\documentclass[a4paper,11pt]{report}

\usepackage{etex}

\usepackage[latin1]{inputenc}
\usepackage[T1]{fontenc}
\usepackage[english]{babel}
\usepackage{mathpazo}	% Garamond	 
\usepackage{eurosym} 	% Symbole euro		
\usepackage{textcomp}	% Bibliothèque de caractères	

\usepackage{soul} 	% Espacement entre lettre + soulignés spéciaux
\usepackage[outline]{contour} 	% Edition du contour de texte
\contourlength{5pt}				% Taille de ce contour
\usepackage{fancybox} 			% Boîtes perso. (généralisation de \fbox)
\usepackage{enumitem} 			% Permet de personnaliser les listes
\usepackage{setspace} 			% Edition de l'interligne dans les §
\usepackage{color} % Définitions des couleurs : fonctionne avec graphics
\usepackage{colortbl} % Couleurs dans les tableaux
\usepackage[svgnames,rgb]{xcolor} % Edition de couleurs avancées (extension de color)

\usepackage{amsthm}		% Outils de théorèmes (AMS)
\usepackage{amsmath}	% Symboles mathématiques (AMS)
\usepackage{amssymb}	% Symboles mathématiques spéciaux (AMS)
\usepackage{mathabx}	% Symboles mathématiques spéciaux (2)
\usepackage{bm} 		% Gras pour les environnements mathématiques
\usepackage{thmtools} 	% Création de théorèmes
\usepackage{cases} 		% Equations multilignes (accolade + numérotation individuelle)
\usepackage{empheq}		% Utilitaire pour équations multilignes
\usepackage{cancel} 	% Simplifications et flèches de résultats

\usepackage{graphicx} 	% Inclusion et travail des images/graphiques
\usepackage{pict2e}		% Complément à graphicx : options avancées
\usepackage{epic}		% Extension des capacités graphiques de LaTeX (1)
\usepackage{eepic}		% Extension des capacités graphiques de LaTeX (2)
\usepackage{wrapfig} 	% Figure dans paragraphe
\usepackage{picins} 	% Figure dans paragraphe avec fonctionnalités étendues de mise en forme
\usepackage{float}		% Interface étendue pour objets flottants
\usepackage{subfig}		% Sous-figures
\usepackage{rotating} 	% Rotation d'images
\usepackage{epstopdf}	% Convertit EPS en PDF en préservant le vectoriel
\usepackage{graphpap}	% Dessine un maillage de papier millimétrique
\usepackage[justification=centering,textfont=it]{caption}	% Légendes d'images

\usepackage{tikz} 								% Lancer TikZ
\usetikzlibrary{decorations.pathmorphing} 		% Traits ondulés
\tikzset{snake it/.style={decorate, decoration=snake}}
\usetikzlibrary{arrows,decorations.markings}	% Jolies flèches
\usetikzlibrary{calc}							% Routines de calcul graphiques
\usetikzlibrary{backgrounds}					% Arrière-plan zone de dessin
\usetikzlibrary{shapes.geometric}				% Formes automatiques
\usepackage{pgfplots}							% Plot library
\usepgfplotslibrary{polar}						% Polar plots
\usetikzlibrary{decorations.text}				% Texte incurvé
\usetikzlibrary{arrows.meta}					% Flèches doubles

\usepackage{multirow} 	% Fusion dans les tableaux
\usepackage{colortbl} 	% Couleur des lignes dans les tableaux
\usepackage{makecell} 	% Editions de bordures dans les tableaux (\Xhline{#\arrayrulewidth})
\usepackage{booktabs}	% Lignes grasses dans les tableaux, et autres joyeusetés :D

\usepackage[top=2.5cm, bottom=2.5cm, left=2.0cm, right=2.0cm]{geometry}
\usepackage{fancyhdr} 	% Personnalisation des en-têtes et pieds de pages
\usepackage{titlesec} 	% Personnalisation de la titraille 
\usepackage{xspace} 	% Espaces intelligents
\usepackage{multicol}	% Texte en multicolonnes
\usepackage[final]{pdfpages} 	% Inclut des pages d'un PDF
\usepackage[pages=some]{background} 
\backgroundsetup{scale=1,color=black,opacity=0.5,angle=0,%
contents={\includegraphics[width=\paperwidth, height=\paperheight]{BackPic.jpg}}}

\usepackage{calc}

\usepackage{xpatch} % Permet la magouille avec les parties
\makeatletter
\renewcommand*\l@part[2]{%
  \ifnum \c@tocdepth >-2\relax
    \addpenalty{-\@highpenalty}%
    \addvspace{2.25em \@plus\p@}%
    \setlength\@tempdima{3em}%
    \begingroup
      \parindent \z@ \rightskip \@pnumwidth
      \parfillskip -\@pnumwidth
      {\leavevmode
       \Large \bfseries #1%
      \leaders\hbox{\normalfont\normalsize$\m@th\mkern\@dotsep mu\hbox{.}\mkern\@dotsep mu$}%\normalfont
       \hfil \hb@xt@\@pnumwidth{\hss #2}}\par
       \nobreak
         \global\@nobreaktrue
         \everypar{\global\@nobreakfalse\everypar{}}%
    \endgroup
  \fi}
\makeatother
\makeatletter
\renewcommand*\l@chapter[2]{%
   \ifnum \c@tocdepth >\m@ne
     \addpenalty{-\@highpenalty}%
     \vskip 1.0em \@plus\p@
     \setlength\@tempdima{1.5em}%
     \begingroup
       \parindent \z@ \rightskip \@pnumwidth
       \parfillskip -\@pnumwidth
       \leavevmode \bfseries
       \advance\leftskip\@tempdima
       \hskip -\leftskip
       #1\nobreak\leaders\hbox{\normalfont$\m@th
         \mkern \@dotsep mu\hbox{.}\mkern \@dotsep
         mu$}\hfill \nobreak\hb@xt@\@pnumwidth{\hss #2}\par
       \penalty\@highpenalty
     \endgroup
   \fi}
\makeatother

\usepackage{tableof} 		% Créer une table à partir d'objets /taggés/.
\usepackage{tocbibind} 		% Affiche les tables créées par l'utilisation dans la TOC

\usepackage{hyperref} 	% Créer des liens et des signets
\hypersetup{
	colorlinks=true, 	% Mise en évidence des liens par coloration
	breaklinks=true, 	% Retour à la ligne dans les liens trop longs
	urlcolor= black, 	% Couleur des liens externes
	linkcolor= black, 	% Couleur des liens internes
	citecolor= black, 	% Couleur des références
}

\usepackage{boiboites} 		% Environnements stylisés (résultats, théorèmes)		
\usepackage[tikz]{bclogo}	% Environnements avec ligne et icône
\usepackage{framed}			% Boîtes brisées sur plusieurs pages
\usepackage{varwidth}		% Minipage à largeur (width) variable

\newenvironment{gbar}[1]{ %
\def \FrameCommand{{\color{#1}\vrule width 1.5pt}\colorbox{gray!0}}%
\MakeFramed {\advance\hsize -\width \FrameRestore }}%
{\endMakeFramed}

\newboxedtheorem[
boxcolor = ForestGreen, 
background = ForestGreen!5, 
titlebackground = ForestGreen!20,
titleboxcolor = ForestGreen]{definition}{Definition}{defi}
\definecolor{GrandVert}{RGB}{0,128,0}

\newboxedtheorem[
boxcolor = red, 
background = red!5, 
titlebackground = red!20,
titleboxcolor = red]{resultat}{Result}{resu}
\definecolor{GrandRouge}{RGB}{192,0,0}

\begin{document}
	% Propriétés du document
		\newcommand{\Titre}{Advanced Lectures on Gravitation}
		\newcommand{\Auteur}{G. COMPÈRE}
		\newcommand{\Annee}{2017}
	% Personnalisation du document
%============================================================
%
% Fichier de personnalisation LaTeX
% ADRIEN FIORUCCI - 25 juin 2017
% 
% Convient pour les fichiers de type syllabus
% ou mémoire/thèse, écrits dans la classe
% /report/. 
%
% Structure :
% -----------
% 1 - Styles de pages
%		1.1.	Déclarations globales
%		1.2.	En-tête et pied de page 
% 2 - Styles des titres
%		2.1.	Mise en forme des titres de chapitres
%		2.2.	Police de la titraille
% 3 - Tables automatiques personnalisées
% 4 - Couverture de document
% 5 - Bibliographie
% 6 - Commandes personnelles
%
% Ce fichier doit être inséré après l'instruction 
% \begin{document} !
%
%============================================================

%============================================================
%
% 1 - Styles de pages
%
%============================================================

% Requiert : package "fancyhdr"

% 1.1.	Déclarations globales
% ---------------------------

% On compile toutes les informations :
\newcommand{\GeneralFooter}{\footnotesize \fontfamily{phv}\selectfont \nouppercase{\leftmark}}
% Numéro de page
\newcommand{\GeneralPageNbr}{\footnotesize \fontfamily{phv}\selectfont \thepage}

% 1.2.	En-tête et pied de page 
% -----------------------------

% Traits
% ------

\renewcommand{\headrulewidth}{1pt}
\renewcommand{\footrulewidth}{1pt}

% Pages de titres
% ---------------

	\fancypagestyle{plain}{%
	\fancyhf{}
	% En-tête
	\lhead{} \chead{} \rhead{}
	% Pied de page
	\lfoot{\GeneralFooter} \cfoot{} \rfoot{\GeneralPageNbr}
	\renewcommand{\headrulewidth}{0pt}
	\renewcommand{\footrulewidth}{1pt} 
	}
	
% Pages de corps de texte
% -----------------------

% Style général
	
	\fancypagestyle{corps}{%
	\fancyhf{}
	% En-tête
	\lhead{}
	\chead{\footnotesize \fontfamily{phv}\selectfont \nouppercase{\rightmark}}
	\rhead{}
	% Pied de page
	\lfoot{\GeneralFooter} \cfoot{} \rfoot{\GeneralPageNbr}
	}
	 		 
% Pages de parties
% ----------------

% Permet de changer le style "plain" par défaut des pages de partie !
\xpatchcmd{\part}{\thispagestyle{plain}}{\thispagestyle{empty}}{}{}
% Requiert le package "xpatch".
	
% Sections spéciales
% ------------------

% Personnalisation des parties sans numéro (*) dans la TOC
% Format du sommaire
	\fancypagestyle{Sommaire}{%
	\fancyhf{}
	% En-tête
	\lhead{}
	\chead{}
	\rhead{}
	% Pied de page
	\lfoot{\GeneralFooter} \cfoot{} \rfoot{\GeneralPageNbr}
	\renewcommand{\headrulewidth}{1pt}
	\renewcommand{\footrulewidth}{1pt} 
	}
	
% Format des parties additionnelles sans numéro
%	\fancypagestyle{AvantPropos}{%
%	\fancyhf{}
%	% En-tête
%	\lhead{}
%	\chead{\footnotesize \fontfamily{phv}\selectfont Avant-Propos}
%	\rhead{}
%	% Pied de page
%	\lfoot{\GeneralFooter} \cfoot{} \rfoot{\GeneralPageNbr}
%	\renewcommand{\headrulewidth}{1pt}
%	\renewcommand{\footrulewidth}{1pt} 
%	}
%	
%	\fancypagestyle{Introduction}{%
%	\fancyhf{}
%	% En-tête
%	\lhead{}
%	\chead{\footnotesize \fontfamily{phv}\selectfont Introduction}
%	\rhead{}
%	% Pied de page
%	\lfoot{\GeneralFooter} \cfoot{} \rfoot{\GeneralPageNbr}
%	\renewcommand{\headrulewidth}{1pt}
%	\renewcommand{\footrulewidth}{1pt} 
%	}
%	
%	\fancypagestyle{Conclusion}{%
%	\fancyhf{}
%	% En-tête
%	\lhead{}
%	\chead{\footnotesize \fontfamily{phv}\selectfont Conclusion}
%	\rhead{}
%	% Pied de page
%	\lfoot{\GeneralFooter} \cfoot{} \rfoot{\GeneralPageNbr}
%	\renewcommand{\headrulewidth}{1pt}
%	\renewcommand{\footrulewidth}{1pt} 
%	}

	\fancypagestyle{Bibli}{%
	\fancyhf{}
	% En-tête
	\lhead{}
	\chead{}
	\rhead{}
	% Pied de page
	\lfoot{\GeneralFooter} \cfoot{} \rfoot{\GeneralPageNbr}
	\renewcommand{\headrulewidth}{1pt}
	\renewcommand{\footrulewidth}{1pt} 
	}
	
	\fancypagestyle{Conventions}{%
	\fancyhf{}
	% En-tête
	\lhead{}
	\chead{\footnotesize \fontfamily{phv}\selectfont $ $}
	\rhead{}
	% Pied de page
	\lfoot{\footnotesize \fontfamily{phv}\selectfont Conventions and notations}
	\cfoot{} \rfoot{\GeneralPageNbr}
	\renewcommand{\headrulewidth}{1pt}
	\renewcommand{\footrulewidth}{1pt} 
	}

%============================================================
%
% 2 - Styles des titres
%
%============================================================	

\renewcommand{\chaptername}{Lecture}
	
% 2.1.	Mise en forme des titres de chapitres
% -------------------------------------------

\titleformat{\chapter}[frame]
{\normalfont \color{NavyBlue}}{\fontfamily{phv}\selectfont \large \enspace 
\chaptertitlename \enspace \large \thechapter \enspace}
{10pt}{\fontfamily{pag}\selectfont \huge \bfseries \filcenter}
% Marges 
\titlespacing{\chapter}{0cm}{-1cm}{1.5cm}
%\titlespacing{<command>}{<left>}{<beforesep>}{<aftersep>}[<right>]
	
% 2.2.	Police de la titraille
% ----------------------------

\titleformat{\section}  
{\fontfamily{lmss}\selectfont \fontsize{14}{16} \bfseries\color{blue}}{\thesection}{1em}{}[{\titlerule[0.8pt]}]
\titleformat{\subsection}
{\color{blue} \fontfamily{lmss}\selectfont \fontsize{12}{15}\bfseries}{\thesubsection}{1em}{}
\titleformat{\subsubsection}
{\color{blue} \fontfamily{lmss}\selectfont \fontsize{11}{11}\bfseries\itshape}{}{1em}{}

%============================================================
%
% 3 - Tables automatiques personnalisées
%
%============================================================

% Table des matières
\renewcommand{\contentsname}{Table of contents}
\setcounter{tocdepth}{2}

% Table des matières par chapitre
%\dominitoc
%\renewcommand{\mtctitle}{Plan du chapitre \thechapter}
	
% Table des illustrations
\renewcommand{\listfigurename}{Table des illustrations}
\renewcommand{\figurename}{Figure}

% Tables des théorèmes et définitions
\newcommand{\ListOfThmAndDef}{
	\chapter*{Table des définitions \& résultats}
	
	\titleformat{\chapter}
	  {\fontfamily{lmss}\selectfont \LARGE \bfseries\color{blue}}{\thesection}{1em}{}[{\titlerule[0.8pt]}]
	
	% Supprime le saut de page entre les chapitres
	\makeatletter 
	    \renewcommand\chapter{\par%
	      \thispagestyle{plain}%
	      \global\@topnum\z@
	      \@afterindentfalse
	      \secdef\@chapter\@schapter}
	\makeatother
	
	%% Table des définitions
	\vspace{20pt} 
	\renewcommand{\listtheoremname}{Table des définitions}
	\pdfbookmark[0]{Table des définitions}{lodefi}{} % Signet sans texte affiché pour hyperref de niveau 0.
	\listoftheorems[ignoreall,show={defn}] % Appel de la liste
	
	% Table des résultats
	\vspace{80pt} % 10*2*n + 40 où n = nbr de def et 2 car 10pt la police et 10pt interligne et 40 de marge avant liste suivante
	
	\renewcommand{\listtheoremname}{Table des résultats}
	\pdfbookmark[0]{Table des résultats}{loresu}{} % Signet sans texte affiché pour hyperref de niveau 0.
	\listoftheorems[ignoreall,show={rslt}] % Appel de la liste
}

%============================================================
%
% 4 - Couverture de document
%
%============================================================

% Page blanche
\newcommand{\blanc}{
	\includepdf[pages=1]{Page_vierge.pdf}}

% Première de couverture
\newcommand{\Couverture}[1]{
	\includepdf[pages=1]{inputs/#1}
	\blanc
	\setcounter{page}{1}
}

% Quatrième de couverture
\newcommand{\BackCouverture}[1]{
	\blanc
	\includepdf[pages=1]{inputs/#1}
}

%============================================================
%
% 5 - Bibliographie
%
%============================================================

\newcommand{\Bibliographie}[1]{
\pagestyle{Bibli}
\bibliography{#1} 
}

%============================================================
%
% 6 - Commandes personnelles
%
%============================================================

% Réduction de l'espace entre les item et enumerate (avec enumitem)
% A utiliser après \begin{itemize} ou \begin{enumerate}
\newcommand{\compresslist}{ 
\setlength{\itemsep}{1pt}
\setlength{\parskip}{0pt}
\setlength{\parsep}{0pt}
}	

% Notations de Dirac	
\newcommand{\bra}[1]{\langle #1 \vert }
\newcommand{\ket}[1]{\vert #1 \rangle }
\newcommand{\braket}[2]{\langle #1 \vert #2 \rangle }

% Opérateurs différentiels dans IR^3
\newcommand{\grad}{\vec{\nabla}}
\newcommand{\divv}[1]{\vec{\nabla} \cdot \vec{#1}}
\newcommand{\rotv}[1]{\vec{\nabla} \times \vec{#1}}
\newcommand{\dive}{\vec{\nabla} \cdot}
\newcommand{\rota}{\vec{\nabla} \times}

% Suppression de l'indentation
\setlength\parindent{0pt}

% Texte en gras et en italique
\newcommand{\textGI}[1]{{\fontseries{sb}\fontshape{it}\selectfont #1}}

% Remplacement du symbole d'ensemble vide
\let\emptyset\varnothing

% Chemin d'accès pour les images
\graphicspath{{SchemasPDF/}}

% Changer les légendes
\newcommand{\scaption}[1]{\caption{\textit{#1}}}

% Raccourcis
\newcommand{\cI}{\mathcal{I}}
\newcommand{\cT}{{\mathcal{T}}}
\newcommand{\Lie}{\mathcal{L}}
\newcommand{\HH}{\mathcal{H}}
\newcommand{\onH}[1]{\left. #1 \right|_{\mathcal{H}}}
\newcommand{\bemph}[1]{{\color{blue} \underline{\textit{#1}}}}
\newcommand{\rmg}{\sqrt{-g}}

%========================================================================================================================

%============================================================

% B - Pages liminaires du document et table des matières
% ------------------------------------------------------

	% Première de couverture
	   %\Couverture{Frontpage_Memoire.pdf}

	% Page de titre
\pagestyle{empty}

\begin{tikzpicture}[overlay,remember picture]
    \draw [line width=1pt,rounded corners=7pt]
        ($ (current page.north west) + (1.3cm,-1.3cm) $)
        rectangle
        ($ (current page.south east) + (-1.3cm,1.3cm) $);
    \draw [xshift=4mm,line width=2.5pt,rounded corners=7pt]
        ($ (current page.north west) + (.5cm,-.5cm) $)
        rectangle
        ($ (current page.south east) + (-.5cm,.5cm) $);
%    \draw [line width=1pt,rounded corners=7pt]
%        ($ (current page.north west) + (1.5cm,-1.5cm) $)
%        rectangle
%        ($ (current page.south east) + (-1.5cm,1.5cm) $);
\end{tikzpicture}

\vspace{-15pt}

\begin{figure}[h!]
\begin{center}
\includegraphics[width=0.3\textwidth]{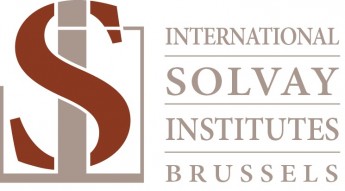}
\end{center}
\end{figure}

\vspace{30pt}

\begin{center}
\textit{Lecture notes prepared for the \textit{Solvay Doctoral School on \\ Quantum Field Theory, Strings and Gravity.} \\ Lectures given in Brussels, \textit{October 2017}. } \\
\end{center}

\vspace{50pt}
\begin{center}
\fbox{
  \begin{minipage}[c]{0.6\textwidth}
    \centering \bfseries
    \vspace{20pt}
    \huge Advanced Lectures on \\
    General Relativity \\
    \setstretch{1.0}$ $
  \end{minipage}
} \\
\vspace{30pt}
%{\textbf{Surface charges, asymptotic symmetries, \textit{AdS} spacetimes features...}}
\end{center}

\vspace{10pt}
\begin{center}
%\setstretch{1.50}
Lecturing \& Proofreading: \\
$ $\\
{\LARGE \textbf{Geoffrey Compère}}\\
\vspace{30pt}
Typesetting, layout \& figures: \\
$ $\\
{\LARGE \textbf{Adrien Fiorucci}}\\
\vspace{30pt}

Fonds National de la Recherche Scientifique (Belgium) \\
Physique Théorique et Mathématique \\
Université Libre de Bruxelles and International Solvay Institutes \\
Campus Plaine C.P. 231, B-1050 Bruxelles, Belgium \\ \vspace{20pt}
{\large Please email any question or correction to: \href{mailto:gcompere@ulb.ac.be}{\textit{gcompere@ulb.ac.be}} }%--- \href{mailto:afiorucc@ulb.ac.be}{\textit{afiorucc@ulb.ac.be}} \\
\end{center}

\newpage
		%\blanc
		
	% Résumé/Abstract

\mbox{}\\

\vspace{10cm}
{\fontfamily{lmss}\selectfont \color{blue} \textbf{Abstract}} --- These lecture notes are intended for starting PhD students in theoretical physics who have a working knowledge of General Relativity. The 4 topics covered are  (1) Surface charges as conserved quantities in theories of gravity; (2) Classical and holographic features of three-dimensional Einstein gravity; (3) Asymptotically flat spacetimes in 4 dimensions: BMS group and memory effects; (4) The Kerr black hole: properties at extremality and quasi-normal mode ringing. Each topic starts with historical foundations and points to a few modern research directions.\newpage

	% Table des matières	
		\pagestyle{Sommaire}	
		\addtocontents{toc}{\protect\setcounter{tocdepth}{-1}}
		\tableofcontents
		\addtocontents{toc}{\protect\setcounter{tocdepth}{3}}
		\newpage
		
	% Conventions
		\thispagestyle{Conventions}		
{\fontfamily{lmss}\selectfont \color{blue} \textbf{Conventions and notations}} --- We employ units such that the speed of light $c=1$ but we will keep $G$ explicit. The spacetime manifold is denoted by the couple $(M,g_{\mu\nu})$. The signature of the Lorentzian metric $g_{\mu\nu}$ obeys the mostly plus convention $(-,+,+,...)$. The dimension of spacetime is generally $n$. If necessary, we will write explicitely $n=d+1$ where $d$ represents the number of spatial dimensions. We use the unit normalized convention for symmetrization and antisymmetrization, $T^{(\mu\nu)} = \frac{1}{2}(T^{\mu\nu}+T^{\nu\mu})$ and $k^{[\mu\nu]} = \frac{1}{2}(k^{\mu\nu}-k^{\nu\mu})$. We employ Einstein's summed index convention: double indices in an expression are implicitly summed over. Finally, we follow the conventions adopted in the textbook \textit{Gravitation} \cite{Wheeler:Gravitation} by Wheeler, Thorne and Misner, concerning the definition of various objects in Relativity. In particular, the Riemann-Christoffel tensor is determined as $R^\mu_{\phantom{\mu}\nu\alpha\beta} = \partial_\alpha \Gamma^\mu_{\phantom{\mu}\nu\beta} - \partial_\beta \Gamma^\mu_{\phantom{\mu}\nu\alpha} + \Gamma^\mu_{\phantom{\mu}\kappa\alpha} \Gamma^\kappa_{\phantom{\kappa}\nu\beta} - \Gamma^\mu_{\phantom{\mu}\kappa\beta} \Gamma^\kappa_{\phantom{\kappa}\nu\alpha}$. In this convention, the $n$-sphere has a positive Ricci curvature scalar $R=R^\alpha_{\phantom{\alpha}\alpha}$ where the Ricci tensor is $R^{\alpha}_{\phantom{\alpha}\mu\alpha\nu}$. \\

The notation of spacetime coordinates is as follows. Greek indexes $\mu,\nu,\dots$ span the full dimension of spacetime $n=d+1$, so $\mu \in \lbrace 0,\dots, d \rbrace$. Often the index $0$ represents a timelike coordinate. Latin indexes will designate the other coordinates $x^a$ with $a \in \lbrace 1,\dots, d \rbrace$. \textit{Capital} latin letters will be used to denote angular coordinates $x^A$ among the spacelike coordinates. \\

Some conventions concerning objects of exterior calculus have also to be detailed. The volume form is denoted by $\epsilon_{\mu_1 \cdots \mu_n}$. It is a tensor so it includes $\sqrt{-g}$. We will keep the notation $\varepsilon_{\mu_1 \cdots \mu_n}$ for the numerically invariant pseudo-tensor with entries $-1,0$ or $1$. We have  $\epsilon_{\mu_1 \cdots \mu_n} = \sqrt{-g}\varepsilon_{\mu_1 \cdots \mu_n}$, see the appendix B of Wald's book \cite{WaldBook:1984} for details. A general $(n-p)$-form ($p \in \mathbb{N}, p \leq n$) is written as boldface $\mathbf{X} = X^{\mu_1 \cdots \mu_p} \sqrt{-g}(d^{n-p} x)_{\mu_1 \cdots \mu_p}$ developed in the base :
\[ (d^{n-p} x)_{\mu_1 \cdots \mu_p} = \frac{1}{p!(n-p)!} \: \varepsilon_{\mu_1 \cdots \mu_p \: \nu_{p+1} \cdots \nu_n} \: dx^{\nu_{p+1}} \wedge \cdots \wedge dx^{\nu_n}. \]
We will always invoke Hodge's duality between $p$-forms and $(n-p)$-forms to define objects in the more convenient way. For example, the Lagrangian density $L$ (equal to $\sqrt{-g}$ times the Lagrangian scalar) will be identified to the $n$-form $\mathbf{L} = L\: d^n x$. A vector field $J^\mu$ will be regarded as a $(n-1)$-form $\mathbf{J} = J^\mu \sqrt{-g}(d^{n-1} x)_\mu$, since $J^\mu$ is the Hodge dual of a  $1$-form. An antisymmetric 2 tensor $k^{\mu\nu}= k^{[\mu\nu]}$ will be identified with its Hodge dual: a $(n-2)$-form $\mathbf{k} = k^{[\mu\nu]} \sqrt{-g}(d^{n-2} x)_{\mu\nu}$ where the antisymetrisation arises naturally from the definition of the natural basis of $(n-2)$-forms. And so on ! (We keep the factors of $\sqrt{-g}$ explicit to easily vary them!)\\

Under an infinitesimal diffeomorphism generated by the vector $\chi^\mu$, a general field $\Phi^i$ with arbitrary index structure summarized by the abstract index $i$ will be modified by the Lie derivative $\delta_{\chi} \Phi^i =+\Lie_\chi \Phi^i$. As a final remark, we note that waved equalities ($\approx$) represent any equation that holds if and only if the Euler-Lagrange equations of motion formulated in the theory of interest are satisfied. 
		%\blanc

%============================================================

% C - Corps de texte
% ------------------

	% Lecture 1 : Charges conservées en Relativité Générale
		\pagestyle{corps}
\chapter{Surface charges in Gravitation}
\label{sec:Symetries}

The main purpose of this first lecture is to introduce the concept of {\bfseries \color{blue} canonical surface charges}  in a generally covariant theory of gravity, whose General Relativity is the most famous representative. \\

As a starter, we will show that a conserved stress tensor can be generated for any classical field theory, simply by coupling it to gravity and using general covariance of the so-enhanced theory. Then we will enter into the main point we have to discuss, and motivate why we cannot define conserved currents and charges in Noether's fashion for generally covariant theories, and more globally, for theories that include gauge transformations. This quite tricky fact will lead us to extend Noether's first theorem to formulate lower degree conservation laws, which will be exploitable for theories such as Einstein's gravity. On the way, we will discuss about the symplectic structure of abstract spaces of fields, and use the covariant phase space formalism to derive a magnificent and powerful result linking this structure and the lower degree conserved forms that we are looking for. We will then be able to compute surface charges associated to these quantities and study their properties and their algebra. Along the text, some pedagogical examples will be provided, namely for pure Einstein's gravity, and Maxwell's electrodynamics, enhanced in a curved background. Finally, we will present another possible definition of these surface charges and use the latter definition as an efficient tool to derive the conserved charges of Chern-Simons theory. We will finally discuss the residual ambiguities of the conserved quantities...\\

\section{Introduction : general covariance and conserved stress tensor}

Before considering a theory of gravity, let us first consider a relativistic field theory of matter. It will allow us to analyse a remarkable relation between the very fundamental concept of \textit{general covariance} of a theory, and the conservation of a \textit{stress tensor} associated to this theory. It is a very nice way to obtain a divergence-free tensor without directly invoking Noether's theorem.  \\

Let us start with the theory of special relativity. There is a background structure in the theory: the flat metric. In Cartesian coordinates $(t,x,y,z)$, the metric takes the most refined form that a Lorentzian metric could take : $\eta_{\mu\nu} = \text{diag}(-1,+1,+1,+1)$, the so-called \textit{metric of Minkowski}. The symmetries of spacetime are defined as vector fields $\xi^\mu$ that preserve this background structure, \textit{i.e.} $\Lie_\xi \eta_{\mu\nu} = 0$. These symmetries preserve all distances and are by right recognized as \textit{isometries}. In components, we have $\xi^\rho \partial_\rho \eta_{\mu\nu} + 2 \eta_{\rho (\mu} \partial_{\nu )} \xi^\rho = 0$. The first term is trivially zero, so it remains $\partial_{(\mu } \xi_{\nu )}   = 0$. The most general solution is given by $\xi^\mu = a^\mu + b^{[\mu \nu]} x_\nu$. The isometries of Minkowski spactime thus depend upon 10 parameters : the 4 components of $a^\mu$ that encode global translations and the 6 matrix elements of $b^{[\mu \nu]}$ associated with Lorentz transformations (rotations and boosts). Under the Lie bracket, these vectors give rise to the \textit{Poincar\'e algebra}. \\

What happens in an arbitrary coordinate system? Thanks to general covariance, we can express all tensorial equations, such as $\Lie_\xi \eta_{\mu\nu} = 0$,  in an arbitrary frame by remplacing partial derivatives $\partial_\mu$ by covariant derivatives $\nabla_\mu$. The covariant derivative is defined with the Levi-Civita connection compatible with the metric. In a general spacetime, although curvature appears, it is still possible to consider free-falling observers, around which a local Lorentzian frame can be constructed. So locally, we find again $\partial_{(\mu } \xi_{\nu )}   = 0$ and general covariance can again be invoked to arrive to 

\begin{resultat}[\textit{Killing equation}]
A necessary and sufficient condition for a vector field to be an \textit{isometry} of spacetime $(M,g)$ is that it verifies the so-called \textit{Killing equation} : $\Lie_\xi g_{\mu\nu} = 0 \Longleftrightarrow \nabla_{( \mu } \xi_{\nu )} = 0$.
\end{resultat}
Since the Lie bracket is a tensorial quantity, the Killing vectors of Minkowski spacetime form the Poincar\'e algebra independently of the coordinates chosen to express the isometries. \\

After obtaining this very crucial formula, we can now show the following theorem :

\begin{resultat}[\textit{Existence of a conserved stress tensor}]
Any relativistic field theory in Minkowski spacetime admits a symmetric stress tensor $T^{\mu\nu}$ that is divergence-free when the equations of motion hold, $\nabla_\mu T^{\mu\nu} \approx 0$. 
\end{resultat}
Let's consider an arbitrary theory of matter fields collectively denoted by $\Phi_M = (\Phi_M^i)_{i\in I}$ (with a totally general index structure $i$ belonging to a set of such structures $I$). The theory is described by a Lagrangian density $L[\Phi_M]$. One can always couple this theory to gravity by introducing a non-flat metric $g_{\mu\nu}$ into the Lagrangian: $L[\Phi_M] \rightarrow L[\Phi_M, g_{\mu\nu}]$. The coupling is said \textit{minimal} when it consists in merely replacing the Minkowski metric by the general metric,  standard derivatives by covariant derivatives and with the necessary mutation of the volume form : $d^n x \rightarrow \sqrt{-g} d^n x$. After that, we can define a natural symmetric tensor :
\begin{equation}
T^{\mu\nu} \triangleq \frac{2}{\rmg} \frac{\delta L}{\delta g_{\mu\nu}}
\label{eq:Tmunu}
\end{equation}
thanks to the \textit{Euler-Lagrange derivative}, rigorously defined as
\begin{equation}
\forall \Phi^i \in \Phi : \frac{\delta L}{\delta \Phi^i} \triangleq \frac{\partial L}{\partial \Phi^i} - \partial_\mu \left( \frac{\partial L}{\partial \: \partial_\mu \Phi^i} \right) + \partial_\mu \partial_\nu \left( \frac{\partial L}{\partial \: \partial_\mu \partial_\nu\Phi^i} \right) + \cdots
\end{equation}
for theories of any order in derivatives. The compact notation $\Phi = \{ (\Phi_M^i)_{i \in I},g_{\mu\nu} \}$ now encompasses at the same time the original matter fields and the metric $g_{\mu\nu}$ of spacetime. We therefore have a natural symmetric candidate stress-tensor in the original field theory, namely $T^{\mu\nu}[\Phi_M,\eta_{\mu\nu}]$ where we substituted back the metric $g_{\mu\nu}$ to $\eta_{\mu\nu}$. What remains to be done is to show that $T^{\mu\nu}[\Phi_M,\eta_{\mu\nu}]$ expressed by (\ref{eq:Tmunu}) is covariantly conserved when the original matter equations hold, \textit{i.e.} when $\frac{\delta L[\Phi_M]}{\delta \Phi_M^i} = 0$. \\

Let's begin by performing an arbitrary variation of the Lagrangian density:
\begin{equation}
\delta L = \delta \Phi^i \frac{\partial L}{\partial \Phi^i} + \partial_\mu \delta \Phi^i\frac{\delta L}{\delta \partial_\mu \Phi^i} + \cdots = \delta \Phi^i \frac{\delta L}{\delta \Phi^i} + \partial_\alpha \Theta^\alpha [\delta \Phi;\Phi] .
\end{equation}
The first term simply contains the Euler-Lagrange equations of motion.  The second one collects the remnants of the inverse Leibniz rule, which was applied in order to factorize the variation of fields $\delta \Phi^i$ without any derivative acting on it. In other words, this second term is nothing else than a boundary term, expressed as a total derivative, or more precisely, the divergence of a vector field density $\Theta^\mu [\delta \Phi^i, \Phi^i]$ named the bare \textit{presymplectic potential}. In the more convenient language of forms, we can rewrite this equation as
\begin{equation}
{\color{blue}\boxed{\delta \mathbf{L} = \delta \Phi^i \frac{\delta \mathbf{L}}{\delta \Phi^i} + d\bm{\Theta} [\delta \Phi;\Phi]}}
.\label{eq:VariationL}
\end{equation}
Here, $\bm{\Theta} = \Theta^\mu (d^{n-1} x)_\mu$ is a $(n-1)$-form, and $\mathbf{L} = L d^n x$ is the $n$-form naturally associated to the Lagrangian density $L$.  The total derivative $d\bm{\Theta}$ is thus also a $n$-form, by virtue of the definition of the exterior derivative $d$ :
\begin{equation}
d\bm{\Theta} = dx^\nu \partial_\nu \left[ \Theta^\mu (d^{n-1} x)_\mu \right] = \partial_\nu \Theta^\mu (d^n x) \delta^\nu_\mu = \partial_\mu \Theta^\mu d^n x.
\end{equation}
Let us now analyse the variation of $\mathbf{L}$ under an infinitesimal diffeomorphism generated by $\xi^\mu$. We get 
\begin{align}
\delta_\xi \mathbf{L} &= \delta_\xi g_{\mu\nu} \frac{\delta \mathbf{L}}{\delta g_{\mu\nu}}+ \delta_\xi \Phi_M^i \frac{\delta \mathbf{L}}{\delta \Phi_M^i}  + d(\cdots) \\
&= \Lie_\xi g_{\mu\nu} \frac{\delta \mathbf{L}}{\delta g_{\mu\nu}} + \delta_\xi \Phi_M^i \frac{\delta \mathbf{L}}{\delta \Phi_M^i}+ d(\cdots) \\
&= 2 \nabla_\mu \xi_\nu \frac{\delta \mathbf{L}}{\delta g_{\mu\nu}} + \delta_\xi \Phi_M^i \frac{\delta \mathbf{L}}{\delta \Phi_M^i}+ d(\cdots) \\
&=d^n x \: \rmg \: T^{\mu\nu} \: \nabla_\mu \xi_\nu+ \delta_\xi \Phi_M^i \frac{\delta \mathbf{L}}{\delta \Phi_M^i} + d(\cdots) .
\end{align}
Let us now substitute $g_{\mu\nu}$ by the original Minkowski metric $\eta_{\mu\nu}$ and let's impose the matter field equations. We are then ``on-shell'' in the original theory. We still have covariant derivatives since we work in arbitrary coordinates. We get
\begin{align}
\delta_\xi \mathbf{L} &\approx d^n x \: \rmg \: \nabla_\mu (T^{\mu\nu} \xi_\nu) - d^n x \: \rmg \: \nabla_\mu T^{\mu\nu} \xi_\nu + d(\cdots) \\
&\approx d^n x \: \partial_\mu \left( \rmg \: T^{\mu\nu} \xi_\nu \right) - d^n x \: \rmg \: \nabla_\mu T^{\mu\nu} \xi_\nu + d(\cdots) \\
&\approx - d^n x \: \rmg \: \nabla_\mu T^{\mu\nu} \xi_\nu + d(\cdots).
\end{align}
Since general covariance requires that the total variation of the Lagrangian density on any diffeomorphism must be a total derivative, including when the equations of motion hold, it implies immediately the conservation of the stress tensor of the original matter theory!
\begin{equation}
{\color{blue} \boxed{\delta_\xi \mathbf{L} = d(\cdots) \Longrightarrow \nabla_\mu T^{\mu\nu} \vert_{g_{\mu\nu} = \eta_{\mu\nu}}\approx 0}}.
\end{equation}

In this relativistic matter theory we can now build a 4-vector from the stress tensor : $J^\mu = T^{\mu\nu} \xi_{\nu}$ which is conserved (or such that its Hodge dual form $\mathbf{J} = J^\mu \sqrt{-g} (d^{n-1}x)_\mu$ is closed), provided that the diffeomorphism $\xi^\mu$ is an isometry of spacetime.
\begin{equation}
\nabla_\mu J^\mu = \nabla_\mu T^{\mu\nu} \xi_\nu + T^{\mu\nu} \nabla_{(\mu} \xi_{\nu )} = 0 + 0 \Longrightarrow d\mathbf{J} = 0. 
\end{equation}

The integral of $\mathbf{J}$ on an arbitrary Cauchy surface\footnote{A Cauchy surface is a subset of $M$ which is intersected by every maximal causal curve exactly \textit{once}. Once the initial data is fixed on such a codimension 1 surface, the field equations lead to the evolution of the system in the entire spacetime.} $\Sigma$ produces a scalar quantity $Q = \int_\Sigma \mathbf{J}$ which is conserved when the system evolves, provided that fields decay sufficiently rapidly at the boundary $\partial\Sigma$. Let us choose a coordinate $x^0 = t$ such that the Cauchy surface is described as the surface $t=0$. Then $(d^{n-1}x)_0 = d\Sigma$ is the volume form on the surface. We have 
\[ \partial_t Q = \partial_t \int_\Sigma  J^0 (d^{n-1}x)_0 = \int_\Sigma d\Sigma\, \partial_0 J^0 = - \int_\Sigma d\Sigma\, \vec{\nabla} \cdot \vec{J} = - \int_{\partial\Sigma} \vec{J} \cdot d\vec{S}= 0. \]
To each isometry corresponds such a conserved charge : 

\begin{center}
\begin{tabular}{c c c}
\toprule
\textit{Isometry} & \textit{Origin} & \textit{Conserved charge} \\ 
\hline 
Translations & Minkowski is homogeneous & $P^{\mu} = \int_\Sigma d\Sigma\,  T^{\mu 0}$ \\ 
Lorentz transformations & Minkowski is isotropic and relativistic & $M^{\mu\nu} = \int_\Sigma d\Sigma\,  (x^\mu T^{\nu 0} - x^\nu T^{\mu 0})$ \\ 
\bottomrule 
\end{tabular} 
\end{center}
\vspace{10pt} 
All these features are not surprising, since there is a fundamental result that permits to deduce immediatly the existence of dynamical invariants when the theory of interest possesses some continuous symmetries. This is the next topic to which we now turn!

\section{Generalized Noether theorem}

\subsection{Gauge transformations and trivial currents}
Let us begin by reviewing one of the most famous statements ever established in modern physics : Noether's first theorem. Proven in 1916, and considered as a "monument of mathematical thought" by Einstein himself, it is not abusive to say that most of modern physical works rely on this result. We will present it without proof, but in a quite modernized form. Let's first clarify the terminology: global symmetries preserve the Lagrangian up to a boundary term. Gauge transformations are global symmetries whose generator arbitrarily depends upon the coordinates.
\begin{resultat}[\textit{Noether's first theorem}]
Take any physical theory described by a Lagrangian density $L$ defined on a spacetime manifold $(M,g)$ that admits global symmetries, some of which might be gauge transformations.
It exists a bijection between :
\begin{itemize}[label=$\rhd$]
\item The equivalence classes of \textit{global} continuous symmetries of $L$, and
\item The equivalence classes of conserved vector fields $J^\mu$, the so-called \textit{Noether currents}.
\end{itemize}
\end{resultat}
On the one hand, we say that two global symmetries of $L$ are equivalent if and only if they differ only by a gauge transformation and another symmetry whose generator is \textit{trivially zero} on shell. On the other hand, we declare that two currents $J_1^\mu$ and $J_2^\mu$ are equivalent if and only if they differ by a trivial current of the form
\begin{equation}
J_2^\mu = J_1^\mu + \partial_\nu k^{[\mu\nu]} + t^\mu
\end{equation}
where $k^{[\mu\nu]}$ is an skew tensor $(2,0)$ %, two times continuously differentiable\footnote{Under this assumption, we can apply the Schwarz theorem : $\partial_\mu\partial_\nu k^{[\mu\nu]} = 0$.} 
and $t^\mu \approx 0$. So we have $\partial_\mu J_2^\mu \approx \partial_\mu J_1^\mu$. 
Using this formulation of Noether's theorem, a thorny problem arises immediately. Imagine that you have a pure gauge theory, \textit{i.e.} a gauge theory with no non-trivial global symmetry at your disposal. From Noether's first theorem, it exists only one equivalence class of conserved currents: the trivial ones. In particular, for generally covariant theories, any transformation like $x^\mu \rightarrow x^\mu + \xi^\mu$ is a gauge transformation, thus the natural symmetries, also called isometries, are associated to trivial currents in a similar way. We can define a charge by integrating on a Cauchy slice $\Sigma$ as we saw before, but it reads simply $Q = \int_\Sigma J^\mu (d^{n-1}x)_\mu \approx \int_{\partial\Sigma} k^{[\mu\nu]} (d^{n-2}x)_{\mu\nu}$ when the equations of motion hold. $Q$ is manifestly completely arbitrary, because $k^{[\mu\nu]}$ is totally unconstrained ! Let us make the issue explicit by computing the Noether current of General Relativity.  \\

We consider the Hilbert-Einstein Lagrangian density coupled to matter $\mathbf{L} = \left( \frac{R\rmg}{16\pi G} + L_M \right)  d^n x$ where $R$ is the scalar Ricci curvature and $L_M$ is the Lagrangian density of matter fields. The conserved stress-tensor built from varying the Lagrangian is
\begin{equation}
T^{\mu\nu} \triangleq \frac{2}{\rmg} \frac{\delta L}{\delta g_{\mu\nu}} = \frac{1}{\rmg} \frac{1}{8\pi G} \frac{\delta (R\rmg)}{\delta g_{\mu\nu}} + \frac{2}{\rmg} \frac{\delta L_M}{\delta g_{\mu\nu}} = -\frac{1}{8\pi G} \left( G^{\mu\nu} - 8\pi G T^{\mu\nu}_M \right) \approx 0
\end{equation}
as we exactly retrieve Einstein's field equations. The Noether current associated to a diffeomorphism $\xi^\mu$ is therefore trivial, $J^\mu = T^{\mu\nu} \xi_\nu \approx 0$.

\subsection{Lower degree conservation laws}

We can sketch a solution to this puzzle simply by considering more carefully the expression of the arbitrary Noether charge $Q = \int_{\partial\Sigma} k^{[\mu\nu]} (d^{n-2}x)_{\mu\nu} = \int_{\partial\Sigma} \mathbf{k}$. We see that $Q$ reduces to the flux of $\mathbf{k}$ through the boundary $\partial\Sigma$\footnote{Remember that $\Sigma$ being a Cauchy slice (and so by definition a $(n-1)$-dimensional volumic object), $\partial\Sigma$ is nothing but a $(n-2)$-surface ! Take $n=4$ to clarify the role of any geometric structure...}, and depends only on the properties of this $(n-2)$-form in the vicinity of $\partial\Sigma$. This suggests to invoke lower degree conservation laws. Indeed, let us imagine that we are able to define uniquely a $(n-2)$-form $\mathbf{k} = k^{[\mu\nu]} (d^{n-2}x)_{\mu\nu}$ such that $d\mathbf{k} =\partial_\nu k^{[\mu\nu]} (d^{n-1}x)_\mu = 0$. Thanks to such an object, we can define an integral charge $Q = \int_{S} \mathbf{k}$ on any surface $S$ which will be conserved when we change surfaces without crossing any singularity (such as the source of the charge!). Seeking for conserved $(n-2)$-forms is the right path to obtain a canonical notion of charges in gauge theories. \\

While the first Noether theorem maps each symmetry to a class a conserved currents (or equivalently closed $(n-1)$-forms $\mathbf{J} = J^\mu (d^{n-1} x)_\mu$), it exists a generalized version of it which precisely focuses on lower degree conserved forms. This result was established by Barnich, Brandt and Henneaux in 1995 \cite{Barnich:2000zw} using cohomological methods, and we present it here without proof.

\begin{resultat}[\textit{Generalized Noether theorem}]
%Let $(M,g)$ be a spacetime manifold and a physical theory described by a Lagrangian density $L$ provided with several symmetries, eventually including gauge invariances. 
Take any physical theory described by a Lagrangian density $L$ defined on a spacetime manifold $(M,g)$ which admits global symmetries, some of which might be gauge transformations. It exists a bijection between :
\begin{itemize}[label=$\rhd$]
\item The equivalence class of gauge parameters $\lambda(x^\mu)$ that are field symmetries, \textit{i.e.} such that the variations of all fields $\Phi^i$ defined on $M$ vanish on shell ($\delta_\lambda \Phi^i \approx 0$).\\ Two gauge parameters are equivalent if they are equal on-shell. 

\item The equivalence class of $(n-2)$-forms $\mathbf{k}$ that are closed on shell ($d\mathbf{k} \approx 0$). \\Two $(n-2)$-forms are equivalent if they differ on-shell by $ d\mathbf{l}$ where $\mathbf{l}$ is a $(n-3)$-form.
\end{itemize}
\label{res:GeneralizedNoether}
\end{resultat}

Note that the expression of conserved $(n-2)$-forms remains ambiguous but the conserved charge $Q = \int_{\partial\Sigma} k^{[\mu\nu]} (d^{n-2}x)_{\mu\nu}$ is not ambiguous! We can always add to $\mathbf{k}$ the divergence of a $(n-3)$-form and a $(n-2)$-form that is trivial on shell. But since the integral of an exact form is zero by Stokes' theorem, it does not modify the conserved charge. We must now understand how we can use this theorem in a general theory with gauge invariance, how we can extract these conserved forms out of any theory and discuss the properties of such conserved charges. \\

As an example, let us show how we can understand the electric charge in classical electrodynamics under the light of this powerful theorem. We denote the 4-potential by $A_\mu$ and the matter current by $J_M^\mu$. Gauge transformations transform $A_\mu \rightarrow A_\mu + \partial_\mu \lambda$. We are interested in the non-trivial field symmetries: the gauge parameters $\lambda \not \approx 0$ such as $\delta_\lambda A_\mu \approx 0$. There is only one set of symmetries, the constant gauge transformations, $\lambda(x^\mu) = c \in \mathbb{R}_0$. As we will derive below, the conserved $(n-2)$-form is given by $\mathbf{k}_c [A] = c F^{\mu\nu} (d^{n-2} x)_{\mu\nu}$. It is conserved on-shell  $d\mathbf{k}_c \approx 0$ outside of matter sources as a consequence of Maxwell's equations, $D_\mu F^{\mu\nu} \approx J_M^\nu$. We can integrate $\mathbf{k}_{c=1}$ on any surface $S$ outside of matter sources, \textit{e.g.} $t,r$ both constant and $r$ large in order to get the electric charge $Q_E = \oint_S \mathbf{k}_{c=1} = \oint_S \vec{E}\cdot \vec{e}_r \: dS$. As an exercice, we can check that it conserved in time, 
\begin{equation}
\frac{d}{dt} Q_E = \oint_S \partial_t k^{tr} \, 2(d^{n-2} x)_{tr} = -\oint_S \partial_A k^{Ar} dS = 0.
\end{equation} 
In the first equation, we evaluated the form on a contant $t$, $r$ slice (the two contributions add up since $k ^{rt}=-k^{tr}$). In the second equation, we developed  the radial component of $d\mathbf{k}=0$, namely $\partial_t k^{tr} + \partial_A k^{Ar}=0$ where $(A=\theta,\phi)$ are the angular coordinates. The last equation follows from the fact that the integration of a closed form on a sphere vanishes (assuming of course that the field strength obeys the free Maxwell equations so without crossing the trajectories of electrons!). Another point to notice is that 
\begin{equation}
\frac{d}{dr} Q_E = \oint_S \partial_r k^{tr} \, 2(d^{n-2} x)_{tr} = -\oint_S \partial_A k^{tA} dS = 0.
\end{equation} 
after using the time component of $d\mathbf{k}=0$, namely $\partial_r k^{tr} + \partial_A k^{tA}=0$ and after assuming again that the field strength is free at $S$. More generally, we obtain the standard Gauss law that only the homology class of the integration surface matter (\textit{i.e.} which sources are included in the surface). 

\subsection{Surface charges in generally covariant theories}

In electrodynamics, we have just seen that it exists exactly one equivalence class of field symmetries, \textit{i.e.} gauge transformations that vanish but such that the gauge parameter itself is non-zero. A representative of this non-trivial symmetry is the global gauge transformation by $1$ everywhere in spacetime. The generalized Noether theorem asserts that it is uniquely associated with the conserved electric charge (we still need to check that explicitly). Now, these symmetries are easily derived because Maxwell theory is linear. General relativity is a non-linear theory and life is more complicated. Gauge transformations in such a generally covariant theory are diffeomorphisms, and the ones that do not transform the metric are the isometries whose generators are the Killing vectors $\delta_\xi g_{\mu\nu} = \Lie_\xi g_{\mu\nu} \approx 0$. However, for a general spacetime, there is no Killing vector since $g_{\mu\nu}$ has no isometries. So the generalized Noether theorem (Result \ref{res:GeneralizedNoether}) cannot be applied to any generally defined diffeomorphism $\xi^\mu$. Correspondingly, it seems hopeless to write a formula describing a conserved $(n-2)$-form for some suitably defined symmetries $\xi^\mu$ in a generally covariant theory. \\

There are however two particular cases where the theorem is just enough: for a family of solutions with shared exact isometries, and for a set of solutions with a shared asymptotic isometry. Both cases make good employ of the linearized theory around a suitably chosen solution. Let us consider a solution $\bar{g}_{\mu\nu}$ of general relativity -- denoted as the \textit{background} field -- which we perturb by adding an infinitesimal contribution $g_{\mu\nu} = \bar{g}_{\mu\nu} + h_{\mu\nu}$. It is not difficult to show that the Lagrangian density linearized around $\bar{g}_{\mu\nu}$ and expanded in powers of $h_{\mu\nu}$ is gauge-invariant under the transformation $\delta_\xi h_{\mu\nu} = \Lie_\xi \bar{g}_{\mu\nu}$, where $\xi^\mu$ is an arbitrary diffeomorphism. Thus, if the background admits some Killing symmetries ($\Lie_\xi \bar{g}_{\mu\nu}=0$), their generators also define a set of symmetries of the linearized theory $\delta_\xi h_{\mu\nu}=0$. We can then apply the Result \ref{res:GeneralizedNoether} to claim the existence of a set of conserved $(n-2)$-forms $\mathbf{k}_\xi [h;\bar g]$ if $h_{\mu\nu}$ satisfies the linearized equations of motion around $\bar{g}_{\mu\nu}$. This is the key to define canonical conserved charges in generally covariant theories! \\

A general method to define symmetries and associated conserved quantities in generic spacetimes consists in introducing boundary conditions in an asymptotic region where the linearized theory can be applied around a reference $\bar{g}_{\mu\nu}$ that admits several Killing vectors. These isometries are only asymptotically defined and relevant, and so are  named \textit{asymptotic symmetries}. In effect, they form the closest analogue in gravity of the group of global symmetries in field theories without gravity. The most obvious illustration of it can be found by looking at the class of asymptotically flat spacetimes. A rough definition of such a spacetimes is provided with metrics that approach the Minkowski metric when some suitably defined radial coordinate $r$ is running to infinity with $g_{\mu\nu} - \eta_{\mu\nu} = \mathcal{O}(1/r)$. Far from the sources of gravitation, the spacetimes becomes approximatively flat : we thus take $\bar{g}_{\mu\nu}=\eta_{\mu\nu}$ as background to linearize the theory. Relevant symmetries $\bar{\xi}^\mu$ include the 10 symmetries of Minkowski spacetime that generate the Poincar\'e algebra, and to each one, a conserved $(n-2)$-form is associated by virtue of the generalized Noether theorem. Then we can integrate these $(n-2)$-forms on a 2-sphere at infinity to get Poincar\'e charges of spacetime! As it turns out, there are even more symmetries leading to additional conserved charges that only exist in gravity, the BMS charges, as we will discuss further in these lectures.  \\

So far we considered boundary conditions where the linearized theory can be directly applied asymptotically. But there are more ways where the linearized theory is useful. Let's formalize the concepts a bit more. The set of metrics that obey the boundary conditions form a set $G$. A particular metric is singled out in this set: the reference or background solution $\bar{g}_{\mu\nu} \in G$. So far we considered the simple case where all asymptotic charges only depend upon $\bar{g}_{\mu\nu}$ and the linearized perturbation $h_{\mu\nu} =g_{\mu\nu}-\bar g_{\mu\nu}$: the charge is then $\oint_S k_\xi [g-\bar{g};\bar{g}]$. But in general, the charge might depend non-linearly on $g_{\mu\nu}$. The way to define the charge is to linearize the theory around each $g_{\mu\nu}$, by considering an abstract field variation $\delta g_{\mu\nu}$. The charge is then defined as 
\begin{equation}\label{ch:def}
Q_\xi[g;\bar g] \equiv \int_{\bar g}^{g} \oint_S k_\xi [\text{d}g';g']
\end{equation}
where we integrate the $(n-2)$-form both on a 2-sphere $S$ and on a path in $G$ joining the reference solution $\bar g_{\mu\nu}$ (\textit{e.g.} Minkowski) to the solution of interest $g_{\mu\nu}$. The charge is conserved as long as $\xi^\mu$ is an asymptotic symmetry in the sense that $d\mathbf{k}_\xi [\delta g;g] \approx 0$ for all $g_{\mu\nu} \in G$ and all variations that are ``tangent'' to $G$. It is not clear at this point if the charge is independent of the path chosen in $G$ to relate $\bar g_{\mu\nu}$ to $g_{\mu\nu}$. In order to define with more care this construction (in particular in which sense $\delta$ can be viewed as an exterior derivative on the field space $G$, how ``tangent to $G$'' can be defined, discuss the independence of the path in $G$, \dots), and in order to compute from first principles the conserved $(n-2)$-forms promised by the result \ref{res:GeneralizedNoether}, we have to develop more formalism...\\

But before, let us mention a last but important conceptual point. The fact that the energy, in particular, is a surface charge in General Relativity can be interpreted as gravity being holographic! Indeed, in quantum gravity the energy levels of all states of the theory can be found by quantizing the Hamiltonian. In the classical limit, the Hamiltonian is a surface charge. If this remains true in the quantum theory (as it does for example in the \textit{AdS/CFT} correspondence) knowing the field on the surface bounding the bulk of spacetime will allow to know all possible states in the bulk of spacetime.

\section{Covariant phase space formalism}

\subsection{Field fibration and symplectic structure}

We work again on a target spacetime $M$ which is a Lorentzian manifold provided with a set of coordinates $\{ x^\mu\} $. Let us set aside the metric tensor $g_{\mu\nu}$ for the moment. On each point $P\in M$, a tangent space $T_P M$ of vectors $v^\mu$ can be constructed, which admits a  natural coordinate basis $\lbrace \partial_\mu \rbrace$. The dual space of $T_P M$ is the so-called cotangent space $T_P^\star M$ which contains $1$-forms $w_\mu$ spanned by a related natural basis $\lbrace dx^\mu \rbrace$. Conversely, vectors are also associated with functions on $1$-forms through the \textit{interior product}. 
\begin{equation}
\begin{array}{rccl}
i : & T_P M & \rightarrow & \text{Linear functions on }T_P^\star M ;\\ 
 & \xi & \mapsto & [ \: i_\xi : T_P^\star M \rightarrow \mathbb{R} : w \mapsto i_\xi w \triangleq \xi^\mu \partial_\mu w \: ] .
\end{array} 
\end{equation}
We can extend this definition to promote the interior product to an operator $i_\xi : \Omega^k  (M)\rightarrow \Omega^{k-1} (M)$ where $\Omega^k (M)$ is the set of $k$-forms defined on $M$, simply by requiring that $i_\xi w \equiv \xi^\mu \frac{\partial}{\partial dx^\mu} w, \: \forall w \in \Omega^k (M)$. We have also at our disposal a differential operator $d = dx^\mu \partial_\mu$, the \textit{exterior derivative} that induces the \textit{De Rham complex}. Starting from scalars (or pedantically $0$-forms), successive applications of $d$ lead to higher order forms :
\begin{equation}
\Omega^0 (M) \rightarrow \Omega^1 (M) \rightarrow \Omega^2 (M) \rightarrow \cdots \rightarrow \Omega^{n-1} (M) \rightarrow \Omega^n (M) \rightarrow 0.
\end{equation}
To be short, we have a first space which is the manifold $M$ of coordinates $\{ x^\mu\} $ equipped with a natural differential operator $d$. Using $d$, we get forms of higher degree, since $d : \Omega^k \rightarrow \Omega^{k+1}$. One can also use $i_\xi$ to ascend the chain of $\Omega$'s. \\

Now we consider the fields only. We designate them by the compact notation  $\Phi = (\Phi^i)_{i\in I}$ where the fields $\Phi^i$ also include the metric field $g_{\mu\nu}$.  Fields are abstract entities without dependence in the coordinates. In order to be complete, the \textit{field space} or \textit{jet space} consists in the fields $\Phi^i$ and a set of ``symmetrized derivatives of fields'' $\lbrace \Phi^i,\Phi_\mu^i, \Phi_{\mu\nu}^i,\dots \rbrace$.  In field space, we can again select a ``point'' $(\Phi^i,\Phi_\mu^i, \Phi_{\mu\nu}^i,\dots )$ and the cotangent space at that point is then defined as $(\delta\Phi^i,\delta\Phi_\mu^i, \delta\Phi_{\mu\nu}^i,\dots )$. The symmetrized derivative is defined such that 
\begin{equation}
\frac{\partial}{\partial  \Phi_{\mu\nu}^i}  \Phi_{\alpha\beta}^j = \delta_\alpha^{(\mu} \delta_\beta^{\nu )} \delta_i^j \: ,\: \text{so in particular }\; \frac{\partial \Phi^i_{xy}}{\partial \Phi_{xy}^i} = \frac{1}{2}.
\end{equation}
The variational operator $\delta$ is defined as
\begin{equation}
\delta = \delta \Phi^i \frac{\partial}{\partial \Phi^i} +\delta \Phi_\mu^i \frac{\partial}{\partial \Phi_\mu^i} +\delta \Phi_{\mu\nu}^i \frac{\partial}{\partial \Phi_{\mu\nu}^i} + \cdots
\end{equation}
It is convenient to use the convention that all $\delta\Phi^i$, $\delta \Phi^i_\mu$, \dots are Grassmann odd. It implies that $\delta^2 = 0$. $\delta$ is then an exterior derivative on the field space and each $\delta\Phi^i,\delta\Phi_\mu^i, \delta\Phi_{\mu\nu}^i,\dots $ is a $1$-form in field space. \\

We can now put the manifold and the field space together and we get the \textit{jet bundle} or \textit{variational bicomplex}. The jet bundle is a manifold with local coordinates $(x^\mu,\Phi^i_{(\mu)})$ where $(\mu)$ stands for any set of symmetrized multi-indices. The fields are all \textit{fibers} above the target manifold. Taking a \textit{section} of the fiber, we obtain the coordinate-dependent fields and their derivatives $\lbrace \Phi^i(x^\mu),\partial_\mu \Phi^i(x^\mu),\partial_\mu \partial_\nu \Phi^i(x^\mu),\dots \rbrace$. The standard differential operator $d$ is still defined on the jet space as $d = dx^\mu \partial_\mu$ but now
\begin{equation}
 \partial_\mu \equiv  \frac{\partial}{\partial x^\mu} + \Phi^i_\mu \frac{\partial}{\partial \Phi^i} +  \Phi^i_{\mu\nu} \frac{\partial}{\partial \Phi_\nu^i}+\cdots
\end{equation}
Thenceforth we have two Grassmann-odd differential operators at our disposal: $d$ and $\delta$ and the formalism ensures that they anti-commute $\{ \delta,d \} = 0$ as you can check. We have a ``\emph{variational bicomplex}''. A form with $p$ $dx^\mu$'s and $q$ $\delta \Phi^i_{(\mu)}$'s is a $(p,q)$-form. 

\begin{figure}[h!]
\centering
\begin{tikzpicture}
% Manifold (M,g)
	\draw[GrandVert,fill=green!25] (0,-5) circle (3 and 1.5);
	\draw (3.5,-5) node[right] {\color{GrandVert} \textbf{Spacetime \textit{M}}};
	\draw[->] (6.3,-5) -- (7,-5);
	\draw (7.5,-5) node[right,draw=black,fill=gray!15,text width=4.5cm,align=center] {$T_P^\star M = \text{Span} \lbrace dx^\mu \rbrace$};
% Fiber above x^\mu
	\def\x{0}; \def\y{-5};
	\draw (\x,\y) node {$\bullet$};
	\draw (\x,\y) node[left]{$x^\mu$} -- (\x,\y+4);
% Fiber above y^\mu
	\def\xx{-2}; \def\yy{-4.5};
	\draw (\xx,\yy) node {$\bullet$};
	\draw (\xx,\yy) node[left]{$y^\mu$} -- (\xx,\yy+4);
	%\draw (\xx,\yy) node {$\bullet$};
	%\draw[->] (\xx,\yy) node[left]{$y^\mu$} -- (\xx,\yy+6) node[above,align=center,text width=1.5cm]{$\Phi(y^\mu)$, $\partial_\alpha \Phi (y^\mu)$, $\cdots$};
	%\draw (\xx,\yy+4) node {\color{orange!50} \Large $\bullet$};
\draw (\xx-0.5,\yy+2) node[above,rotate=90,align=center,text width=4cm]{Field fibration};
% Fiber above z^\mu
	\def\xxx{2.2}; \def\yyy{-5.5};
	\draw (\xxx,\yyy) node {$\bullet$};
	\draw (\xxx,\yyy) node[left]{$z^\mu$} -- (\xxx,\yyy+4);
	%\draw (\xxx,\yyy) node {$\bullet$};
	%\draw[->] (\xxx,\yyy) node[left]{$z^\mu$} -- (\xxx,\yyy+6) node[above,align=center,text width=1.5cm]{$\Phi(z^\mu)$, $\partial_\alpha \Phi (z^\mu)$, $\cdots$};
	%\draw (\xxx,\yyy+4) node {\color{orange!50} \Large $\bullet$};
% Section in the fiber
	\draw[orange,fill=orange!25,rotate=-13.5,fill opacity=0.6] (\x+0.35,\y+4) circle (3 and 0.8);
	\draw (3.5,-1.7) node[right] {\color{orange} \textbf{Section }$ = \bm{ \lbrace \Phi^i (x^\mu), \partial_\alpha \Phi^i (x^\mu),...\rbrace}$};
	% Complete arrows
	\draw[->] (\x,\y+4) -- (\x,\y+7) node[above,align=center,text width=2cm]{$\Phi^i , \Phi_\mu^i ,\dots$};
	\draw[->] (\xx,\yy+4) -- (\xx,\yy+7) node[above,align=center,text width=2cm]{$\Phi^i ,  \Phi_\mu^i ,\dots$};
	\draw[->] (\xxx,\yyy+4) -- (\xxx,\yyy+7) node[above,align=center,text width=2cm]{$\Phi^i ,  \Phi_\mu^i ,\dots$};
	\fill[fill=orange] (\x,\y+4) circle (0.1);
	\fill[fill=orange] (\xx,\yy+4) circle (0.1);
	\fill[fill=orange] (\xxx,\yyy+4) circle (0.1);
% Field space
	\draw (3.5,1.8) node[right] {\color{black} \textbf{Field space \textit{J}}};
	\draw[->] (6.3,1.8) -- (7,1.8);
	\draw (7.5,1.8) node[right,draw=black,fill=gray!15,text width=4.5cm,align=center] {$T_P^* J = \text{Span} \lbrace \delta \Phi^i_I \rbrace$ };

% Jet space
	%\draw (9.8,-4.2) -- (9.8,0.7);
	\draw[blue,decorate,decoration={brace,amplitude=10pt}] (3.0,-7.0) -- (-3.0,-7.0);
	\def\z{-8.3}
	\draw (0.0,\z) node[blue,draw=blue] {$\bm{ \text{\textbf{Jet bundle}} = \lbrace (x^\mu,\Phi^i_{(\mu)} )\rbrace}$};
	\draw (4,\z) node[right,fill=gray!15,text width=8cm,draw=black] {Horizontal derivative = exterior derivative $d$ Vertical derivative = variational operator $\delta$};
	\draw[->] (2.65,\z) -- (3.45,\z);
	
% Frame
\draw (-3.8,-9.6) -- (13,-9.6) -- (13,3.7) -- (-3.8,3.7) -- cycle;
\end{tikzpicture}
\caption{Elements from the variational bicomplex structure.}
\label{fig:Bicomplex}
\end{figure}
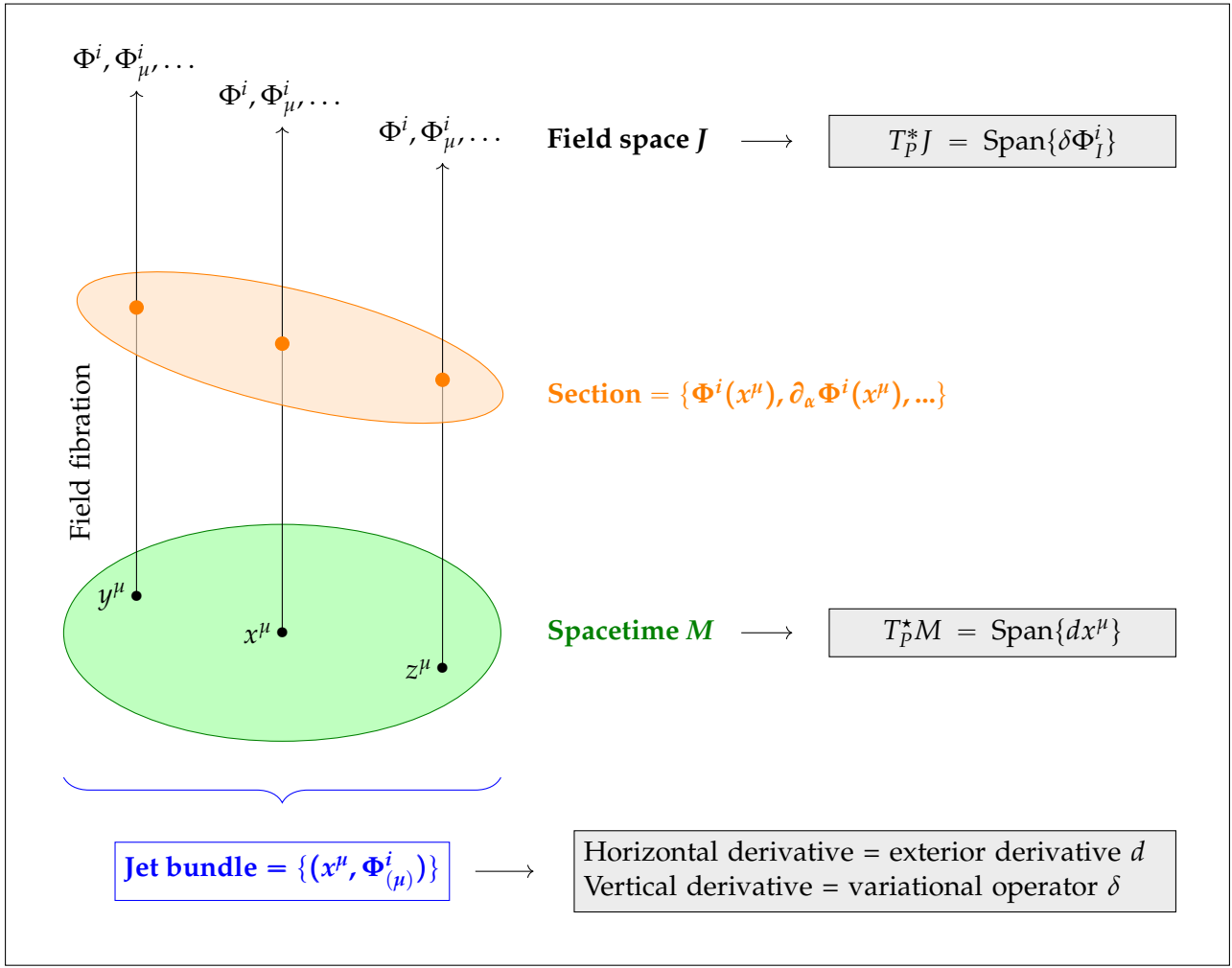

\newpage
The classical physics of the fields is encoded into a Lagrangian density $n$-form $\mathbf{L}$ and a set of boundary conditions. Since the Lagrangian is a $n$-form and depends on the fields, it is a natural object in this variational bicomplex structure. Let us now revise the formula giving an arbitrary variation of the Lagrangian density (\ref{eq:VariationL}). Remember that the boundary terms arise after iterative applications of inverse Leibniz rule.  Now, $\delta$ has been defined as a $1$-form that anticommutes with $dx^\mu$ so we should now write 
\begin{equation}
{\color{blue}\boxed{\delta \mathbf{L} = \delta \Phi^i \frac{\delta \mathbf{L}}{\delta \Phi^i} - d\bm{\Theta} [\delta \Phi;\Phi]}.}
\label{eq:VariationLa}
\end{equation}
We can get back to (\ref{eq:VariationL}) by contracting each side of \eqref{eq:VariationLa} with the inner product $i_{\delta_a}$ where 
\begin{equation}
i_{\delta_a} \triangleq \delta_a \Phi^i_I \frac{\partial}{\partial \delta \Phi^i_I}
\end{equation}
and we take $\delta_a \Phi_I$ Grassmann even by definition. The minus sign is compensated by the fact that $\delta$ needs to anticommute with $d$ to reach $\bm{\Theta}$. \\
\newpage	
We name $\bm{\Theta}[\delta \Phi;\Phi]$ the \textit{presymplectic potential}. It depends by definition on the fields and their variations, but not explicitly on the coordinates. It is a $(n-1,1)$-form! As a consequence, $\delta \bm{\Theta}$ is a $(n-1,2)$-form! It is the so-called \textit{presymplectic form} :
\begin{equation}
{
\color{blue} \boxed{ \bm\omega [\delta \Phi,\delta \Phi;\Phi] = \delta \bm\Theta[\delta \Phi;\Phi]. }
}
\end{equation}
In order to go back to a notation where variations are more familiar Grassmann even quantities, one can contract both sides of the equation with the inner product $i_{\delta_2}i_{\delta_1}$ where $i_{\delta_a}$ is defined as above. The operator $i_{\delta_1}$ hits either the first or second $\delta$ so there are two terms; in each case the remaining $\delta$ is replaced by $\delta_2$. Taking into account the sign obtained by anticommuting $\delta$ with $d$ we obtain
\begin{equation}
i_{\delta_2}i_{\delta_1}\bm\omega \triangleq  \bm\omega[\delta_1 \Phi,\delta_2 \Phi;\Phi] = \delta_1 \bm\Theta [\delta_2 \Phi;\Phi] - \delta_2 \bm\Theta [\delta_1 \Phi;\Phi].
\end{equation}

Our main goal consists in linking the symplectic form that we have just defined on the jet space to conserved $(n-2)$-forms that we announced before. But before that, let us make a necessary intermezzo about the second Noether theorem on continuous symmetries, which we will use afterwards as a lemma ! 

\subsection{Noether's second theorem : an important lemma}

Each gauge symmetry of a Lagrangian gives rise to an identity among its equations of motion. This fundamental property of gauge theories leads to Noether's second theorem: 

\begin{resultat}[\textit{Noether's second theorem}]
Given a generally covariant Lagrangian $n$-form $\mathbf{L}=L d^n x$ and an arbitrary infinitesimal diffeomorphism $\xi^\mu$, one has 
\[ \frac{\delta \mathbf{L}}{\delta \Phi^i} \delta_\xi \Phi^i = d\mathbf{S}_\xi \left[ \frac{\delta L}{\delta \Phi} ; \Phi \right] \]
where $\mathbf{S}_\xi$ is a $n-1$ form proportional to the equations of motion and its derivatives. The equality also holds for other types of gauge transformations where $\xi^\mu$ is then replaced by an arbitrary gauge parameter of the other type. 
\label{res:SecondNoetherThm}
\end{resultat}
Instead of giving a formal proof of this relation, we prefer verify it for two famous gauge theories !

\subsubsection{Einstein's gravity}
Let us focus first on the Einstein-Hilbert Lagrangian density $L = \frac{1}{16\pi G} R \rmg $, where $R$ is the Ricci curvature associated to the metric field $g_{\mu\nu}$. 
\begin{align}
\frac{\delta {\bf L}}{\delta g_{\mu\nu}} \delta_\xi g_{\mu\nu} &= \frac{1}{16\pi G} \: d^n x \: \sqrt{-g} \: \left( \frac{1}{\sqrt{-g}} \frac{\delta\: (\sqrt{-g}\:R)}{\delta g_{\mu\nu}} \right) \delta_\xi g_{\mu\nu} \\
 &= \frac{1}{16\pi G}\:  d^n x \: \sqrt{-g} \: (- G^{\mu\nu}) \mathcal{L}_\xi g_{\mu\nu} \\
 &= -\frac{1}{8\pi G} \: d^n x \: \sqrt{-g} \: G^{\mu\nu} \nabla_\mu \xi_\nu \\
 &= -\frac{1}{8\pi G} \: d^n x \: \sqrt{-g} \: \nabla_\mu ( G^{\mu\nu} \xi_\nu) + \frac{1}{8\pi G} \: d^n x \: \sqrt{-g} \: \nabla_\mu G^{\mu\nu} \xi_\nu \\
 &= d^n x \: \partial_\mu \left( -\frac{1}{8\pi G} \: \sqrt{-g} \: G^{\mu\nu} \xi_\nu \right) \\
 &\Longrightarrow \boxed{ \mathbf{S}_\xi = -\frac{1}{8\pi G} \: (d^{n-1} x)_\mu \: \sqrt{-g} \: G^{\mu\nu} \xi_\nu  } \label{eq:Noether2ndThmEinstein}
\end{align}
and the second Noether theorem is proven for this case. In the crucial fourth step, we used Bianchi's identities $\nabla_\mu G^{\mu\nu} =0$, which is the identity among the equations of motion directly related to general covariance. 

\subsubsection{Einstein-Maxwell electrodynamics}
As an exercice, we can also show a similar result for classical electrodynamics that is minimally coupled to Einstein's gravity. The field is the 4-vector potential $A_\mu$. We thus consider the Maxwell field into a curved spacetime manifold described by its metric tensor $g_{\mu\nu}$. They are two gauge symmetries in the game : the classical invariance of electrodynamics $A_\mu \rightarrow A_\mu + \partial_\mu \lambda$, and also the invariance under diffeomorphisms, guaranteed by the generally covariant property of a theory coupled to gravity. The minimal coupling assumption leads us to the Lagrangian $n$ form $\mathbf{L} = \mathbf{L}^{G} + \mathbf{L}^{EM} = \frac{1}{16\pi G} R \rmg d^n x -\frac{1}{4} \rmg F^{\alpha \beta} F_{\alpha\beta} d^n x$, where $F_{\alpha \beta} = \partial_\alpha A_\beta - \partial_\beta A_\alpha$ is the antisymmetrical Faraday tensor, which is gauge invariant and contains the physical electric and magnetic fields. \\

Let us show as a little lemma that the electromagnetic stress tensor is conserved on-shell. We have :
\begin{equation}
T_{EM}^{\mu\nu} = \frac{2}{\rmg} \frac{\delta L^{EM}}{\delta g_{\mu\nu}} = F^{\mu \alpha} F^{\nu}_{\phantom{\nu}\alpha} - \frac{1}{4} F^{\alpha \beta} F_{\alpha\beta} g^{\mu\nu} .
\end{equation}
To do this calculation, the useful formulas are :
\begin{equation}
\delta \rmg = \frac{1}{2} \rmg g^{\mu\nu} \delta g_{\mu\nu} \quad ; \quad \delta g^{\mu\nu} = - g^{\mu\alpha} g^{\nu\beta} \delta g_{\alpha\beta}.
\end{equation}

Recalling that $\partial_{[\alpha} F_{\beta\gamma ]} = 0$ or equivalently $\nabla_{[\alpha} F_{\beta\gamma ]} = 0$ since all Christoffel symbols cancel out by antisymmetry, it can be checked that $\nabla_{[\mu} F^\nu_{\phantom{\nu}\alpha]} = -\frac{1}{2} \nabla^{\nu} F_{\alpha \mu}$. We can now check:
\begin{align}
\nabla_\mu T_{EM}^{\mu\nu} &= \nabla_\mu F^{\mu\alpha} F^{\nu}_{\phantom{\nu}\alpha} + F^{{\color{orange} [}\mu\alpha{\color{orange}]}} \nabla_{{\color{orange} [}\mu} F^\nu_{\phantom{\nu}\alpha {\color{orange}]}} - \frac{1}{2} F^{\alpha\beta} \nabla^\nu F_{\alpha\beta} \\
&= \nabla_\mu F^{\mu\alpha} F^{\nu}_{\phantom{\nu}\alpha}   - \frac{1}{2} F^{\mu\alpha} \nabla^{\nu} F_{\alpha \mu} - \frac{1}{2} F^{\alpha\beta} \nabla^\nu F_{\alpha\beta} \\
&= \nabla_\mu F^{\mu\alpha} F^{\nu}_{\phantom{\nu}\alpha}  + \frac{1}{2} F^{\mu\alpha} \nabla^{\nu} F_{\mu \alpha} - \frac{1}{2} F^{\alpha\beta} \nabla^\nu F_{\alpha\beta} \\
&=\nabla_\mu F^{\mu\alpha} F^{\nu}_{\phantom{\nu}\alpha} ,\label{id4}
\end{align}
which is indeed zero on-shell. \\

Let us now compute the left-hand-side of the second Noether theorem. Pay attention to the fact that in the case of Einstein-Maxwell theory, the gauge parameter is a couple $(\xi^\mu,\lambda)$ (where $\xi^\mu$ is a diffeomorphism, and $\lambda$ a gauge transformation for $A_{\mu}$).
\begin{align}
(1) &\quad \frac{\delta L}{\delta A_{\mu}} = \frac{\delta L^{EM}}{\delta A_{\mu}} = -\partial_\nu \frac{\partial L^{EM}}{\partial\partial_\nu A_{\mu}} = \partial_\nu \left( \rmg F^{\nu \mu} \right) =  \rmg \nabla_\nu F^{\nu\mu}; \\
(2) &\quad \frac{\delta L}{\delta g_{\mu\nu}} = \frac{\delta L^{EM}}{\delta g_{\mu\nu}} + \frac{\delta L^G}{\delta g_{\mu\nu}} = \frac{\rmg}{2} T^{\mu\nu}_{EM} - \frac{1}{16\pi G}  \sqrt{-g}  G^{\mu\nu}.
\end{align}
And thus, since $\Phi^i = \{ A_\mu,g_{\mu\nu}\}$ :
\begin{equation}
\frac{\delta L}{\delta \Phi^i} \delta_{(\xi,\lambda)} \Phi^i = \frac{\delta L}{\delta A_{\mu}} \delta_{(\xi,\lambda)} A_\mu + \frac{\delta L}{\delta g_{\mu\nu}} \delta_{(\xi,\lambda)} g_{\mu\nu}.
\label{eq:deltaLforEinsteinMax}
\end{equation}
The potential field $A_\mu$ varies under the two gauge transformations : $\delta_{(\xi,\lambda)} A_\mu = \Lie_\xi A_\mu + \delta_\lambda A_\mu = \Lie_\xi A_\mu  + \partial_\mu \lambda$, while $g_{\mu\nu}$ is only affected by diffeomorphisms $\delta_{(\xi,\lambda)} g_{\mu\nu} = \delta_\xi g_{\mu\nu} = \Lie_\xi g_{\mu\nu}$. It remains to compute the Lie derivative of $A_\mu$ on the flow of $\xi^\mu$ :
\begin{align}
\Lie_\xi A_\mu &\triangleq \xi^\rho \partial_\rho A_\mu + A^\rho \partial_\mu \xi_\rho = \xi^\rho F_{\rho \mu} + \xi^\rho \partial_\mu A_\rho + A^\rho \partial_\mu \xi_\rho = \xi^\rho F_{\rho \mu} + \partial_\mu (\xi^\rho A_\rho).
\end{align} 
Inserting all these expressions into (\ref{eq:deltaLforEinsteinMax}) leads us to
\begin{align}
\frac{\delta L}{\delta \Phi^i} \delta_{(\xi,\lambda)} \Phi^i &= \rmg \nabla_\nu F^{\nu\mu} \left[ \xi^\rho F_{\rho \mu} + \partial_\mu (\xi^\rho A_\rho + \lambda ) \right] + 2 \rmg \left( \frac{1}{2} T^{\mu\nu}_{EM} - \frac{1}{16\pi G}  G^{\mu\nu} \right) \nabla_\mu \xi_\nu \\
&= \rmg \nabla_\mu T^{\mu\nu}_{EM} \xi_\nu + \rmg \nabla_\nu F^{\nu\mu} \nabla_\mu (\xi^\rho A_\rho + \lambda ) + \rmg \left( T^{\mu\nu}_{EM} - \frac{1}{8\pi G}  G^{\mu\nu} \right) \nabla_\mu \xi_\nu\label{id7}
\end{align}
where we used \eqref{id4} in the first term. 
We now apply the inverse Leibniz rule on the third term, and we remember that $G^{\mu\nu}$ is divergence-free, to obtain :
\begin{align}
\rmg \left( T^{\mu\nu}_{EM} - \frac{1}{8\pi G}  G^{\mu\nu} \right) \nabla_\mu \xi_\nu &= \rmg \nabla_{\mu} \left(  T^{\mu\nu}_{EM} - \frac{1}{8\pi G} G^{\mu\nu} \right) - \rmg \nabla_\mu T^{\mu\nu}_{EM} \xi_\nu \\
&= \partial_{\mu} \left(  \rmg T^{\mu\nu}_{EM} - \frac{1}{8\pi G} \rmg G^{\mu\nu} \right) - \rmg \nabla_\mu T^{\mu\nu}_{EM} \xi_\nu \\
&= \partial_{\mu} \left( 2 \frac{\delta L}{\delta g_{\mu\nu}} \xi_\nu \right)- \rmg \nabla_\mu T^{\mu\nu}_{EM} \xi_\nu .
\end{align}
The last term is compensated as expected by the first term of \eqref{id7}. We then have 
\begin{align}
\frac{\delta L}{\delta \Phi^i} \delta_{(\xi,\lambda)} \Phi^i&= \rmg \nabla_\nu F^{\nu\mu} \nabla_\mu (\xi^\rho A_\rho + \lambda ) + \partial_{\mu} \left( 2 \frac{\delta L}{\delta g_{\mu\nu}} \xi_\nu \right) \\
&= \rmg \nabla_\mu \left[ \nabla_\nu F^{\mu\nu}   (\xi^\rho A_\rho + \lambda ) \right] + \partial_\mu  \left( 2 \frac{\delta L}{\delta g_{\mu\nu}} \xi_\nu \right) \\
&= \partial_\mu \left[ \rmg \nabla_\nu F^{\mu\nu}   (\xi^\rho A_\rho + \lambda ) \right] + \partial_\mu  \left( 2 \frac{\delta L}{\delta g_{\mu\nu}} \xi_\nu \right) \\
&= \partial_\mu \left[ \frac{\delta L}{\delta A_{\mu}}   (\xi^\rho A_\rho + \lambda ) + 2 \frac{\delta L}{\delta g_{\mu\nu}} \xi_\nu \right] \triangleq \partial_\mu S_{(\xi,\lambda)}^\mu . 
\end{align}
We get the second line thanks to the Bianchi identity $\nabla_{\mu}\nabla_{\nu} F^{\mu\nu} = 0$, and the third one by virtue of the property $\rmg \nabla_\mu (\cdots) \equiv \partial_\mu (\rmg \cdots )$. We are thus left with a $(n-1)$-form $\mathbf{S}_{(\xi,\lambda)}$ that satisfies the second Noether theorem :
\begin{equation}
\boxed{ \mathbf{S}_{(\xi,\lambda)} = \left[ \frac{\delta L}{\delta A_{\mu}}   (\xi^\rho A_\rho + \lambda ) + 2 \frac{\delta L}{\delta g_{\mu\nu}} \xi_\nu \right] (d^{n-1} x)_\mu .}
\end{equation}
We have just proven that the second Noether theorem was valid for both diffeomorphisms and electromagnetic gauge transformations! 

\subsection{Fundamental theorem of the covariant phase space formalism}

\subsubsection{Cartan's magic formula}
Since we are considering generally covariant theories, we can always identify the variation along a diffeomorphism and the Lie derivative along its flow : $\delta_\xi \equiv \Lie_\xi$ when acting on tensors. In turn, the Lie derivative of a tensor can subdivided into several operations using Cartan's magic formula 
\begin{equation}
\Lie_\xi (\cdots) = d\: i_\xi (\cdots) + i_\xi d(\cdots),
\label{eq:Cartan}
\end{equation}
which makes it useful for deriving algebraic relations.
Here, recall that the involution along a vector $\xi^\mu$ is defined as $i_\xi = \xi^\mu \partial_\mu$ and $d = dx^\mu \partial_\mu$. We can easily prove it when the Lie derivative acts on a scalar field $\phi$, since it reduces to the directional derivative on the integral curves of $\xi^\mu$. So $\Lie_\xi \phi = \xi^\mu \partial_\mu \phi = i_\xi d\phi$ which is correct because $i_\xi \phi = 0$ (the space of $(-1)$-forms is empty !). The proof that Cartan magic's formula holds for all forms follows by induction from this observation. Indeed, one can show that, for any integer $k$, $\Omega^k$ is generated by scalars, their exterior derivative, and some exterior products. So we can accept that it is true for all tensors without exhaustively do the proof here !
 
\subsubsection{Noether-Wald surface charge}
Let us now take the variation of $\mathbf{L}$ along any infinitesimal diffeomorphism $\xi^\mu$ :
\begin{align}
\delta_\xi \mathbf{L} = \Lie_\xi \mathbf{L} &\stackrel{(\ref{eq:Cartan})}{=} d(i_\xi \mathbf{L}) + i_\xi d\mathbf{L} = d(i_\xi \mathbf{L}) + 0 \\
&\stackrel{(\ref{eq:VariationL})}{=} \frac{\delta \mathbf{L}}{\delta \Phi} \Lie_\xi \Phi + d \bm\Theta[\Lie_\xi \Phi;\Phi].
\end{align}
By virtue of Noether's second theorem (Result \ref{res:SecondNoetherThm}), we get :
\begin{equation}
d(i_\xi \mathbf{L}) = d\mathbf{S}_\xi \left[ \frac{\delta L}{\delta \Phi}; \Phi \right] + d \bm\Theta[\Lie_\xi \Phi;\Phi] \Longrightarrow \partial_\mu \left( \xi^\mu L - \Theta^\mu [\Lie_\xi \Phi;\Phi] - S_\xi^\mu \left[ \frac{\delta L}{\delta \Phi}; \Phi \right] \right) = 0.
\end{equation}
The standard Noether current of field theories is the Hodge dual of the conserved $n-1$ form 
\begin{equation}
\mathbf{J}_\xi \triangleq i_\xi \mathbf{L} - \bm\Theta [\Lie_\xi \Phi;\Phi] \quad \text{with} \quad d\mathbf{J}_\xi = d\mathbf{S}_\xi \Rightarrow d\mathbf{J}_\xi \approx 0.
\label{eq:LeeWaldCurrent}
\end{equation}
Now, a fundamental property of the covariant phase space is that a closed form that depends linearly on a vector $\xi^\mu$ and its derivatives is locally exact. Therefore, this Noether current can be written as $\mathbf{J}_\xi = \mathbf{S}_\xi + d\mathbf{Q}_\xi$.  The proof is simple. It relies on the existence of an operator $I_\xi$ such that 
\begin{equation}
d\: I_\xi + I_\xi \: d = 1. 
\end{equation}
Acting with $I_\xi$ on $d(\mathbf{J}_\xi -\mathbf{S}_\xi) $ we get 
\begin{equation}
0 = I_\xi d(\mathbf{J}_\xi -\mathbf{S}_\xi) = \mathbf{J}_\xi -\mathbf{S}_\xi - d I_\xi (\mathbf{J}_\xi -\mathbf{S}_\xi)
\end{equation}
so we deduce that $\mathbf{Q}_\xi = I_\xi (\mathbf{J}_\xi -\mathbf{S}_\xi)$. The operator $I_\xi$ is in fact given by 
\begin{equation}
\forall \bm\omega_\xi \in \Omega^k(M), \:\;\; I_\xi \bm\omega_\xi = \frac{1}{n-k} \: \xi^\alpha \frac{\partial}{\partial \partial_\mu \xi^\alpha} \: \frac{\partial}{\partial dx^\mu} \bm\omega_\xi + (\text{Higher derivative terms}).
\end{equation}
Since only terms proportional to at least one derivative of $\xi^\alpha$ matter and neither $\mathbf{S}_\xi $ nor $i_\xi \mathbf{L}$ do contain derivatives of $\xi^\mu$ we have $I_\xi\mathbf{S}_\xi =I_\xi i_\xi \mathbf{L}=0$ and we have more simply 
\begin{equation}
\mathbf{Q}_\xi [\Phi] = -I_\xi \bm\Theta [\delta_\xi \Phi;\Phi].
\label{eq:NoetherWaldCharge}
\end{equation}
We will call this $(n-2)$-form the \textit{Noether-Wald surface charge}. \\

We are now ready to state and prove the fundamental theorem: 
\begin{resultat}[\textit{Fundamental theorem of the covariant phase space formalism}]
\label{res:FundamentalThmPSF}
In the Grassmann odd convention for $\delta$, contracting the presymplectic form with a gauge transformation $\delta_{\xi} \Phi^i$, it exists a $(n-2,1)$-form $\mathbf{k}_\xi [\delta\Phi;\Phi]$ that satisfies the identity
\[ 
\boxed{ \bm\omega[ \delta_\xi\Phi,\delta \Phi;\Phi] \approx d\mathbf{k}_\xi [\delta \Phi;\Phi] }
\]
where $\Phi^i$ solves the equations of motion, and $\delta\Phi^i$ solves the linearized equations of motion around the solution $\Phi^i$. The infinitesimal surface charge $\mathbf{k}_\xi [\delta\Phi;\Phi]$ is unique, up to a total derivative that does not affect the equality above, and it is given in terms of the Noether-Wald surface charge and the presymplectic potential by the following relation :
\[
\mathbf{k}_\xi [\delta\Phi;\Phi] = -\delta \mathbf{Q}_\xi [\delta \Phi;\Phi] + i_\xi \bm\Theta [\delta\Phi;\Phi] + \text{total derivative}
\]
\end{resultat}

\subsubsection{Proof}
We are considering the jet space where $\delta$ is Grassmann odd and anticommutes with the exterior derivative $d$. Let us compute 
\begin{align}
\delta \mathbf{S}_\xi \left[ \frac{\delta L}{\delta \Phi}; \Phi \right] &= \delta \mathbf{J}_\xi [\delta \Phi;\Phi] - \delta d\mathbf{Q}_\xi [\delta \Phi;\Phi] \\
&= \delta i_\xi \mathbf{L}[\Phi] - \delta\bm\Theta [\Lie_\xi \Phi;\Phi] + d\delta \mathbf{Q}_\xi [\delta \Phi;\Phi] \\
&=- i_\xi \delta \mathbf{L}[\Phi] - \delta\bm\Theta [\Lie_\xi \Phi;\Phi] + d\delta \mathbf{Q}_\xi [\delta \Phi;\Phi] \\
&=- i_\xi \left( \frac{\delta \mathbf{L}[\Phi]}{\delta\Phi^i} \delta \Phi^i - d\bm\Theta [\delta\Phi;\Phi] \right) - \delta\bm\Theta [\Lie_\xi \Phi;\Phi] + d\delta \mathbf{Q}_\xi [\delta \Phi;\Phi] \\
&\approx i_\xi d\bm\Theta [\delta\Phi;\Phi] - \delta\bm\Theta [\Lie_\xi \Phi;\Phi] + d\delta \mathbf{Q}_\xi [\delta \Phi;\Phi].
\end{align}
Cartan's magic formula implies that $\Lie_\xi \bm\Theta [\delta\Phi;\Phi] = di_\xi \bm\Theta [\delta\Phi;\Phi] + i_\xi d\bm\Theta [\delta\Phi;\Phi]$, so we get
\begin{align}
\delta \mathbf{S}_\xi \left[ \frac{\delta L}{\delta \Phi}; \Phi \right] &\approx \Lie_\xi \bm\Theta [\delta\Phi;\Phi] - d i_\xi \bm\Theta [\delta\Phi;\Phi] - \delta\bm\Theta [\Lie_\xi \Phi;\Phi] + d\delta \mathbf{Q}_\xi [\delta \Phi;\Phi] \\
&= \delta_\xi \bm\Theta [\delta\Phi;\Phi] - \delta\bm\Theta [\delta_\xi \Phi;\Phi] + d \left( \delta \mathbf{Q}_\xi [\delta \Phi;\Phi] - i_\xi \bm\Theta [\delta\Phi;\Phi] \right) \\
&\triangleq \bm\omega [\delta_\xi \Phi,\delta\Phi;\Phi] - d\mathbf{k}_\xi [\delta \Phi;\Phi]
\end{align}
where $\mathbf{k}_\xi [\delta \Phi;\Phi] = -\delta \mathbf{Q}_\xi [\delta \Phi;\Phi] + i_\xi \bm\Theta [\delta\Phi;\Phi] + d(\cdots)$. Now we are about to conclude : the form $\mathbf{S}_\xi$ vanishes identically by definition on shell, and if $\delta \Phi^i$ solves the linearized equations of motion, its variation vanishes too. So we have proven the fundamental theorem of the covariant phase space formalism.

\subsubsection{Some residual ambiguities}

The fundamental theorem allows to uniquely define (up to an exact form) the infinitesimal surface charge $\mathbf{k}_\xi$ from the presymplectic form. Now, is the definition of the presymplectic form unambiguous? \\

First notice that the presymplectic potential $\bm\Theta$ is ambiguous. If we add a boundary term $d\mathbf{M}$ to the Lagrangian density $\mathbf{L} \rightarrow \mathbf{L} + d\mathbf{M}$, we get exactly $\bm\Theta \rightarrow \bm\Theta +\delta \mathbf{M}$. However, since $\bm\omega = \delta\bm\Theta$, this transformation has no effect on the presymplectic form because $\delta^2 = 0$. Second, we defined $\bm\Theta$ from an integration by part prescription, which gives the canonical definition of $\bm\Theta$ but our derivation goes through by modifying $\bm\Theta \rightarrow \bm\Theta - d\mathbf{B}$ and therefore $\bm\omega \rightarrow \bm\omega - \delta d\mathbf{B} = \bm\omega + d \delta \mathbf{B} \triangleq \bm\omega + d\bm\omega_B$. This ambiguity reflects our ignorance on how to select the boundary terms in the presymplectic form. In principle, most of these ambiguities should be related to the so-called ``corner terms''  in the action principle, but a generic derivation has not been proven (see one specific example in \cite{Compere:2008us}). Fortunately, this ambiguity is irrelevant for exact symmetries of the fields (Killing symmetries in the case of Einstein's theory) as can be shown quickly:
\begin{equation}
\bm\omega[\delta_\xi \Phi,\delta \Phi;\Phi] \rightarrow \bm\omega[\delta_\xi \Phi,\delta \Phi;\Phi] + d\bm\omega_B[\delta_\xi \Phi,\delta \Phi;\Phi] \Longrightarrow \mathbf{k}_\xi \rightarrow \mathbf{k}_\xi + \bm\omega_B[{\Lie_\xi \Phi},\delta \Phi;\Phi] = \mathbf{k}_\xi + 0. 
\end{equation}

\section{Conserved surface charges}

Now we will show how the Result \ref{res:FundamentalThmPSF} can help us defining surface charges for generally covariant and other gauge theories. 

\subsection{Definition of the charges}

We defined so far  a $(n-2)$-form $\mathbf{k}_\xi [\delta\Phi;\Phi]$ with special properties. We now integrate $\mathbf{k}_\xi$ on a closed surface $S$ of codimension 2 (\textit{e.g.} a sphere at time and radius fixed). Doing it, we are left with the \textit{local variation of charge} between the two solutions $\Phi^i$ and $\Phi^i+\delta \Phi^i$, where $\Phi^i$ satisfies the equations of motion, and $\delta\Phi$ their linearized counterpart around $\Phi$. We denote this by
\begin{equation}
\slash\hspace{-5pt}\delta H_\xi [\delta\Phi;\Phi] = \oint_S \mathbf{k}_\xi [\delta\Phi;\Phi].
\end{equation}
We denote $\slash\hspace{-5pt}\delta$ instead of $\delta$ in order to emphazise that the right-hand-side is not necessarily an exact differential on the space of fields. If it is the case, the charge will be said \textit{integrable}, otherwise it is not.

\subsection{Integrability condition}
Let us comment a bit on this very important concept of integrability. $\slash\hspace{-5pt}\delta H_\xi$ is a functional depending on the fields $\Phi^i$ and their variations $\delta \Phi^i$. It is obviously a $1$-form from the point of view of the fields, and a scalar on the manifold. But nothing tells us that this $1$-form is \textit{exact} for the exterior derivative $\delta$, \textit{i.e.} we are not sure that it exist some $H_\xi[\Phi]$ such as $\slash\hspace{-5pt}\delta H_\xi [\delta\Phi;\Phi] = \delta (H_\xi[\Phi])$. A necessary condition for allowing the existence of a Hamiltonian generator $H_\xi$ associated with $\xi$ is the so-called \textit{integrability condition} :
\begin{equation}
\delta_1 \oint_S \mathbf{k}_\xi [\delta_2 \Phi;\Phi] - \delta_2 \oint_S \mathbf{k}_\xi [\delta_1 \Phi;\Phi] =0, \quad \forall \delta_1\Phi ,\, \delta_2 \Phi \in \mathcal{T}[\Phi].
\label{eq:Integrability}
\end{equation}
It is also a sufficient condition if the space of fields does not have any topological obstruction, which is most often the case. \\

If the charge is integrable, $H_\xi$ exists. In order to define it, we denote by $\bar{\Phi}^i$ some reference field configuration, and we continue to denote by $\Phi^i$ our target configuration. Then we select a path $\gamma$ linking $\bar{\Phi}^i$ and $\Phi^i$ in field space, and we perform a path integration along $\gamma$ 
\begin{equation}
{\color{blue} \boxed{ H_\xi [\Phi;\bar{\Phi}] = \int_\gamma \oint_S \mathbf{k}_\xi [\delta \Phi;\Phi] + N_\xi [\bar{\Phi}] .} }\label{defH}
\end{equation}
Here $N_\xi [\bar{\Phi}]$ is a charge associated with the reference $\bar{\Phi}^i$ that is not fixed by this formalism (it can be fixed in other formalisms, \textit{e.g.} the counterterm method in $AdS/CFT$). The definition of $H_\xi$ does not depend on the path $\gamma$ chosen precisely because the integrability condition is obeyed.

\subsection{Conservation criterion}
Let us suppose from now on that the integrability condition (\ref{eq:Integrability}) is obeyed. The surface charge $H_\xi[\Phi,\bar \Phi]$ is clearly conserved on shell under continuous deformations of $S$ if and only if $d\mathbf{k}_\xi [\delta\Phi;\Phi] \approx 0$ or, equivalently, 
\begin{equation}
{\color{blue} \boxed{ H_\xi [\Phi;\bar{\Phi}] \text{ is conserved } \Longleftrightarrow \:\: \bm\omega[ \delta_\xi\Phi,\delta\Phi;\Phi] \approx 0. }}
\end{equation}
We repeat again that ``on shell'' means here : ``$\Phi^i$ solves the equations of motion and $\delta \Phi^i$ solves the linearized equations of motion around $\Phi^i$''. In many cases, asking for conservation in the entire spacetime is too stringent, but one at least requires conservation at spatial infinity, far from sources and radiation. The conservation condition implies that the difference of charge between two surfaces $S_1$ and $S_2$ vanishes,
\begin{equation}
\left. H_\xi \right|_{S_1} - \left. H_\xi \right|_{S_2} = \int_\gamma \oint_{S_1} \mathbf{k}_\xi - \int_\gamma \oint_{S_2} \mathbf{k}_\xi \stackrel{(\text{Stokes})}{=}  \int_\gamma \int_\mathcal{C} d\mathbf{k}_\xi \stackrel{(\text{\ref{res:FundamentalThmPSF}})}{\approx}  \int_\gamma \int_\mathcal{C} \bm\omega \approx 0
\end{equation}
where $\mathcal C$ is the codimension one surface whose boundary is $S_1 \cup S_2$. \\

In gravity, we get conserved charges in two famous cases that we have already discussed :
\begin{itemize}[label=$\rhd$]
\item If $\xi^\mu$ is an exact (Killing) symmetry, we know that $\delta_\xi g_{\mu\nu} = \Lie_\xi g_{\mu\nu} = 0$ so $\bm\omega[\delta_\xi g, \delta g ;g] = 0$. Therefore, any Killing symmetry is associated with a conserved surface charge in the bulk of spacetime. 

\item For asymptotic symmetries, the Killing equation $\Lie_\xi g_{\mu\nu} = 0$ is only verified in an asymptotic sense when $r \rightarrow \infty$, so $\bm\omega[\delta_\xi g, \delta g ;g] \rightarrow 0$ only in an asymptotic region. As a consequence, the charges associated to $\xi^\mu$ will be conserved only in the asymptotic region.
\end{itemize}
Morever, we mention a third particular case: the so-called \textit{symplectic symmetries}, which are vectors $\xi^\mu$ that are no longer isometries of $g_{\mu\nu}$ but still lead to a vanishing presymplectic form. They also lead to conserved charges in the bulk of spacetime (see examples in \cite{Compere:2014cna,Compere:2015knw}). 

\subsection{Charge algebra}

In special relativity, we have 10 Killing vectors and a bracket between these Killing vectors: the Lie bracket. Under the Lie bracket, the 10 Killing vector form the Poincar\'e algebra. Moreover, the charges associated with these vectors represent the algebra of symmetries and also form the Poincar\'e algebra under a suitably defined Poisson bracket between the charges.  What we want to do now is to derive this representation theorem for gravity, and for more general gauge theories.\\

We only consider the most important case of asymptotic symmetries. Let us consider a set $G$ of field configurations that obeys some boundary conditions.  A vector $\xi^\mu$ is said to be an \textit{allowed diffeomorphism} if its action is tangential to $G$. In other words, the infinitesimal Lie variation of the fields is a tangent vector to $G$, which therefore preserves the boundary conditions that define $G$. The set $\lbrace \xi^\mu_a \rbrace$ of such vectors fields form an algebra for the classical Lie bracket $[\xi_a,\xi_b]^\mu = C_{ab}^c \xi^\mu_c$. One can integrate these infinitesimal transformations to obtain their global counterparts, which form a group of \textit{allowed transformations}, always preserving $G$. \\

We assume that the boundary conditions are chosen such that any allowed vector $\xi^\mu$ asymptotically solves the Killing equation. We can then define a conserved charge $H_\xi$, which we assume is integrable and finite\footnote{If the charge is not finite, it means that the boundary conditions constraining the fields on which we are defining the charges are too large and not physical. If the charge is not integrable, one could attempt to redefine $\xi^\mu$ as a function of the fields to solve the integrability condition.}. Two cases can be distinguished :
\begin{itemize}[label=$\rhd$]
\item If $H_\xi$ is non-zero for a generic field configuration, the action of $\xi^\mu$ on fields is considered to have physical content. For example, boosting or rotating a configuration changes the state of the system. 
\item If $H_\xi$ is zero, the diffeomorphism $\xi^\mu$ is considered to be a gauge transformation, and does nothing more than a change of coordinates. These diffeomorphisms are also called trivial gauge transformations. 
\end{itemize}
We define the asymptotic symmetry group as the quotient
\[
{\color{blue}\boxed{ \text{Asymptotic symmetry group} = \frac{\text{Allowed diffeomorphisms}}{\text{Trivial gauge transformations}} }}
\]
which extracts the group of state-changing transformations. This is the closest concept in gravity to the group of global symmetries of field configurations obeying a given set of boundary conditions. 

Let us now derive the representation of the asymptotic algebra obeyed by the charges themselves !

\subsubsection{Representation theorem}
First, we need a Lie bracket for the charges. The definition is the following : for any infinitesimal diffeomorphisms $\chi^\mu,\, \xi^\mu$, we define
\begin{equation}
\lbrace H_\chi , H_\xi \rbrace \triangleq \delta_\xi H_\chi = \oint_S \mathbf{k}_\chi [\delta_\xi\Phi;\Phi]. 
\end{equation}
The last equality directly follows from the definition of the charge \eqref{defH}, in the same way as $\frac{d}{dt}\int_0^t dt' f(t') = f(t)$. To derive the charge algebra, we have to express the right-hand-side as a conserved charge for some yet unknown diffeomorphism. The trick is to use again a reference field $\bar{\Phi}^i$ to re-introduce a path integration :
\begin{eqnarray}
\lbrace H_\chi , H_\xi \rbrace &=& \left( \oint_S \mathbf{k}_\chi [\delta_\xi\Phi;\Phi] - \oint_S \mathbf{k}_\chi [\delta_\xi \bar{\Phi}; \bar{\Phi}] \right) + \oint_S \mathbf{k}_\chi [\delta_\xi \bar{\Phi}; \bar{\Phi}] \\
&=& \left( \int_\gamma \oint_S \delta \mathbf{k}_\chi [\delta_\xi\Phi;\Phi] \right) + \oint_S \mathbf{k}_\chi [\delta_\xi \bar{\Phi}; \bar{\Phi}]
\end{eqnarray}
by virtue to the fundamental theorem of integral calculus. The first term needs some massaging. After using the integrability condition, one can show that 
\begin{equation}
\int_\gamma \oint_S \delta \mathbf{k}_\chi [\delta_\xi\Phi;\Phi] = \int_\gamma \oint_S \mathbf{k}_{[\chi,\xi]} [\delta \Phi;\Phi]\label{rel4}
\end{equation}
and with all fields $\Phi^i$ and their variations $\delta\Phi^i$ on shell. The proof will be given below for the interested reader. So we are left with :
\begin{eqnarray}
\lbrace H_\chi , H_\xi \rbrace &=& \int_\gamma \oint \mathbf{k}_{[\chi,\xi]} [\delta \Phi;\Phi]  + \oint \mathbf{k}_\chi [\delta_\xi \bar{\Phi}; \bar{\Phi}] \\
&=& H_{[\chi,\xi]} + \mathcal{K}_{\chi,\xi} [\bar{\Phi}]
\end{eqnarray}
where we defined 
\begin{equation}
\mathcal{K}_{\chi,\xi} [\bar{\Phi}] \triangleq \oint \mathbf{k}_\chi [\delta_\xi \bar{\Phi}; \bar{\Phi}] - N_{[\chi,\xi]} [\bar{\Phi}].
\end{equation}
The charge algebra is now determined. It reproduces the diffeomorphism algebra up to an extra functional $\mathcal{K}_{\chi,\xi} [\bar{\Phi}]$ that depends only on the reference $\bar{\Phi}^i$. For this reason, it commutes with any surface charge $H_\xi$ under the Poisson bracket, and so it belongs to the center of this algebra. Thus we obtain a \textit{central extension} when we consider the charges instead of the associated vectors. We can show that the central extension is antisymmetric under the exchange $\chi^\mu \leftrightarrow \xi^\mu$, and 
\begin{equation}
\mathcal{K}_{[\chi_1,\chi_2],\xi} [\bar{\Phi}] + \mathcal{K}_{[\xi,\chi_1],\chi_2} [\bar{\Phi}] + \mathcal{K}_{[\chi_2,\xi],\chi_1} [\bar{\Phi}] = 0, \quad \forall \xi^\mu_1,\, \xi^\mu_2,\, \chi^\mu.
\end{equation}
In other words, $\mathcal{K}_{\chi,\xi} [\bar{\Phi}]$ forms a 2-\textit{cocycle} on the Lie algebra of diffeomorphisms, and furthermore confers to $\lbrace \cdot, \cdot \rbrace$ a rightful structure of Lie bracket, since the presence of the central extension affects neither the properties of antisymmetry nor Jacobi's identity. A central extension $\mathcal{K}_{\chi,\xi}$ which cannot be absorbed into a normalization of the charges $N_{[\chi,\xi]}$ is said to be non-trivial. So we have proved the representation theorem :
\begin{resultat}[\textit{Charge representation theorem}]
Assuming integrability, the conserved charges associated to a Lie algebra of diffeomorphisms also form an algebra under the Poisson braket $\lbrace H_\chi , H_\xi \rbrace \triangleq \delta_\xi H_\chi$, which is isomorphic to the Lie algebra of diffeomorphisms up to a central extension
\[ \boxed{ \lbrace H_\chi , H_\xi \rbrace = H_{[\chi,\xi]} + \mathcal{K}_{\chi,\xi} [\bar{\Phi}] .} \]
\end{resultat}

It remains to prove the remaining equality \eqref{rel4}, which we provide here for the interested reader. For that purpose, we need some algebra on the variational bicomplex. We define the operator $\delta_Q = Q_I^i \frac{\partial}{\partial\Phi^i_I}+ \delta Q_I^i \frac{\partial}{\partial\delta\Phi^i_I}$ and we recall that $i_Q = Q_I^i \frac{\partial}{\partial \delta\Phi^i_I}$. One can check that 
\begin{eqnarray}
 [\delta_Q, d] &=& 0,\qquad [\delta_Q, \delta] = 0, \qquad \{ i_Q, I_\xi \} = 0, \\
 \{ i_Q, \delta \} &=& \delta_Q, \qquad [i_{Q_1},\delta_{Q_2}] = i_{[Q_1,Q_2]},\\
\mbox{} [ \delta_{Q_1},\delta_{Q_2}] &=& -\delta_{[Q_1,Q_2]},\qquad [Q_1,Q_2]^i \equiv  \delta_{Q_1} Q_2^i - \delta_{Q_2} Q_1^i . \label{rel5}
\end{eqnarray}

In particular for gravity, we are interested in the operator $\delta_{\mathcal L_\xi \Phi}$ that acts on tensor fields as a Lie derivative, $\delta_{\mathcal L_\xi \Phi}= +\mathcal L_\xi$ in our conventions. Note that the commutator in \eqref{rel5} is consistent with the standard commutator of Lie derivatives: 
\begin{eqnarray}
 [\delta_{\mathcal L_{\xi_1} \Phi} , \delta_{\mathcal L_{\xi_2} \Phi}]\Phi^i &=&  \delta_{\mathcal L_{\xi_1}\Phi } \mathcal L_{\xi_2} \Phi^i - (1\leftrightarrow 2) \\
 &=&   \mathcal L_{\xi_2} \delta_{\mathcal L_{\xi_1}\Phi} \Phi^i - (1\leftrightarrow 2) \\
  &=&   \mathcal L_{\xi_2} \mathcal L_{\xi_1} \Phi^i - (1\leftrightarrow 2) \\
    &=& -  \mathcal L_{[\xi_1,\xi_2]}  \Phi^i - (1\leftrightarrow 2) \\
       &=& - \delta_{ \mathcal L_{[\xi_1,\xi_2]}  \Phi} \Phi^ i - (1\leftrightarrow 2) .
\end{eqnarray}
With these tools in mind, let us start the proof. Applying the operator $I_\xi$ on the fundamental relation of Result 6, we obtain the definition of the surface charge form from the presymplectic form,
\begin{eqnarray}
I_\xi \bm\omega[ \delta_\xi\Phi,\delta\Phi;\Phi] & \approx & I_\xi  d\mathbf{k}_\xi [\delta \Phi;\Phi] \\
&\approx & \mathbf{k}_\xi [\delta \Phi;\Phi]  + d(\dots). \label{defk3}
\end{eqnarray}
Contracting with  $i_{\delta_\chi\Phi }$ we further obtain 
\begin{eqnarray}
\mathbf{k}_\xi [\delta_\chi \Phi;\Phi]   & \approx & I_\xi \bm\omega[ \delta_\chi\Phi,\delta_\xi\Phi;\Phi] + d(\dots). 
\end{eqnarray}
We would like to compute 
\begin{eqnarray}
\delta \mathbf{k}_\xi [\delta_\chi \Phi;\Phi] &\approx& \delta I_\xi \bm\omega[ \delta_\chi \Phi,\delta_\xi\Phi;\Phi]+ d(\dots)  \\
&\approx& -I_\xi \delta  \bm\omega[ \delta_\chi \Phi,\delta_\xi\Phi;\Phi]+ d(\dots),\\
&\approx&- I_\xi \delta i_{\delta_\xi \Phi} i_{\delta_\chi  \Phi}  \bm  \omega[ \delta \Phi,\delta \Phi;\Phi]+ d(\dots). 
\end{eqnarray}
We would like to use the fact that the presymplectic structure is $\delta$-exact, $\delta  \omega[ \delta \Phi,\delta \Phi;\Phi] =0$, so we will (anti-)commute the various operators as 
\begin{eqnarray}
\delta  i_{\delta_\xi \Phi} i_{\delta_\chi \Phi}  &=& -i_{\delta_\xi \Phi} \delta i_{\delta_\chi \Phi}-\delta_{\delta_\xi \Phi} i_{\delta_\chi \Phi}\\
&=&i_{\delta_\xi \Phi} i_{\delta_\chi \Phi}\delta + i_{\delta_\xi \Phi}\delta_{\delta_\chi \Phi}-\delta_{\delta_\xi \Phi} i_{\delta_\chi \Phi}\\
&=& i_{\delta_\xi \Phi} i_{\delta_\chi \Phi}\delta +\delta_{\delta_\chi \Phi} i_{\delta_\xi \Phi}+i_{[\delta_\xi \Phi,\delta_\chi \Phi]}-\delta_{\delta_\xi \Phi} i_{\delta_\chi \Phi}\\
&=& i_{\delta_\xi \Phi} i_{\delta_\chi \Phi}\delta +\delta_{\delta_\chi \Phi} i_{\delta_\xi \Phi}-i_{\delta_{[\xi,\chi]} \Phi }-\delta_{\delta_\xi \Phi} i_{\delta_\chi \Phi}.
\end{eqnarray}
The first term does not contribute as announced. The second and fourth term in fact combine to a contraction of the integrability condition \eqref{eq:Integrability} after using the definition \eqref{defk3} of the surface charge in terms of $\boldsymbol \omega$. We refer to \cite{Barnich:2007bf} for this piece of the proof. We are then left with 
\begin{eqnarray}
\delta \mathbf{k}_\xi [\delta_\chi \Phi;\Phi] &\approx&  I_\xi   \bm  \omega[ \delta_{[\xi,\chi]} \Phi,\delta \Phi;\Phi]+ d(\dots) \\
&\approx&  \mathbf{k}_{[\xi,\chi]} [\delta \Phi;\Phi]+ d(\dots) 
\end{eqnarray}
which proves \eqref{rel4}. \\

This closes our presentation of the covariant phase space formalism. What we have discussed is in fact one general method to derive canonical conserved charges in gauge theories. Let us now make some explicit calculations in General Relativity, to illustrate a bit all these formulae that we have just written...

\subsection{Conserved charge formula for General Relativity}
Let us consider the Hilbert-Einstein Lagrangian density $L = \frac{1}{16\pi G} \: \rmg \: R$. The only field $\Phi^i$ to take into account is the metric tensor $g_{\mu\nu}$, whose local variation will be denoted as $h_{\mu\nu} = \delta g_{\mu\nu}$ (convention: $\delta$ is Grassmann even). First, we need the expression of a general perturbation of $L$ : the calculation is straightforward and you should already performed it during your gravitation classes, in particular when you extracted the Einstein's equations from the variational principle, so it is left as an exercise :
\begin{align}
\delta L &= -\frac{\rmg}{16\pi G} G^{\mu\nu} h_{\mu\nu} + \partial_\mu\Theta^\mu [h;g] ;\\
\Theta^\mu [h;g] &= \frac{\rmg}{16\pi G} \left( \nabla_\nu h^{\mu\nu} - \nabla^\mu h^\nu_{\phantom{\nu}\nu} \right)
\end{align}
where $\nabla_\alpha$ is the Levi-Civita connection compatible with $g_{\mu\nu}$. If the variation is contracted with the action of a diffeomorphism $\xi^\mu$, we are able to explicit the presymplectic superpotential :
\begin{equation}
\Theta^\mu [\Lie_\xi g;g] = \frac{\rmg}{16\pi G} \left(2 \nabla_\nu \nabla^{(\mu} \xi^{\nu )} - 2 \nabla^\mu \nabla_\nu \xi^\nu \right).
\end{equation}
Recalling the definition of Riemann's curvature tensor, one gets easily that $\nabla^\mu \nabla_\nu \xi^\nu = \nabla_\nu \nabla^\mu \xi^\nu + R^{\nu\phantom{\alpha}\mu\phantom{\nu}}_{\phantom{\nu}\alpha\phantom{\mu}\nu} \xi^\alpha \approx \nabla_\nu \nabla^\mu \xi^\nu$ because the last term is proportional to the Ricci tensor which vanishes on shell for pure gravity without matter. So :
\begin{equation}
\Theta^\mu [\Lie_\xi g;g] \approx \frac{\rmg}{16\pi G} \nabla_\nu \left(  \nabla^\nu \xi^\mu - \nabla^\mu \xi^\nu \right).
\end{equation}
Knowing the symplectic prepotential gives us access to the Noether-Wald charge (\ref{eq:NoetherWaldCharge}) after some derivations :
\begin{equation}
\mathbf{Q}_\xi =- I_\xi \bm\Theta [\delta_\xi g;g] = \frac{\rmg}{16 \pi G} \left(  \nabla^\mu \xi^\nu - \nabla^\nu \xi^\mu \right) \: (d^{n-2} x)_{\mu\nu} = \frac{\rmg}{8 \pi G} \nabla^\mu \xi^\nu \: (d^{n-2} x)_{\mu\nu}
\label{eq:Komar}
\end{equation}
which is often called \textit{Komar's term}, in reference to the Komar's integrals that give the mass and angular momentum of simple spacetimes when (\ref{eq:Komar}) is evaluated on the asymptotic 2-sphere. The last ingredient we need is 
\begin{equation}
-i_\xi \bm\Theta = -\xi^\nu \frac{\partial}{\partial dx^\nu} \Theta^\mu (d^{n-1} x)_\mu = \left( \xi^\mu \Theta^\nu - \xi^\nu \Theta^\mu \right) \: (d^{n-2} x)_{\mu\nu}.
\end{equation}
The total surface charge is $\mathbf{k}_\xi [h;g] = -\delta \mathbf{Q}_\xi [g]- i_\xi \bm\Theta [h;g]$ where the last minus sign is valid in the Grassmann even convention for $h_{\mu\nu}$ since $i_\xi$ is Grassmann odd! Finally, after some tensorial algebra, we are left with :
\begin{equation}
{\color{blue} \boxed{ \mathbf{k}_\xi [h;g] = \frac{\rmg}{8\pi G} \: (d^{n-2} x)_{\mu\nu} \: \left( \xi^\mu \nabla_\sigma h^{\nu\sigma} - \xi^\mu \nabla^\nu h + \xi_\sigma \nabla^\nu h^{\mu\sigma} + \frac{1}{2} h \nabla^\nu \xi^\mu - h^{\rho \nu} \nabla_\rho \xi^\mu \right).\label{def:ch1}
}}
\end{equation}
One can explicitly prove that this object is conserved when $g_{\mu\nu}$ and $h_{\mu\nu}$ are on shell and $\xi^\mu$ is a Killing vector of $g_{\mu\nu}$:
\begin{equation}
d\mathbf{k}_\xi [h;g] \approx 0 \Longleftrightarrow \partial_\nu k^{[\mu\nu]}_\xi (d^{n-2}x)_{\mu\nu} \approx 0.
\end{equation}

Don't forget that it remains an ambiguity on this surface charge, which appears when we attempt to add a boundary term to the presymplectic form. If we impose that this term is only made up of covariant objects, the form of this term is highly constrained. Indeed, one can be convinced that the only boundary symplectic form constituted from $g_{\mu\nu}$ is :
\begin{equation}
\mathbf{E}[\delta g,\delta g ;g]= \frac{1}{16\pi G} (\delta g)^\mu_{\phantom{\mu}\sigma} \wedge (\delta g)^{\sigma \nu} (d^{n-2} x)_{\mu\nu}.
\label{eq:AmbiguityForRG}
\end{equation}
When the variations are generated by an infinitesimal diffeomorphism $\xi^\mu$, (\ref{eq:AmbiguityForRG}) results in 
\begin{equation}
\mathbf{E}[\delta_\xi g,\delta g ;g] = \frac{1}{16 \pi G}  \left( \nabla^\mu \xi_\sigma + \nabla_\sigma \xi^\mu \right) (\delta g)^{\sigma \nu} (d^{n-2} x)_{\mu\nu}.
\end{equation}
It is not surprising to obtain a contribution proportional to the Killing equation, since we have already shown that charges associated to exact symmetries do not suffer from any ambiguity ! The charge $\mathbf k_\xi[h;g] + \alpha \,\mathbf{E}[\delta_\xi g,h ;g]$ is the Iyer-Wald charge \cite{Iyer:1994ys} when $\alpha = 0$ and  the Abbott-Deser charge \cite{Abbott:1981ff} when $\alpha = 1$.  \\

Let us conclude this section by performing a concrete calculation on the most simple black hole metric: the Schwarzchild metric. In spherical coordinates $(t,r,\theta,\phi)$, we can describe the region outside the horizon by
\begin{equation}
g_{\mu\nu} [m] = -\left(1 - \frac{2m}{r}\right) dt^2 + \left(1 - \frac{2m}{r}\right)^{-1} dr^2 + r^2 d\Omega^2 \quad \text{with }\quad d\Omega^2 = d\theta^2+ \sin^2\theta d\phi.
\end{equation}
Only the mass parameter $m$ labels the family of metrics. Therefore, $h_{\mu\nu} \triangleq \delta g_{\mu\nu} [\delta m, m] = \frac{\partial g_{\mu\nu}}{\partial m} \delta m$. We find  $\delta g_{\mu\nu} dx^\mu dx^\nu = \frac{2\delta m}{r} dt^2 + \frac{2 \delta m}{r} \left(1 - \frac{2m}{r}\right)^{-2} dr^2$. Choosing a 2-surface $S$ on which both $t$ and $r$ are constant and fixing $\xi = \partial_t$, a direct evaluation of \eqref{def:ch1} with the natural orientation $\epsilon_{tr\theta\phi} = +1$ shows that :
\begin{equation}
\slash\hspace{-5pt}\delta H_\xi = \oint_S d\Omega \frac{\delta m}{4 \pi G} = \int_0^{2\pi} d\phi \: \int_0^\pi d\theta \: \sin \theta \frac{\delta m}{4 \pi G} = \frac{\delta m}{G} = \delta M.
\end{equation}
where $M = m/G$ is the total mass of spacetime. So the charge is trivially integrable and, after a simple path integration between the Minkowski metric ($m=0$) and a target metric with given $m> 0$, we get the right result according to which $M$ is the total energy of the Schwarzschild black hole !

\section{Conserved charges from the equations of motion}

In this section, we quickly discuss another way to define conserved charges through $(n-2)$-forms. This will lead to a particular prescription to fix the boundary ambiguity in the presymplectic form. This method is sensitively the same as the Iyer-Wald's one, and it also relies on the link between the symplectic structure of the space of fields and lower degree conserved currents. 

\subsection{Anderson's homotopy operator}

We first introduce a more formal procedure for performing integration by parts on expressions that do not necessarily involve $\xi^\mu$ but must involve the fields $\Phi^i$. It involves the fundamental operator, called \textit{Anderson's homotopy operator} $I^p_{\delta\Phi}$, which bears some ressemblance with the operator $I_\xi$ constructed and used above. Using the Grassmann odd convention for $\delta$ its constitutive relations are 
\begin{align}
- d I^n_{\delta\Phi} + \delta \Phi^i \frac{\delta}{\delta\Phi^i} &= \delta \quad \text{ when acting on }n\text{-forms ;} \label{eq:Anderson} \\
- d I^p_{\delta\Phi} + I^{p+1}_{\delta\Phi} d &= \delta \quad \text{ when acting on }p\text{-forms }(p<n). \label{eq:Anderson2}
\end{align}
As an exercise, the reader can convince him/her-self that the correct definition is
\begin{align}
I^n_{\delta\Phi} &= \left[ \delta \Phi^i \frac{\partial}{\partial \partial_\mu \Phi^i} - \delta \Phi^i \partial_\nu \frac{\partial}{\partial \partial_\mu \partial_\nu \Phi^i} + \partial_\nu  \delta \Phi^i \frac{\partial}{\partial \partial_\mu \partial_\nu \Phi^i} + \cdots \right] \frac{\partial}{\partial dx^\mu}, \\
I^{n-1}_{\delta\Phi} &= \left[ \frac{1}{2} \delta \Phi^i \frac{\partial}{\partial \partial_\mu \Phi^i} - \frac{1}{3} \delta \Phi^i \partial_\nu \frac{\partial}{\partial \partial_\mu \partial_\nu \Phi^i} + \frac{2}{3} \partial_\nu  \delta \Phi^i \frac{\partial}{\partial \partial_\mu \partial_\nu \Phi^i} + \cdots \right] \frac{\partial}{\partial dx^\mu}\label{Ander2}
\end{align}
where higher derivative terms are omitted. 

\subsection{Invariant presymplectic current}
Recalling that the Lagrangian density can be promoted to a $n$-form $\mathbf{L}$, we can use (\ref{eq:Anderson}) on it :
\begin{equation}
\delta \mathbf{L} = \delta \Phi^i \frac{\delta \mathbf{L}}{\delta \Phi^i} - d I^n_{\delta\Phi} \mathbf{L} \triangleq \delta \Phi^i \frac{\delta \mathbf{L}}{\delta \Phi^i} - d\bm\Theta[\delta \Phi ; \Phi].
\label{eq:deltaL2}
\end{equation}
So the definition $\bm\Theta = I^n_{\delta \Phi} \mathbf{L}$ fixes the boundary term ambiguity in $\bm\Theta$. Note the global sign in front of $\bm\Theta$, because $d$ and $\delta$ are both Grassmann-odd and anticommute. We always define the Iyer-Wald presymplectic current as $\bm\omega[\delta\Phi,\delta\Phi ; \Phi] = \delta\bm\Theta[\delta\Phi ; \Phi]$, and using (\ref{eq:Anderson2}), we can apply $I_{\delta\Phi}^n$ on both sides of (\ref{eq:deltaL2}) to get :
\begin{align}
I^n_{\delta\Phi} \delta \mathbf{L} &= I^n_{\delta\Phi} \left( \delta \Phi^i \frac{\delta \mathbf{L}}{\delta \Phi^i} \right) - I_{\delta\Phi}^n d I_{\delta\Phi}^n \mathbf{L} \\
&= I^n_{\delta\Phi} \left( \delta \Phi^i \frac{\delta \mathbf{L}}{\delta \Phi^i} \right) - \delta I_{\delta\Phi}^n \mathbf{L} - dI^{n-1}_{\delta\Phi}I^n_{\delta\Phi} \mathbf{L} \\
\Longrightarrow I^n_{\delta\Phi} \delta \mathbf{L} + \delta I_{\delta\Phi}^n \mathbf{L} &= I^n_{\delta\Phi} \left( \delta \Phi^i \frac{\delta \mathbf{L}}{\delta \Phi^i} \right) - dI^{n-1}_{\delta\Phi}I^n_{\delta\Phi} \mathbf{L}.
\end{align}
Since $[\delta,I^n_{\delta\Phi}] = 0$ because $\delta^2=0$, the left-hand-side is nothing but $2\: \delta I^n_{\delta\Phi} \mathbf{L} = 2\: \bm\omega[\delta\Phi,\delta\Phi ; \Phi]$, and so :
\begin{equation}
{\color{blue} \boxed{
\bm\omega[\delta\Phi,\delta\Phi ; \Phi] = \mathbf{W}[\delta\Phi,\delta\Phi ; \Phi] + d\mathbf{E}[\delta\Phi,\delta\Phi ; \Phi]
}}\label{ea:W4}
\end{equation}
where we have isolated and defined the \textit{invariant presymplectic current} :
\begin{equation}
\mathbf{W}[\delta\Phi,\delta\Phi ; \Phi] \triangleq \frac{1}{2} I^n_{\delta\Phi} \left( \delta \Phi^i \frac{\delta \mathbf{L}}{\delta \Phi^i} \right).\label{defW1}
\end{equation}
It differs from the Iyer-Wald one by a boundary term that reads as :
\begin{equation}
\mathbf{E}[\delta\Phi,\delta\Phi ; \Phi] \triangleq -\frac{1}{2} I^{n-1}_{\delta\Phi}I^n_{\delta\Phi} \mathbf{L}.
\label{eq:AmbiguityE_BarnichBrandt}
\end{equation}
This formulation allows us to choose $\mathbf{W}$ instead of $\bm\omega$ as symplectic form to build conserved surface charges. It is called invariant because it is defined in terms of the equations of motion and does not depend upon the boundary terms added to the action. \\

Let us consider again an infinitesimal diffeomorphism $\xi^\mu$. In order to compute $\mathbf{W}[\delta_\xi\Phi,\delta\Phi;\Phi] $ we need to contract either of the two $\delta \Phi$ on the right-hand side of \eqref{defW1}. There are therefore two terms. Now, it is a mathematical fact of the variational bicomplex that these two terms are equal, 
\begin{equation}
 I^n_{\Lie_\xi \Phi} \left( \frac{\delta \mathbf{L}}{\delta \Phi^i} \delta \Phi^i \right) = - I^n_{\delta\Phi} \left( \frac{\delta \mathbf{L}}{\delta \Phi^i} \Lie_\xi \Phi^i \right) 
\end{equation}
The proof is given in the Appendix of \cite{Barnich:2007bf} (denoted as Proposition 13). Therefore, 
\begin{equation}
\mathbf{W}[\Lie_\xi \Phi,\delta\Phi ; \Phi] \triangleq i_{\Lie_\xi \Phi} \mathbf{W} \approx - I^n_{\delta\Phi} \left( \frac{\delta \mathbf{L}}{\delta \Phi^i} \Lie_\xi \Phi^i \right) .\label{antiss}
\end{equation}
The trick to progress is to consider Noether's second theorem
\begin{equation}
d\mathbf{S}_\xi = \frac{\delta \mathbf{L}}{\delta \Phi^i} \Lie_\xi \Phi^i . 
\end{equation}
and apply Anderson's operator $I^n_{\delta\Phi}$ to both sides :
\begin{align}
I^n_{\delta\Phi} d\mathbf{S}_\xi &= I^n_{\delta\Phi} \left( \frac{\delta \mathbf{L}}{\delta \Phi^i} \Lie_\xi \Phi^i \right) \label{eq:FirstTerm} \\
&= \delta\mathbf{S}_\xi + d I^{n-1}_{\delta\Phi} \mathbf{S}_\xi.
\end{align}
If $\Phi^i$ is on shell and $\delta \Phi^i$ is also on shell in the linearized theory, the variation $\delta \mathbf{S}_\xi$ vanishes. Using \eqref{antiss} we are left with a familiar formula :
\begin{equation}
\mathbf{W}[\Lie_\xi \Phi, \delta\Phi ; \Phi] \approx d\mathbf{k}^{BB}_\xi [\delta\Phi;\Phi]
\end{equation}
where now the \emph{invariant surface charge form} or Barnich-Brandt charge form is 
\begin{equation}
{\color{blue}\boxed{
\mathbf{k}^{BB}_\xi [\delta\Phi;\Phi] = I^{n-1}_{\delta\Phi} \mathbf{S}_\xi \left[ \frac{\delta L}{\delta \Phi};\Phi \right].
}}\label{formBB}
\end{equation}
The surface charges are obtained by integration on a 2-surface and on a path in field space, as before.

\subsection{Expression of Barnich-Brandt's charge for Einstein's gravity}

The computation of the Barnich-Brandt's charge for General Relativity can be performed thanks to the formula \eqref{formBB}, and with the mere knowledge of $\mathbf{S}_\xi$ already derived, see (\ref{eq:Noether2ndThmEinstein}). But it is not necessary, since using \eqref{ea:W4} we have $\mathbf{k}^{BB}_\xi [\delta\Phi ;\Phi] = \mathbf{k}^{IW}_\xi [\delta\Phi ;\Phi] + \mathbf{E}[\delta_\xi\Phi,\delta \Phi ; \Phi]$. Therefore, the two formulations differ by this ambiguous term, which can be computed explicitly with (\ref{eq:AmbiguityE_BarnichBrandt}). In doing so we get exactly (\ref{eq:AmbiguityForRG}), so the Barnich-Brandt's local charge for Einstein's theory reads as follows
\begin{equation}
{\color{blue} \boxed{ \mathbf{k}^{\mu\nu}_\xi [h;g] = \frac{\rmg}{8\pi G} \left( \xi^\mu \nabla_\sigma h^{\nu\sigma} - \xi^\mu \nabla^\nu h + \xi_\sigma \nabla^\nu h^{\mu\sigma} + \frac{1}{2} h \nabla^\nu \xi^\mu -\frac{1}{2}  h^{\rho \nu} \nabla_\rho \xi^\mu + \frac{1}{2} h^{\nu}_{\phantom{\nu}\sigma} \nabla^{\mu} \xi^{\sigma } \right).\label{def:ch2}
}}
\end{equation}
This formula was also obtained by Abbott and Deser by a similar procedure involving integrations by parts, without using  formal operators \cite{Abbott:1981ff,Deser:2002rt,Deser:2002jk}. \\

When we will show that 3-dimensional Einstein's gravity can be reduced to a couple of Chern-Simons theories, we will use this formulation of conserved charges (instead of the Iyer-Wald one) to compute the charges in that alternative formalism, simply because it is faster.

\section*{References}
\addcontentsline{toc}{section}{References}

Many textbooks on \textit{QFT}s explain {\color{blue} Noether's theorem} in detail, as \textit{e.g.} the book of di Franscesco et al. \cite{DiFrancesco:1997nk}. \\

The {\color{blue} covariant phase space formalism} was developed in \cite{Lee:1990nz,Iyer:1994ys}. Several introductions to the formalism can be found in research articles including \textit{e.g.} Section 3 of \cite{Compere:2009dp} and Appendix A of \cite{Compere:2015bca}. For a list of references on the definition of the symplectic structure for Einstein gravity, see \textit{e.g.} Section 4.4 of \cite{Compere:2011ve}. \\

The {\color{blue} cohomological formalism} for defining the surface charges (and in particular the proof of uniqueness or ``Generalized Noether theorem") was developed in \cite{Barnich:2000zw,Barnich:2001jy,Barnich:2007bf}. For additional details on the variational bicomplex, see \textit{e.g.} Appendix C of \cite{Azeyanagi:2009wf} or Section I.1. of \cite{Compere:2007az}. \\

The {\color{blue} Hamiltonian formalism}, which also leads to a complete theory of conserved charges and which is equivalent to the covariant phase space formalism, was developed in \cite{Arnowitt:1959ah,Regge:1974zd,Brown:1986ed}. Due to a lack of time, it was not covered in these lectures. \\

The proof of the {\color{blue} representation theorem} for the algebra of (integrable) charges can be found in the case of the Hamiltonian formalism in \cite{Brown:1986ed} and for the covariant formalism in \cite{Barnich:2007bf}. \\

The {\color{blue} definition of surface charges using the equations of motion} was developed for Einstein gravity and higher curvature theories in \cite{Abbott:1981ff,Deser:2002rt,Deser:2002jk}. Only the Abbott-Deser formula for Einstein gravity was covered here. These definitions are equivalent to the cohomological formalism. \\

If you need to use surface charges for various theories of second order in derivatives (Einstein, Maxwell, Chern-Simons, scalars), the {\color{blue} explicit formulae for the surface charges} with all signs and factors right up to my knowledge can be found in Section 4.4. of \cite{Compere:2009dp}. For those interested, a {\color{blue} Mathematica package} is also available to compute surface charges in several theories with tutorial on my \href{http://www.ulb.ac.be/sciences/ptm/pmif/gcompere/package.html}{\color{ForestGreen} homepage}.

	% Lecture 2 : Gravité à 3D et AdS3
\chapter{Three dimensional Einstein's gravity}
\label{sec:3DGravity}

General Relativity is the commonly accepted modern paradigm of gravitation, and is supported nowadays by compelling experimental evidence, which consolidates its position as the unavoidable model of gravitation. Yet, it is a very complex theory presenting a lot of puzzles to physicists, both at the classical and quantum level. Classically, one has to deal with laborious analytic or numerical calculations and some conceptual issues that still remain unresolved. Quantum mechanically, Einstein's theory is not well-behaved and after several years of intense searches, quantum gravity is still an elusive theory without direct experimental prospect.  \\

Now, it exists a reduced version of Einstein's general relativity where the physical objects are more under control and where its quantization, if it is consistent, is within our reach : \textbf{\color{blue} 3-dimensional Einstein's gravity} to which this lecture is dedicated. As a toy-model, it is a very useful framework thanks to which we can experiment some techniques and derive features, some of which extend to the physical $4d$ case. As one removes a spatial dimension, the solution space is reduced as well, in fact to constant curvature solutions in the absence of matter. But by considering topological and asymptotic properties, Einstein's solutions still contain black holes and infinite-dimensional asymptotic symmetries and so save the theory from its apparent triviality. \\

Anti-de Sitter spacetime, one of the three constant curvature spacetimes, was studied in the early 80's. The seminal paper of Brown and Henneaux in 1986 \cite{Brown:1986nw} gave a definition of asymptotically $AdS_3$ spacetimes together with an analysis of its infinite dimensional symmetries. The paper was ignored for long, until the discovery of the $AdS/CFT$ correspondence by Maldacena in 1997 \cite{Maldacena:1997re,Aharony:1999ti}. Notably in 1988, Witten attempted a quantization of $3d$ gravity \cite{Witten:1988hc} by using its equivalent Chern-Simons representation, constructed by Achúcarro and Townsend in 1986 \cite{Achucarro:1987vz}. Most of the community considered the theory as trivial, until a black hole solution, the $BTZ$ black hole, was derived in 1992 by Bañados, Teitelboim and Zanelli \cite{Banados:1992wn,Banados:1992gq}, which shares some features with the $4d$ Kerr solution. In string theory, many higher dimensional supersymmetric black holes contain the $BTZ$ black hole in their near-horizon geometry, and $3d$ Einstein gravity becomes a universal tool to understand black hole entropy, as emphasized by Strominger in 1997 \cite{Strominger:1997eq}. The black hole entropy of any $BTZ$ black hole can be ``holographically'' computed using Cardy's formula of a dual putative $CFT_2$ thanks to an extension of Maldacena's  $AdS/CFT$ correspondence.  Today, it is not yet clear whether pure $3d$ Einstein gravity makes sense quantum mechanically without a string theory embedding. Several attempts have been made but are not conclusive \cite{Witten:2007kt,Maloney:2007ud}. \\

In the last years, the holography community has paid a particular attention to $4d$ asymptotically flat spacetimes where another infinite dimensional symmetry group, the $BMS$ group, plays a particular role. There is also a $3d$ analogue of this $BMS$ group, and again it is useful to understand the toy model in parallel to the physical $4d$ case. \\

Through this course, we will review the typical properties of $3d$ gravity, which are mostly due to the vanishing of the Weyl curvature. Next we will turn to the $AdS_3$ phase space: we will describe global features of $AdS_3$ itself, and then give several elements on the Brown-Henneaux boundary conditions and the resulting asymptotic group, without forgetting a long discussion on $BTZ$ black holes. We will show that the asymptotically flat phase space can be obtained from the flat limit of the $AdS_3$ phase space. Finally, we will shortly present the Chern-Simons formulation of $3d$ gravity, which reduces the theory to the one of two non-abelian gauge vector fields!

\section{Overview of typical properties}

\subsection{A theory without bulk degrees of freedom}

Beyond the fact that the analytic complexity of General Relativity is reduced in $2+1$ dimensions, it occurs a particular phenomenon that heavily constraints the gravitational field. To see how $3d$ gravity is special, let us consider a $n$-dimensional spacetime manifold $(M,g)$. On $M$, the intrinsic curvature is encoded in the Riemann tensor $R_{\mu\nu\alpha\beta}$, which natively possesses $n^4$ components. This number can be reduced to $\frac{n^2}{12} (n^2-1)$ independent components, after taking into account several well-known indicial symmetries :
\begin{itemize}[label=$\rhd$]
\item Antisymmetry on the pair $(\mu,\nu)$ $\rightarrow$ $N = n(n-1)/2$ independent choices of $(\mu,\nu)$ ;
\item Antisymmetry on the pair $(\alpha,\beta)$ $\rightarrow$ $N = n(n-1)/2$ independent choices of $(\alpha,\beta)$ ; 
\item Symmetry under permutation of the pairs $\rightarrow$ reduces the number of independent choices of pairs from $N^2$ to $N(N+1)/2$ ;
\item Identity $R_{\mu (\nu\alpha\beta )}=0$. Any repeated index would lead to zero after using the above three properties. Indeed, for a repeated index in the last three symmetrized indices,  $R_{1(233)} = R_{1332}+R_{1323}= 0$, and for repeated mixed indices, $R_{1(123)}=R_{1231}+R_{1312}=R_{1231}+R_{1213} = 0$. Each identity with distinct indices (up to reschuffling the indices) brings new constraints. The number of new constraints brought by this last identity is therefore equal to $C=n(n-1)(n-2)(n-3)/4!$. We see that if $n<4$, it brings no additional restriction.
\end{itemize}
Finally we are left with the correct number of independent components : $F(n) = N(N+1)/2- C = \frac{n^2}{12} (n^2-1)$. Now we recall that the Riemann tensor can be written with an explicit decomposition between the trace-part (the \textit{Ricci tensor} $R_{\mu\nu} = R^{\alpha}_{\phantom{\alpha}\mu\alpha\nu}$) and the traceless conformally invariant part (the \textit{Weyl tensor} $W_{\mu\nu\alpha\beta}$, sometimes also denoted as $C_{\mu\nu\alpha\beta}$) :
\begin{align}
R_{\mu\nu\alpha\beta} = W_{\mu\nu\alpha\beta} + \frac{2}{n-2} \left( g_{\alpha[\mu} R_{\nu] \beta} + R_{\alpha[\mu} g_{\nu] \beta}\right) - \frac{2}{(n-1)(n-2)} R g_{\alpha [\mu} g_{\nu ] \beta}. 
\label{eq:RiemannGeneral}
\end{align}
According to Einstein's equations $G_{\mu\nu} = 8\pi G T_{\mu\nu}$ the source of curvature that governs the Ricci tensor is the local distribution of matter energy-momentum and stresses. Outside the sources, the Ricci curvature vanishes ($R_{\mu\nu} = 0$). Yet, the Weyl curvature might not vanish. It carries the gravitational information of the sources on the local observer (gravitational waves, Newtonian potential and additional Einsteinian potentials). 

But now in $3d$, a simplification occurs, because the number of independent components in the Riemann tensor $\frac{n^2}{12} (n^2-1) = \frac{9}{12}\times (9-1) = 6$ perfectly matches with the number of independent components of the Ricci tensor, which is symmetrical, and thus owns $\frac{1}{2}n(n+1) = \frac{3}{2}\times 4 = 6$ independent components ! So the Weyl tensor is \textit{identically zero}. That implies that \textit{there is no gravitational degree of freedom} in $3d$ Einstein's gravity because gravitational information cannot propagate. Also, there is no Newtonian potential, and so \textit{the masses do not attract} ! This astounding fact allows to extend the mass spectrum of the theory to negative masses (as long as the spectrum is bounded from below for stability reasons): indeed negative masses cannot give rise to repulsion since two masses do not ``feel'' each other in $3d$ gravity !

\subsection{Einstein-Hilbert action and homogeneous spacetimes}
Until now we didn't use the equations of motion. Now let us write the Einstein-Hilbert action, and obtain another characterization of the local triviality of $3d$ gravity :
\begin{equation}
S[g] = \frac{1}{16\pi G} \int_M d^3 x \: \rmg (R+2\Lambda )
\end{equation}
where $R$ is the Ricci scalar curvature, and $\Lambda$ the cosmological constant. In the case of pure gravity, the equations of motion are $G_{\mu\nu} +\Lambda g_{\mu\nu} = R_{\mu\nu} +(\Lambda-\frac{R}{2}) g_{\mu\nu} =0$. Taking the trace we find that $R = 6\Lambda$ (since $g^{\mu\nu}g_{\mu\nu} = n = 3$). On shell, the Ricci tensor is completely determined by the metric tensor $R_{\mu\nu} \approx 2\Lambda g_{\mu\nu}$. Inserting this expression into (\ref{eq:RiemannGeneral}) after deleting the Weyl tensor, one gets :
\begin{equation}
R_{\alpha\beta\mu\nu} \approx \Lambda \left( g_{\alpha\mu}g_{\beta\nu} - g_{\alpha\nu}g_{\beta\nu} \right). 
\end{equation}
This is the curvature tensor of an \textit{homogeneous} spacetime or constant curvature spacetime. The solutions are thus distinguished by the sign of the cosmological constant $\Lambda$ :
\begin{itemize}[label=$\rhd$]
\item If $\Lambda > 0$, the solutions of Einstein's field equations are locally the \textit{de Sitter} space ;
\item If $\Lambda = 0$, the solutions of Einstein's field equations are locally Minkowski (flat) ;
\item If $\Lambda < 0$, the solutions of Einstein's field equations are locally the \textit{anti-de Sitter} space.
\end{itemize}
Remark that the spacetime being locally homogeneous everywhere is consistent with the non-interaction between masses. Indeed, if we consider a set of test masses, the region that separates them is locally an homogeneous spacetime where no preferential direction can be chosen, and in particular there is no direction of attraction. 
The solution space of $3d$ gravity is however not trivial! There are particles (local defects), black holes (due to topological properties) and asymptotic symmetries, as discussed later on!

\newpage
\section{Asymptotically anti-de Sitter phase space}

\subsection{Global properties of \textit{AdS}\textsubscript{3}}

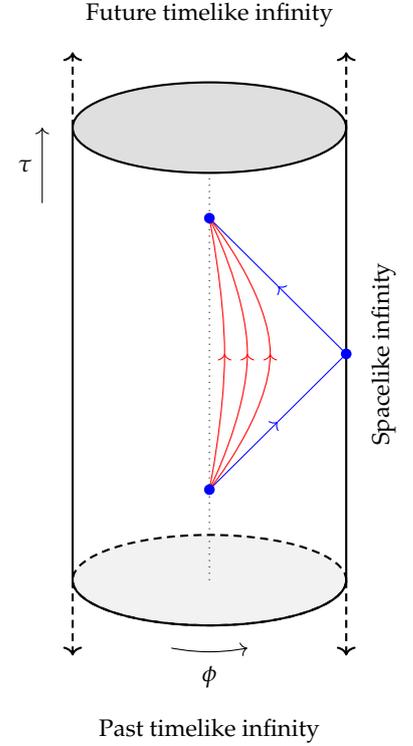
\begin{wrapfigure}{r}{0.3\textwidth}
\begin{tikzpicture}[scale=1]
% Carré externe
%\draw[blue] (-3.5,-3.5) -- (-3.5,3.5) -- (3.5,3.5) -- (3.5,-3.5) -- cycle;
\coordinate (top) at (0,3);
\coordinate (bottom) at (0,-3);
\coordinate (centre) at (0,0);
\draw[densely dashed,thick,fill=gray!10] (bottom) ellipse (1.8 and 0.6);
\draw[dotted] (bottom) -- (top);
\draw[thick,fill=gray!25] (top) ellipse (1.8 and 0.6);
\draw[thick] ( $(bottom)+(-1.8,0)$ ) -- ( $(top)+(-1.8,0)$ );
\draw[thick] ( $(bottom)+( 1.8,0)$ ) -- ( $(top)+( 1.8,0)$ );
\draw[thick] (bottom)+(1.8,0) arc (0:-180:1.8 and 0.6);
\path (bottom) + (0,1.2) coordinate (A);
\draw[blue,->] (A) -- ( $(A)+(0.9,0.9)$ );
\draw[blue] ( $(A)+(0.9,0.9)$ ) -- ( $(A)+(1.8,1.8)$ );
\node[below,rotate=90] at ( $(A)+(2.0,1.8)$ ) {{\footnotesize \color{black} Spacelike infinity}};
\draw[blue,->] ( $(A)+(1.8,1.8)$ ) -- ( $(A)+(0.9,2.7)$ );
\draw[blue] ( $(A)+(0.9,2.7)$ ) -- ( $(A)+(0,3.6)$ );
\draw[red] plot [smooth,tension=0.8] coordinates { (A) ( $(A)+(0.8,1.8)$ ) ( $(A)+(0,3.6)$ )};
\draw[red,->] ( $(A)+(0.8,1.8)$ );
\draw[red] plot [smooth,tension=0.8] coordinates { (A) ( $(A)+(0.5,1.8)$ ) ( $(A)+(0,3.6)$ )};
\draw[red,->] ( $(A)+(0.5,1.8)$ );
\draw[red] plot [smooth,tension=0.8] coordinates { (A) ( $(A)+(0.2,1.8)$ ) ( $(A)+(0,3.6)$ )};
\draw[red,->] ( $(A)+(0.2,1.8)$ );
\fill[blue] (A) circle [radius=2pt];
\fill[blue] (A)+(1.8,1.8) circle [radius=2pt];
\fill[blue] ( $(A)+(0,3.6)$ ) circle [radius=2pt];
\draw[thick,densely dashed,->] ( $(bottom)+(-1.8,0)$ ) -- ( $(bottom)+(-1.8,-1)$ );
\draw[thick,densely dashed,->] ( $(bottom)+( 1.8,0)$ ) -- ( $(bottom)+( 1.8,-1)$ );
\draw[thick,densely dashed,->] ( $(top)+(-1.8,0)$ ) -- ( $(top)+(-1.8, 1)$ );
\draw[thick,densely dashed,->] ( $(top)+( 1.8,0)$ ) -- ( $(top)+( 1.8, 1)$ );
\coordinate (legend) at ( $(bottom)+(0,-1)$ );
\node[below] at (legend) {\footnotesize $\phi$};
\draw[->] plot [smooth,tension=1] coordinates {($(legend)+(-0.5,0.1)$) ($(legend)+(0,0.05)$) ($(legend)+(0.5,0.1)$)};
\draw[<-] ($(top)+(-2.2,0)$) -- ($(top)+(-2.2,-1)$);
\node[left] at ($(top)+(-2.2,-0.5)$) {\footnotesize $\tau$};
\node[] at ($(top)+(0,1.5)$){\footnotesize Future timelike infinity};
\node[] at ($(bottom)+(0,-2.0)$){\footnotesize Past timelike infinity};		
\end{tikzpicture}
\caption{Penrose diagram of global $AdS_3$ spacetime.}
\label{fig:PenroseAdS3}
\end{wrapfigure}

Let us focus on negative curvature solutions. We introduce a length scale $\ell\in \mathbb{R}_0^+$ such as $\Lambda = -1/\ell^2$. Since we are first interested in the global homogeneous solution, we provide ourselves with \textit{global coordinates} $(t,r,\phi)$, in the sense that any complete geodesic can be maximally extended in this coordinate system. $t$ is a timelike coordinate, while $r,\phi$ are spacelike coordinates, $r \in \mathbb{R}^+$ is the luminosity distance and $\phi \in [0,2\pi ]$ is the angular coordinate. The maximally symmetric solution of Einstein's equation with negative curvature reads as :
\begin{equation}
ds^2 =  - \left( 1+ \frac{r^2}{\ell^2} \right) dt^2 +\left( 1+ \frac{r^2}{\ell^2} \right)^{-1} dr^2 + r^2 d\phi^2 .
\label{eq:AdS3Metric}
\end{equation}
This spacetime is called the \textit{anti-de Sitter spacetime} ($AdS_3$). The length scale $\ell$ determines the distance at which the curvature sets in. At shorter distances ($r\ll \ell$), the metric (\ref{eq:AdS3Metric}) is simply Minkowski. Near infinity ($r\gg \ell$), we get $ds^2 \sim \frac{\ell^2}{r^2} dr^2 + \frac{r^2}{\ell^2} (-dt^2 + \ell^2 d\phi^2 )$ so the asymptotic boundary is a cylinder, with the $\phi$ circle as a base and extending along the timelike coordinate $t$. Given that there is a potential barrier to reach infinity, one may think about $AdS_3$ as a ``spacetime in a box'', and its Penrose diagram can be easily obtained by compactifying the luminosity distance to bring the asymptotic cylinder to a finite distance. Before to do that, let us introduce a new coordinate system in which the luminosity distance $r$ is remplaced by another (dimensionless) radial coordinate $\rho$ as $r \triangleq \ell \sinh \rho$. The line element takes the form
\begin{equation}
ds^2 = \ell^2  \left( d\rho^2 - \cosh^2 \rho \: d\tau^2 + \sinh^2 \rho \: d\phi^2 \right)
\end{equation}
where we have rescaled $\tau = t/\ell$ for later convenience. We wish to compactify the radial direction given by $\rho$. So let us define a "conformal radial coordinate" $q$ such as $dq = d\rho / \cosh \rho$, or $\cosh \rho = 1/\cos q$. Since $\rho \geq 0$, $\cosh \rho \in [1,\infty [$ and it is mapped to $q \in [0,\pi/2 [$. In the patch $(q,\tau,\phi)$ the metric is brought into the form
\begin{equation}
ds^2 = \frac{\ell^2}{\cos^2 q} (dq^2 - d\tau^2 + \sin^2 q \: d\phi^2) = \frac{\ell^2}{\cos^2 q} \left(-d\tau^2 + d\Omega_{D}^2 \right).
\end{equation}
Surfaces of constant $\tau$ are half-spheres or disks (since the colatitude $q$ must be lower than $\pi/2$) with boundary at $q = \pi/2$ ($\rho = \infty$). At this boundary, the conformal metric is simply $(-d\tau^2 + d\phi^2)$ as expected : one can thus actually visualize $AdS_3$ as an infinite cylinder as represented by the figure \ref{fig:PenroseAdS3}. But note that since the range of the timelike coordinate $\tau$ remains infinite while that of $q$ is finite, there is no way to compress $AdS_3$ into a finite range of coordinates for both $\tau$ and $q$ if we want to preserve the condition that lightrays
are diagonal. If we attempt to perform another conformal transformation to reduce $\tau$ in a finite range, the circles generating the conformal boundary will be mapped to points, since the interval of the conformal radius $q$ will be squeezed to a single point. So we admit that the conformal diagram is that we have drawn : it is an infinite cyclinder, whose upper and lower boundary, rejected to infinity, are respectively future and past timelike infinities. \\

We have still more to learn from the metric written in the new global coordinates $(\rho,\tau,\phi)$. First we clearly remark that $AdS_3$ has the topology of $\mathbb{R}^3$. It is also clear that $AdS_3$ is \textit{static}, because $\partial_\tau$ is a trivial Killing vector of $g_{\mu\nu}$ (since no component depends on $\tau$) that is also orthogonal to the constant time slices ($g_{\tau\, a} = 0$). Since the staticity is manifest in this coordinate system, it bears the name of \textit{global static coordinates}. We can calculate the trajectories of particles in $AdS_3$, which is also more easy in this system. Let us focus only on radial null geodesics $x^\mu (\lambda )$ which have to verify
\begin{equation}
ds^2 = 0 \Rightarrow d\rho^2 = \cosh^2 \rho \: d\tau^2 \Rightarrow \left(\frac{d\rho}{d\lambda}\right)^2 = \cosh^2 \rho \left(\frac{d\tau}{d\lambda}\right)^2.
\end{equation}
To integrate this equation on $\lambda$, we can use a dynamical invariant which is the energy of the null ray:
\begin{equation}
E \triangleq - \partial_\tau \cdot u \triangleq - g_{\mu\nu} (\partial_\tau )^\mu \frac{dx^\nu}{d\lambda} = - g_{\tau\tau} \frac{d\tau}{d\lambda} = \cosh^2 \rho \: \frac{d\tau}{d\lambda}.
\end{equation}
So we have to integrate now 
\begin{equation}
\left(\frac{d\rho}{d\lambda}\right)^2 = \frac{E^2}{\cosh^2 \rho} \Longrightarrow \frac{d}{d\lambda} \sinh \rho = {\color{blue} +} E \Longleftrightarrow \sinh \rho(\lambda) = E (\lambda-\lambda_0)
\end{equation}
for outgoing lightrays. We find thus that spatial infinity $\rho \rightarrow \infty$ is reached when the affine parameter $\lambda \rightarrow \infty$. But for the coordinate time $\tau$, we directly integrate 
\begin{equation}
\frac{d\tau}{d\lambda} = \frac{E}{\cosh^2 \rho(\lambda)} = \frac{E}{1 + E^2 (\lambda-\lambda_0)^2} \Longrightarrow \tau = \arctan \left[ E (\lambda-\lambda_0) \right]
\end{equation} 
and thus when $\lambda \rightarrow \infty$, $\tau = \pi/2$, so null geodesics \textit{reach infinity after a finite coordinate time interval} (see the blue curve on figure \ref{fig:PenroseAdS3}) ! $AdS_3$ spacetime is said not to be globally hyperbolic: one needs boundary conditions at spatial infinity to arbitrarily extend the dynamics in coordinate time in the bulk of spacetime. Equivalently, Cauchy surfaces do not exist. \\

Concerning timelike geodesics, it can be shown thanks to a cautious examination of the $AdS_3$ geodesic equation that they cannot reach the boundary, and are in fact \textit{reflected} at large distances. The Christoffel symbols contain a potential proportional to $r^2$ that blows up at infinity and confines timelike objects to the center of spacetime. Some timelike geodesics are drawn in red on the conformal diagram in Figure \ref{fig:PenroseAdS3}.  \\

To study the symmetries of $AdS_3$, the most convenient way is to first realize that one can immerse it into the $4$-dimensional manifold $\mathbb{R}^{(2,2)}$, the space $\mathbb{R}^4$ provided with the pseudo-riemannian metric $\zeta_{ab} = \text{diag}(-1,-1,+1,+1)$. On this manifold, we denote by $X_0$ and $\tilde X_0$ the timelike coordinates, and ($X_1, X_2$) the spacelike ones. $AdS_3$ can be viewed as an hyperboloid isometrically immersed into $\mathbb{R}^{(2,2)}$:
\begin{equation}
AdS_3 \equiv \mathcal{H} \triangleq \lbrace X^\mu \in \mathbb{R}^{(2,2)} \: | \: -X_0^2  -\tilde X_0^2 + X_1^2 + X_2^2 = \zeta_{ab} X^a X^b = -\ell^2 \rbrace .
\end{equation}
We immediately see that a natural parametrization of this hypersurface is the following\footnote{Note that the application defined by the system above, and which sends $AdS_3$ into $\mathbb{R}^{(2,2)}$, is not injective, because it contains $2\pi$-periodic functions of $\tau$. So it wraps $AdS_3$ around $\mathcal{H}$ an infinity of times. But locally the injectivity is however ensured, so we  talk about \textit{immersion} rather than \textit{embedding}.}
 :
\begin{equation}
X^\mu \in \mathcal{H} \Longleftrightarrow
\left\lbrace
\begin{array}{ccc}
X_0  &=& \ell \cosh \rho \cos \tau \\
\tilde X_0 &=& \ell \cosh \rho \sin \tau \\
X_1  &=& \ell \sinh \rho \cos \phi \\
X_2  &=& \ell \sinh \rho \sin \phi
\end{array}\right. \quad (\rho\in  \mathbb{R}^+ ,\;\; \tau,\phi \in [0,2\pi[).
\end{equation}
The application is isometric because the pull-pack of the ambient metric $dS^2_{(2,2)} = \zeta_{ab} dX^a dX^b$ on $\mathcal{H}$ exactly reproduces the metric of $AdS_3$ in global coordinates :
\begin{equation}
\onH{dS^2_{(2,2)}} = \ell^2  \left( d\rho^2 - \cosh^2 \rho \: d\tau^2 + \sinh^2 \rho \: d\phi^2 \right).
\end{equation}
We can now easily analyze the isometries of $AdS_3$. The host space $\mathbb{R}^{(2,2)}$ is 4-dimensional flat space with 2 timelike directions. It possesses the maximal number of symmetries allowed in $4d$, namely $10$. The isometries consist of the 4 translations along each direction (because the metric does not depend on the coordinates $X^a$) and the $6$ matricial transformations that preserve the metric $\zeta$ : $M^{-1} \zeta M = \zeta \Rightarrow M \in SO(2,2)$ by definition. All these symmetries cannot survive on $\mathcal{H}$ : in particular, the hypersurface $\mathcal{H}$ is clearly not invariant under translations $X^a \rightarrow X^a + A^a$, but acting with $SO(2,2)$ still preserves the condition $\zeta_{ab} X^a X^b = -\ell^2$, so we have obtained 6 exact symmetries of $AdS_3$. Since it cannot have more than 6 global isometries, we have proven that $AdS_3$ admits $SO(2,2)$ as Killing isometry group. The generators of $so(2,2)$ are given by :
\begin{equation}
J_{ab} = X_b \frac{\partial}{\partial X^a} - X_a \frac{\partial}{\partial X^b}
\end{equation}
where $X_a = \zeta_{ab}X^b$. After expanding in global coordinates, one can see that $J_{01} = \partial_\tau$ generates time evolution on $\mathcal{H}$, whereas $J_{12} = \partial_\phi$ generates rotations. The most general Killing vector is naturally given by $\frac{1}{2}\omega^{ab}J_{ab}$, and is thus determined by an antisymmetric $\omega^{ab} = \omega^{[ab]}$ tensor in $4d$.

\subsection{Asymptotically \textit{AdS}\textsubscript{3} black holes}

We now set aside the global homogeneous case to consider more evolved geometries, but which asymptote nevertheless to $AdS_3$ when approaching spatial infinity in a sense that we will make precise below. Such spacetimes are called \textit{asymptotically} $AdS_3$. The space of such spacetimes will be quite rich, in particular because of the presence of black hole solutions. There is no contradiction with our earlier derivation that all solutions to Einstein's equations are locally $AdS_3$. A $3d$ black hole does not possess any curvature singularity. It is a black hole because it admits an event horizon, which turns out to protect a ``causal singularity''. We will first present the solution and discuss its properties, and afterwards  we will see how to obtain such a solution starting from $AdS_3$ itself !

\subsubsection{\textit{BTZ} black holes and more\dots}

Let us consider a set of Boyer-Lindquist-like coordinates $(t,r,\phi)$, where $t$ is an asymptotically timelike coordinate, $r$ is an asymptotically radial coordinate, and $\phi$ is a polar angle identified as $\phi \sim \phi + 2\pi$. Imposing that the spacetime behaves like $AdS_3$ when $r \rightarrow \infty$, and that the spacetime is stationary and axisymmetric, a natural ansatz to consider is 
\begin{equation}
ds^2 = -N^2 (r) dt^2 + \frac{dr^2}{N^2(r)} + r^2 \left( d\phi + N^\phi (r) dt \right)^2 . 
\label{eq:BTZ}
\end{equation}
This solution for $N(r)$ and $N^\phi(r)$ was found for first time in 1992 by Bañados, Teitelboim, and Zanelli, and describes in some range of the parameters, as we are about to show, the so-called \textit{BTZ black hole}. The boundary conditions are fixed such as 
\begin{align}
N^2 (r) &= -8MG + \frac{r^2}{\ell^2} + \frac{16 G^2 J^2}{r^2} &&(\text{Lapse function}) ;\\
N^\phi (r) &= -\frac{4GJ}{r^2} &&(\text{Angular dragging}).
\end{align}
We can explicitly verify that the \textit{BTZ} solution has 2 Killing vectors $\partial_t$ and $\partial_\phi$ since the metric coefficients depend only on $r$. But the solution is not static because the term $dtd\phi$ is not invariant under time reversal. The quantities $M$ and $J$ which naturally appears in the metric are respectively the surface charges associated to $\partial_t$ and $\partial_\phi$ evaluated on the circle at infinity $S = \lbrace x^\mu \in M \: | \: t = \text{Cst}, r = \text{Cst} \rightarrow \infty \rbrace$ (or actually any circle homotopic to it) :
\begin{align}
\oint_{S} \mathbf{k}_{\partial_t} [\delta g ;g] = \delta M &\quad\rightarrow\quad M = \text{total mass of the black hole} ;\\
\oint_{S} \mathbf{k}_{-\partial_\phi} [\delta g ;g] = \delta J &\quad\rightarrow\quad J = \text{total angular momentum of the black hole}.
\end{align}
The minus sign in the definition of $J$ is conventional. We have a 2-parameter family of solutions,  with the scale length $\ell$ being fixed. In fact, it is a black hole for special values of the parameters, but it is also more than that as we will now show. 

\subsubsection{Main properties}

Global $AdS_3$ spacetime is recovered if $N^\phi = 0 \Leftrightarrow J = 0$ and $N = 0 \Leftrightarrow M = -1/8G$. In that case, $r=0$ is the origin of polar coordinates and there is no singularity there. \\

More generally, we can check that the solution is asymptotically $AdS_3$ simply by taking the limit $r\rightarrow \infty$ in (\ref{eq:BTZ}). One can compute the curvature scalar $R$ and show that it is equal to $-6/\ell^2$ everywhere. So again, the solution does not contain any curvature singularity. Starting from infinity and going inwards, the first particular surface we encounter is the \textit{limit of staticity} $g_{tt} (r_{erg}) = -N^2 (r_{erg}) + r^2 [N^{\phi}(r_{erg})]^2 = 0$ below which $\partial_t$ has positive norm, and so a static observer with a 4-velocity colinear to $\partial_t$ cannot exist since $t$ is no more a timelike coordinate. The critical "radius" is given by $r_{erg} = \ell \sqrt{8GM}$, and this surface, called \textit{ergocircle} in analogy to the Kerr metric, exists if and only if $M > 0$. Another set of critical values of $r$ are the roots of the lapse function $N^2(r) = 0$. The latter equation is quadratic equation in $r^2$, so we will find two roots in terms of $r^2$. We are interested in positive $r$ so we choose the positive roots, which are 
\begin{equation}
r_\pm = \ell \sqrt{4GM} \sqrt{1 \pm \sqrt{1 - \left(\frac{J}{M\ell}\right)^2}}.
\end{equation} 
Let us denote by $\mathcal{H}_\pm$ the surfaces $\lbrace x^\mu \: | \: r = r_\pm \rbrace$. They exist if and only if
\begin{equation}
|J| \leq M\ell \,\,\,;\,\, M > 0.\label{BTZrange}
\end{equation}
This limits the spectrum of allowed black holes. They are said \textit{extremal} if $|J| = M\ell$ or equivalently $r_+ = r_-$. Another special place is $r=0$ beyond which $\partial_\phi$ becomes timelike: this is the causal singularity.  Let us now demonstrate that $\mathcal{H}_+$ is a rightful (outer) event horizon. Taking advantage of our knowledge about the Kerr black hole, we claim that the surface $r=r_+$ is in fact a Killing horizon ruled by integral curves of the helicoidal vector $\xi = \partial_t + \Omega_H \partial_\phi$ for a certain $\Omega_H$. To find $\Omega_H$, we use the trick to solve $g_{\mu\nu}\xi^\nu = 0$ at $r=r_+$ which gives $\Omega_H = -g_{tt}/g_{t\phi}|_{r=r_+} =-g_{t\phi}/g_{\phi\phi} |_{r=r_+}= -N^\phi (r_+) = 4GJ/r_+^2 = r_-/r_+ \ell$. This is the angular velocity of the horizon. By construction, $\xi^\mu \xi_\mu = 0$ on $\mathcal{H}_+$. As an exercise, you can show that $\xi^\mu$ actually generates the horizon $\xi^\mu D_\mu \xi^\nu = \kappa \xi^\nu$ where the surface gravity is given by $\kappa = (r_+^2 - r_-^2)/\ell^2 r_+$. These two expressions make sense in the range \eqref{BTZrange}. So we have proven that $\mathcal{H}_+$ was the rightful outer event horizon, and from now we can use the name "black hole" without abuse. The event horizon is found to be a Killing horizon, so the rigidity theorem is obeyed, since \textit{BTZ} spacetime is stationary and axisymmetric. Moreover, note also that $r_- \leq r_+ \leq r_{erg}$. So there is a non-trivial region beyond the ergosphere and still outside the horizon where the observers experience some frame dragging due to the rotation of the black hole. This \textit{ergoregion} is a supplementary feature that \textit{BTZ} black hole shares with the Kerr solution in $4d$. \\

Now let us look at the thermodynamical properties. Hawking's temperature reads simply as
\begin{equation}
T_H = \frac{\kappa}{2\pi} = \frac{r_+^2 - r_-^2}{2\pi\ell^2 r_+}.
\end{equation}
Note that $T_H = 0$ for extremal cases where also $\Omega_H = 1/\ell$. Now we can define the entropy of the \textit{BTZ} black hole :
\begin{align}
S_H = \frac{1}{4G} (\text{Perimeter of the horizon}) = \frac{1}{4G} \int_0^{2\pi} d\phi \: \sqrt{g_{ind}}
\end{align}
where $g_{ind}$ is the determinant of the induced metric on the horizon at constant value of time, so $ds^2_{ind} = ds^2 [r=r_+, t = \text{Cst}] = r_+^2 d\phi^2$, so
\begin{equation}
S_H = \frac{1}{4G} \int_0^{2\pi} d\phi \: \sqrt{g_{ind}} = \frac{\pi r_+}{2G}\label{SBTZ} .
\end{equation}
We can check the first law $T_H \delta S_H = \delta M - \Omega_H \delta J$, which is quite obvious here :
\begin{equation}
8GM = \frac{r_+^2 - r_-^2}{\ell^2},\, 4GJ = \frac{r_+ r_-}{\ell} \Longrightarrow \delta M - \Omega_H \delta J = \frac{r_+^2 - r_-^2}{2\pi\ell^2 r_+} \frac{\pi}{2G} \delta r_+ = T_H \delta S_H .
\end{equation}

\subsubsection{Penrose diagrams}

To analyze the causal features of the \textit{BTZ} spacetime, there is nothing like a Penrose diagram ! The procedure is sensitively the same as for the Kerr black hole. The idea is to introduce a set of \textit{Kruskal coordinates} $(U,V,r,\phi)$ in the vicinity of each root of the lapse function $N^2(r) = 0$. On each Kruskal patch, we wish to write the line element as 
\begin{equation}
ds^2 = \Omega^2(U,V) (dU^2-dV^2) + r^2 \left( N^\phi(U,V) dt + d\phi \right)^2.
\end{equation}
where $t=t(U,V)$. If $J=0$, $r_- = 0$ and there is only one non-trivial root $r_+$: in this case the Kruskal patch around $r_+$ actually covers all the spacetime. Let us start with $r_+$ : we leave as an exercise (see the original \cite{Banados:1992gq}) to show that Kruskal coordinates around $r_+$ are defined by the patch $K_+$ :
\begin{equation}
\text{If } r \in \: ]r_- , r_+] : \left\lbrace
\begin{array}{ccc}
U_+ & = & \sqrt{ \left( \frac{-r + r_+}{r + r_+} \right) \left( \frac{r + r_-}{r - r_-} \right)^{r_-/r_+ } } \sinh [\kappa t(U_+,V_+) ]; \\ 
V_+ & = & \sqrt{ \left( \frac{-r + r_+}{r + r_+} \right) \left( \frac{r + r_-}{r - r_-} \right)^{r_-/r_+ } } \cosh [\kappa t(U_+,V_+) ] ;
\end{array} 
\right.
\end{equation}
\begin{equation}
\text{If } r \in \: [r_+ , \infty [ : \left\lbrace
\begin{array}{ccc}
U_+ & = & \sqrt{ \left( \frac{r - r_+}{r + r_+} \right) \left( \frac{r + r_-}{r - r_-} \right)^{r_-/r_+ } } \cosh [\kappa t(U_+,V_+) ]; \\ 
V_+ & = & \sqrt{ \left( \frac{-r + r_+}{r + r_+} \right) \left( \frac{r + r_-}{r - r_-} \right)^{r_-/r_+ } } \sinh [\kappa t(U_+,V_+) ] ;
\end{array} 
\right.
\end{equation}
and we recall that $\kappa = (r_+^2-r_-^2)/\ell^2 r_+$. Within the patch $K_+$, the angular coordinate $\phi_+$ is chosen such as $N^\phi (r_+) = 0$ to ensure that the metric element $N^\phi dt$ remains regular at $r_+$. Up to this change of coordinates, the \textit{BTZ} metric takes the right form in $K_+$ with 
\begin{equation}
\Omega_+^2 (r) = \frac{(r^2-r_-^2)(r+r_+)^2}{\kappa^2 r^2 \ell^2} \left( \frac{r-r_-}{r+r_-} \right)^{r_-/r_+} \text{  for } r \in \: ]r_-, \infty [ \; .
\end{equation}
We can define a similar Kruskal patch $K_-$ around $r_-$. The expressions are quite similar, up to some permutations $(+ \rightarrow -)$. The two patches have a non-trivial overlap $K = K_- \cap K_+$. Just as in the $3+1$ Kerr metric, one may maximally extend the geometry by gluing together an infinite number of copies of patches $K_+, K_-$ through their overlap $K$. Since we have at our disposal a set of Kruskal-like coordinates, the Penrose compactification is quite straightforward. We introduce another change of coordinates $(U,V) \rightarrow (p,q)$ such as
\begin{equation}
U+V = \tan \left( \frac{p+q}{2} \right) \quad ; \quad U-V = \tan \left( \frac{p-q}{2} \right).
\label{eq:PenroseCoord}
\end{equation}
This transformation is a bijection for the usual determination of the arctangent function, which lies between $-\pi/2$ and $\pi/2$. Let us start with the non-rotating case $J=0$. Only the outer horizon at $r_+$ exists, since $r_- = 0$ coincides with the causal singularity. In this case,  spacelike infinity $r = \infty$ is mapped to the (vertical) lines $p = \pm \pi/2$, the singularity $r=0$ is mapped to the (horizontal) lines $q= \pm \pi/2$, and finally the horizon lies at $p = \pm q$. This gives the following Penrose diagram: \vspace{3cm}

\begin{figure}[!]
\centering
\begin{tikzpicture}[scale=1]
\draw[thick,red] (-2,-2) -- (-2,2);
\node[above,rotate=90,red] at (-2,0) {\footnotesize $r=\infty$};
\draw[thick,red] ( 2,-2) -- ( 2,2);
\node[below,rotate=90,red] at ( 2,0) {\footnotesize $r=\infty$};
\draw[thick,GrandVert] (-2, 2) -- ( 2, 2);
\node[above,GrandVert] at (0, 2) {\footnotesize $r=0$};
\draw[thick,GrandVert] (-2,-2) -- ( 2,-2);
\node[below,GrandVert] at (0,-2) {\footnotesize $r=0$};
\draw[blue] (-2,-2) -- (2, 2);
\node[above,blue,rotate= 45] at ( 1, 1) {\footnotesize $r=r_+$};
\draw[blue] (-2, 2) -- (2,-2);
\node[above,blue,rotate=-45] at (-1, 1) {\footnotesize $r=r_+$};
\draw[gray,->] (-2.5,-2.5) -- (-2.5,-1.8) node[above] {\footnotesize $q$};
\draw[gray,->] (-2.5,-2.5) -- (-1.8,-2.5) node[right] {\footnotesize $p$};
\end{tikzpicture}
\caption{Penrose diagram for a static \textit{BTZ} black hole ($J=0$).}
\end{figure}
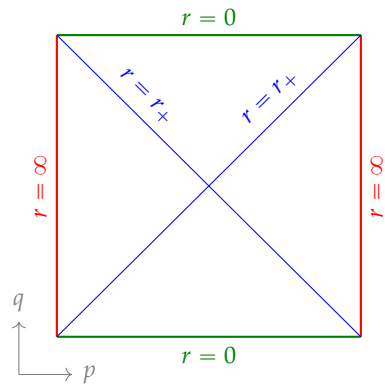

In the rotating case, we have to perform the change of coordinates $(U,V) \rightarrow (p,q)$ in each Kruskal patch, so we find the following Penrose diagram: 

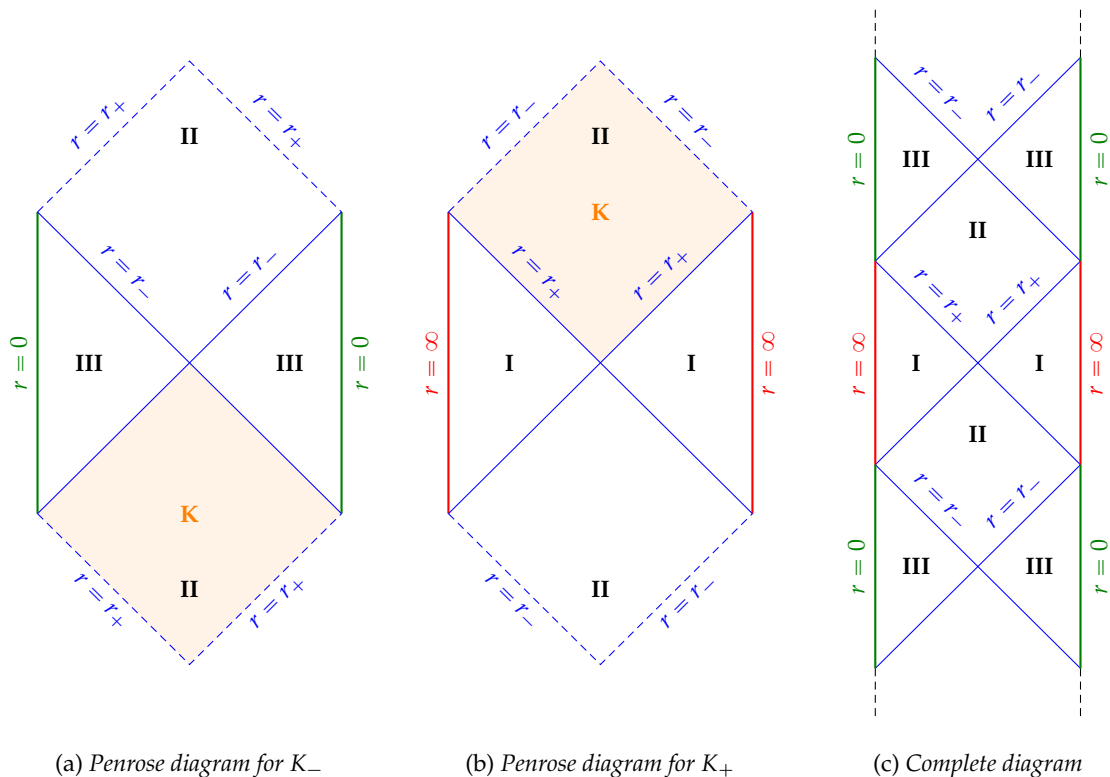
\begin{figure}[!ht]
  \centering
  \subfloat[Penrose diagram for $K_-$]{
  	\begin{tikzpicture}[scale=1]
  		\draw[white] (-2.6,-5) -- (-2.6, 5) -- ( 2.6, 5) -- ( 2.6,-5) -- cycle;
		\draw[thick,GrandVert] (-2,-2) -- (-2,2);
		\node[above,rotate=90,GrandVert] at (-2,0) {\footnotesize $r=0$};
		\draw[thick,GrandVert] ( 2,-2) -- ( 2,2);
		\node[below,rotate=90,GrandVert] at ( 2,0) {\footnotesize $r=0$};
		\fill[orange!10] (-2,-2) -- (0,-4) -- (2,-2) -- (0,0) -- cycle;
		\node[orange] at (0,-2) {\footnotesize \textbf{K}};
		\node at ( 0,-3) {\footnotesize \textbf{II}};
		\node at ( 0, 3) {\footnotesize \textbf{II}};
		\node[left] at (-1, 0) {\footnotesize \textbf{III}};
		\node[right] at ( 1, 0) {\footnotesize \textbf{III}};
		\draw[blue,densely dashed] (-2, 2) -- (0,4) -- ( 2, 2);
		\node[above,blue,rotate=-45] at ( 1,3) {\footnotesize $r=r_+$};
		\node[above,blue,rotate= 45] at (-1,3) {\footnotesize $r=r_+$};
		\draw[blue,densely dashed] (-2,-2) -- (0,-4) -- ( 2,-2);
		\node[below,blue,rotate= 45] at ( 1,-3) {\footnotesize $r=r_+$};
		\node[below,blue,rotate=-45] at (-1,-3) {\footnotesize $r=r_+$};
		\draw[blue] (-2,-2) -- (2, 2);
		\node[above,blue,rotate= 45] at ( 1, 1) {\footnotesize $r=r_-$};
		\draw[blue] (-2, 2) -- (2,-2);
		\node[above,blue,rotate=-45] at (-1, 1) {\footnotesize $r=r_-$};
	\end{tikzpicture}
   } 
   \subfloat[Penrose diagram for $K_+$]{
  	\begin{tikzpicture}[scale=1]
  		\draw[white] (-2.6,-5) -- (-2.6, 5) -- ( 2.6, 5) -- ( 2.6,-5) -- cycle;
		\draw[thick,red] (-2,-2) -- (-2,2);
		\node[above,rotate=90,red] at (-2,0) {\footnotesize $r=\infty$};
		\draw[thick,red] ( 2,-2) -- ( 2,2);
		\node[below,rotate=90,red] at ( 2,0) {\footnotesize $r=\infty$};
		\fill[orange!10] (-2, 2) -- (0, 4) -- (2, 2) -- (0,0) -- cycle;
		\node[orange] at (0, 2) {\footnotesize \textbf{K}};
		\node at ( 0,-3) {\footnotesize \textbf{II}};
		\node at ( 0, 3) {\footnotesize \textbf{II}};
		\node[left] at (-1, 0) {\footnotesize \textbf{I}};
		\node[right] at ( 1, 0) {\footnotesize \textbf{I}};
		\draw[blue,densely dashed] (-2, 2) -- (0,4) -- ( 2, 2);
		\node[above,blue,rotate=-45] at ( 1,3) {\footnotesize $r=r_-$};
		\node[above,blue,rotate= 45] at (-1,3) {\footnotesize $r=r_-$};
		\draw[blue,densely dashed] (-2,-2) -- (0,-4) -- ( 2,-2);
		\node[below,blue,rotate= 45] at ( 1,-3) {\footnotesize $r=r_-$};
		\node[below,blue,rotate=-45] at (-1,-3) {\footnotesize $r=r_-$};
		\draw[blue] (-2,-2) -- (2, 2);
		\node[above,blue,rotate= 45] at ( 1, 1) {\footnotesize $r=r_+$};
		\draw[blue] (-2, 2) -- (2,-2);
		\node[above,blue,rotate=-45] at (-1, 1) {\footnotesize $r=r_+$};
	\end{tikzpicture}
   } 
   \subfloat[Complete diagram]{
  	\begin{tikzpicture}[scale=1.35]
  		\draw[white] (-1.6,-3.703) -- (-1.6, 3.703) -- ( 1.6, 3.703) -- ( 1.6,-3.703) -- cycle;
		\draw[blue] (-1,-3) -- ( 1,-1) -- (-1, 1) -- ( 1, 3);
		\draw[blue] ( 1,-3) -- (-1,-1) -- ( 1, 1) -- (-1, 3);
		\draw[densely dashed] (-1,-3) -- (-1,-3.5);
		\draw[densely dashed] ( 1,-3) -- ( 1,-3.5);
		\draw[densely dashed] (-1, 3) -- (-1, 3.5);
		\draw[densely dashed] ( 1, 3) -- ( 1, 3.5);
		\draw[thick,red] (-1,-1) -- (-1, 1);
		\draw[thick,red] ( 1,-1) -- ( 1, 1);
		\draw[thick,GrandVert] (-1,-3) -- (-1,-1);
		\draw[thick,GrandVert] ( 1,-3) -- ( 1,-1);
		\draw[thick,GrandVert] (-1, 1) -- (-1, 3);
		\draw[thick,GrandVert] ( 1, 1) -- ( 1, 3);
		\node[above,rotate=90,red] at (-1, 0) {\footnotesize $r=\infty$};
		\node[below,rotate=90,red] at ( 1, 0) {\footnotesize $r=\infty$};
		\node[above,rotate=90,GrandVert] at (-1, 2) {\footnotesize $r=0$};
		\node[below,rotate=90,GrandVert] at ( 1, 2) {\footnotesize $r=0$};
		\node[above,rotate=90,GrandVert] at (-1,-2) {\footnotesize $r=0$};
		\node[below,rotate=90,GrandVert] at ( 1,-2) {\footnotesize $r=0$};
		\node[above,blue,rotate= 45] at ( 0.5,-1.5) {\footnotesize $r=r_-$};
		\node[above,blue,rotate=-45] at (-0.5,-1.5) {\footnotesize $r=r_-$};
		\node[above,blue,rotate= 45] at ( 0.5, 0.5) {\footnotesize $r=r_+$};
		\node[above,blue,rotate=-45] at (-0.5, 0.5) {\footnotesize $r=r_+$};
		\node[above,blue,rotate= 45] at ( 0.5, 2.5) {\footnotesize $r=r_-$};
		\node[above,blue,rotate=-45] at (-0.5, 2.5) {\footnotesize $r=r_-$};
		\node[] at (-0.6,-2.0) {\footnotesize \textbf{III}};
		\node[] at (-0.6, 0.0) {\footnotesize \textbf{I}};
		\node[] at (-0.6, 2.0) {\footnotesize \textbf{III}};
		\node[] at ( 0.6,-2.0) {\footnotesize \textbf{III}};
		\node[] at ( 0.6, 0.0) {\footnotesize \textbf{I}};
		\node[] at ( 0.6, 2.0) {\footnotesize \textbf{III}};
		\node[] at ( 0.0,-0.7) {\footnotesize \textbf{II}};
		\node[] at ( 0.0, 1.3) {\footnotesize \textbf{II}};
	\end{tikzpicture}
	\label{fig:PenroseBTZ}
   }
\caption{Penrose diagrams for non-extremal \textit{BTZ} black holes}
\end{figure}

As we have already seen, the $K$ parts are identified because they represent the overlap between the patches $K_-$ and $K_+$. In the $K_-$ patch, the original black hole coordinates covered $K$ and one region labeled by \textit{III}, while in the $K_+$ patch, they covered $K$ and one region labeled by \textit{I}. To obtain a``maximal causal extension" (\textit{i.e.} where all causal curves extend maximally), we must include the others regions in each diagram, and then glue together an \textit{infinite} sequence of them. The resulting Penrose diagram bears strong ressemblence with the Kerr one but the asymptotics and singularities differ.\\

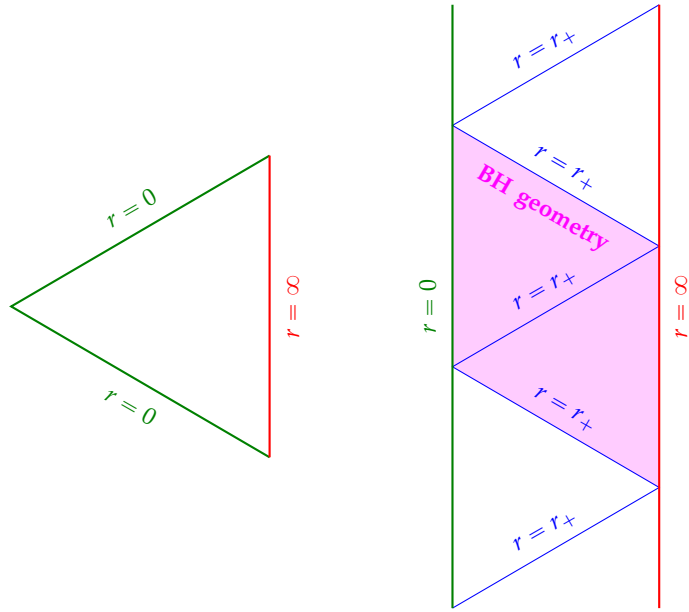
\begin{figure}[h!]
  \centering
  \subfloat[Penrose diagram for $M=J=0$]{
  	\begin{tikzpicture}[scale=1]
  		\draw[white] ( 1.0,-4.4) -- ( 1.0, 4.4) -- (-4.0, 4.4) -- (-4.0,-4.4) -- cycle;
  		\draw[thick,GrandVert] ( 0, 2) -- ( -3.4, 0) -- ( 0, -2);
  		\draw[thick,red] ( 0, 2) -- ( 0, -2);
  		\node[above,rotate= 30,GrandVert] at (-1.7, 1.1) {\footnotesize $r=0$};
  		\node[below,rotate=-30,GrandVert] at (-1.7,-1.1) {\footnotesize $r=0$};
  		\node[below,rotate= 90,red] at (0.1,0) {\footnotesize $r=\infty$};
	\end{tikzpicture}
   } 
   \subfloat[Penrose diagram for $|J| = M\ell$]{
  	\begin{tikzpicture}[scale=0.8]
  		\draw[white] (-4.9,-4.5) -- (-4.9,6.5) -- (1.5,6.5) -- (1.5,-4.5) -- cycle;
  		\fill[magenta!20] ( 0,-2) -- ( -3.4, 0) -- ( -3.4, 4) -- ( 0, 2) -- cycle;
  		\draw[blue] (-3.4,-4) -- ( 0,-2) -- ( -3.4, 0) -- ( 0, 2) -- ( -3.4, 4) -- (0,6);
  		\draw[thick,red] ( 0, -4) -- ( 0, 6);
  		\draw[thick,GrandVert] ( -3.4, -4) -- ( -3.4, 6);
  		\node[above,rotate= 30,blue] at (-1.7,-3.0) {\footnotesize $r=r_+$};
  		\node[above,rotate= 30,blue] at (-1.7, 1.0) {\footnotesize $r=r_+$};
  		\node[above,rotate=-30,blue] at (-1.7,-1.0) {\footnotesize $r=r_+$};
  		\node[above,rotate= 30,blue] at (-1.7, 5.0) {\footnotesize $r=r_+$};
  		\node[above,rotate=-30,blue] at (-1.7, 3.0) {\footnotesize $r=r_+$};
  		\node[below,rotate= 90,red] at (0.1,1) {\footnotesize $r=\infty$};
  		\node[above,rotate= 90,GrandVert] at (-3.5,1) {\footnotesize $r=0$};
  		\node[above,rotate=-30,magenta,align=center,text width=2cm] at (-2.1, 2.3) {\footnotesize \textbf{BH geometry}};
	\end{tikzpicture}
   }
\caption{Penrose diagrams for extremal \textit{BTZ} black holes.}
\label{fig:ExtremalBTZ}
\vspace{30pt}
\end{figure}

Let us now comment on the extremal case. For the non-massive case $M=0$ (which is also non-rotating), the metric reduces to $ds^2 = -(r/\ell)^2 dt^2 + (r/\ell)^{-2} dr^2 + r^2 d\phi^2$, and we can directly define null dimensionless coordinates $U = (t/\ell) - (\ell/r), \: V = -(t/\ell) - (\ell/r)$ such that $ds^2 = r^2 dU dV + r^2 d\phi^2$. This allows to directly write the compact Penrose coordinates (\ref{eq:PenroseCoord}), in which the metric reads as
\begin{equation}
ds^2 = \frac{\ell^2}{\sin^2 p} (dp^2 - dq^2) + r^2 d\phi^2 \quad ; \quad r = -\ell \frac{\cos p + \cos q}{\sin p}.
\end{equation}
So the origin $r=0$ is mapped to the segment of the line $p = \pi \pm q$ running from $p=0$ to $p=\pi$. On the other hand, spacelike infinity is mapped to the segment of the line $p=\pi$, and the Penrose diagram is simply a closed triangle. Now we finish with the extremal rotating case $|J| = M\ell$. In this case only the outer horizon survives, and the metric can be written as a function of $r_+$ instead of $M$. The lapse function becomes simply $N^2 (r) = \frac{(r^2-r_+^2)^2}{r^2\ell^2}$. The appropriate null coordinates are $U = t+r^\star$ and $V = -t+r^\star$ where $r^\star$ is the so-called \textit{tortoise coordinate} 
\begin{equation}
r^\star = \int \frac{dr}{N^2 (r)} = -\frac{r\ell^2}{2(r^2-r_+^2)} + \frac{\ell^2}{4r_+} \ln \left| \frac{r-r_+}{r+r_+} \right| . 
\end{equation}
Defining the Penrose coordinate again as before, we get
\begin{equation}
ds^2 = \frac{4 N^2 (r) \ell^2 (dp^2 - dq^2)}{(\cos p + \cos q )^2} + r^2 (N^\phi dt + d\phi )^2 \quad \text{where \textit{r} is solution of } \frac{\sin p}{\cos p + \cos q} = \frac{r^\star}{\ell}. 
\end{equation}
We see that the horizon $r=r_+$ is represented by lines at $\pm 45°$, whereas $r=0$ lies at $p \rightarrow k \pi$ (the limit is taken from above values) and $r=\infty$ at $p \rightarrow k \pi$ (the limit is this time taken from below values). The region beyond the horizon ($0 < r < r_+$) is mapped onto a triangle bounded by the lines $p=0$ (which is $r=0$) and $p=q=\pi, \: p-q = \pi$ which is quite similar to the non-rotating case. The outer region $r>r_+$ is obtained thanks to another determination of the \textit{arctangent} function, and is a symmetric triangle too. To obtain the full spacetime, we have to glue the two triangles along the common edge $r=r_+$ at $45^\circ$. Once this is done, we can go safely accross the gluing edge because the root of $N(r)$ is compensated by the cancellation of the denominator in the $(p,q)$ term in the metric. As before, the maximal extension is build by including an infinite sequence of triangles (see Figure \ref{fig:ExtremalBTZ}). \\

We conclude here the description of the \textit{BTZ} geometry. What remains to be done is see how to reach the black hole solution from the global $AdS_3$ spacetime. Since there is no topological defect for $M>0$, the process at work is in fact a fold of $AdS_3$ on itself after that some points have been identified. We discuss this in more details in the next paragraph.

\newpage
\subsubsection{Identifications}

\begin{wrapfigure}{r}{0.4\textwidth}
\begin{tikzpicture}[scale=1.2]
	\coordinate (A1) at (0, 0);
    \coordinate (A2) at (0, 4);
    \coordinate (A3) at (4, 4);
    \coordinate (A4) at (4, 0);
    \coordinate (B1) at (1.2, 1.2);
    \coordinate (B2) at (1.2, 5.2);
    \coordinate (B3) at (5.2, 5.2);
    \coordinate (B4) at (5.2, 1.2);
    \draw[thick] (A1) -- (A2) -- (A3) -- (A4) -- cycle;
    \draw[thick] (A2) -- (B2) -- (B3) -- (B4) -- (A4);
    \draw[thick] (A3) -- (B3);
    \draw[dashed] (A1) -- (B1);
    \draw[dashed] (B1) -- (B2);
    \draw[dashed] (B1) -- (B4);
	\draw[orange,fill=orange,opacity=0.3] (0.7,1) -- (1.3,1.6) -- (1.3,3.6) -- (0.7,3) -- cycle;
	\draw[orange] (0.7,1) -- (1.3,1.6) -- (1.3,3.6) node[right] {\footnotesize $S(0)$} -- (0.7,3) -- cycle;
	\draw[black!50,thick,densely dashed] (1.0,2.3) -- (3.0,2.3);
	\fill[fill=blue] (1.0,2.3) circle (0.05);
	\node[above,blue] at (1.0,2.3) {\footnotesize $P$};
	\draw[blue,thick,->] (1.0,2.3) -- (1.7,2.3) node[below] {\footnotesize $\xi$};
	\draw[orange,fill=orange,opacity=0.3] (2.7,1) -- (3.3,1.6) -- (3.3,3.6) -- (2.7,3) -- cycle;
	\draw[orange] (2.7,1) -- (3.3,1.6) -- (3.3,3.6) node[right] {\footnotesize $S(s)$} -- (2.7,3) -- cycle;
	\fill[fill=blue] (3.0,2.3) circle (0.05);
	\node[above,blue] at (3.0,2.3) {\footnotesize $e^{s\xi} P$};
	\draw[red,<->] (0.7,0.9) to [out=-30,in=-150] (2.7,0.9);
	\node[red] at (1.7,0.3) {\footnotesize Identification};
	\node[below] at (3.1, 5.2) {\footnotesize Asymptotic box};
\end{tikzpicture}
\caption{Identifications in $AdS_3$.}
\end{wrapfigure}
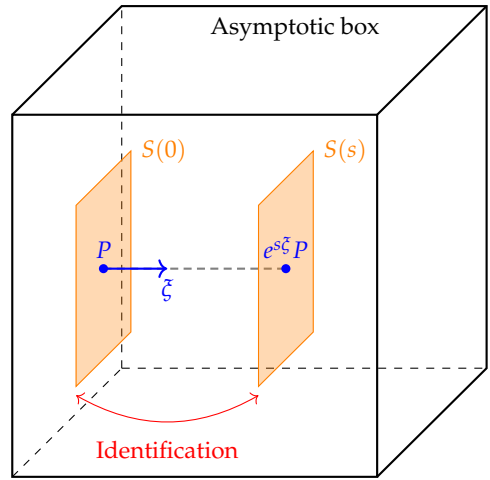

Any Killing vector $\xi^\mu$ generates a $1$-parameter subgroup of isometries of $AdS_3$. It acts on points as $P \rightarrow e^{s\xi} P$ where $s$ is a continuous parameter, and locally as $P = (x^\mu) \rightarrow P' = (x^\mu + \xi^\mu )$. If we restrict $s$ to discrete values $k\: \Delta s$ where $k \in \mathbb{Z}$ and $\Delta s$ is a basic step conventionally fixed as $2\pi$, we are left with the so-called \textit{identification subgroup}. The identified quotient space is obtained by identifying points that belong to a given orbit of the identification subgroup. Geometrically, we can view this operation as follows. Let us consider a surface $S_0$ in $AdS_3$. Each point of this surface belongs to an orbit of $\xi^\mu$, and we can apply an element of the identification subgroup to get another surface $S(2\pi k)$ whose points have coordinates $e^{2\pi k \xi} P, \: P\in S_0$, $k \in \mathbb Z$. The quotient process consists in gluing the surface $S_0 = S(0)$ with $S(2\pi)$, $S(-2\pi)$, $S(4\pi)$, and so on ! \\

Since $\xi^\mu$ is a Killing vector, the gluing leads to a continuous spacetime and the quotient space inherits from anti-de Sitter space a well defined metric which has constant negative curvature. Indeed, let us choose a coordinate system $(t,r,\varphi)$ in which $\xi = \partial_\varphi$ and $\varphi \in \mathbb R$. So $\Lie_\xi g_{\mu\nu} = \xi^\rho \partial_\rho g_{\mu\nu} + 0 = \partial_\varphi g_{\mu\nu} = 0 \Rightarrow g_{\mu\nu}(t,r,\varphi) = g_{\mu\nu}(t,r)$. When we perform identifications along $\varphi$, $g_{\mu\nu}$ is not a multivalued function of the $\varphi$ coordinate. So the metric remains locally smooth and is still a solution of Einstein's equations. The identification process makes the curves joining two points that are on the same orbit to be closed in the quotient space. In order to preserve causality in the quotient space or, equivalently, prevent the appearance of closed causal curves, a necessary condition (but not sufficient in general) is that the Killing vector $\xi^\mu$ must be spacelike : $\xi^\mu \xi_\mu > 0$. Let us see what happens particularly for the \textit{BTZ} black hole. \\

One can show that the non-extremal \textit{BTZ} black hole solutions are obtained by making identifications in $AdS_3$ by the discrete group generated by the Killing vector  
\begin{equation}
\xi = \frac{r_+}{\ell} J_{12} - \frac{r_-}{\ell} J_{03}\label{defph}
\end{equation}
where the $J_{ab}$ are the Killing vectors of $AdS_3$, belonging to the $so(2,2)$ algebra \cite{Banados:1992gq}. To see geometrically in what consist these identifications, the most ``simple'' thing to do is to find a coordinate system $(t,r,\phi)$ in which $\xi = \partial_\phi$. The answer is simply that $AdS_3$ takes the form (\ref{eq:BTZ}) but with $\phi \in \mathbb{R}$.  
Recognizing that $\partial_\phi$ is given by \eqref{defph} takes more effort, which is narrated in \cite{Banados:1992gq}. It follows that the non-extremal $BTZ$ black hole is obtained by realizing periodic identifications along \eqref{defph} in $AdS_3$! \\

But all is not resolved, since $\xi^\mu$ is not spacelike everywhere in $AdS_3$. In fact, before performing  identifications, we better remove the regions where $\xi^2 \leq 0$ in global $AdS_3$ in order to avoid closed timelike curves. We are left with a spacetime denoted by $\overline{AdS}_3$ which is geodesically incomplete, since before the removal, some geodesics traveled from regions $\xi^2 > 0$ to regions $\xi^2 \leq 0$. The critical surface $\xi^2 = 0$ appears as a singularity in the causal structure of spacetime, since continuing beyond it would produce closed timelike curves. We retrieve a familiar feature of $3+1$ black holes : the only incomplete geodesics are those that hit the singularity. In $4d$ it is a curvature singularity but here it is a causal singularity. Remember that in $2+1$ dimensions the Weyl tensor is identically zero while the Ricci tensor is determined by the cosmological constant by the vacuum Einstein's equations, so curvature singularities cannot appear. It turns out however that a horizon prevents the asymptotic observer to detect the causal singularities. This is a form of ``cosmic censorship''.\\ 

Going back to $\overline{AdS}_3$ without identifications, we can recognize its boundary $\xi^2 = 0$ as a surface. Explicitly, we have
\begin{equation}
\xi^2 = \xi^a \xi_a = \zeta_{ab} \xi^a \xi^b = \frac{r_+^2}{\ell^2} (X_0^2 - X_1^2) + \frac{r_-^2}{\ell^2} (\tilde X_0^2 - X_2^2). 
\end{equation}
Therefore, $\overline{AdS}_3$ is bounded by the planes $X_0 = \pm X_1$ and $\tilde X_0 = \pm X_2$ intersecting the conformal cylinder. The locus $\xi^2 = 0$ is therefore a connected, diamond-shaped region depicted below. \\

\begin{figure}[h!]
\centering
\begin{tikzpicture}[scale=1.2]
\coordinate (top) at (0,3);
\coordinate (bottom) at (0,-3);
\coordinate (centre) at (0,0);
\coordinate (leftp) at (-1.8,0);
\coordinate (rightp) at ( 1.8,0);
\coordinate (edgebotbot) at ( $(bottom)-(0.3,0.59)$ );
\coordinate (edgebottop) at ( $(bottom)+(0.3,0.59)$ );
\coordinate (edgetopbot) at ( $(top)-(0.3,0.59)$ );
\coordinate (edgetoptop) at ( $(top)+(0.3,0.59)$ );
\draw[densely dashed] (edgebottop) -- (edgetoptop);
\draw[densely dashed] (bottom) ellipse (1.8 and 0.6);
\draw[thick] ( $(bottom)+(-1.8,0)$ ) -- ( $(top)+(-1.8,0)$ );
\draw[thick] ( $(bottom)+( 1.8,0)$ ) -- ( $(top)+( 1.8,0)$ );
\draw[thick] (bottom)+(1.8,0) arc (0:-180:1.8 and 0.6);
\draw[red,fill=red,fill opacity=0.2] (edgebotbot) to [bend left=15] (leftp) to [bend left=5] (edgebottop);
\draw[red,fill=red,fill opacity=0.2] (edgebotbot) to [bend right=15] (rightp) to [bend right=5] (edgebottop);
\draw[red] (edgebotbot) -- (edgebottop);
\draw[densely dashed] (centre) ellipse (1.8 and 0.6);
\draw[] (centre)+(1.8,0) arc (0:-180:1.8 and 0.6);
\draw[red,fill=red,fill opacity=0.2] (edgetopbot) to [bend left=15] (leftp) to [bend left=5] (edgetoptop);
\draw[red,fill=red,fill opacity=0.2] (edgetopbot) to [bend right=15] (rightp) to [bend right=5] (edgetoptop);
\draw[] (edgetopbot) -- (edgetoptop);
\draw[] (edgebotbot) -- (edgetopbot);
\draw[thick] (top) ellipse (1.8 and 0.6);
\coordinate (legend) at ( $(bottom)+(0,-1)$ );
\node[below] at (legend) {\footnotesize $\phi$};
\draw[->] plot [smooth,tension=1] coordinates {($(legend)+(-0.5,0.1)$) ($(legend)+(0,0.05)$) ($(legend)+(0.5,0.1)$)};
\draw[<-] ($(top)+(-2.2,0)$) -- ($(top)+(-2.2,-1)$);
\node[left] at ($(top)+(-2.2,-0.5)$) {\footnotesize $t$};
\end{tikzpicture}
\caption{A finite-time section of the conformal representation of $AdS_3$. \\
The red sections are the null surfaces on which $\xi^2 = 0$.}
\label{fig:BarAdS3}
\end{figure}
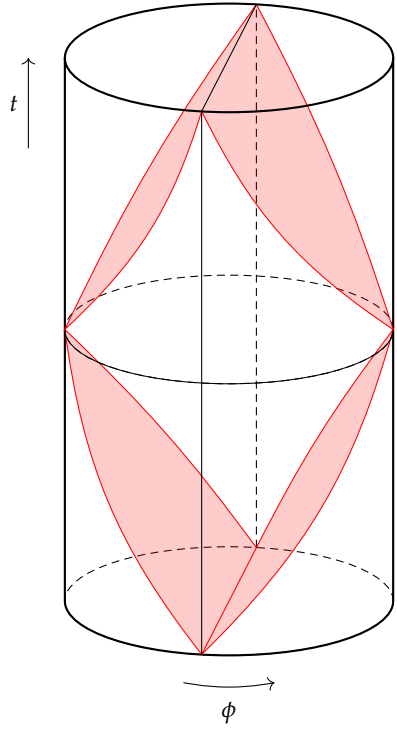

\subsubsection{Symmetries of the quotient space}
The \textit{BTZ} solution is stationary and axisymmetric: it has 2 commuting Killing vectors $\partial_t$ and $\partial_\phi$. One may ask whether there are any other independent Killing vectors, since it comes from an identification of a maximally symmetric spacetime that has 4 more isometries. We will show that the two aforementioned Killing vectors are the only two isometries of the $BTZ$ solution. \\

Before any identification, the spacetime has 6 independent Killing vectors which are the generators of $so(2,2)$. After identification, some of these vectors become multivalued, and so are no longer proper symmetries of the quotient space. A necessary and sufficient condition for a vector $\eta^\mu \in so(2,2)$ to induce a well-defined vector field on the quotient space is that $\eta^\mu$ be invariant under the identification subgroup $(e^{2\pi \xi})^\star \eta = \eta$. Here we consider $\eta^\mu$ and $\xi^\mu$ as $so(2,2)$ matrices. Since $\eta$ is a Killing vector, this condition is equivalent to $(e^{2\pi \xi}) \eta (e^{2\pi \xi})^{-1} = \eta$ or $[e^{2\pi \xi},\eta] = 0$. 
A theorem due to Chevalley and Jordan states that any matrix $\eta$ can be fragmented in two commuting parts $s$ and $n$ : $\xi = s+n$. $s$ is semi-simple with real eigenvalues, and $n$ is nilpotent. Any matrix commuting with $e^{2\pi \xi}$ must also commute separately with $e^{2\pi s}$ and $e^{2\pi n}$, so we get $[s,\eta ] = [n,\eta ] = 0$. This implies $[\xi,\eta]=0$.\\

So the problem of finding all Killing vectors of the \textit{BTZ} solution is equivalent to  finding all the $so(2,2)$ matrices that commute with $\xi$. Similarly to $so(4) = so(3) \oplus so(3)$, we know that $so(2,2) = so(2,1) \oplus so (2,1)$ ($so(2,1)$ is the Lorentz algebra in $2+1$ dimensions) and we can decompose $\xi = \xi^+ + \xi^-, \eta = \eta^+ + \eta^-$, where the "$+$" parts are the self-dual parts, and the "$-$" the anti-self-dual ones. Since the sum between the algebras is direct, we are left with $[\xi^+, \eta^+] = [\xi^-, \eta^-] = 0$. Now, recall that the only elements of $so(2,1)$ that commute with a given non-zero element of $so(2,1)$ are only the multiples of that element. After decomposing \eqref{defph} as $\xi^+ + \xi^-$, one finds that both $\xi^+$ and $\xi^-$ are non-zero. We conclude that $\xi^\pm \: \propto \: \eta^\pm$ and there is no additional Killing vector. The identification kills 4 out of the 6 Killing isometries of $AdS_3$ and only $\partial_t$ and $\partial_\phi$ survive.

\subsubsection{\textit{BTZ} phase space}
We conclude this quite long section by a summary of the different spacetimes that are described by the line element (\ref{eq:BTZ}). At fixed $\ell$, it is only described by two parameters, which are the total mass $M$ and the total angular momentum $J$, so we can represent the phase space in a $(J,M)$-plane.
\begin{itemize}[label=$\rhd$]
\item For $M > 0$ and $|J| \leq M\ell$, we find all \textit{BTZ} black holes whose main features were explained. They are identifications of $AdS_3$ that possess a Killing horizon. The latter is a rightful event horizon that shields a causal (but not a curvature) singularity. These solutions are delimited by the extremal ones, which obey $|J| = M \ell$. The special $BTZ$ black hole with zero mass has $M = J = 0$. 
\item For $M > 0$ and overspin $|J| > M\ell $, the spacetime leaves exposed the chronological singularity (closed timelike curve or C.T.C.) at $r = 0$. This is a naked singularity, which is usually considered unphysical. 
\item Solutions with $M < 0$ and $J = 0$ represent \textit{particles} in $AdS_3$ sitting at $r=0$. If $M > -1/8G$, they produce a conical defect around it\footnote{We save more detailed explanations for the lecture on $3d$ asymptotically flat spacetimes where conical defects also appear. Doing so we do not develop the same explanations twice !}. 
 When $M = -1/8G$, the conical singularity disappears and the spacetime, smooth everywhere, is nothing but global $AdS_3$. When $M < -1/8G$, the angular defect becomes an excess. The energy spectrum of these solutions is not bounded from below and has a mass lower than the natural ground state, global $AdS_3$. These solutions are therefore also usually considered unphysical and discarded. If $J \neq 0$, the spacetime is identified with a twist in time around the conical defect/excess: we thus find \textit{spinning particles} with mass $M$ and angular momentum $J$. Again when $ |J| > -M \ell$ the spacetime contains naked closed timelike curves... 
\end{itemize}

\begin{figure}[h!]
\centering
\begin{tikzpicture}[scale=1]
\draw[white] (-4,-4) -- (-4, 4) -- (4,4) -- (4,-4);
\coordinate (A)   at (-3.0,-2.6);
\coordinate (B)   at (-3.0, 2.6);
\coordinate (C)   at ( 3.0, 2.6);
\coordinate (D)   at ( 3.0,-2.6);
\coordinate (O)   at ( 0.0, 0.0);
\coordinate (AdS) at ( 0.0,-1.3);
\coordinate (AdSL)at (-1.5,-1.3);
\coordinate (AdSR)at ( 1.5,-1.3);
\fill[GrandVert!30] (B) -- (O) -- (C);
\fill[orange!30] (AdSL) -- (O) -- (AdSR);
\fill[red!30] (A) -- (AdSL) -- (AdSR) -- (D) -- cycle;
\draw[red,thick] (A) -- (C) node[above]{\footnotesize $M\ell = |J|$};
\draw[red,thick] (B) node[above]{\footnotesize $M\ell = |J|$} -- (D);
\draw[->,thick] (0,-3.5) -- (0,3.5) node[above]{\footnotesize $M$};
\draw[->,thick] (-3.5,0) -- (3.5,0) node[right]{\footnotesize $J$};
\node[GrandVert,text width=2.5cm, fill=GrandVert!30,align=center] at (0,2.0) {\footnotesize \textbf{\textit{BTZ} black holes}};
\node[fill=orange!30,text width=0.5cm] at (0,-0.7) {$\phantom{A}$};
\node[orange,align=center] at (0,-0.7) {\footnotesize \textbf{Defects}};
\node[red,align=center,fill=red!30,text width=2cm] at (0,-1.3-0.7) {\footnotesize \textbf{Excesses}};
\draw[densely dashed] (AdS) -- ( $(AdSL)-(1.5,0)$ );
\draw[<->] ( $(AdSL)-(1.5,0)$ ) -- ( $(AdSL)-(1.5,-1.3)$ );
\node[left] at ( $(AdSL)-(1.5,-0.65)$ ) {$-\frac{1}{8G}$};
\fill[fill=blue] (AdS) circle (0.1) node[blue,right]{\footnotesize $AdS_3$};
\node[] at ( 2.7, 0.8) {\footnotesize \textit{Naked sing.}};
\node[] at (-2.7, 0.8) {\footnotesize \textit{Naked sing.}};
\node[] at ( 2.7,-0.8) {\footnotesize \textit{C.T.C.}};
\draw[->] (1.05,0) to [bend right=15] (0.8,0.7);
\node[] at (1.2,0.4) {\footnotesize $\beta$};
\end{tikzpicture}
\caption{Solutions described by the \textit{BTZ} metric. \\
The slope of extremal lines is $\tan \beta = \ell^{-1}$.\\ \tiny{Figure adapted with permission from \cite{Barnich:2012aw}.Copyrighted by the American Physical Society.}}
\label{fig:BTZPhaseSpace}
\end{figure}
\newpage
\subsection{Asymptotically \textit{AdS}\textsubscript{3} spacetimes}
In the previous section, we have studied the phase space of stationary and axisymmetric solutions of Einstein's $3d$ gravity that also asymptote to $AdS_3$ spacetime. We have found many kinds of solutions, including \textit{BTZ} black holes and spinning particles... Is it possible to formalize a phase space that includes all these solutions, and contains even more solutions?

\subsubsection{Boundary conditions}
Let us denote by $\mathcal{P}$ this phase space. Any point of $\mathcal{P}$ is a solution of $3d$ Einstein's equations that  is asymptotically $AdS_3$ at spatial infinity. To define precisely the assumptions under which a given metric $g_{\mu\nu}$ belongs or not to $\mathcal{P}$, we have to construct a set of \textit{boundary conditions} that tell us how $g_{\mu\nu}$ behaves when a suitable spacelike coordinate runs to infinity. The problem of fixing the asymptotic behaviour of fields in gravity is not straightforward because the choice of appropriate fall-offs at infinity is not unique at all ! We will consider the boundary conditions that obey the following criteria: 
\begin{itemize}[label=$\rhd$]
\item $\mathcal{P}$ has to contain the solutions we have already analyzed : \textit{BTZ} black holes, spinning particles,... and of course $AdS_3$ itself. So the boundary conditions must not to be too restrictive ;
\item $\mathcal{P}$ needs to lead to finite and integrable surfaces charges. It restricts the possibilities for $\mathcal{P}$. For example, a geometry $g_{\mu\nu}$ that has infinite energy is not allowed, and must be avoided by the boundary conditions;
\item The asymptotic symmetry group must at least contain $SO(2,2)$, the group of exact symmetries of global $AdS_3$.
\end{itemize}

These three requirements lead to the boundary conditions of \textit{Brown} and \textit{Henneaux} \cite{Brown:1986nw}. However, keep in mind that several other choices of boundary conditions exist if one for example changes the third requirement (see \cite{Compere:2013bya} and further developments for alternatives).\\

Remember that the asymptotic symmetry group is the quotient group between ``large" diffeomorphisms, associated to non-vanishing canonical charges, and the ``gauge" diffeomorphisms, which act trivially on the phase space and are not associated with any charge. In their original work,  \textit{Brown} and \textit{Henneaux} started with fall-off conditions for all components of the metric. We find more pedagogical and illuminating to first reduce the coordinate system by removing all gauge diffeomorphisms. It indeed allows to isolate in detail the structure of the asymptotic symmetry group which then coincides with all remaining (non-trivial) diffeomorphisms that obey the boundary conditions. Gauge diffeomorphisms are typically removed by fixing a coordinate system such that no diffeomorphism depending arbitrarily on all coordinates is still  allowed. Such a coordinate system exists for asymptotically $AdS_3$ spacetimes, as stated in the \textit{Fefferman-Graham theorem}: 
\begin{resultat}[\textit{Fefferman-Graham coordinates}]
Any asymptotically $AdS_3$ spacetime can be written in the neighborhood of the boundary as
\begin{align}
ds^2 	&= \ell^2 d\rho^2 + e^{2\rho} g_{(0)ab} dx^a dx^b + \mathcal{O}(e^{\rho}) \label{eq:FeffermanGraham}\\
		&= \ell^2  \frac{dr^2}{r^2} + \frac{r^2}{\ell^2} g_{(0)ab} dx^a dx^b + \mathcal{O}(r) \quad (r = \ell e^\rho ).
\end{align}
The coordinates $(\rho,x^a)$ are called the \textit{Feffermann-Graham} coordinates : $\rho$ is a spacelike coordinate such as $\rho \rightarrow +\infty$ represents the boundary, and $(x^a)$ are the coordinates defined on this boundary, $x^1 = t$ being a timelike coordinate and $x^2 = \phi$ is a spacelike coordinate with closed periodic orbits (an angle). The geometry at the boundary is dictated by the $2d$ metric $g_{(0)ab}$, whose associated covariant derivative will be denoted by $\mathcal{D}_a$.
\end{resultat}
In this framework, the Brown-Henneaux boundary conditions are simply Dirichlet boundary conditions for the boundary metric:
\begin{equation}
g_{(0)ab} = \eta_{ab} \text{ is fixed } 
\end{equation}
where $\eta_{ab}dx^a dx^b = -dt^2 + \ell^2 d\phi^2$. Imposing Fefferman-Graham gauge, the 3 diffeomorphisms depending on arbitrary functions of the 3 coordinates are reduced to the residual diffeomorphisms which are 3 diffeomorphisms depending on arbitrary functions of 2 coordinates. In that sense, the trivial bulk diffeomorphisms are removed, and only the more interesting residual diffeomorphisms remain. The boundary conditions then reduce this set of residual diffeomorphisms to the asymptotic symmetries. In summary, the asymptotic symmetry group can be computed simply by inspecting the set of non-trivial diffeomorphisms that preserve both the Fefferman-Graham asymptotic expansion and the Brown-Henneaux boundary conditions. Let us consider an arbitrary diffeomorphism $\xi = \xi(\rho,x^a)$. The preservation of (\ref{eq:FeffermanGraham}) leads us to
\begin{equation}
\Lie_\xi g_{\rho\rho} = 0 \Rightarrow 2 g_{\rho \mu} \partial_\rho \xi^\mu = 2 g_{\rho \rho} \partial_\rho \xi^\rho = 2 \partial_\rho \xi^\rho = 0 \Leftrightarrow \xi^\rho (\rho,x^a) = R(x^a);
\end{equation}
\begin{equation}
\Lie_\xi g_{\rho a} = 0 \Rightarrow g_{\mu a} \partial_\rho \xi^\mu  + g_{\mu\rho} \partial_a \xi^\mu = g_{ab} \partial_\rho \xi^b  + \partial_a R = 0.
\end{equation}
If we denote by $g^{ab}$ the inverse of $g_{ab}$, we get
\begin{equation}
\partial_\rho \xi^b = - g^{ab} \partial_a R \Rightarrow \xi^b (\rho,x^a ) = V^b (x^a) - \int d\rho \: g^{ab} \partial_a R.
\end{equation}
So we are left with the most general diffeomorphism that preserve the Fefferman-Graham gauge, which depends upon 3 functions of 2 coordinates as announced. Now we impose also the boundary conditions :
\begin{align}
\Lie_\xi g_{ab} = \mathcal{O}(e^\rho) &\Rightarrow \xi^\rho \partial_\rho g_{ab} + \xi^c \partial_c g_{ab} + g_{ac} \partial_b \xi^c + g_{bc} \partial_a \xi^c = \mathcal{O}(e^\rho)\\
&\Leftrightarrow 2 e^{2\rho} g_{(0)ab} R + e^{2\rho} \left( V^c \partial_c g_{(0)ab} + g_{(0)ac} \partial_b V^c + g_{(0)bc} \partial_a V^c \right) = \mathcal{O}(e^\rho) \\
&\Leftrightarrow 2  g_{(0)ab} R +   \Lie_V g_{(0)ab} = 0.
\end{align}
Taking the trace of this latter equation, and recalling that the boundary metric is flat, we get $4R(x^a) = g^{ab}_{(0)} (\mathcal{D}_a V_b + \mathcal{D}_b V_a )$ so 
\begin{equation}
R(x^a) = -\frac{1}{2} \mathcal{D}_c V^c \Longrightarrow {\color{blue}\boxed{ \mathcal{D}_a V_b + \mathcal{D}_b V_a = \mathcal{D}_c V^c \eta_{ab} .}}
\label{eq:CKEBoundary}
\end{equation}
In conclusion, the asymptotic vectors are determined at leading order by a boundary vector field $V^a$ which is in fact a \textit{conformal Killing vector} on the boundary. To solve the conformal Killing equation, let us introduce lightcone coordinates on the boundary $x^{\pm} = (t/\ell) \pm \phi$, such as $ds^2_{(0)} = g_{(0)ab} dx^a dx^b = -dt^2 + \ell^2 d\phi^2 = - \ell^2 dx^+ dx^-$. From (\ref{eq:CKEBoundary}), we get
\begin{equation}
2 \partial_+ V_+ = 0 \Rightarrow V_+ = V_+ (x^-) \Leftrightarrow V^- = g_{(0)}^{-+} V_+ = -\frac{2}{\ell^2} V_+ \Rightarrow V^- = V^- (x^-)
\end{equation}
and symmetrically $V^+ = V^+ (x^+)$. The conformal Killing vectors on the boundary are naturally divided into left-moving and right-moving fields on the cylinder, and we obtain thus 2 infinite families of generators:
\begin{align}
\xi^{(+)} = V^+ (x^+) \partial_+ - \frac{1}{2} \partial_+ V^+ \partial_\rho + \int d\rho \: g^{+-} \partial_+ \partial_+ V^+ \partial_-  ; \label{xiplus} \\
\xi^{(-)} = V^- (x^-) \partial_- - \frac{1}{2} \partial_- V^- \partial_\rho +\int d\rho \: g^{+-} \partial_+ \partial_- V^- \partial_- .
\end{align}
The first term contains the independent function that defines the generator. The second term is completely fixed in terms of the boundary conformal vectors in order to preserve the Feffermann-Graham gauge and it is also leading. The third term is subleading, since $g^{ab} = e^{-2\rho} g_{ab}^{(0)} + \dots$. Therefore, both the first and second term will bring the leading contribution to the conserved charges. 

\subsubsection{Asymptotic symmetry algebra}

Let us look at this asymptotic symmetry algebra of vector fields ! One word of caution however: the asymptotic symmetry algebra requires finite conserved charges, which we will have to check next! First, we develop each vector field into Fourier modes $\xi^{(+)}_m = \xi^{(+)} (V^+ = e^{imx^+}), \: \xi^{(-)} = \xi^{(-)}_n (V^- = e^{inx^-})$ and compute their usual Lie bracket. First, $\xi^{(+)}$ and $\xi^{(-)}$ always commute since they depend at leading order on opposite boundary lightcone coordinates, and the subleading orders trivially follow by the Fefferman-Graham gauge,
\begin{equation}
\left[\xi^{(+)}_m, \xi^{(-)}_n \right]^\mu = \left(\xi^{(+)}_m \right)^\alpha \partial_\alpha \left(\xi^{(-)}_n \right)^\mu - \left(\xi^{(-)}_n \right)^\alpha \partial_\alpha \left(\xi^{(+)}_m \right)^\mu = 0. 
\end{equation}
The asymptotic algebra is thus a direct sum of two chiral subalgebras. Moreover
\begin{align}
\left[\xi^{(+)}_m, \xi^{(+)}_n \right] &= \left(\xi^{(+)}_m \right)^\alpha \partial_\alpha \left(\xi^{(+)}_n \right) - \left(\xi^{(+)}_n \right)^\alpha \partial_\alpha \left(\xi^{(+)}_m \right) \\
&= e^{imx^+} \partial_+ \left( e^{inx^+} \right) - e^{inx^+} \partial_+ \left( e^{imx^+} \right) + o(1) \\
&= i(n-m) \xi^{(+)}_{m+n} \Rightarrow {\color{blue} \boxed{ i \left[\xi^{(+)}_m, \xi^{(+)}_n \right] = (m-n) \xi^{(+)}_{m+n}. }}
\end{align}
The same calculation can be performed for the vectors $\xi^{(-)}$. We have just proven that each chiral subalgebra is isomorphic to the \textit{Witt algebra}, which is the centerless algebra of circle diffeomorphisms. We can now verify that the asymptotic algebra contains the generators of $AdS_3$ exact symmetries. Let us consider simply the subset $\lbrace \xi^{(+)}_{-1} , \xi^{(+)}_0 , \xi^{(+)}_1 \rbrace$. They form a closed subalgebra under the Lie bracket
\begin{equation}
i\left[ \xi^{(+)}_1 , \xi^{(+)}_0 \right] = \xi^{(+)}_1 \quad ; \quad i\left[ \xi^{(+)}_1 , \xi^{(+)}_{-1} \right] = 2 \xi^{(+)}_0 \quad ; \quad i\left[ \xi^{(+)}_0 , \xi^{(+)}_{-1} \right] = \xi^{(+)}_{-1}
\end{equation}
which we recognize as the $sl(2,\mathbb{R})$ algebra. But since it also holds for $\lbrace \xi^{(-)}_{-1} , \xi^{(-)}_0 , \xi^{(-)}_1 \rbrace$ and we know that $sl(2,\mathbb{R}) \simeq so(2,1)$, the asymptotic algebra contains a set of 6 generators which form a subalgebra isomorphic to $sl(2,\mathbb{R}) \oplus sl(2,\mathbb{R}) \simeq so(2,1) \oplus so(2,1) \simeq so(2,2)$, so the asymptotic symmetry group is the natural extension of the exact symmetry group of the asymptotic $AdS_3$ space ! \\

Now, we would like to compute the charges. While the boundary conditions are sufficient to compute the charges, it is useful to first make a detour to the solution space, which will allow us to make the integral on fields in 
\eqref{ch:def} totally explicit! 

\subsubsection{Phase space}

Now that the asymptotic symmetries have been characterized, let us go back to the description of $\mathcal{P}$ itself. Recall that $\mathcal{P}$ is the set of Einstein's solutions which can be written as (\ref{eq:FeffermanGraham}) in the neighborhood of the boundary and obey Brown-Henneaux boundary conditions. In Feffermann-Graham gauge, the remaining metric coefficients are given as an asymptotic expansion in terms of $r = \ell e^\rho$ with $r \rightarrow \infty$. Solving order by order Einstein's equations, one finds that the expansion miraculously \textit{stops} at second order: 
\begin{equation}
ds^2 = \ell^2 \frac{dr^2}{r^2} + \frac{r^2}{\ell^2} \left( g_{(0)ab} + \frac{\ell^2 }{r^2} g_{(2)ab} + \frac{\ell^4}{r^4} g_{(4)ab} \right) dx^a dx^b . 
\end{equation}
Moreover, the second order $g_{(4)ab}$ is completely fixed by the leading orders
\begin{equation}
g_{(4)ab} = \frac{1}{4} g_{(2)ac} g_{(0)}^{cd} g_{(2)db}
\end{equation}
and finally the trace of $g_{(2)ab}$ and its covariant divergence $\mathcal{D}^a g_{(2)ab}$ are also fixed by the equations of motion and the boundary conditions to be zero ! So the form of the metric is nearly totally fixed! In fact the only metric components that are left over to vary independently are the traceless part and divergence-free part of $g_{(2)ab}$, which is naturally interpreted as a boundary stress tensor. To make these remaining boundary degrees of freedom explicit, we use again lightcone coordinates at the boundary. The vanishing trace condition reads as $g_{(2)+-} = 0$ and the conservation condition gives
\begin{equation}
\mathcal{D}^a g_{(2)ab} = 0 \Rightarrow \partial^+ g_{(2)++} = 0 \Leftrightarrow \partial_- g_{(2)++} = 0 \Leftrightarrow g_{(2)++} (x^a) \equiv  \ell^2 L_+ (x^+)
\end{equation}
and immediately $g_{(2)--} (x^a) \equiv   \ell^2 L_- (x^-)$ (the $\ell^2$ factor is conventional). Therefore, each metric in $\mathcal{P}$ can be written into the form
\begin{equation}
ds^2 = \ell^2 \frac{dr^2}{r^2} - \left( r dx^+ -  \ell^2 \frac{L_- (x^-)}{r} dx^- \right) \left( r dx^- -  \ell^2 \frac{L_+ (x^+)}{r} dx^+ \right). 
\label{eq:EinsteinSolFGBH}
\end{equation}
This is the most general (analytic) Einstein solution which obeys the Brown-Henneaux boundary conditions. One may check that all the solutions we have discussed in the previons section can be brought to this form. For example, the \textit{BTZ} black hole \eqref{eq:BTZ} can be written as (\ref{eq:EinsteinSolFGBH}) with constant $L_+$ and $L_-$ such as $M = (L_+ + L_-)/(4G)$ and $J = \ell (L_+ -L_-)/(4G)$. In this patch the outer horizon lies at $r=0$ ! \\

Let us finally complete our discussion by computing the charges associated to this phase space, and check the representation theorem. The set of metrics is now defined by \eqref{eq:EinsteinSolFGBH} and an arbitrary variation of the metric can be written as $h_{\mu\nu} \equiv \delta g_{\mu\nu} = \frac{\partial g_{\mu\nu}}{\partial L_+}\delta L_+ + \frac{\partial g_{\mu\nu}}{\partial L_-}\delta L_-$. We can therefore compute the infinitesimal surface charge using either formalism \eqref{def:ch1} or \eqref{def:ch2}. In fact, the annoying supplementary term $\boldsymbol E$ \eqref{eq:AmbiguityForRG} identically vanishes in Fefferman-Graham coordinates so the surface charge is uniquely defined!  As we have done for the vectors themselves, we can develop the charges into Fourier modes and we obtain 
\begin{align}
\delta \mathcal{L}^{(+)}_m &= \oint_{S^1} \mathbf{k}_{\xi^{(+)}_m} [\delta g ;g] = \frac{\ell}{8\pi G} \int_0^{2\pi} d\phi \: \delta L_+ (x^+) e^{i m x^+} ;\label{eq:ChargePlus} \\
\delta \mathcal{L}^{(-)}_n &= \oint_{S^1 } \mathbf{k}_{\xi^{(-)}_n} [\delta g ;g] = \frac{\ell}{8\pi G} \int_0^{2\pi} d\phi \: \delta L_- (x^-) e^{i n x^-} .
\end{align}
The integration is performed on any circle, either at infinity or at finite $r$. The charges are clearly integrable, 
\begin{align}
\mathcal{L}^{(+)}_m = \frac{\ell}{8\pi G} \int_0^{2\pi} d\phi \:  L_+ (x^+) e^{i m x^+},\qquad \mathcal{L}^{(-)}_m = \frac{\ell}{8\pi G} \int_0^{2\pi} d\phi \:  L_- (x^-) e^{i m x^-}
\label{ch0}
\end{align}
so the representation theorem holds! When we integrated the charges, we chose to define all charges of the zero mass $BTZ$ black hole (with $L_+ = L_- = 0$) as zero. Let us cross-check that these charges form an algebra under the Poisson bracket
\begin{equation}
\left\lbrace  \mathcal{L}^{(+)}_m,  \mathcal{L}^{(+)}_n \right\rbrace = \delta_{\xi^{(+)}_n} \mathcal{L}^{(+)}_m = \oint_{S^1_\infty} \mathbf{k}_{\xi^{(+)}_m} \left[\delta_{\xi^{(+)}_n} g ,g \right] = \frac{\ell}{8\pi G} \int_0^{2\pi} d\phi \: \delta_{\xi^{(+)}_n} L_+ (x^+) e^{i m x^+} . 
\end{equation}
The variation of the first component of the stress tensor $L_+$ can be deduced as follows. We can compute the Lie derivative of $g_{ab}$ on the flow of $\xi^{(+)}$. Since this vector is a generator of the asymptotic group, it must preserve the expansion (\ref{eq:EinsteinSolFGBH}) then at the linear level
\begin{equation}
\Lie_{\xi^{(+)}} g_{ab} = g_{ab} [L_+ + \delta_{\xi^{(+)}} L_+ , L_- + \delta_{\xi^{(+)}} L_- ] - g_{ab} [L_+ , L_- ] .
\end{equation}
Simply by inspecting this relation component by component, one finds that
\begin{align}
\delta_{\xi^{(+)}} L_+ &= V^+ \partial_+ L_+ + 2 L_+ \partial_+ V_+ - \frac{1}{2} \partial^3_+ V_+; \\
\delta_{\xi^{(+)}} L_- &= 0.
\end{align}
The first relation implies that $L_+$ transforms as an element of the coadjoint representation of the Witt algebra, while the second one indicates to us that $\left\lbrace  \mathcal{L}^{(+)}_m,  \mathcal{L}^{(-)}_n \right\rbrace = 0$ $\forall m,n \in \mathbb{Z}$, so the chiral fragmentation also holds at the level of the charge algebra. Let us compute explicitly the Poisson bracket :
\begin{align}
i \left\lbrace  \mathcal{L}^{(+)}_m,  \mathcal{L}^{(+)}_n \right\rbrace &= \frac{i \ell}{8\pi G} \int_0^{2\pi} d\phi \: e^{imx^+} \left( e^{inx^+} \partial_+ L_+ + 2 L_+ \partial_+ e^{inx^+} - \frac{1}{2} \partial_+^3 e^{inx^+} \right) \\
&= \frac{i \ell}{8\pi G} \int_0^{2\pi} d\phi \: \left[ -i(m+n) e^{i(m+n)x^+}L_+  + 2 i n L_+ e^{i(m+n)x^+} - \frac{1}{2} (-i) n^3 e^{i(m+n)x^+} \right] \\
&= \frac{\ell}{8\pi G} (m-n) \left[ \int_0^{2\pi} d\phi \: L_+ e^{i(m+n)x^+} \right]   + \delta_{m+n,0} \frac{\ell m^3}{8G} \\
&= (m-n) \mathcal{L}^{(+)}_{m+n} + m^3 \delta_{m+n,0} \frac{\ell}{8G}
\end{align}
where the second equality was obtained by performing an integration by parts on $\partial_+$ or equivalently on $\partial_\phi$ since the charges are computed at $t=\text{Cst}$. The third equality comes from the integral representation of the discrete $\delta$-function. We see that the representation theorem is obeyed, and the central charge is given by $m^3 \delta_{m+n,0} \frac{\ell}{8G} \equiv \frac{c}{12} m^3 \delta_{m+n,0}$ where the dimensionless \textit{Brown-Henneaux central charge} is:
\begin{equation}
{\color{blue} \boxed{ c= \frac{3\ell}{2G} .}}
\end{equation}
The central extension is obviously zero for the zero-mode $m=0$, but by  shifting the zero mode of the charges \eqref{ch0}, it is possible to cancel it for $m = -1,0,+1$ :
\begin{equation}
\tilde{\mathcal{L}}^{(+)}_m = \mathcal{L}^{(+)}_m + \delta_{m,0} N, \: N \in \mathbb{R} \Rightarrow i \left\lbrace \tilde{\mathcal{L}}^{(+)}_m , \tilde{\mathcal{L}}^{(+)}_n \right\rbrace = (m-n) \left( \tilde{\mathcal{L}}^{(+)}_{m+n} - \delta_{m+n,0} N\right) + \frac{c}{12} m^3 \delta_{m+n,0}.
\end{equation}
We choose $N=c/24$ in order to get a centerless subalgebra of $AdS_3$ exact symmetries:
\begin{equation}
{\color{blue} \boxed{ i \left\lbrace \tilde{\mathcal{L}}^{(+)}_m , \tilde{\mathcal{L}}^{(+)}_n \right\rbrace = (m-n) \tilde{\mathcal{L}}^{(+)}_{m+n} + \frac{c}{12} \: (m^2 - 1)  m \: \delta_{m+n,0} .}}\label{Virch}
\end{equation}
We need to also shift the $(-)$ sector. We also have $\left\lbrace \tilde{\mathcal{L}}^{(+)}_m , \tilde{\mathcal{L}}^{(-)}_n \right\rbrace = 0$, so the check is achieved. The shift $c/24$ of both sectors amounts to a shift of the mass of $c/12=1/(8G)$, which is nothing else than the difference of mass between the zero mass $BTZ$ black hole and global $AdS_3$, as shown in Figure \ref{fig:BTZPhaseSpace}. Everything fits in nicely!\\

This result is a strong hint that $AdS_3$ gravity is deeply related to a $2d$ \textit{CFT}. It took many years to unravel a deeper connection, through one instance of Maldacena's $AdS/CFT$ correspondence \cite{Maldacena:1997re}, which requires much more structure. $AdS_3$ is in this case embedded in $10d$ supergravity itself the low energy limit of string theory. It is remarkable that a relatively simple semi-classical analysis of pure gravity already hints at a holographically dual $2d$ conformal field theory! \\

A more detailed connection with \textit{CFT}s can be also made with what has been presented together with one additional ingredient. The entropy of a high energy state in a $2d$ \textit{CFT} is given by a universal formula known as Cardy's formula, 
\begin{equation}
{\color{blue} \boxed{ S_{CFT} = 2\pi \left( \sqrt{\frac{c_L E_L}{6}} + \sqrt{\frac{c_R E_R}{6}} \right) .}}
\label{eq:CardyMicro} 
\end{equation}
Here, $E_L$ and $E_R$ are the eigenvalues  in the high energy state of $\mathcal{L}_0$ and $\bar{\mathcal{L}}_0$, the zero-modes of the Virasoro algebra, and $c_L$, $c_R$ are the left and right Virasoro central charges. Using the Brown-Henneaux central charge for both the left and right moving Virasoro's and the zero modes of the $BTZ$ black hole, you will find that $S_{CFT}$ is exactly the geometrical $BTZ$ black hole entropy \eqref{SBTZ}! In exact instances of the $AdS/CFT$ correspondence, a black hole can be described as a high energy state in a $CFT$ and its microscopic entropy can be exactly computed from field theory degrees of freedom without gravity! (Now this is not all: finding instances of $AdS/CFT$ correspondences and finding the exact microscopic degrees of freedom for non-supersymmetric black holes are hard problems, which are equivalent to quantifying gravity! This has been done only in a few cases in string theory starting from Maldacena's work).

\section{Asymptotically flat phase space}

\subsection{Flat limit and the \textit{BMS}\textsubscript{3} group}
\begin{figure}[h!]
    \begin{center}
        \begin{tikzpicture}[scale=1]
        	% Couleur violette Powerpoint
       			\definecolor{VioletPPT}{RGB}{204,0,153};
       		% Coordonnées temps/espace conformes
        		\def\x{4.5}; \def\y{-0.375}; \def\ddelta{0.75};
            	\draw[->] (\x,\y) -- (\x+\ddelta,\y) node[right] {$R$};
            	\draw[->] (\x,\y) -- (\x,\y+\ddelta) node[above] {$T$};
            	\fill (\x, \y) circle [radius=1.5pt];
            %\draw[blue] (\x,\y) -- (0,\y); \draw[blue] (\x,\y+\ddelta) -- (0,\y+\ddelta); %(Guides pour le centrage)
            % Coordonnées-lumière conformes
				\def\offset{0.75}; \def\base{2.75};
				\draw[->] (\base,-0.5-\offset) node[left] {$V$} -- (\base+0.5,-0-\offset);
				\draw[<-] (\base,0.5+\offset) node[left] {$U$} -- (\base+0.5,0+\offset);
            	\draw[thick] (3,0) node[right] {$i^0$} -- (0,3) node[above] {$i^+$} -- (0,-3) node[below] {$i^-$} -- cycle;
            % Infinis de genre lumière             
	            \node[right] (1) at (2.5, 2.5) {$\cI^+$};
    	        \node[right] (2) at (2.5,-2.5) {$\cI^-$};
        	    \draw[->,bend right = 15] (1) to (1.5, 1.75);
            	\draw[->,bend left  = 15] (2) to (1.5,-1.75);
            % Géodésiques lumière
            	\draw[-latex, blue] (1,-2) -- (0,-1) -- (2,1);
            	\draw[-latex, cyan] (2,-1) -- (0,1) -- (1,2); 
            % Géodésique temporelle
				\draw[-latex, bend right = 25, red] (0,-3) to (0,3); 
			% Points à l'infini
				\fill (0, 3) circle [radius=2pt];
            	\fill (0,-3) circle [radius=2pt];
            	\fill (3, 0) circle [radius=2pt];
            % Légendes
            	\node [rotate=90] at (-0.3,0) {$r = 0$};
        \end{tikzpicture}
    \end{center}
    \caption{Penrose conformal diagram of Minkowski spacetime $\mathbb{R}^{(3,1)}$. Radial null geodesics are represented in blue, while a radial timelike geodesic is represented in red.}
    \label{fig:PenroseMinkowski}
\end{figure}
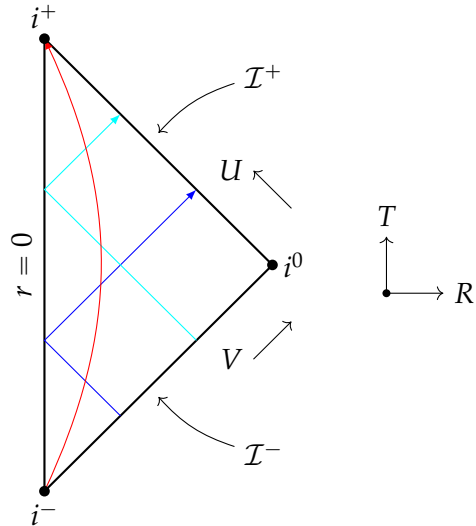

Let us give some elements about the asymptotically flat case. Now the cosmological constant $\Lambda$ vanishes and Einstein's equation in the vacuum are simply $R_{\mu\nu} = 0$. The asymptotic structure of Minkowski spacetime is very different than $AdS$ spacetime. We refer the reader to the Penrose diagram depicted in Figure \ref{fig:PenroseMinkowski},  which also applies in $3d$. We focus on spacetimes that are locally asymptotically flat, \textit{i.e.} whose metric tensor reduce to $\eta_{\mu\nu}$ when some \emph{null} coordinate $r$ reaches infinity. As in the asymptotically $AdS_3$ case, a set of boundary conditions can be formulated \cite{Barnich:2006av}, and the most general solution of Einstein's equations can be \textit{exactly} derived \cite{Barnich:2010eb}. Instead of developing here the full deduction of suitable boundary conditions and the analytical derivation of the solutions, we take advantage of our knowledge about  asymptotically $AdS_3$ spacetimes, and we simply take the limit $\ell \rightarrow \infty$ (which is equivalent to $\Lambda \rightarrow 0$) \cite{Barnich:2012aw}. Geometrically, this process rejects the boundary cylinder to infinity and as the length scale become infinite, the entire bulk looks like the previous "center" of $AdS_3$, and thus is locally flat. Expressed in terms of \textit{Bondi coordinates} $(u,r,\phi)$, where $u$ is the retarded time, $r$ is the luminosity distance, and $\phi$ the angle on the circle at infinity, the solution reads as
\begin{equation}
ds^2 = \Theta (\phi ) du^2 - 2dudr + 2\left[ \Xi(\phi) + \frac{u}{2} \partial_\phi \Theta (\phi) \right] dud\phi + r^2 d\phi^2.
\label{eq:FlatPS}
\end{equation}
The phase space is also parametrized by 2 arbitrary functions on the boundary (here the circle at infinity) but here they depend only on $\phi$ because $x^\pm = (t/\ell ) \pm \phi \rightarrow \phi$ when $\ell \rightarrow \infty$. One can check that this metric is Ricci-flat, and clearly, we retrieve Minkowski spacetime for $\Theta = -1$ and $\Xi = 0$. Again, instead of computing the asymptotic symmetry group that preserves the phase space, we can directly take the``flat limit" of the asymptotic symmetry group of the $AdS_3$ phase space. To do this, we first define 
\begin{equation}
\xi^{(+)}_m = \frac{1}{2} (\ell P_m + J_m ) \quad ; \quad \xi^{(-)}_n = \frac{1}{2} (\ell P_{-n} - J_{-n} ).
\end{equation}
A straightforward computation shows that
\begin{equation}
i[P_m,P_n] = \frac{1}{\ell^2} (m-n) J_{m+n} \quad ; \quad i[J_m,J_n] = (m-n) J_{m+n} \quad ; \quad i[J_m,P_n] = (m-n) P_{m+n}.
\end{equation}
So when we take the flat limit $\ell \rightarrow \infty$, the $P_m$ commute, the $J_m$ form a Witt algebra, and act non-trivially on the $P_m$.
\begin{equation}
{\color{blue} \boxed{ i[P_m,P_n] = 0 \quad ; \quad i[J_m,J_n] = (m-n) J_{m+n} \quad ; \quad i[J_m,P_n] = (m-n) P_{m+n} .}}
\end{equation}
These commutation relations define the so-called $bms_3$ algebra ! Let us first note that it extends the Poincaré algebra $iso(2,1)$, in the same way that the asymptotic symmetry algebra of the $AdS_3$ phase space extended the $so(2,2)$ algebra of exact symmetries of $AdS_3$ : the first modes $m,n = -1,0,+1$ form a subalgebra of $bms_3$ containing (check it as an exercise) :
\begin{itemize}[label=$\rhd$]
\item $2+1$ translations : $\partial_t = P_0 , \partial_x = P_{+1} + P_{-1}, \partial_y = i(P_{+1} - P_{-1})$ ;
\item $1$ rotation $\partial_\phi = R_0$ and two boosts $x\partial_t - t \partial_x = J_{+1} + J_{-1}$, $y\partial_t - t \partial_y = i(J_{+1} - J_{-1})$.
\end{itemize}
We see that the asymptotic symmetry group of asymptotically flat spacetimes is also larger than the exact symmetry group of the flat spacetime itself ! Instead of $3$ translations, we get an abelian subalgebra of $bms_3$,  usually denoted as $vect(S^1)_{ab}$, that contains the so-called \textit{supertranslations} $P_m$. The Lorentz algebra $so(2,1)$ is also enhanced into an infinite-dimensional algebra of diffeomorphisms on the circle, $vect(S^1)$, which now contains the so-called \textit{superrotations} $J_m$. The last commutation relations tell us that the supertranslations form an ideal as do the translations in the Poincar\'e subalgebra. Since $vect(S^1)$ acts on $vect(S^1)_{ab}$ as the adjoint representation, we can write
\begin{equation}
bms_3 = vect(S^1) \oright_{ad} vect(S^1)_{ab}.
\end{equation}
The asymptotic symmetry group is the integral version of this algebra, and reads as a semi-direct product $BMS_3 = \text{Diff}(S^1) \ltimes_{Ad} \text{Vect}(S^1)_{ab}$ between the group of diffeomorphisms on the circle, and its own Lie algebra, seen here as a abelian normal subgroup. \\

We can obtain by a fairly similar process the $bms_3$ charge algebra. Before taking $\ell \rightarrow \infty$, we again define
\begin{equation}
\mathcal{L}^{(+)}_m = \frac{1}{2} (\ell \mathcal{P}_m + \mathcal{J}_m ) \quad ; \quad \mathcal{L}^{(-)}_n = \frac{1}{2} (\ell \mathcal{P}_{-n} - \mathcal{J}_{-n} ).
\end{equation}
Recalling the Virasoro charge algebra \eqref{Virch}, one can check that
\begin{align}
i[\mathcal{P}_m,\mathcal{P}_n] &= \frac{1}{\ell^2} (m-n) \mathcal{J}_{m+n} \quad ; \\
i[\mathcal{J}_m,\mathcal{J}_n] &= (m-n) \mathcal{J}_{m+n} \quad ; \\
i[\mathcal{J}_m,\mathcal{P}_n] &= (m-n) \mathcal{P}_{m+n} + \frac{1}{4G} m(m^2-1) \delta_{m+n,0}.
\end{align}
The central charge is now free of any $\ell$. The charges remain finite in the flat limit, and their algebra reads as 
\begin{align}
i[\mathcal{P}_m,\mathcal{P}_n] &= 0 \quad ; \\
i[\mathcal{J}_m,\mathcal{J}_n] &= (m-n) \mathcal{J}_{m+n} \quad ; \\
i[\mathcal{J}_m,\mathcal{P}_n] &= (m-n) \mathcal{P}_{m+n} + \frac{1}{4G} m(m^2-1) \delta_{m+n,0}.
\end{align}
The centerless part of this algebra forms an algebra isomorphic to $iso(2,1)$, so the lowest modes $m,n = -1,0,+1$ are the Poincaré charges (energy, linear and angular momentum and Lorentz charges). By direct analogy, the $\mathcal{P}_m$ are called \textit{supermomenta} while the $\mathcal{J}_m$ receive the name of \textit{super-Lorentz charges} or \textit{superrotation charges}. \\

Again using the flat limit process, it is possible to compute the explicit form of the $bms_3$ vectors. What we have to do is simply take the expressions of $\xi^+,\xi^-$, compute the transformation to pass into $P,J$ vectors, then take the flat limit and express the result in Bondi coordinates (see again \cite{Barnich:2012aw}) :
\begin{equation}
\xi_{Y,T} = \xi^\mu \partial_\mu : \left\lbrace 
\begin{array}{rcl}
\xi^u &=& T(\phi) + u \partial_\phi R(\phi) + \mathcal{O}(r^{-1}); \\ 
\xi^r &=& -r \partial_\phi R(\phi) + \mathcal{O}(r^{0})  ;\\ 
\xi^\phi &=& R(\phi) - \frac{u}{r} \partial_\phi^2 R (\phi) + \mathcal{O}(r^{-1}).
\end{array} 
\right.
\label{eq:GeneratorsBMS3}
\end{equation}
The generators depend on two arbitrary functions on the circle, $T(\phi)$ representing arbitrary supertranslations, and $R(\phi)$ representing arbitrary superrotations! We obtain $P_m$ with $(T=e^{im\phi},R=0)$ and $J_n$ with $(T=0,R=e^{in\phi})$. The corresponding surface charges are 
\begin{align}
\mathcal P_n &= \frac{1}{16 \pi G} \int_0^{2\pi} d\phi (\Theta(\phi) + 1) e^{i n \phi} ,\\
\mathcal J_n &= \frac{1}{8 \pi G} \int_0^{2\pi} d\phi \Xi(\phi) e^{i n \phi}. 
\end{align}
We can transform (\ref{eq:FlatPS}) along the flow of $\xi_{T,R}$ to deduce the transformation laws of the metric fields $\Theta,\Xi$ :
\begin{align}
\delta_{T,R} \Theta &= R \partial_\phi \Theta + 2 \partial_\phi R \Theta - 2 \partial_\phi^3 R, \label{eq:VarTheta} \\
\delta_{T,R} \Xi &= R \partial_\phi \Xi + 2 \partial_\phi R \Xi + \frac{1}{2} T \partial_{\phi} \Theta + \partial_\phi T \Theta - \partial_\phi^3 T .\label{eq:VarXi}
\end{align}
We can now infer the ``boundary field content'' of the phase space. By virtue of (\ref{eq:VarTheta}), $\Theta$ belongs to the coadjoint representation of $\text{Diff}(S^1)$. We can introduce what we call the superrotation field $\Psi(\phi)$ which is invariant under supertranslations and which transforms under superrotations as
\begin{align}
\delta_{T,R}\Psi = R \partial_\phi \Psi + \partial_\phi R. \label{tr5}
\end{align}
The transformation \eqref{eq:VarTheta} allows to recognize $\Theta = (\partial_\phi \Psi)^2-2\partial_\phi^2\Psi +8 G M e^{2\Psi}$. When $\Psi = 0$, one is left with a zero mode which cannot be generated by a diffeomorphism $\Theta_0 = 8GM$. The mass $M$ is recognized after computing the charges, as the canonical conjugated charge to $P_0 = \partial_t$. To untangle the second transformation law \eqref{eq:VarXi} a second fundamental boundary field is necessary which we call the supertranslation field $C(\phi)$. After some algebra, we find convenient to define $\Xi = \Theta \partial_\phi C-\partial^3_\phi C+4 G J e^{2\Psi}+\frac{1}{2}\partial_\phi \Theta C$. The transformation property \eqref{eq:VarXi} is reproduced from \eqref{tr5} and
\begin{align}
\delta_{T,R}C = T+R \partial_\phi C - C \partial_\phi R. 
\end{align}
The zero mode $\Xi_0 = 4GJ$ is recognized after computing the charges as determined by the angular momentum conjugated to $-\partial_\phi$. 
In summary, the field space is parameterized by the supertranslation field $C(\phi)$, the superrotation field $\Psi(\phi)$ and the two zero modes $M,J$. This description is slightly redundant, because not all the modes of $C$ and $\Psi$ lead to distinct metrics (for example the lowest 3 harmonics of $C$ are annihilated by $\partial_\phi (\partial_\phi^2 +1) $ and therefore they do not modify $\Xi$). Studying these subtleties is called studying the orbits of the $BMS_3$ group and it has been done in detail, we refer the reader to \cite{Oblak:2016eij}! \\

Let us finally discuss how Minkowski spacetime transforms under supertranslations and superrotations. First, it does not transform under the Poincar\'e subgroup since these are isometries. In the language that we just developped, $\Theta=-1$ and $\Xi=0$ are left invariant under Poincar\'e transformations, because $\partial_\phi (\partial_\phi^2 +1)R=\partial_\phi (\partial_\phi^2 +1)T=0$. Now, acting with general supertranslations and superrotations, the metric changes and the canonical charges also change. If one only acts with supertranslations, the energy remains zero, so the vacuum is degenerate and parameterized by its superrotation charge! The field $C(\phi)$ is precisely the Goldstone boson which comes from the spontaneous breaking of supertransation invariance. 
When also acting with superrotations, the energy now changes and one finds new (classical) states with supertranslation charges, which are related to Minkowski by the action of superrotations.

\subsection{Constant representatives : spinning particles and flat cosmologies}

To conclude our discussion of the flat case, let us focus on constant representatives of the phase space. As we did for the asymptotically $AdS_3$ phase space, let us reduce $\Theta,\Xi$ to their zero modes in terms of $M,J$ and represent the phase diagram. If we look at the previous phase space (Figure \ref{fig:BTZPhaseSpace}), we see that the flat limit $\ell \rightarrow \infty$ cancels the slope of the extremal lines. A second effect is that the upper half plane is also not anymore filled by black holes in the flat case: the reason is that by sending $\ell \rightarrow \infty$ we zoom in the interior of the black holes so we are left with cosmological spacetimes without horizon! This is consistent with a theorem due to Ida \cite{Ida:2000jh} that says that black holes cannot exist in $2+1$ dimensions when the cosmological constant is not negative assuming reasonable matter (obeying the null energy condition). Below the massless line, we find again spinning particles that we will describe a bit more here ! \\

For later convenience, let us note $\bar{M} = 8GM$ and $\bar{J} = 4GJ$. For constant representatives, the metric reads simply as
\begin{equation}
ds^2 = \bar{M} du^2 - 2 du dr + \bar{J} du d\phi + r^2 d\phi^2.
\end{equation}
When $\bar{M}>0$ and $\bar{J}$ is arbitrary, we get an expanding spacetime (or a contracting spacetime after time reversal) enclosing an (unphysical) time machine hidden by a cosmological horizon! Let us define $\alpha^2 = \bar{M}$ and $r_0 = |\bar{J}/\alpha|$. The spacetime is clearly divided in two distinct parts:
\begin{itemize}[label=$\rhd$]
\item If $r < r_0$ we can perform the change of coordinate $r^2 \rightarrow \bar{r}^2 = (r_0^2 -r^2)/\alpha^2$ to get  the line element $ds^2 = (-\alpha dt + r_0 d\phi)^2 + d\bar{r}^2 - \alpha^2 \bar{r}^2 d\phi^2$. In this inner region, $\partial_\phi$ is always timelike, and generates closed timelike curves. This unphysical ``time machine'' is shielded from the rest of the spacetime by the cosmological horizon $r=r_0$ \cite{Barnich:2012xq,Bagchi:2012xr} ;
\item In the outer region ($r > r_0$), one can define new coordinates $T^2 = (r^2 -r_0^2)/\alpha^2$ and $X = (\alpha t/r_0)+\phi$ such that $(X,\phi) \sim (X + 2\pi,\phi + 2\pi)$. The metric becomes $ds^2 = -dT^2 + r_0^2 dX^2 + \alpha^2 T^2 d\phi^2$, which describes a spacetime expanding with growing $T$ with $T = 0$ as the big bang. The closed timelike curves of the inner region are enclosed in the ``pre big-bang'' era, which we need to cut out. This kind of cosmological spacetime is called a \textit{flat cosmology} !
\end{itemize}
Let us consider the subset of metrics for which $\bar{M}<0$. Let us denote $\alpha^2 = -\bar{M} > 0$ ($\alpha=\sqrt{-M}$) and $r_0 = \bar{J}/\alpha$. The change of coordinate $\bar{r}^2 = (r^2 + r_0^2)/\alpha^2$ puts the line element into the form
\begin{equation}
ds^2 = -\left(\alpha dt - \frac{4GJ}{\alpha} d\phi \right)^2 + d\bar{r}^2 + \alpha^2 \bar{r}^2 d\phi^2
\end{equation}
which is nothing but a \textit{spinning particle} found in 1984 by Deser, 't Hooft and Jackiw \cite{Deser:1983tn}. We can give a more geometrical interpretation of this line element by performing a second change of coordinates $\bar{t} = \alpha t - r_0 \phi$, $\bar \phi = \alpha \phi$. In these coordinates, $ds^2 = -d\bar{t}^2 + d\bar{r}^2 + \bar{r}^2 d\bar \phi^2$ : it is now manifest that the spacetime is locally flat ! But the difference between a spinning particle and the Minkowskian vacuum lies in the periodic identification :
\begin{equation}
(t,\phi) \sim (t,\phi+2\pi) \rightarrow (\bar{t},\bar{\phi}) \sim \left( \bar{t} - \frac{8\pi G J}{\alpha} , \bar{\phi} + 2\pi\alpha \right).
\end{equation}
For $J \neq 0$, there is a twist in the time identification, which leads to spin.  Let us discuss the static case $J=0$. The phase space is summarized in Figure \ref{fig:FlatPhaseSpace}.  \\

\begin{figure}[h!]
\centering
\begin{tikzpicture}[scale=1]
\draw[white] (-4,-4) -- (-4, 4) -- (4,4) -- (4,-4);
\coordinate (A)   at (-3.0,-2.6);
\coordinate (B)   at (-3.0, 2.6);
\coordinate (C)   at ( 3.0, 2.6);
\coordinate (D)   at ( 3.0,-2.6);
\coordinate (O)   at ( 0.0, 0.0);
\coordinate (Mink)at ( 0.0,-1.3);
\coordinate (MinL)at (-3.0,-1.3);
\coordinate (MinR)at ( 3.0,-1.3);
\fill[magenta!30] (B) -- (-3.0, 0) -- ( 3.0, 0) -- (C);
\fill[orange!30] (MinL) -- (-3.0, 0) -- ( 3.0, 0) -- (MinR);
\fill[red!30] (A) -- (MinL) -- (MinR) -- (D) -- cycle;
\draw[->,thick] (0,-3.5) -- (0,3.5) node[above]{\footnotesize $M$};
\draw[->,thick] (-3.5,0) -- (3.5,0) node[right]{\footnotesize $J$};
\node[magenta,text width=2.5cm, fill=magenta!30,align=center] at (0,2.0) {\footnotesize \textbf{Flat cosmologies}};
\node[fill=orange!30,text width=0.5cm] at (0,-0.7) {$\phantom{A}$};
\node[orange,align=center] at (0,-0.7) {\footnotesize \textbf{Conical defects}};
\node[red,align=center,fill=red!30,text width=4cm] at (0,-1.3-0.7) {\footnotesize \textbf{Conical excesses}};
\fill[fill=blue] (Mink) circle (0.1) node[blue,right]{\footnotesize \textit{Minkowski}};
\draw[densely dashed] (Mink) -- (MinL);
\draw[<->] (MinL) -- ( $(MinL)-(0,-1.3)$ );
\node[left] at ( $(AdSL)-(1.5,-0.65)$ ) {$-\frac{1}{8G}$};
\end{tikzpicture}
\caption{Constant representatives of the asymptotically flat phase space.\\ \tiny{Figure adapted with permission from \cite{Barnich:2012aw}.Copyrighted by the American Physical Society.} }
\label{fig:FlatPhaseSpace}
\end{figure}

\begin{itemize}[label=$\rhd$]
\item For $\bar{M} = -1$ or, equivalently, $\alpha = 1$, one has Minkowski spacetime. 

\item For $-1 < \bar{M} < 0$ ($\alpha^2 < 1$), we find a conical defect. The spacetime can be created from a cut and paste procedure. Cut a wedge of angle $2\pi( 1 - |\alpha|)$ out of the plane and glue the remaining edges. This is a conical defect. 

\item When $\bar{M} < -1$ ($\alpha^2 > 1$), we find a conical excess. This is equivalent to incise the plane along a half-line and then introduce an excendentary section of angle $2\pi( |\alpha| - 1)$ between the two edges of the incision. In this case, the energy spectrum is not bounded from below, so we often discard these solutions.
\end{itemize}

Relaxing the hypothesis of analyticity, multi-particle solutions can be found \cite{Deser:1983tn}, and each of them may carry a $BMS$ representation! Such a general metric has not yet been described. As already mentioned, these particles will not attract, as we have already stressed before; the Newtonian potential does not exist in $3d$ gravity ! On this remark, we close the presentation of the asymptotically flat phase space, with the hope that through it, you got more  intuition about the intriguing properties of $3d$ gravity !

%\newpage 
\section{Chern-Simons formulation}

The last topic that we will discuss concerning $3d$ gravity  is a reformulation of the theory. Einstein's theory with $\Lambda < 0 $ is equivalent (at least classically) to the difference of two Chern-Simons actions of non-abelian gauge fields which both transform under the adjoint representation of $SO(2,1)$. This is the Chern-Simons formulation of $3d$ Einstein gravity, first discovered by Achúcarro and Townsend in 1986 \cite{Achucarro:1987vz}. \\

First, let us reset our notations. Spacetime indices will still be denoted as $\mu,\nu,...$ but now latin indices will represent Lorentz indices in the local triad frame. We will denote the Levi-Civita connection compatible with $g_{\mu\nu}$ as $\nabla_\mu$ and reserve $D$ as the covariant derivative on objects that transform under the gauge group! We will write forms in bold, except the triad and spin connection, which is conventional. 

\subsection{Local Lorentz triad}

At each point of the manifold $M$, we can find a local change of frame in which the metric is locally flat (this is the equivalence principle). The natural basis of this frame is given by a \textit{triad} of Lorentz vector-valued 1-forms $\lbrace e^a = e^{a}_{\phantom{a}\mu} dx^\mu \rbrace$ which obey $g_{\mu\nu} dx^\mu dx^\nu = \eta_{ab} e^a e^b$. Since the metric admits an inverse, we can also define the inverse of $e^a_{\;\mu}$ which we denote by $e_a^{\; \mu}$ ($e_a^{\; \mu} e^b_{\; \mu} = \delta_a^b$ and $e_a^{\; \mu} e_{\; \nu}^a = \delta_\nu^\mu$). With respect to this orthonormal local basis, the connection coefficients are given by $e^{\; \mu}_a \nabla_\mu e_b\equiv \nabla_a e_b = e_c \omega^c_{\phantom{c}ab}$. Recall that for a vector $\chi = \chi^\alpha e_{\alpha}$ we have $\nabla_\mu \chi = (\partial_\mu \chi^\alpha + \Gamma^\alpha_{\mu\nu} \chi^\nu ) e_\alpha$ so we can inverse the previous relation to obtain the equation linking the connection coefficients $\Gamma^\alpha_{\mu\nu}$ in the coordinate basis with the connection coefficients $\omega^c_{\phantom{c}ab}$ in the orthonormal basis :
\begin{equation}
\omega^c_{\phantom{c}ab} = e^{c}_{\phantom{c}\nu} e_{a}^{\phantom{a}\mu} (\partial_\mu e_{b}^{\phantom{b}\nu} + e_{b}^{\phantom{b}\alpha} \Gamma^\nu_{\mu\alpha} ) = e^{c}_{\phantom{c}\nu} e_{a}^{\phantom{a}\mu} \nabla_\mu e_{b}^{\phantom{b}\nu}.
\label{eq:EqConnection}
\end{equation}
We also define the \textit{spin connection} $\omega^a_{\phantom{a}b} \triangleq \omega^a_{\phantom{a}bc} e^c$ which are tensor-valued 1-forms on $M$ that belong to the adjoint representation of the local Lorentz group. Indeed, under a Lorentz rotation of the triad, 
\begin{equation}
\omega'^a_{\phantom{a}b} = \Lambda^a_{\phantom{a}c} \omega^c_{\phantom{c}d} (\Lambda^{-1})^d_{\phantom{d}b} + \Lambda^a_{\phantom{a}c} (d\Lambda^{-1})^c_{\phantom{c}b} .
\end{equation}

%We can rewrite (\ref{eq:EqConnection}) in a more economical way by contracting both sides with the triad vector basis $e^a$ (and use $dx^\mu = e_a^{\phantom{a}\mu} e^a$) :
%\begin{align}
%&\omega^c_{\phantom{c}b} = e^c_{\phantom{c}\nu} (d e_b^{\phantom{b}\nu} + e_b^{\phantom{b}\alpha} \Gamma^{\nu}_{\rho\alpha} dx^\rho ) \quad \Leftrightarrow 
%\quad d e^a_{\phantom{a}\nu} + \omega^a_{\phantom{a}b} e^b_{\phantom{b}\nu} = e^a_{\phantom{a}\mu} \Gamma^{\mu}_{\alpha\nu} dx^\alpha .
%\end{align}
We can also use the flat metric $\eta_{ab}$ and its inverse $\eta^{ab}$ to lower and raise local Lorentz indices, $\omega_{ab}=\eta_{ac}\omega^c_{\phantom{c}b} $. The metric-compatibility of the connection (\textit{i.e.} $\nabla_\alpha g_{\mu\nu}= 0$) reads as $\omega_{ab} = -\omega_{ba}$ in this formalism, so the spin connection can be seen as antisymmetric $3\times 3$ matrices of one-forms.\\

In $3d$ spacetime, the special feature that leads to a further reformulation is that \textit{antisymmetric matrices $\omega_{ab}$ are dual to vectors}, since the completely antisymmetric tensor $\varepsilon^{abc}$ exists in $3d$ (with $\varepsilon^{012}=1$). So we are able to construct a local Lorentz vector from the spin connection :
\begin{equation}
\omega^a \triangleq \frac{1}{2} \varepsilon^{abc}\omega_{bc} \Leftrightarrow \omega_{ab} = -\varepsilon_{abc}\omega^c
\end{equation}
or, equivalently, a Lorentz one-form $\omega_a = \eta_{ab}\omega^b$. Therefore, it is possible to treat the spin connection $\omega_a$ and the triad $e_a$ on an equal footing!

\subsection{Chern-Simons action}

Using the existence of the cosmological length scale $\ell$, $\Lambda=-1/\ell^2$, we can introduce the dimensionally consistent \textit{connections}
\begin{equation}
\mathbf A^a = \omega^a + \frac{e^a}{\ell} \quad ; \quad \bar{\mathbf{A}}^a = \omega^a - \frac{e^a}{\ell}.\label{defAAb}
\end{equation}
Let us now denote by $J_a$ a set of matrices of $so(2,1)$ that obey the algebra $[J_a,J_b] = \varepsilon_{abc} \eta^{cd} J_d \equiv \varepsilon_{ab}^{\phantom{ab}c} J_c$. Since $\mathbf A^a$ carries one Lorentz index, it can be understood as the components of a $so(2,1)$ vector in the matricial base $\lbrace J_a \rbrace$. We will consider a matrix representation of the connection:
\begin{equation}
\mathbf A = \mathbf A^a J_a = \omega^a J_a + \frac{e^a J_a}{\ell} \equiv \omega + \frac{e}{\ell}
\end{equation}
and the same for the other connection $\bar{\mathbf A} = \bar{\mathbf A}^a J_a$. We can say that $\mathbf A$ and $\bar{\mathbf A}$ transform under the \textit{adjoint representation} of $SO(2,1)$ ! \\

Since we are working with a local orthonormal triad, we can make profit of Cartan's calculus. In particular, the \textit{second Cartan structure equation} links curvature with the spin connection:
\begin{equation}
\mathbf R^a_{\phantom{a}b} = d\omega^a_{\phantom{a}b} + \omega^a_{\phantom{a}c} \wedge \omega^c_{\phantom{c}b}. 
\end{equation}

From $R^a_{\phantom{a}b}$ we can reconstruct the components of the Riemann tensor, given that
\begin{equation}
\mathbf R^a_{\phantom{a}b} = \frac{1}{2} R^a_{\phantom{a}bcd} e^c \wedge e^d = \frac{1}{2} R^a_{\phantom{a}b\mu\nu} dx^\mu \wedge dx^\nu \Rightarrow R^\alpha_{\phantom{\alpha}\beta \mu\nu} = e_a^\alpha e^b_\beta R^a_{\phantom{a}b\mu\nu}.
\end{equation}
We can peform a contraction between $\mathbf R_{ab}$ and $\varepsilon^{abc}$ to get a vector in the Lorentz frame
\begin{equation}
\mathbf R^a \triangleq \frac{1}{2} \varepsilon^{abc} \mathbf R_{bc} = d\omega^a + \frac{1}{2} \varepsilon^a_{\phantom{a}bc} \omega^b \wedge \omega^c .
\end{equation}
Since $\mathbf R^a$ also transforms under the adjoint representation of the local Lorentz group, we can build a matrix representation of it:
\begin{equation}
\mathbf{R} \triangleq \mathbf R^a J_a = d\omega + \frac{1}{2} [J_c,J_f] \omega^c \wedge \omega^f  = d\omega + \omega \wedge \omega.
\end{equation}
The two-form $\mathbf{R}$ should not be confused with the Ricci scalar $R$. \\

Now that we have defined all the necessary geometrical objects, we are ready to show that the Einstein-Hilbert action for $3d$ gravity is equivalent (up to a boundary term) to a couple of Chern-Simons actions, one for each connection. The latter is build up from a $3$-form, because the spacetime is $3$-dimensional.  The scalar action is obtained by integration of this $3$-form on $M$. The most simple $3$-forms that one can construct from $\mathbf A$ are $\mathbf A\wedge d\mathbf A$ and $\mathbf A \wedge \mathbf A \wedge \mathbf A$. These are the only two terms which appear in the \textit{Chern-Simons form}
\begin{equation}
I[\mathbf A] = \mathbf A \wedge d\mathbf A + \frac{2}{3} \mathbf A \wedge \mathbf A \wedge \mathbf A . 
\end{equation}
After a straightforward computation that can be done without any subtlety, we obtain
\begin{align}
\text{tr} (I[\mathbf A] - I[\bar{\mathbf A}]) = \frac{2}{\ell} \text{tr} \left[ 2 e \wedge \mathbf{R} + \frac{2}{3\ell^2} e \wedge e \wedge e - d(\omega \wedge e) \right].
\end{align}
where the trace is taken on the representation of $so(2,1)$ to which belong the matrices $\mathbf A$, $\mathbf R$,... On the $so(2,1)$ algebra we can define a Killing product such as $k(J_a,J_b) = \frac{2}{N} \text{tr} (J_a J_b) \equiv \eta_{ab}$. The normalisation $N$ is representation-dependent, and as we will work with the natural $2d$ representation, we pick the value $N = 1$ which is the standard normalisation for this representation. So the explicit computation of the trace gives
\begin{equation}
\text{tr} \left[ 2 e \wedge \mathbf{R} + \frac{2}{3\ell^2} e \wedge e \wedge e - d(\omega \wedge e)  \right] = \frac{1}{2} \varepsilon_{abc} \left[ e^a \wedge \mathbf R^{bc} + \frac{1}{3\ell^2}  e^a \wedge e^b \wedge e^c \right]- d(\omega^a \wedge e_a).
\end{equation}
Recalling the definition of a determinant, one has $\det(e) = \det(e^a_\mu) = \sqrt{-g}$ and furthermore
\begin{equation}
\varepsilon_{abc} \: e^a \wedge \mathbf R^{bc} = \sqrt{-g} \: R \: d^3 x \quad ; \quad \varepsilon_{abc} \: e^a \wedge e^b \wedge e^c = 3! \: \rmg \: d^3 x
\end{equation}
where $R$ is here the Ricci scalar associated to $g_{\mu\nu}$. Finally
\begin{equation}
\text{tr} (I[\mathbf A] - I[\bar{\mathbf A}]) = \frac{2}{\ell} \left( \frac{1}{2} \sqrt{-g} R + \frac{1}{\ell^2} \rmg \right) d^3 x - \frac{2}{\ell} d(\omega^a \wedge e_a ).
\end{equation}
We have thus showed that the difference of two Chern-Simons actions:
\begin{equation}
{\color{blue} \boxed{ S_{CS} [\mathbf A,\bar{\mathbf A}] = \frac{k}{4\pi} \int_M \text{tr} (I[\mathbf A] - I[\bar{\mathbf A}])}} ,\quad\quad k \triangleq \frac{\ell}{4G} \text{ (dimensionless)} 
\label{eq:ChernSimons}
\end{equation}
is equivalent to the $3d$ Einstein-Hilbert action, up to a boundary term that has no effect on the equations of motion:
\begin{equation}
S_{CS} [\mathbf A,\bar{\mathbf A}] = \frac{1}{16\pi G} \int_M d^3 x \: \rmg\:  ( R + 2\Lambda ) - \frac{1}{16\pi G} \int_{\partial M} \omega^a \wedge e_a = S_{EH} [g] + \text{Boundary term}.
\end{equation}
So we can use the Chern-Simons formulation of $3d$ gravity, which presents the very nice advantage to be based on gauge fields which belong to the adjoint  representation of $so(2,1)$! \\

It is a simple matter of tensorial calculus to get the equations of motion associated to (\ref{eq:ChernSimons}). First let us develop all exterior products in components. We are allowed to use spacetime indices to write explicitly $I[\mathbf A], I[\bar{\mathbf A}]$ since all objects are tensorial...
\begin{equation}
L_{CS} [\mathbf A,\bar{\mathbf A}] = \frac{k}{4\pi} \text{tr} (I[\mathbf A] - I[\bar{\mathbf A}]) = \frac{k}{4\pi} \text{tr} \left[ - \varepsilon^{\mu\alpha\beta} \left( A_\mu \partial_\alpha A_\beta + \frac{2}{3} A_\mu A_\alpha A_\beta \right)  \right] - (\bar{\mathbf A}\text{-part}).
\end{equation}
The Euler-Lagrange equations for the gauge field $\mathbf A$ then read as
\begin{align}
\frac{\delta L_{CS}}{\delta A_\mu} &= \frac{\partial L_{CS}}{\partial A_\mu} - \partial_\alpha \frac{\partial L_{CS}}{\partial\partial_\alpha A_\mu} \\
&= -\frac{k}{4\pi} \varepsilon^{\mu\alpha\beta} \left( \partial_\alpha A_\beta + 2  A_\nu A_\beta \right) - \frac{k}{4\pi} \left( -\varepsilon^{\beta\alpha\mu} \partial_\alpha A_\beta \right) \\
&= -\frac{k}{2\pi} \varepsilon^{\mu\alpha\beta} \left( \partial_\alpha A_\beta + A_\alpha A_\beta \right) \\
&= -\frac{k}{2\pi} \varepsilon^{\mu\alpha\beta} F_{\alpha \beta} 
\end{align}
where we have defined the curvature tensor $\mathbf F$ associated to the connection $\mathbf A$ as $\mathbf F = d\mathbf A + \mathbf A \wedge \mathbf A \Rightarrow F_{\mu\nu} = \partial_\mu A_\nu - \partial_\nu A_\mu + [A_\mu , A_\nu ]$. $F_{\mu\nu}$ is antisymmetric by construction. The equations of motion exactly state that \textit{the connections $\mathbf A$ and $\bar{\mathbf A}$ are flat} !
\begin{equation}
{\color{blue} \boxed{
\mathbf F = d\mathbf A + \mathbf A\wedge \mathbf A = 0 \quad\text{\&}\quad \bar{\mathbf F} = d\bar{\mathbf A} + \bar{\mathbf A}\wedge \bar{\mathbf A} = 0.
}}
\end{equation}
Any flat solution is locally pure gauge, \textit{i.e.} of the form
\begin{equation}
\mathbf A = g^{-1} dg,
\end{equation}
as we can check straighforwardly
\begin{align}
d\mathbf A + \mathbf A\wedge \mathbf A &= dg^{-1} \wedge dg + g^{-1} dg  \wedge g^{-1} dg  \\
&= -g^{-1} dg \wedge g^{-1} dg + g^{-1} dg  \wedge g^{-1} dg = 0.
\end{align}

\subsection{General covariance and charges}

We now know the relationship between Chern-Simons theory and $3d$ Einstein gravity.  The latter theory is by construction invariant under arbitrary diffeomorphisms, so a natural question is: how is a diffeomorphism represented in terms of gauge transformations of the fundamental fields $\mathbf A$ and $\bar{\mathbf A}$ ? First recall that an infinitesimal gauge transformation with parameter $\lambda = \lambda^a J_a$ acts on the connection field as $\delta_\lambda A_\mu = D_\mu \lambda$ where $D_\mu$ is the gauge-covariant derivative defined by $D_\mu \lambda = \partial_\mu \lambda + [A_\mu , \lambda ]$. The second term is necessary for non-abelian gauge groups. \\

%As a first hint, we can study the action of diffeomorphisms on the Chern-Simons action: $A$ is not only a gauge field, it is a $1$-form for the manifold $M$, so the variation under a diffeomorphism $\xi^\mu$ is given by the Lie derivative $\delta_\xi A = \Lie_\xi A$. By virtue of Cartan's magic formula (\ref{eq:Cartan}) we get
%\begin{align}
%\delta_\xi A &= d(i_\xi A) + i_\xi dA 
%= d(i_\xi A) + i_\xi F - i_\xi (A \wedge A) \\
%&= d(i_\xi A) + [A,i_\xi A] + i_\xi F 
%= D(i_\xi A) + i_\xi F 
%\approx D(i_\xi A) \equiv D(\lambda)
%\end{align}
%where we denote by $D$ the gauge-covariant exterior derivative and we have explicitly used the equations of motion $F \approx 0$. So on-shell the diffeomorphisms are represented by gauge parameters $\lambda^a = A^a_\mu \xi^\mu$. Similarly, they act simultaneously on $\bar A$ with gauge parameters $\bar \lambda^a = \bar A^a_\mu \xi^\mu$. 

If we perform a gauge transformation with parameters $\lambda^a$ and $\bar \lambda^a$, does it correspond to a diffeomorphism of the metric obtained from the Chern-Simons dictionary? We need to compute
\begin{align}
\delta e_\mu &= \frac{\ell}{2} (\delta_\xi A_\mu - \delta_\xi \bar A_\mu )\\
&=  \frac{\ell}{2} (D_\mu \lambda - \bar D_\mu \bar \lambda)\\
&= \frac{\ell}{2} (\partial_\mu (\lambda - \bar \lambda) + [\omega_\mu,\lambda - \bar\lambda])+\frac{1}{2}[e_\mu,\lambda+\bar\lambda]\\
&= \partial_\mu ( e_\nu \xi^\nu )+ [\omega_\mu, e_\nu \xi^\nu]+\frac{1}{2}[e_\mu,\lambda+\bar\lambda].
\end{align}
In the last step, we introduced our ansatz for the diffeomorphism: $e^a_\mu \xi^\mu = \frac{\ell}{2}(\lambda^a-\bar \lambda^a)$. We then have 
\begin{align}
\delta e_\mu &= e_\nu \partial_\mu \xi^\nu + (\partial_\mu e_\nu+ [\omega_\mu , e_\nu]) \xi^\nu +\frac{1}{2}[e_\mu,\lambda+\bar\lambda]\\
&= J_a (e^a_\nu \partial_\mu \xi^\nu + (\partial_\mu e^a_\nu+\omega^a_{\mu b}e^b_\nu)\xi^\nu)+\frac{1}{2}[e_\mu,\lambda+\bar\lambda]\\
&=J_a (e^a_\nu \partial_\mu \xi^\nu + e^a_\nu \Gamma_{\mu\lambda}^\nu \xi^\lambda) +\frac{1}{2}[e_\mu,\lambda+\bar\lambda]\\
&= e_\lambda \nabla_\mu \xi^\lambda +\frac{1}{2}[e_\mu,\lambda+\bar\lambda].
\end{align}
The first term is responsible for the transformation of the metric under the infinitesimal diffeomorphism parametrized by $\xi^\mu$. The second term is responsible for a local Lorentz transformation which does not appear in the variation of the metric. We can check that it is so by evaluating
\begin{align}
\delta g_{\mu\nu} &= \delta \left(e^a_\mu e^b_\nu \eta^{ab}\right) \\
&= \delta \left(e^a_\mu e^b_\nu tr(J_a J_b) \right) \\
&= \text{tr} (\delta e_\mu e_\nu + e_\mu \delta e_\nu)\\
&= \text{tr} \left( e_\lambda \nabla_\mu \xi^\lambda +\frac{1}{2}[e_\mu,\lambda+\bar\lambda] \right)e_\nu + e_\mu \left( e_\lambda \nabla_\nu \xi^\lambda +\frac{1}{2}[e_\nu,\lambda+\bar\lambda]) \right)\\
&= \text{tr} (e_\lambda e_\nu \nabla_\mu \xi^\lambda + e_\mu e_\lambda \nabla_\nu \xi^\lambda)\\
&= \nabla_\nu \xi_\mu + \nabla_\mu \xi_\nu\, .
\end{align}
That proves it! Since we are not interested in the local Lorentz transformations, we can fix the ambiguity in our definition of $\xi^\mu$ by selecting 
\begin{align}
\lambda^a = \xi^\mu A_\mu^a,\qquad \bar \lambda^a = \xi^\mu \bar A_\mu^a\, .
\end{align}
Using \eqref{defAAb}, we have $\lambda^a - \bar\lambda^a = \frac{2}{\ell} \xi^\mu e^a_{\; \mu}$ which matches our definition. \\

Let us now quickly compute the surface charges associated to these diffeomorphisms. In the Chern-Simons theory, we can obtain a general expression of these quantities (which are conserved if $\xi^\mu$ is a symmetry), which is furthermore very simple and elegant ! The path we take is the Barnich-Brandt method. We begin by computing the conserved current $S^\mu_\xi$ defined in the second Noether theorem (Result \ref{res:SecondNoetherThm}):
\begin{align}
d\mathbf{S}_\xi &= \frac{\delta \mathbf{L}_{CS}}{\delta A_\mu^a} \delta_\lambda A_\mu^a+\frac{\delta \mathbf{L}_{CS}}{\delta \bar A_\mu^a} \delta_{\bar \lambda} \bar A_\mu^a ,\quad\quad (\lambda = A_\mu \xi^\mu ,\qquad \bar \lambda = \bar A_\mu \xi^\mu ) \\
&=  \frac{\delta \mathbf{L}_{CS}}{\delta A_\mu^a}  (\partial_\mu \lambda^a + [A_\mu ,\lambda]^a ) + (\text{barred sector})\\
&= \partial_\mu \left(  \frac{\delta \mathbf{L}_{CS}}{\delta A_\mu^a} \lambda^a \right) -\partial_\mu  \frac{\delta \mathbf{L}_{CS}}{\delta A_\mu^a} \lambda +
 \frac{\delta \mathbf{L}_{CS}}{\delta A_\mu^a}  [A_\mu ,\lambda]^a+ (\text{barred sector})\\
 &= \partial_\mu \left(  \frac{\delta \mathbf{L}_{CS}}{\delta A_\mu^a} \lambda^a \right) + (\text{barred sector}).
\end{align}
In the last step, we replaced $ \frac{\delta \mathbf{L}_{CS}}{\delta A_\mu^a} =-\frac{k}{2\pi}\varepsilon^{\mu\alpha\beta}F^a_{\alpha\beta}$ and used that $\varepsilon^{\mu\alpha\beta}\partial_\mu F^a_{\alpha\beta} = 0$ and $\varepsilon^{\mu\alpha\beta}F^a_{\alpha\beta} [A_\mu,\lambda]^a= \text{tr} (\varepsilon^{\mu\alpha\beta}F_{\alpha\beta} [A_\mu,\lambda]) =\text{tr} (\varepsilon^{\mu\alpha\beta}[F_{\alpha\beta} ,A_\mu ]\lambda)=0$ by cyclic property of the trace and antisymmetry. Therefore, 
\begin{align}
\mathbf{S}^\mu_\xi &= -\frac{k}{2\pi}\varepsilon^{\mu\alpha\beta} (F^a_{\alpha\beta} \lambda^a + \bar F^a_{\alpha\beta} \bar \lambda^a ). 
\end{align}
A simple application of Anderson's operator \eqref{Ander2} separates us from the charge formula. Let us look only at the unbarred sector. Since there is only one derivative acting on the field $A_\mu$ in $\mathbf{S}^\mu_\xi$, there is only one term to compute:
\begin{align}
\mathbf{k}_\xi^{BB} &= I^2_{\delta A} \mathbf{S}_\xi = \frac{1}{2} \delta A_\alpha \frac{\partial}{\partial \partial_\nu A_\alpha} \frac{\partial}{\partial dx^\nu} \left( S^\mu_\xi (d^2 x)_\mu \right) \\
&= \frac{1}{2} \delta A_\alpha \frac{\partial}{\partial \partial_\nu A_\alpha} \left[ -\frac{k}{2\pi} \text{tr} \left\lbrace  \varepsilon^{\mu\rho \sigma} (\partial_\rho A_\sigma + A_\rho A_\sigma ) \lambda \right\rbrace \right] \frac{\partial}{\partial dx^\nu} \left[ \frac{1}{2} \varepsilon_{\mu\beta\gamma} dx^\beta dx^\gamma \right] \\
&= -\frac{k}{4\pi} \text{tr} \left\lbrace  \delta A_\alpha \varepsilon^{\mu\rho\sigma} (\delta_{\rho}^{\nu} \delta^\sigma_\alpha + 0) \lambda \right\rbrace \varepsilon_{\mu\nu\gamma} dx^\gamma = -\frac{k}{4\pi} \text{tr} (  \delta A_\alpha \lambda ) \varepsilon^{\mu\nu\alpha} \varepsilon_{\mu\nu\gamma} dx^\gamma  \\
&= -\frac{k}{4\pi} \text{tr}(  \delta A_\alpha \lambda ) (-2 \delta^\alpha_\gamma ) dx^\gamma = \frac{k}{2\pi} \text{tr} ( \delta A_\alpha \lambda) dx^\alpha = \frac{k}{2\pi} \text{tr}(  \lambda \delta A_\alpha  ) dx^\alpha.
\end{align}
The last equality is obtained thanks to the cyclicity of the trace. If we incorporate the contribution of the $\bar{\mathbf A}$-part and substitute $\lambda$, $\bar\lambda$ in terms of $\xi^\mu$, we have just found that
\begin{equation}
{\color{blue} \boxed { \mathbf{k}_\xi [\delta \mathbf A, \mathbf A] = \frac{k}{2\pi} \text{tr} \left[ (i_\xi \mathbf{A}) \delta \mathbf{A} - (i_\xi \mathbf{\bar A})  \delta \mathbf{\bar{A}} \right]. }}
\end{equation}
This is our final infinitesimal charge formula! Note that it is not clearly integrable, as in other non-linear theories such as Einstein gravity: compare with \eqref{def:ch2}! One needs to specify either a specific vector, or boundary conditions to deduce the (integrated) surface charge.

\subsection{\textit{AdS}\textsubscript{3} phase space in the Chern-Simons formalism}
We wish to close this section, and at the same time this chapter by deriving the connections $\mathbf A$, $\bar {\mathbf A}$ that describe the $AdS_3$ phase space in the Chern-Simons formulation. Let us begin by the most simple case which is global $AdS_3$ itself. Recall that in global coordinates $(\rho,t,\phi)$, the metric reads as
\begin{equation}
ds^2 = \ell^2 \left( d\rho^2 - \cosh^2 \rho \: dt^2 + \sinh^2 \rho \: d\phi^2 \right).
\end{equation}
Since the line element is diagonal, one choice of the orthonormal basis is quite trivial:
\begin{equation}
e^0 = \ell \cosh \rho \: dt \quad ; \quad e^1 = \ell \sinh \rho \: d\phi \quad ; \quad e^2 = \ell d\rho.
\end{equation}
Consequently, the non-vanishing components of the spin connection are
\begin{equation}
\left. 
\begin{array}{ccc}
\omega_{20} &=& \sinh \rho \: dt \\ 
\omega_{12} &=& \cosh \rho \: d\phi
\end{array} \right\rbrace
\Longrightarrow \left\lbrace
\begin{array}{ccccc}
\omega^0 &=& \omega_{12} &=& \cosh \rho \: d\phi \\ 
\omega^1 &=& \omega_{20} &=& \sinh \rho \: dt
\end{array} 
\right.
\end{equation}
together with the other components related by indicial symmetries. The components of the connections $\mathbf A = \mathbf A^a J_a, \: \bar{\mathbf A} = \bar{\mathbf A}^a J_a$ evaluate to
\begin{equation}
\begin{array}{r|l} 
\mathbf A^0 = \cosh \rho \: (dt+d\phi) \phantom{\quad} & \quad \bar{\mathbf A}^0 = -\cosh \rho \: (dt-d\phi) ;\\
\mathbf A^1 = \sinh \rho \: (dt+d\phi) \phantom{\quad} & \quad \bar{\mathbf A}^1 = \sinh \rho \: (dt-d\phi); \\
\mathbf A^2 = d\rho  \phantom{\quad} & \quad \bar{\mathbf A}^2 = -d\rho .
\end{array}
\end{equation}
We can clearly simplify the notations by introducing again lightcone coordinates $x^\pm = t \pm \phi$, so we get :
\begin{equation}
\left\lbrace
\begin{array}{ccl}
\mathbf A & = & \left(\cosh \rho \: J_0 + \sinh \rho \: J_1 \right) dx^+ + J_2 d\rho ; \\ 
\bar{\mathbf A} & = & \left(-\cosh \rho \: J_0 + \sinh \rho \: J_1 \right) dx^- - J_2 d\rho.
\end{array} 
\right.
\end{equation}
We claim that the connections are locally pure gauge, $\mathbf A = g^{-1} dg$, with $g = e^{x^+ J_0} e^{\rho J_2}$. Trivially we get $A_\rho = g^{-1} \partial_\rho g = J_2$ and $A_- = g^{-1} \partial_- g = 0$. The last component is more tricky to compute since $[J_0,J_2] \neq 0$. To progress we have to invoke Hadamard's lemma
\begin{equation}
\text{Adj}_X Y = e^{\text{adj}_X} Y \Leftrightarrow e^X Y e^{-X} = Y + [X,Y] + \frac{1}{2!} [X,[X,Y]] + \frac{1}{3!} [X,[X,[X,Y]]] + \cdots
\end{equation}
We find $A_+ = g^{-1} \partial_+ g = e^{-\rho J_2} J_0 e^{\rho J_2} = \cosh \rho \: J_0 + \sinh \rho \: J_1$ after summation of MacLaurin's series of $\cosh,\sinh$. As an exercise, show that $\bar{A}$ is also pure gauge, and associated to $\bar{g} = e^{-x^- J_0} e^{-\rho J_2}$. To make connection with the more general form that we will derive in a few moments, let us write the gauge fields in terms of the generators $\lbrace L_{-1}, L_0, L_{+1} \rbrace$ defined by
\begin{equation}
J_0 = \frac{1}{2}(L_{+1} + L_{-1}) \quad ; \quad J_1 = \frac{1}{2}(L_{+1} - L_{-1}) \quad ; \quad J_2 = L_0
\end{equation}
and which satisfy the $sl(2,\mathbb{R})$ algebra (isomorphic to the Lorentz algebra, or the centerless part of the Witt algebra):
\begin{equation}
[L_{+1},L_0] = +L_{+1} \quad ; \quad [L_{-1},L_0] = -L_{-1} \quad ; \quad [L_{+1},L_{-1}] = 2 L_0.
\end{equation}
The global $AdS_3$ solution is thus perfectly reproduced by the Chern-Simons connections
\begin{equation}
\left\lbrace
\begin{array}{ccl}
\mathbf A & = & +\frac{1}{2}\left(e^\rho \: L_{+1} + e^{-\rho} \: L_{-1} \right) dx^+ + L_0 d\rho ,\\ 
\bar{\mathbf A} & = & -\frac{1}{2}\left(e^\rho \: L_{-1} + e^{-\rho} \: L_{+1} \right) dx^- - L_0 d\rho ,
\end{array} 
\right.
\end{equation}
or, if we want to specify these in a manifestly pure gauge fashion $\mathbf A = g^{-1} dg$, $\bar{\mathbf A} = \bar{g}^{-1} d\bar{g}$ where
\begin{equation}
g = e^{\frac{1}{2}\left(L_{+1} + L_{-1} \right) x^+} e^{\rho L_0} \quad ; \quad \bar{g} = e^{-\frac{1}{2}\left(L_{+1} + L_{-1} \right) x^-} e^{-\rho L_0}.
\end{equation}
As a side note, remember that the Killing symmetry algebra of $AdS_3$ is $so(2,2)$. The fact that the Chern-Simons action contains two chiral terms makes manifest the chiral decomposition $so(2,2) = so(2,1)\oplus so(2,1)$... \\

Let us extend the discussion now to the entire $AdS_3$ phase space. We recall that under the Brown-Henneaux boundary conditions, any asymptotically $AdS_3$ metric can be written in the Fefferman-Graham gauge as (\ref{eq:EinsteinSolFGBH}). After setting $r = e^\rho$ and rewriting the null boundary fields as $L_\pm (x^\pm) \rightarrow \frac{1}{k}\mathcal L_\pm (x^\pm)$ we get
\begin{equation}
ds^2 = \ell^2 \left[ d\rho^2 + \frac{1}{k} \left( \mathcal L_+ (dx^{+})^2 + \mathcal L_- (dx^{-})^2 \right) - \left(e^{2\rho} + \left( \frac{1}{k}\right)^2 \mathcal L_+ \mathcal L_- e^{-2\rho} \right) dx^+ dx^- \right].
\end{equation}
The Lorentz triad 
\begin{equation}
e^{(\rho)} = \ell d\rho \: ; \:\;\; e^{(+)} = \ell \left( e^\rho dx^+ - \frac{1}{k} \mathcal L_- e^{-\rho} dx^- \right) \: ; \: \;\; e^{(-)} = \ell \left( e^\rho dx^- - \frac{1}{k} \mathcal L_+ e^{-\rho} dx^+ \right)
\end{equation}
is chosen to bring the metric into the form $ds^2 = e^{(\rho)} \otimes e^{(\rho)} - \frac{1}{2}e^{(+)} \otimes e^{(-)}-\frac{1}{2}e^{(-)} \otimes e^{(+)} $ where indices in brackets are the local Lorentz indices.
We leave as an exercise to the reader to compute the connections
\begin{equation}
\left\lbrace
\begin{array}{ccl}
\mathbf A & = & +\left(\frac{1}{2}e^\rho \: L_{+1} - \frac{2}{k} \mathcal L_+ (x^+) e^{-\rho} \: L_{-1} \right) dx^+ + L_0 d\rho , \\ 
\bar{\mathbf A} & = & -\left(\frac{1}{2}e^\rho \: L_{-1} - \frac{2}{k} \mathcal L_- (x^-) e^{-\rho} \: L_{+1} \right) dx^- - L_0 d\rho . 
\end{array} 
\right.
\label{eq:AForBHBC}
\end{equation}
This is a generalization of global $AdS_3$ spacetime, which is recovered when $\mathcal L_+ =\mathcal L_-= -k/4$. \\

As a very last check, we can show that the Chern-Simons charge formula reproduces the charges that we have derived in the metric formalism. Using (\ref{eq:AForBHBC}), we get immediately
\begin{equation}
\delta\mathbf   A = - e^{-\rho} \frac{2}{k} \delta \mathcal L_+ L_{-1} dx^+ \quad\text{and}\quad \delta \bar{\mathbf  A} = + e^{-\rho} \frac{2}{k} \delta \mathcal L_- L_{+1} dx^- . 
\end{equation}
The charge formula is 
\begin{align}
\delta H_{\xi} &= \frac{k}{2\pi} \int_S \text{tr} [\lambda \delta \mathbf  A - \bar{\lambda} \delta \bar{\mathbf  A}] .
\end{align}
Since $\delta A \sim L_{-1}$, the trace will be non-zero only for $\lambda \sim L_{+1}$, since $\text{tr}(L_{+1}L_{-1}) = -1$ and otherwise zero. With the same reasoning, we can concentrate on the $\bar \lambda \sim L_{-1}$ part only. The gauge parameters which capture the $n$-th harmonic of the diffeomorphism $\xi^{(+)}$ defined in \eqref{xiplus} are
\begin{equation}
\lambda = \xi^{(+)\mu}_n A_\mu = (\cdots ) L_0 + \frac{1}{2}e^{inx^+} e^{\rho} L_{+1} \quad\text{and}\quad \bar{\lambda} =  \xi^{(+)\mu}_n \bar{A}_\mu = (\cdots) L_0+0.
\end{equation}
The charge formula then evaluates to 
\begin{align}
\delta H_{\xi^{(+)}_n} &= \frac{k}{2\pi} \int_S  \text{tr}(L_{+1}L_{-1}) \frac{1}{2} e^{i n x^+}e^{\rho} \left(-e^{-\rho} \frac{2}{k} \delta \mathcal L_+ \right) dx^+ \\
&= \delta \left( \frac{1}{2\pi} \int_S e^{i n x^+} \mathcal L_+(x^+) d\phi\right)
\end{align}
since $dx^+ = d\phi$ on the boundary circle. The charge is clearly integrable and it reproduces our expectations. 
The null boundary fields $\mathcal L_\pm$ being related to the previous ones by $\mathcal L_\pm =k L_\pm = \frac{\ell}{4G} L_\pm$, we have in fact proven that
\begin{equation}
H_{\xi^{(+)}_n} = \frac{\ell}{8\pi G}\int_S d\phi \: L_+ e^{inx^+} 
\end{equation}
which is precisely the surface charge \eqref{ch0} obtained in the metric formalism. This last cross-check ends up our trip into the marvellous world of $3d$ gravity!

\section*{References}
\addcontentsline{toc}{section}{References}

{\color{blue} Spinning particles} as conical defects and excesses in $3d$ Minkowski spacetime and $AdS_3$ were described in 1984 by Deser, Jackiw and 't Hooft \cite{Deser:1983tn,Deser:1983nh}. \\

The  {\color{blue} Chern-Simons formulation} of $3d$ gravity was found by Achúcarro and Townsend in 1986 \cite{Achucarro:1987vz}. In this lectures, we followed the presentation of unpublished notes of Juan Jottar (partly published in \cite{deBoer:2013gz}) who we gratefully thank. \\

The same year Brown and Henneaux performed their  {\color{blue} asymptotic symmetry group} analysis of $AdS_3$ \cite{Brown:1986nw} using the Hamiltonian formalism. We followed here the equivalent covariant approach and used the notation of \cite{Compere:2015knw}.\\

The Bañados, Teitelboim and Zanelli  {\color{blue}  $BTZ$ black hole} was found in 1992 \cite{Banados:1992wn}. We followed the subsequent paper \cite{Banados:1992gq} in which Henneaux joined, where most of the geometrical properties of the $BTZ$ black hole were found. \\

The  {\color{blue} $BMS_3$ asymptotic symmetry algebra} was found at null infinity in \cite{Ashtekar:1996cd,Barnich:2006av} and at spatial infinity in \cite{Compere:2017knf}. Understanding the limit from $AdS_3$ to Minkowski for the symmetry algebra was described in \cite{Barnich:2006av}, and for the full phase space in \cite{Barnich:2012aw}. The  {\color{blue} flat spacetime cosmologies} and their thermodynamical properties were understood only in 2012 in \cite{Barnich:2012xq,Bagchi:2012xr}.

		\blanc % Pour avoir debut en page impaire!
	% Lecture 3 : BMS
\chapter{Asymptotically flat spacetimes}
\label{sec:BMS}

For the next lecture, we go back to more realistic gravitational models: indeed we consider four-dimensional asymptotically flat spacetimes, which are the solutions of General Relativity with localised energy-momentum sources. We are obviously not going to make an exhaustive overview of this rich and deep topic. We will start with a review of the work of Penrose on the conformal compactification of asymptotically flat spacetimes in order to get a global view on the asymptotic structure. We will then concentrate on the properties of radiative fields by reviewing the work of van der Burg, Bondi, Metzner and Sachs of 1962. One may think at first that the group of asymptotic symmetries of radiative spacetimes is the Poincar\'e group, but a larger group appears, the {\color{blue} \textbf{\textit{BMS} group}}  which contains so-called supertranslations. Additional symmetries, known as superrotations, also play a role and we shall briefly discuss them too.\\

This enhanced $BMS$ symmetry group has been the focus of much recent work. In particular, it is related to the so-called  {\color{blue} \textbf{displacement memory effect}} of General Relativity whose various facets where independently discovered in the 70s, 80s and 90s and that we will review here, an independent subleading spin memory effect, and to soft graviton theorems that we will not cover in these lectures.  \\

Finally, we will give some comments on the {\color{blue} \textbf{scattering problem}} in General Relativity, show that the extended asymptotic group gives conserved quantities once junction conditions are fixed at spatial infinity. This analysis is still under development by the international community at the time of writing these lecture notes and brings fascinating insights into the infrared properties of gravity!

\section{A definition of asymptotic flatness}

\subsection{Asymptotic structure of Minkowski spacetime}

The easiest way to introduce the various notions of asymptotic infinities of Minkowski spacetime is to introduce the Penrose compactification, which conformally maps the spacetime to another non-physical Lorentzian manifold with finite extent and boundary that is differentiable \emph{almost everywhere}. We first recall fundamentals about the conformal compactification of Minkowski spacetime, and then derive from it a geometric definition of asymptotic flatness. Even though the construction is simple, it leads to a rich asymptotic structure with a lot of subtetlies on the order of limits and non-geometric properties at the boundary of the unphysical spacetime, so let's be careful! 

\subsubsection{Conformal compactification of Minkowski spacetime}

Let us begin with a quick review of the conformal compactification of Minkowski spacetime, written in spherical coordinates $(t,r,x^A)$ as $ds^2 = -dt^2+dr^2 + r^2 \gamma_{AB} dx^A dx^B$, where $\gamma_{AB}$ is the unit round metric on the $2$-sphere. To represent the whole spacetime on a finite portion of a sheet of paper, we declare that every point on this sheet actually represent a $2$-sphere (except the space origin which is a spacetime line), so we consider for the moment the $1+1$ dimensional metric $-dt^2+dr^2$. To study causal motion on that spacetime, we introduce null coordinates $u = t-r$ and $v = t+r$ such that the metric becomes $-dudv$ and radial outgoing (resp. ingoing) null geodesics are simply labeled by constant values of $u$ (resp. $v$). We compactify the support of these coordinates thanks to a coordinate transformation $u = \tan U,\: v = \tan V$, whose action remains diagonal on $(u,v)$. Now $U,V \in ]-\pi/2,\pi/2[$, but the metric $-dUdV/(\cos U \cos V)^2$ still diverges at ``infinities'' mapped on the line segments $|U| = \pi/2$ and $|V| = \pi/2$. Let us now perform a Weyl transformation with conformal factor $\Omega(U,V) = \cos U \cos V$ to delete the diverging prefactor. The spacetime does not obey Einstein's equations anymore and does not faithfully represent the physical distances but the causal structure of the original metric is preserved. Reintroducing the angular contributions, we are left with 
\begin{equation}
ds^2 = - dudv + \frac{(u-v)^2}{4} \gamma_{AB} dx^A dx^B \Longrightarrow \Omega^2 ds^2 = -dUdV + \frac{\sin^2 (U-V)}{4} \gamma_{AB} dx^A dx^B
\end{equation}
and now we can extend the spacetime to $|U| = \pi/2$ and $|V| = \pi/2$. 
Going back to space- and timelike coordinates $T = U+V\in [-\pi,\pi]$ and $R = V-U \in [0,\pi]$, the conformal metric reads as
\begin{equation}
\Omega^2 ds^2 = -dT^2 + dR^2 + \frac{\sin^2 R}{4} \gamma_{AB} dx^A dx^B
\end{equation}
On the sheet of paper, the Minkowski spacetime has been compressed into a triangle with finite extent. Each point of this triangle, including the null boundaries but excluding the left vertical spacetime origin line and the boundary points $i^+$, $i^0$, $i^-$, represents a $2$-sphere whose radius varies with $R$ in a non-monotone way, since it equals $\sin R / 2$. At the end of the day, we get thus the following diagram, on which the geometrical structure of infinity is highlighted.

\newpage	
We now detail this structure:
\begin{enumerate}
\item \textit{Past timelike infinity} ($i^-$) : at $(R,T) = (0,-\pi)$, it represents the asymptotic sphere reached when $t\rightarrow -\infty$ while keeping $r$ fixed. It is also the starting point of any (maximally extended) timelike geodesic;
\item \textit{Future timelike infinity} ($i^+$) : at $(R,T) = (0,+\pi)$, it represents the asymptotic sphere reached when $t\rightarrow +\infty$ while keeping $r$ fixed. It is also the ending point of any (maximally extended) timelike geodesic;
\item \textit{Spacelike infinity} ($i^0$) : at the right-hand vertex of the triangle $(R,T) = (+\pi,0)$, it represents the asymptotic sphere reached when $r \rightarrow +\infty$ at fixed $t$;
\item \textit{Past null infinity} ($\cI^-$) : the line segment $R+T=\pi$ in the conformal diagram, it represents the $3$-surface formed by the starting points of ingoing null geodesics (the region reached when $r\rightarrow \infty$ and $u$ is fixed);
\item \textit{Future null infinity} ($\cI^+$) : the line segment $R-T=\pi$ in the conformal diagram is the future counterpart of $\cI^-$, and contains the terminal points of outgoing null geodesics (the region reached when $r\rightarrow +\infty$ and $v$ is fixed).
\end{enumerate}

What is fantastic is that we can easily read off the causal structure of spacetime in a glance! The radial null geodesics (or equivalently the lightcones of observers) point in the directions $+\pi/4$ and $-\pi/4$. Such a (maximally prolonged) geodesic always starts at some point of $\cI^-$, continues perpendicularly to the center of spacetime, is ``reflected'' by the segment $r=0$ and ends its journey at some other point of $\cI^+$. At the null infinities $\cI^\pm$ the null direction leads to an induced metric $ds^2=0 du^2 + d\Omega^2$ of zero determinant, while the topology is $S^2 \times \mathbb{R}$. 

\subsubsection{Singular points}

Now, don't get fooled. We didn't solve the asymptotic structure of Minkowski. The points/spheres at $i^0$, $i^+$, $i^-$ are singular in the conformal description. For example, fields propagating on Minkowski get multivalued there! More precisely, a propagating field will get a different limit to either of these points, depending on the order of limits between large distances and future or past. We therefore need to resolve these singular points. \\

Resolving the structure around $i^0$ amounts to introduce a foliation of spacetime around $i^0$ with well chosen $3$-surfaces. A useful foliation is the \textit{hyperbolic} one $(\tau,\rho,x^A)$ where $t = \rho \sinh \tau$, $r = \rho \cosh \tau$. By definition, $\tau$ is timelike, and $\rho$ is spacelike. The Minkowski metric becomes
\begin{equation}
ds^2 = d\rho^2 + \rho^2 \left( -d\tau^2 + \cosh^2 \tau \: \gamma_{AB} dx^A dx^B \right)
\end{equation}
outside the origin lightcone (centered at $r=t = 0$) which is the domain where this set of hyperbolic coordinates is well-defined. It is clear that the spacetime is now foliated by $dS_3$ (hyperboloids) of constant $\rho$. Spatial infinity is now defined as  $\rho \rightarrow \infty$. The boundary hyperbolic metric is now a smooth codimension 1 manifold which resolves $i^0$. It intersects null infinity at two spheres denoted by $\mathcal{I}^+_-$ and $\mathcal{I}^-_+$ which are respectively the \textit{past limit of future null infinity} and  the \textit{future limit of past null infinity}. In the hyperbolic description, $\mathcal{I}^+_-$ coincides with the sphere at the future time $\tau \rightarrow \infty$ of the boundary hyperboloid, and $\mathcal{I}^-_+$ is the sphere at the past $\tau \rightarrow -\infty$ of the boundary hyperboloid. \\

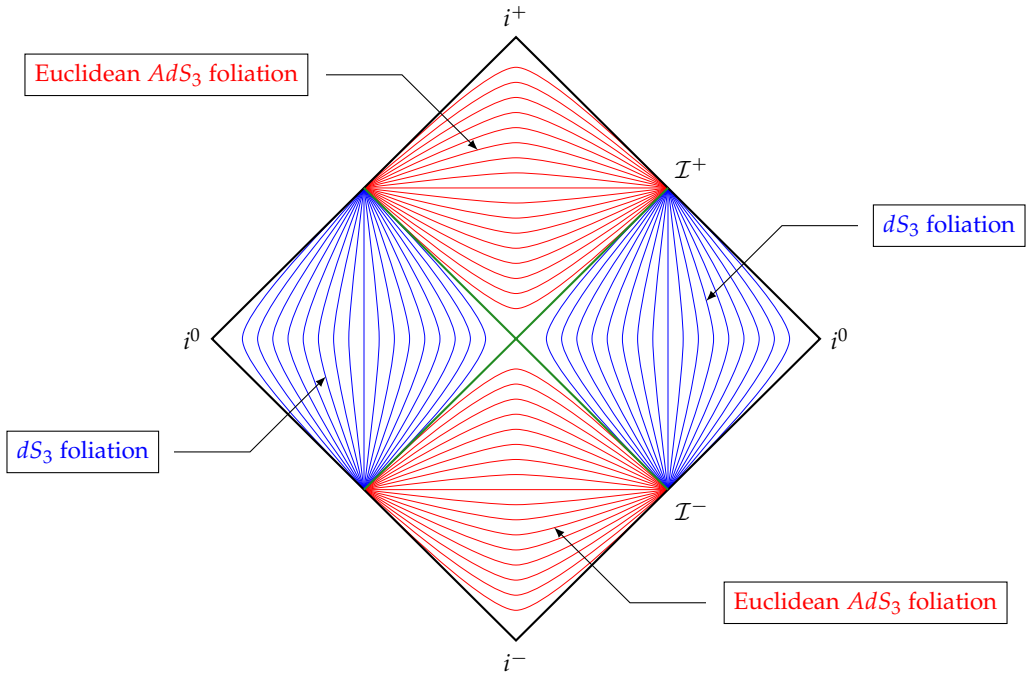
\begin{figure}[h!]
\begin{center}
\begin{tikzpicture}[scale=1]
% Foliations
\def\step{0.2};
\foreach \i in {-8,...,8}
{
\draw[very thin,blue] plot[smooth,tension=0.4] coordinates{(2,2) (2+\i*\step,0) (2,-2)};
\draw[very thin,blue] plot[smooth,tension=0.4] coordinates{(-2,2) (-2+\i*\step,0) (-2,-2)};
\draw[very thin,red] plot[smooth,tension=0.4] coordinates{(2,2) (0,2+\i*\step,0) (-2,2)};
\draw[very thin,red] plot[smooth,tension=0.4] coordinates{(2,-2) (0,-2+\i*\step,0) (-2,-2)};
}
% Diamond
\draw[thick,black] (-4,0) node[left]{\footnotesize $i^0$} -- (0,-4) node[below]{\footnotesize $i^-$} -- (4,0) node[right]{\footnotesize $i^0$} -- (0,4) node[above]{\footnotesize $i^+$} -- cycle;
% Origin lightcone
\draw[thick,ForestGreen] (-2,-2) -- (2,2) (2,-2) -- (-2,2);
\node[above] at (2, 2) {\footnotesize \qquad $\cI^+$};
\node[below] at (2,-2) {\footnotesize \qquad $\cI^-$};
% Labels
%\draw[-latex] (-3, 2) -- (-3, 3) -- (-1.5, 3);
%\draw[-latex] (-3,-2) -- (-3,-3) -- (-1.5,-3);
\draw[latex-] (2.5,0.5) -- (3.5, 1.5) -- (4.5, 1.5);
\node[fill=white,draw=black] at (5.7,1.5) {\color{blue} \footnotesize $dS_3$ foliation};
\draw[latex-] (-0.5,2.5) -- (-1.5, 3.5) -- (-2.5, 3.5);
\node[fill=white,draw=black] at (-4.6,3.5) {\color{red} \footnotesize Euclidean $AdS_3$ foliation};
\draw[latex-] (-2.5,-0.5) -- (-3.5, -1.5) -- (-4.5, -1.5);
\node[fill=white,draw=black] at (-5.7,-1.5) {\color{blue} \footnotesize $dS_3$ foliation};
\draw[latex-] (0.5,-2.5) -- (1.5, -3.5) -- (2.5, -3.5);
\node[fill=white,draw=black] at (4.6,-3.5) {\color{red} \footnotesize Euclidean $AdS_3$ foliation};
\end{tikzpicture}
\end{center}
\caption{\textit{Hyperbolic foliations which blow up the geometry near the singular points $i^+,i^-,i^0$. The origin lightcone (centered at $r = t= 0$) is drawn in green.}}
\label{fig:BlowUp}
\end{figure}

One can also blow up the geometry around $i^+$ and $i^-$, by introducing a second set of coordinates $(\hat\tau,\hat\rho,x^A)$, this time covering the patch inside the origin lightcone, with $t = \hat\tau\cosh\hat\rho$, $r = \hat\tau\sinh\hat\rho$. The slicing will be realised on the timelike coordinate $\hat\tau$. Each hypersurface is conformal to the Euclidean version of $AdS_3$, as we can directly see from the line element
\begin{equation}
ds^2 = -d\hat\tau^2+ \hat\tau^2 \left( d\hat\rho^2 + \sinh^2 \hat\rho \: \gamma_{AB} dx^A dx^B \right).
\end{equation}
The three foliations are represented on Figure \ref{fig:BlowUp}. The union of the foliations inside and outside the origin lightcone provides a manifold whose ``corners'' are smooth and have a differentiable structure.

\subsection{Gravity in Bondi gauge}

Let us now introduce gravity. Let $g_{\mu\nu}$ be the metric. We would like to define a notion of asymptotic flatness. There are two ways to do so: 
\begin{enumerate}
\item Using covariant objects but involving unphysical fields such as a conformal factor (a scalar field) used to do a Penrose compactification of spacetime;
\item Using an adapted coordinate system and specifying fall-off conditions.
\end{enumerate}
We will follow the second route which more easily allows to analyse the details of the asymptotic structure. \\

We would like to define asymptotically flat spacetimes which approach a notion of future null infinity $\cI^+$. Physically, this describes the so-called ``radiation zone'' where gravitational waves and other null wave phenomena leave their imprint on spacetime far from the sources. This problem has been addressed by Bondi, van der Burg, Metzner and Sachs in the 60ies, which we now review. We consider a family of null hypersurfaces labeled by a constant $u$ coordinate. The normal vector of these hypersurfaces $n^\mu = g^{\mu\nu} \partial_\nu u$ is null by construction, so we fix $g^{uu} = 0$. We define angular coordinates $x^A = (\theta,\phi)$ such that the directional derivative along the normal $n^\mu$ is zero, $n^\mu \partial_\mu x^A = 0 \Rightarrow g^{uA} = 0$. We finally select the radial coordinate $r$ to be the luminosity distance, \textit{i.e.} we fix $\partial_r \det (g_{AB}/r^2) = 0$. The coordinates $x^\mu = (u,r,x^A)$ so defined are known in the literature as the \textit{Bondi-Sachs} coordinate system or \textit{Bondi gauge}. After lowering the indices, we find $g_{rr} = g_{rA} = 0$. The  $4d$ line element takes the form
\begin{equation}
ds^2 = g_{\mu\nu} dx^\mu dx^\nu = g_{uu} du^2 +2 g_{ur} dudr +2 g_{uA}du dx^A + g_{AB} dx^A dx^B .
\end{equation}
We can now define the notion of asymptotic flatness. We would like to obtain Minkowski spacetime in the limit $r \rightarrow \infty$ at constant $u,x^A$, which is written in retarded coordinates as $ds^2 = -du^2 - 2 dudr + r^2 \gamma_{AB} dx^A dx^B$ where $\gamma_{AB}$ is the unit round metric on the $2$-sphere. Therefore, we demand
\begin{equation}
\lim_{r\rightarrow \infty}  g_{uu} = \lim_{r\rightarrow \infty}  g_{ur} = -1 \quad ; \quad \lim_{r\rightarrow \infty} g_{uA} = 0 \quad ; \quad \lim_{r\rightarrow \infty}  g_{AB} = r^2 \gamma_{AB}. 
\end{equation}
Boundary conditions are in fact more restrictive. We need to ensure that we define a phase space with well defined charges. We cannot be too restrictive, since we need to keep all physical spacetimes, such as black hole mergers for example. After analysis, it was proposed to consider
\begin{align}
\begin{array}{llll}
g_{uu} = -1 + \mathcal{O}(r^{-1}), & g_{ur} = -1 + \mathcal{O}(r^{-2}), & g_{uA} = \mathcal{O}(r^{0}), & g_{AB} = r^2 \gamma_{AB} + \mathcal{O}(r) . 
\end{array}
\label{eq:BondiBoundaryConditions}
\end{align}
The class of allowed metrics for these fall-off conditions can be derived:
\begin{align}
ds^2 	&= -du^2 - 2dudr + r^2 \gamma_{AB} dx^A dx^B \quad\quad (\text{Minkowski}) \nonumber\\
		&+ \frac{2 m}{r} du^2 + r C_{AB} dx^A dx^B + D^B C_{AB} du dx^A \nonumber\\
		&+ \frac{1}{16 r^2} C_{AB} C^{AB} du dr 
		+ \frac{1}{r} \left[ \frac{4}{3} (N_A + u \partial_A m_B) - \frac{1}{8} \partial_A (C_{BC} C^{BC}) \right] du dx^A+ \frac{1}{4} \gamma_{AB} C_{CD} C^{CD} dx^A dx^B \nonumber\\
		& + (\text{Subleading terms}).
\end{align}
Here all indices are raised with $\gamma^{AB}$ and $\gamma^{AB}C_{AB} = 0$. In Bondi gauge, the metric defines a hierarchy of several physically relevant fields which we now explicit:
\begin{itemize}[label=$\rhd$]
\item $m(u,x^A)$ is the \textit{Bondi mass aspect}. It gives the angular density of energy of the spacetime as measured from a point at $\cI^+$ labeled by $u$ and in the direction pointed out by the angles $x^A$. The \emph{Bondi mass} is obtained after performing an integration of $m$ on the sphere: $M (u) = \oint_{S^2_\infty} d^2\Omega \, m (u,x^A)$. One can show that $\partial_u M(u) \leq 0$ for pure gravity or gravity coupled to matter obeying the null energy condition. Physically, radiation carried by gravitational waves or null matter such as electromagnetic fields escapes through $\mathcal I^+$ and lowers the energy of spacetime when the retarded time $u$ evolves. At $u \rightarrow -\infty$, the Bondi mass equates the ADM energy, or total energy of a Cauchy slice of spacetime. 
\item At the first subleading order, we find another field: the field $C_{AB} (u,x^A)$ which is traceless ($\gamma^{AB} C_{AB} = 0$) and symmetric. It therefore contains two polarization modes. It contains all the information about the gravitational radiation around $\cI^+$. Its retarded time variation is the \textit{Bondi news} tensor $N_{AB} = \partial_u C_{AB}$. This is the analog of the Maxwell field for gravitational radiation and its square is proportional to the energy flux across $\cI^+$ as we will see a bit later.
\item At second subleading order, one finds $N_A(u,x^A)$ the \textit{angular momentum aspect}. It is closely related to the angular density of angular momentum with respect to the origin defined as the zero luminosity distance $r=0$.  Its integration on $S^2$ contracted with the generator of rotations is related to the total angular momentum of the spacetime, evaluated at $\cI^+$ at retarded time $u$.
\end{itemize}

The metric as written so far does not obey Einstein's equations. One finds two additional constraints upon pluging this consistent ansatz into Einstein's equations: 
\begin{equation}
\partial_u m = \frac{1}{4} D^A D^B N_{AB} - T_{uu} \quad \text{ with } T_{uu} = \frac{1}{8} N_{AB} N^{AB} + 4\pi \lim_{r\rightarrow \infty} (r^2 T_{uu}^M)
\label{eq:BondiMassVariation}
\end{equation}
and 
\begin{align}
\partial_u N_A &= -\frac{1}{4} D^B \left( D_B D^C C_{AC} - D_A D^C C_{BC} \right) + u \partial_A \left(T_{uu} - \frac{1}{4} D^B D^C N_{BC} \right) - T_{uA} \label{eq:EvolutionForNA} \\
&\text{with } T_{uA} = 8\pi \lim_{r\rightarrow\infty} (r^2 T_{uA}^M) - \frac{1}{4} \partial_A (C_{BC} N^{BC}) + \frac{1}{4} D_B (C^{BC} N_{CA}) - \frac{1}{2} C_{AB} D_C N^{BC}. 
\end{align}
Here we denote by $T_{\mu\nu}^M$ the stress tensor of matter, and $D_A$ is the covariant derivative associated to $\gamma_{AB}$. Of course, since we are performing an expansion close to $\cI^+$, the only relevant matter is the null matter. The gravitational wave contributions naturally add up to the null matter contributions. \\

Because of these constraints, a generic initial data on $\mathcal I^+$ is specified by $m$, $C_{AB}$ and $N_{A}$ at initial retarded time and $N_{AB}$ at all retarded times, in addition of course with all the subleading fields that we ignored so far. Given that there is an infinite tower of subleading multipoles at spatial infinity, there will also be a tower of subleading terms around null infinity. 

\subsection{Initial and late data}

Let us now study in more details the initial and final data at $\mathcal I^+_-$ and $\mathcal I^+_+$, the early and late retarded times of $\mathcal I^+$ at first and second subleading order, which are the relevant orders to study mass and angular momentum conservation. This initial or late data depends upon the class of spacetimes that we are studying.  Let us restrict our analysis to solutions that start from the vacuum in the far past, and revert to it in the far future (in particular, this assumption rules out black hole formation). Such spaces have been defined in a rigorous way by Christodoulou and Klainerman \cite{Christodoulou:1993uv} and subsequent authors. They showed that it exists a class of Cauchy data which decays sufficiently fast at spatial infinity such that the Cauchy problem leads to a smooth geodesically complete solution. In fact they proved the non-linear stability of Minkowski spacetime. In such analyses, the Bondi news falls off as 
\begin{equation}
N_{AB} = \mathcal{O} \left( |u|^{-(1+\varepsilon)} \right), \qquad  (\varepsilon > 0)
\end{equation}
when $ u \rightarrow \pm \infty$, while $m$ and $N_A$ remain finite in the two limits. Now, even if black holes form in the spacetime, we don't expect that these quantities will behave differently since they don't emit radiation at early or late retarded times, so we just assume that all asymptotically flat spacetimes obey these conditions. \\

Since $\partial_u N_A \rightarrow 0$ and $T_{uA} \rightarrow 0$ when one approaches $\cI^+_-$, the evolution equation for $N_A$ gives
\begin{equation}
\left. D^B (D_B D^C C_{AC} - D_A D^C C_{BC}) \right|_{\cI^+_-} = 0
\label{eq:CequationAtIPM}
\end{equation}
which constraints the initial value of $C_{AB}$. Note that the divergent term $u\partial_A (\cdots)$ in (\ref{eq:EvolutionForNA}) has cancelled thanks to the fall off condition of the Bondi news. A symmetric traceless tensor on the $2$-sphere like $C_{AB}$ forms a representation of $SO(3)$ and takes the general form
\begin{equation}
C_{AB} = -2 D_A D_B C + \gamma_{AB} D^2 C + \varepsilon_{C(A} D_{B)} D^C \Psi. 
\label{eq:Cgeneral}
\end{equation}
The first term defines a scalar field $C(u,x^A)$ on the sphere. The second term is parity-violating, and depends on a pseudo-scalar field, $\Psi(u,x^A)$. Now, the equation \eqref{eq:CequationAtIPM} implies that $D^2 (D^2+2)\Psi(u,x^A)=0$ at $\cI^+_-$ which implies $\Psi = 0$\footnote{Strictly speaking it implies that $\Psi = 0$ up to the lowest $l = 0,1$ spherical harmonics. However, these harmonics are exactly zero modes of the differential operator $\varepsilon_{C(A} D_{B)} D^C \Psi$ defined in $C_{AB}$. Therefore we can set them to zero.}, but $C(u,x^A)\vert_{\cI^+_-} = C(x^A)$ can be non-vanishing. We will call $C(x^A)$ the \textit{supertranslation memory field} for reasons that will be clear in a few moments.  \\

A similar construction can be performed at $\cI^-$ up to switching $u$ into $v$ and $+$ into $-$, so we are left with two sets of radiative data on both null infinities:
\begin{itemize}[label=$\rhd$]
\item Radiation at past null infinity : $\left\lbrace \left. C(x^A) \right|_{\cI_+^-}, \left. m (x^A) \right|_{\cI_+^-}, \left. N_A (x^A) \right|_{\cI_+^-}, N_{AB} (v,x^A) , \dots \right\rbrace$ ;
\item Radiation at future null infinity : $\left\lbrace \left. C(x^A) \right|_{\cI_-^+}, \left. m (x^A) \right|_{\cI_-^+}, \left. N_A (x^A) \right|_{\cI_-^+}, N_{AB} (u,x^A) ,\dots \right\rbrace$.
\end{itemize} 
These quantities form a set of initial data at null infinity at first and second subleading order in the luminosity distance expansion since Einstein's equations and the gauge conditions provide with all the metric components from this set of data.  This closes our discussion on the set of physical solutions, and we now turn to the asymptotic symmetries of asymptotically flat spacetimes.

\section{Asymptotic symmetries : the \textit{BMS}\textsubscript{4} group}

Let us discuss the vector fields that preserve the Bondi gauge and the boundary conditions. As discussed in Lecture 2, such infinitesimal diffeomorphisms are either pure gauge, or belong to the non-trivial set of asymptotic symmetries. This distinction requires to compute the conserved charges associated with the infinitesimal diffeomorphisms. At this point, let us just enumerate the vector fields that change the leading fields in the asymptotic expansion of the metric and therefore are candidates to be asymptotic symmetries. Preserving the Bondi gauge requires
\begin{align}
\begin{array}{lll}
\Lie_\xi g_{rr} = 0, &  \Lie_\xi g_{rA} = 0 , & \Lie_\xi \partial_r \det (g_{AB}/r^2) = 0 .\label{Bg}
\end{array}
\end{align}
Preserving the boundary conditions (\ref{eq:BondiBoundaryConditions}) further requires
\begin{align}
\begin{array}{llll}
\Lie_\xi g_{uu} = \mathcal{O}(r^{-1}) ,& \Lie_\xi g_{ur} = \mathcal{O}(r^{-2}) ,& \Lie_\xi g_{uA} = \mathcal{O}(r^{0}), &\Lie_\xi g_{AB} = \mathcal{O}(r). \label{BC1}
\end{array}
\end{align}
One first solves the constraints \eqref{Bg} exactly which allows to express the 4 components of $\xi^\mu$ in terms of 4 functions of $u,x^A$. One can then solve \eqref{BC1} to reduce these 4 functions to only 3 functions on the $2$-sphere, namely $T(x^A)$ and $R^A(x^B)$. The resulting vector is 
\begin{align}
\xi_{T,R} = &\left[ T(x^C) + \frac{u}{2} D_A R^A (x^C) + o(r^0) \right] \partial_u \\
&+ \left[ R^A (x^C) - \frac{1}{r} D^A T(x^C) + o(r^{-1}) \right] \partial_A \\
&+ \left[ - \frac{r+u}{2} D_A R^A (x^C) + \frac{1}{2} D_A D^A T(x^C) + o(r^0) \right] \partial_r 
\end{align}
where $T(x^A)$ is unconstrained. The last constraint in \eqref{BC1} imposes that $R^A$ obeys the conformal Killing equation on the $2$-sphere, 
 \begin{align}
D_A R_B + D_B R_A = \gamma_{AB} D_C R^C. \label{cKe}
\end{align}
These vectors are known as the \textit{BMS} generators, in honour to Bondi, Metzner, Sachs and van der Burg who were the pioneers in studying the asymptotic behaviour of the gravitational radiation field. We will see later that the canonical charges associated with these generators are non-trivial and therefore these diffeomorphisms acquire the name of asymptotic symmetries. Analogously to the $3d$ case, we immediately see that the asymptotic algebra is larger than the Poincaré algebra! The explicit computation of the algebra under the Lie bracket is not quite difficult and ends with
\begin{equation}
\xi_{\hat T, \hat R} = [\xi_{T,R}, \xi_{T',R'}] \Longrightarrow \left\lbrace
\begin{array}{ccl}
\hat T &=& R^A D_A T' + \frac{1}{2} T D_A R'^A - R'^A D_A T - \frac{1}{2} T' D_A R^A \; ;\\
\hat R^A &=& R^B D_B R'^A - R'^B D_B R^A\; .
\end{array}
\right. 
\label{eq:BMSCommu}
\end{equation}
These relations define the $bms_4$ algebra. Trivial boundary diffeomorphisms $T = R^A = 0$ form an ideal of this algebra. Taking the quotient by this ideal, we are left with the asymptotic algebra of asymptotically flat spacetimes compatible with the Bondi-Sachs boundary conditions. The generators can be divided into 2 categories, the vectors generated by $T$ known as supertranslations and the vectors generated by $R^A$ the Lorentz transformations or their extension: the superrotations. 

\subsection{Supertranslations}

The relations (\ref{eq:BMSCommu}) imply that $\xi_{T,0} \equiv \xi_T$ form an abelian ideal of the $bms_4$ algebra. They generalize the translations and receive for these reasons the name of \textit{supertranslations}
\begin{equation}
\xi_{T} = T(x^C) \partial_u - \frac{1}{r} D^A T(x^C) \partial_A + \frac{1}{2} D_A D^A T(x^C) \partial_r + \cdots
\end{equation}
Note that supersymmetry was found 10 years later, so this denomination has nothing to do with supersymmetry! Since $T(x^A)$ is a completely arbitrary scalar field on the sphere, the exponentiation of these vectors gives rise to an abelian subgroup $\mathcal{S}$ of the asymptotic symmetry group, which is infinite-dimensional. $\mathcal{S}$ admits one unique normal finite subgroup that reproduces exactly the Poincaré translations. The associated generators $\xi_T$ are built from the first 4 spherical harmonics in the decomposition of $T$, which verify $D_A D_B T + \gamma_{AB} D^2 T = 0$, namely $T(x^A) = a_0 Y^0_0 (x^A) + a_{m+2} Y^m_1 (x^A), \: a_\mu \in \mathbb{R}, m \in \lbrace -1,0,+1 \rbrace$. As an example,  $\partial_z$ amounts to $T = Y^1_1(x^A)$. Indeed, we have $\partial_z =  \cos \theta \partial_r - \frac{1}{r} \sin \theta \partial_\theta$ in spherical static coordinates $(t,r,\theta,\phi)$, so in retarded  coordinates $\partial_z = - \cos \theta \partial_u +\cos \theta \partial_r - \frac{1}{r} \sin \theta \partial_\theta$. 

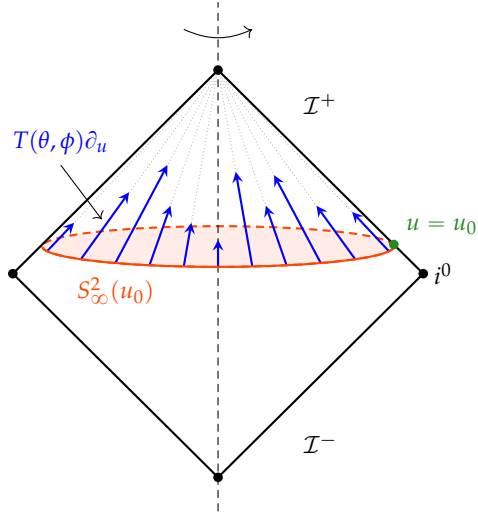
\begin{figure}[h!]
    \begin{center}
        \begin{tikzpicture}[scale=0.9]
        	% Sphère à l'infini
        		\coordinate (centre) at (0,0.4);
        		\draw[OrangeRed,densely dashed,thick,fill=OrangeRed!10] (centre) ellipse (2.57 and 0.3);
			% Axe de rotation
        		\draw[densely dashed] (0, -3.5) -- (0,4);
        		\path[->, bend right = 25] (-0.5,3.6) edge (0.5,3.6);       
			% Guides radiants
				\draw[densely dotted,line width=0.1mm,gray!75] (-0.5,0.11) -- (0,3);
				\draw[densely dotted,line width=0.1mm,gray!75] (-1.0,0.13) -- (0,3);
				\draw[densely dotted,line width=0.1mm,gray!75] (-1.5,0.16) -- (0,3);
				\draw[densely dotted,line width=0.1mm,gray!75] (-2.0,0.22) -- (0,3);
				\draw[densely dotted,line width=0.1mm,gray!75] (-2.5,0.33) -- (0,3);
				\draw[densely dotted,line width=0.1mm,gray!75] (0.5,0.11) -- (0,3);
				\draw[densely dotted,line width=0.1mm,gray!75] (1.0,0.13) -- (0,3);
				\draw[densely dotted,line width=0.1mm,gray!75] (1.5,0.16) -- (0,3);
				\draw[densely dotted,line width=0.1mm,gray!75] (2.0,0.22) -- (0,3);
				\draw[densely dotted,line width=0.1mm,gray!75] (2.5,0.33) -- (0,3);
			% Super-translations
				\draw[->,>=stealth,blue,thick] (-0.5,0.11) -- (-0.39,0.75);
				\draw[->,>=stealth,blue,thick] (-1.0,0.13) -- (-0.695,1.0);
				\draw[->,>=stealth,blue,thick] (-1.5,0.16) -- (-0.74,1.60);
				\draw[->,>=stealth,blue,thick] (-2.0,0.22) -- (-1.30,1.2);
				\draw[->,>=stealth,blue,thick] (-2.5,0.33) -- (-2.105,0.75);
				\draw[->,>=stealth,blue,thick] (0,0.1) -- (0,0.5);
				\draw[->,>=stealth,blue,thick] (0.5,0.11) -- (0.26,1.5);
				\draw[->,>=stealth,blue,thick] (1.0,0.13) -- (0.70,1.0);
				\draw[->,>=stealth,blue,thick] (1.5,0.16) -- (0.82,1.45);
				\draw[->,>=stealth,blue,thick] (2.0,0.22) -- (1.44,1.0);
				\draw[->,>=stealth,blue,thick] (2.5,0.33) -- (1.96,0.91);
			% Arc de face
        		\draw[OrangeRed,thick] (2.57,0.4) arc (0:-180:2.57 and 0.3);
			% Diamant
        		\draw[thick] (0, 3) -- (3, 0) node[right] {\footnotesize $i^0$} -- (0,-3) -- (-3,0) -- cycle;
            % Infinis de genre lumière             
	            \node[] (1) at (1.5, 2.5) {\footnotesize $\cI^+$};
    	        \node[] (2) at (1.5, -2.5) {\footnotesize $\cI^-$};
			% Points à l'infini
				\fill[black] ( 3, 0) circle [radius=2pt];
				\fill[black] ( 0, 3) circle [radius=2pt];
				\fill[black] ( 0,-3) circle [radius=2pt];
				\fill[black] (-3, 0) circle [radius=2pt]; 
			% Légendes
				\node[] at (-1.5,-0.25) {{\color{OrangeRed} {\footnotesize $S^2_\infty(u_0)$}}};
				\node[above] at (2.57,0.43) {{\color{ForestGreen} {\footnotesize $\qquad \qquad u=u_0$}}};
				\fill[ForestGreen] (2.57,0.43) circle [radius=2pt];
				\node[above] at (-2.3,1.6) {{\color{blue} {\footnotesize $T(\theta,\phi) \partial_u$}}};
				\draw[->,black] (-2.3,1.6) -- (-1.7,0.8);
        \end{tikzpicture}
    \end{center}
    \caption{\textit{BMS supertranslations}}
    \label{fig:Supertranslations}
\end{figure} 

The associated conserved charges are the supermomenta. The infinitesimal canonical charges are finite, non-vanishing, not integrable but their non-integrable piece is related to the flux passing through null infinity,
\begin{equation}
\delta Q_T = \delta \left[ \frac{1}{4 \pi G} \int d^2\Omega \: \sqrt{\gamma} \: T\: m\:  \right] + \frac{1}{32 \pi G} \int d^2\Omega \: \sqrt{\gamma} \: N^{AB} \delta C_{AB}. 
\end{equation}
As non-trivial diffeomorphisms, the supertranslations act on the asymptotically flat phase space, transforming a geometry into another one, physically inequivalent. The Bondi news is transformed following $\delta_T N_{AB} = \Lie_T N_{AB} = T \partial_u  N_{AB}$ so supertranslations have a relationship with gravitational radiation as we will see later. Other fields also vary non-trivially under $\xi_T$, for example
\begin{align}
\Lie_T C_{AB} = T \partial_u C_{AB} - 2 D_A D_B T + \gamma_{AB} D_C D^C T ,\label{eq:STField} \\
\Lie_T m_B = T \partial_u m_B + \frac{1}{4} \left[ N^{AB} D_A D_B T + 2 D_A N^{AB} D_B T \right]. 
\end{align}
From these relations, it is obvious that a supertranslation cannot create inertial mass, or gravitational radiation. Indeed, if we apply a supertranslation on the Minkowski global vacuum $m = N_{AB} = C_{AB} = 0$, we get $\Lie_T m = \Lie_T N_{AB} = 0$. The only field that can be shifted is $C_{AB}$, but since the Bondi news remain zero, we are left with $\Lie_T C_{AB} = - 2 D_A D_B T + \gamma_{AB} D^2 T$. Recalling that $C_{AB} = - 2 D_A D_B C + \gamma_{AB} D^2 C$ in a non-radiative configuration, we deduce that the supertranslation memory field $C$ is shifted as 
\begin{align}
\delta_T C(x^A) = T(x^A)
\end{align}
under a supertranslation. This explains half of the name of the field! The fixation of $C$ is equivalent to a spontaneous breaking of the supertranslation invariance between gravitational vacua, and as a consequence $C$ is the Goldstone boson which accompanies this breaking. It is noteworthy that the 4 Poincaré translations are not concerned by this breaking, because they consist in the 4 lowest spherical harmonics of $T(x^A)$, which are annihilated by the differential operator $-2D_A D_B + \gamma_{AB} D^2$. So $C$, up to the first 4 harmonics, labels the various degeneracies of the gravitational field. Moreover, since supertranslations commute with the time translation, their associated charges will commute with the Hamiltonian which means that all these degenerate states have the same energy. This remarkable feature of asymptotically flat gravity was only found in 2013 by Strominger \cite{Strominger:2013jfa}\footnote{Similarly to the construction of the $AdS_3$ phase space of stationary field configurations with boundary fields turned on that was described in \eqref{eq:EinsteinSolFGBH} or for the asymptotically flat analogue \eqref{eq:FlatPS}, one can construct gravitational vacua and Schwarzschild black holes that carry a supertranslation field, see \cite{Compere:2016jwb,Compere:2016hzt}.}.

\subsection{Lorentz algebra and its extensions}
The second set of generators of the $bms_4$ algebra is constituted by the generators with $T=0$, denoted by $\xi_R$ and given asymptotically by
\begin{equation}
\xi_R = R^A (x^C)\partial_A - \frac{r+u}{2} D_A R^A (x^C) \partial_r + \frac{u}{2} D_A R^A (x^C) \partial_u + \cdots
\end{equation}
Remember that the fixation of the boundary sphere metric as a part of the boundary conditions imposes the conformal Killing equations \eqref{cKe}. The easiest way to solve the conformal Killing equation on $S^2$ is to introduce complex \textit{stereographic coordinates} on the sphere $z = e^{i\phi} \cot (\theta/2),\: \bar{z} = z^*$. The unit round metric on $S^2$ is simply the off-diagonal line element $ds^2_{S^2} = 4(1+z\bar{z})^{-2} dzd\bar{z}$. The $(z,\bar z)$ component of $D_A R_B + D_B R_A = D_C R^C \gamma_{AB}$ is identically obeyed. The $(z,z)$ and $(\bar{z},\bar{z})$ components  simply reduce to the holomorphicity conditions $\partial_{\bar{z}} R^z = 0$ and $\partial_z R^{\bar{z}} = 0$. So $R^z (z)$ is an holomorphic function (and $R^{\bar{z}} (\bar{z})$ is its antiholomorphic counterpart) which can be expanded in Laurent's series, and appears so as a sum of monomial terms $R^z = z^k, \: k\in \mathbb{Z}$. Considering $v_k = z^k \partial_z$, we claim that $v_k$ is globally well-defined only when $k = 0,1,2$. Indeed, when $k<0$, $z^k$ is singular at the origin $z=0$ (south pole $\theta = \pi$), and when $k > 2$, $z^k$ is singular at the point at infinity $z = \infty$ (the north pole $\theta = 0$), since under the transformation $z \rightarrow w = z^{-1}$, $v^k$ becomes $-w^{2-k} \partial_w$. Three globally well-defined (complex) vectors also come from the antiholomorphic part, so we are left with a subalgebra of 6 well-defined conformal isometries of the sphere. The real part of this algebra gives 6 asymptotic Killing vectors. These are nothing but the \textit{Lorentz generators}. Indeed, the complexification of $so(3,1)$ is isomorphic to the direct sum $sl(2,\mathbb R) \oplus sl(2,\mathbb R)$. In stereographic retarded coordinates, a simple exercise can convince us that the Lorentz $so(3,1)$ vectors of Minkowski read as
\begin{equation}
\xi_R \in so(3,1) \Longleftrightarrow \xi_R = \left( 1 + \frac{u}{2r} \right) R^z \partial_z - \frac{u}{2r} D^{\bar{z}} D_z R^z \partial_{\bar{z}} - \frac{u+r}{2} D_z R^z \partial_r + \frac{u}{2} D_z R^z \partial_u + c.c.
\end{equation}
with $R^z \in \lbrace 1,z,z^2,i,iz,iz^2 \rbrace$. So the globally defined $\xi_R$ are simply the asymptotic Lorentz transformations. If we discard the singular generators, we finally get after exponentiation the historical form of the $BMS_4$ group:
\begin{equation}
{\color{blue} \boxed{ BMS_4 = SO(3,1) \ltimes \mathcal{S} = \text{Lorentz} \ltimes \text{Supertranslations} }}
\end{equation}
It reproduces the semi-direct structure of the Poincaré group: the Lorentz group acts non-trivially on the abelian factor $\mathcal{S}$ as it does on the global translations. The only difference, and a crucial one, is that the translational part is enhanced, which implies the degeneracy of the gravitational Poincaré  vacua. \\

One can argue that there is no obvious reason to restrict $R^z(z)$ to be a globally well-defined function on the sphere. After all, conformal field theories exist on a two-sphere and singular local conformal transformations play an important role. The proposition of Barnich and Troessaert \cite{Barnich:2010eb} (and before them of de Boer and Solodukhin \cite{deBoer:2003vf}) is to allow the full range of $k$ in the Laurent spectrum of $R^z(z)$, generalizing the global conformal transformations to \emph{meromorphic superrotations} defined from a  \textit{meromorphic} function with a finite set of poles on the sphere. The conformal Killing vectors are now obeyed locally, except at the poles. For example, if we pick $R^z = (z-z_0)^{-1}$, we get $\partial_{\bar{z}} R^z = 2 \pi \delta^2 (z-z_0)$. The boundary conditions are therefore slightly enhanced, since singularities of the boundary metric $\gamma_{AB}$ are now allowed on the two-sphere. Physically, one can interpret these singularities as cosmic strings that reach out to null infinity as recently emphasized by Strominger and Zhiboedov \cite{Strominger:2016wns} following an earlier construction of Penrose \cite{Penrose:1972xrn}. Several problems however arise: if one insists in defining a consistent asymptotic symmetry algebra, the commutators \eqref{eq:BMSCommu} imply that one needs to generalise the supertranslations to \emph{supertranslations with poles as well}. Then, one can show that all these singular supertranslations admit infinite conserved charges for the Kerr black hole \cite{Barnich:2010eb}.  Another issue is that the standard definition of energy is not bounded from below if the phase space is enhanced consistently with the action of meromorphic superrotations \cite{Compere:2016jwb}. These problems can be resolved, but at the cost of renormalizing the symplectic structure \cite{Compere:2018ylh}. The meromorphic superrotations are therefore not on the same footing as the supertranslation asymptotic symmetries.\\

Another asymptotic symmetry group was also proposed \cite{Campiglia:2015yka} with a distinct extension of the Lorentz group, the entire group of diffeomorphisms on the 2-sphere, Diff$(S^2)$, leading to the total asymptotic symmetry group
\begin{equation}
{\color{blue} \boxed{ \text{Extended } BMS_4 = \text{Diff}(S^2) \ltimes \text{Supertranslations} }}
\end{equation}
The argument is based on the equivalence \cite{Campiglia:2014yka} of the Ward identities of Diff$(S^2)$ symmetry with the subleading soft graviton theorem \cite{Cachazo:2014fwa}. In this case, the leading order metric $g_{AB}$ is allowed to fluctuate, except its determinant. A set of consistent boundary conditions which leads to the definition of surface charges associated with these symmetries can be worked out \cite{Compere:2018ylh} but it requires again a renormalization procedure.

\subsection{Gravitational memory effects}

In this short section, we want to provide evidence that $BMS$ symmetries are more than mathematical artefacts acting on the phase space, but are intrinsically linked to physical phenomena, known as \textit{gravitational memory effects}.

\subsubsection{Displacement memory effect} 

\begin{figure}[h!]
    \begin{center}
        \begin{tikzpicture}[scale=1.0]
			% Observateurs
				%\draw[ForestGreen] plot [smooth,tension=0.5] coordinates { (0,-3) (0.1,-2.6) (2.0,-0.5) (2.5,0) (2.0,0.5)};
				\draw[ForestGreen] (0,-3) [bend left=30] to (0.1,-2.5) -- (2.0,-0.5) -- plot [smooth,tension=0.5] coordinates { (2.0,-0.5) (2.37,0) (2.0,0.5)} decorate [decoration={snake, segment length=7.2pt, amplitude=1pt}] {(2.0,0.5) -- (1.25,1.25)} (1.25,1.25) -- (0.1,2.5) [bend left=30] to (0,3);
				\draw[ForestGreen] (0,-3) [bend left=30] to (0.15,-2.5) -- (2.05,-0.5) -- plot [smooth,tension=0.5] coordinates { (2.05,-0.5) (2.42,0) (2.03,0.53)} decorate [decoration={snake, segment length=7.2pt, amplitude=1pt}] {(2.03,0.53) -- (1.28,1.28)} (1.28,1.28) -- (0.13,2.5) [bend left=30] to (0,3);
				\draw[red,thick,-latex] (1.2,0.05) -- (2.35,1.20);
				\draw[red,thick,-latex] (0.8,0.45) -- (1.95,1.60);
				\draw[red,thick,-latex] (1.0,0.25) -- (2.15,1.40);
				\node[below,red,rotate=-45] at (1.0,0.25) {\footnotesize Radiation};
			% Diamant
        		\draw[thick] (0, 3) -- (3, 0) node[right] {\footnotesize $i^0$} -- (0,-3) -- (-3,0) -- cycle;
            % Infinis de genre lumière             
	            \node[] (1) at (1.25,  2.75) {\footnotesize $\cI^+$};
    	        \node[] (2) at (1.25, -2.75) {\footnotesize $\cI^-$};
			% Points à l'infini
				\fill[black] ( 3, 0) circle [radius=2pt];
				\fill[black] ( 0, 3) circle [radius=2pt] node[above] {\footnotesize $i^+$};
				\fill[black] ( 0,-3) circle [radius=2pt] node[below] {\footnotesize $i^-$};
				\fill[black] (-3, 0) circle [radius=2pt]; 
			% Légendes
				\fill[blue] (2.25,0.75) circle [radius=2pt];
				\fill[blue] (1.50,1.50) circle [radius=2pt];
				\draw[blue] (1.75,0.25) -- (2.50,1.00) node[right] {\footnotesize $u=u_i$};
				\draw[blue] (1.00,1.00) -- (1.75,1.75) node[right] {\footnotesize $u=u_f$};
        \end{tikzpicture}
    \end{center}
    \caption{\textit{Displacement memory effect.}}
    \label{fig:Memory}
\end{figure}
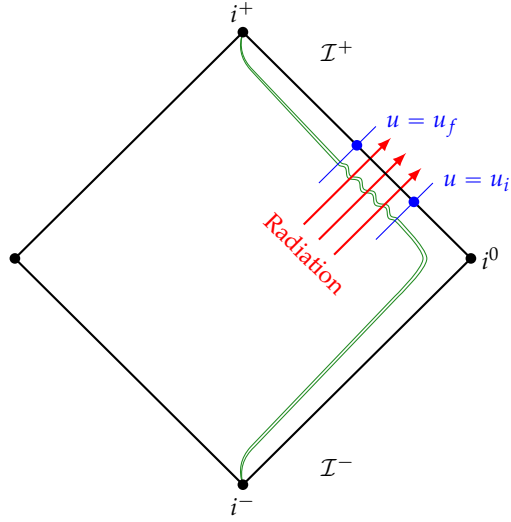

Let us consider a couple of inertial observers (that we will refer to as the "detector") travelling near future null infinity $\cI^+$. The detector is localized in a region with no gravitational radiation or more generally no null signal at both late and early (retarded) times. Let us declare that the radiation is turned on at $u = u_i$ and stops at $u=u_f$. For any value of retarded time excluded from the interval $[u_i,u_f]$, the Bondi news tensor and the matter stress-tensor are identically zero by hypothesis. The detector, which moves on a timelike trajectory in the far region, experiences null radiation only during the time interval $\Delta u = u_f - u_i$ which we suppose $\Delta u \ll r$. We will show that their constant separation will be \textit{permanently shifted} due to the null radiation in a precise way. For the setup, see Figure \ref{fig:Memory}. The leading shift is an angular displacement which is not visible on the Penrose diagram.  

The two inertial observers forming the detector follow a timelike geodesics in the vicinity of $\cI^+$, characterised by a $4$-velocity $v^\mu$. Since their trajectories are located near $\cI^+$, we can admit that $v^\mu \partial_\mu = \partial_u$ up to subleading corrections (necessary for $v^\mu$ to verify $v^\mu v_\mu = -1 + \mathcal{O}(r^{-2})$ in Bondi gauge). The separation between the two geodesic trajectories is given by the deviation vector $s^{\mu}$ which is solution of the famous \textit{equation of geodesic deviation} :
\begin{equation}
\nabla_v \nabla_v s^{\mu} = R^\mu_{\phantom{\mu} \alpha\beta\gamma} V^\alpha V^\beta s^\gamma
\end{equation}
where $\nabla_v = v^\mu \nabla_\mu$ is the directional derivative along $v^\mu$. 
We suppose that both detectors move on the same celestial sphere so $s^r = 0$. We get
\begin{equation}
r^2 \gamma_{AB} \partial_u^2 s^B = R_{uAuB} s^B \Longleftrightarrow \gamma_{AB} \partial_u^2 s^B = \frac{1}{2r} \partial_u^2 C_{AB} s^B
\end{equation}
after using $R_{uAuB}  = -\frac{r}{2} \partial_u^2 C_{AB} + \mathcal{O}(r^0)$ as one can check using the metric in Bondi gauge. Let us introduce the perturbation of the deviation vector as $s^B = s^B_i + r^{-1} s_{sub}^B$. Integrating once in $u$ we obtain an integration constant, which we set to zero assuming that the velocity is zero if the news is zero. Integrating once more, we get
\begin{equation}
{\color{blue} \boxed{ \gamma_{AB} \Delta s_{sub}^B = \frac{1}{2r} \Delta C_{AB} s_i^B + \mathcal{O}(r^{-2}) }}. 
\end{equation}
where $\Delta s_{sub}^A = s_{sub}^A(u=u_f)-s_{sub}^A(u=u_i)$ and $\Delta C_{AB} = C_{AB}(u=u_f) - C_{AB}(u=u_i) $.
Therefore, if the field $\Delta C_{AB}$ is non-zero during the time interval $[u_i,u_f]$, the deviation between the two inertial observers of the detector will be irreversibly shifted : this is the displacement memory effect ! This is a $DC$ effect. Flashing a light between these two observers will measure the shift, which is therefore detectable. This was first observed by Zeldovich and Polnarev in 1974 \cite{1974SvA....18...17Z}. \\

Let us now study the causes of such a displacement. Any process that can change the tensor $C_{AB}$ will lead to the displacement. We integrate the variation of the Bondi mass aspect (\ref{eq:BondiMassVariation}) between $u_i$ and $u_f$ to get simply
\begin{equation}
\Delta m = \frac{1}{4} D^A D^B \Delta C_{AB} - \int_{u_i}^{u_f} du \: T_{uu}.
\end{equation}
If the spacetime is stationary before $u_i$ and after $u_f$ :
\begin{equation}
\Delta C_{AB} = C_{AB,f} - C_{AB,i} = -2 D_A D_B \Delta C + \gamma_{AB} D^2 \Delta C.
\end{equation}
Injecting this in the previous relation, we obtain that the shift of supertranslation memory field $\Delta C$ obeys a quartic elliptic equation which is sourced by 3 qualitatively distinct terms :
\begin{equation}
-\frac{1}{4} (D^2 +2) D^2 \Delta C = \Delta m + \int_{u_i}^{u_f}  du \: \left[ \frac{1}{8} N_{AB} N^{AB} + 4\pi \lim_{r\rightarrow\infty} (r^2 T_{uu}^M) \right].
\label{eq:SourcingC}
\end{equation}
The displacement memory detector will trigger for each of the following causes:  
\begin{itemize}[label=$\rhd$]
\item If the Bondi mass aspect varies between $u_i$ and $u_f$. This is sometimes called \textit{ordinary memory}, but it is not ordinary to our common sense! For example, a single massive body containing a string that suddently separate into two parts due to a trigger will modify the Bondi mass aspect $m$ because the mass will suddenly possess a strong dipolar component. What Einstein gravity tells us is that a signal is sent at null infinity with that information, and the memory effect follows.
\item If null matter reaches $\cI^+$ between $u_i$ and $u_f$. This is sometimes called the \textit{null memory effect}. For example, electromagnetic radiation causes the displacement memory effect. 
\item And finally, if gravitational waves pass through $\cI^+$ between $u_i$ and $u_f$. This is sometimes called the \textit{Christodoulou effect} \cite{Christodoulou:1991cr,Thorne:1992sdb}, even though one could argue that it was found earlier by Blanchet and Damour in the post-Newtonian formalism \cite{Blanchet:1987wq,Blanchet:1992br}. 
\end{itemize}

The displacement memory effect was never observed at the time of writing, but it may be observed in the close future by gravitational wave detectors \cite{Lasky:2016knh} or pulsar timing arrays. 

\subsubsection{Spin memory effect}

Remember the intimate relationship between the $BMS$ supertranslation charge (the local energy on the celestial sphere or Bondi mass) and the displacement memory effect: the change of Bondi mass between initial and final stationary states sources the displacement memory effect -- see equation (\ref{eq:SourcingC}). Given the existence of a local angular momentum on the celestial sphere, the Bondi angular momentum aspect, and the associated Diff$(S^2)$ extended symmetry, Pasterski, Strominger and Zhiboedov raised and proposed an answer in 2015 \cite{Pasterski:2015tva} to the following question: is there a gravitational effect sourced by the Bondi angular momentum aspect? Yes, they called it the \textit{spin memory effect}. \\

%{\color{blue} How does $\Psi$ transforms under Lorentz transformations?}

The starting point is that in the decomposition of $C_{AB}$ \eqref{eq:Cgeneral}, $\Psi$ is zero at retarded times without radiation, but $\int_{u_i}^{u_f} du \: \Psi$ can still be measured\footnote{The attentive reader might already notice that such an observable is not clearly the difference between a quantity in the initial and final state. Instead $\int_{u_i}^{u_f} du \: \Psi $ is non-local in retarded time! Therefore, though it bears analogy with the displacement memory effect, the spin memory effect is not (yet?) proven to be a memory effect at the first place! For a discussion of memory effects associated with Diff$(S^2)$ symmetries, see \cite{Compere:2018ylh}.}. Note that the polarisation mode $\Psi$ is a pseudo-scalar which flips under parity. In order to measure it, let's consider the following thought experiment. We consider light rays which orbit a circle $\mathcal{C}$ of radius $L$ (like a toroid optic fiber, or a circular wall of mirrors) in the vicinity of $\cI^+$. The photons are allowed to travel clockwise or counterclockwise in this system. Again we assume that no null radiation passes through the system at early and late times. The system experiences the passage of radiation which carries non-trivial angular momentum ($N_A \neq 0$) during a finite range of retarded time $[u_i,u_f]$, as before. Let us denote by $\Delta u$ the relative time delay between the clockwise and counterclockwise light rays. For $u<u_i$, we set the system such that $\Delta u=0$. The spin memory effect resides in the fact the $\Delta u$ is no more zero at late time after the null radiation has passed. After a computation explained in \cite{Pasterski:2015tva}, one gets
\begin{equation}
\Delta u = \frac{1}{2\pi L} \int du \: \int_{\mathcal{C}} \left( D^A C_{AB} dx^B \right). 
\end{equation}
This integral is independent of $C$ and therefore only depends upon $\Psi$. In order to see it, compute $D^A C_{AB}$ using the decomposition  \eqref{eq:Cgeneral}. The terms involving $C$ lead to  $D^A C_{AB} = -2D_B (D^2+2)C$ which is a total derivative on the circle, and therefore leads to a vanishing integral on the closed circle. Therefore, $\Delta u$ is only function of $\int_{u_i}^{u_f} du \: \Psi$. \\

The sources of spin memory can be obtained by integrating the conservation law of the angular momentum density \eqref{eq:EvolutionForNA} between $u_i$ and $u_f$. The definite integral of the first term on the right-hand side does not depend upon $C$ as one can check. It leads to some differential operator acting on $\int_{u_i}^{u_f} du \: \Psi$, symbolically denoted by $\mathcal D \int_{u_i}^{u_f} du \: \Psi$. The rest of the right-hand side brings a total contribution of the null matter and gravitational wave sources: $- \int_{u_i}^{u_f} du \: \hat T_{uA}$. We therefore find
\begin{equation}
\mathcal D \int_{u_i}^{u_f} du \: \Psi = \Delta N_A +  \int_{u_i}^{u_f} du \: \hat T_{uA}.
\end{equation}
There is a decomposition of the origin of the spin memory effect into 3 qualitative classes similar to the displacement memory effect: a change of angular momentum aspect, angular momentum flux from null matter and angular momentum flux from gravitational waves. \\

The spin memory effect is subleading with respect to the displacement memory effect and is probably not observable by the current 2G technology of gravitational wave detectors \cite{Nichols:2017rqr}. 

\section{Scattering problem and junction conditions}

We close this lecture about asymptotically flat spacetimes by a little glance at the very fundamental \textit{scattering problem} in General Relativity.  Let us first consider scattering of null radiation (null matter fields or gravitational waves) around Minkowski spacetime. We are interested in relating the out states at $\cI^+$ to the in states at $\cI^-$. Are there universal constraints among these $\mathcal{S}$-matrix elements? This is the question that Strominger asked and answered in his 2013 paper \cite{Strominger:2013jfa}. \\

So far we found an asymptotic symmetry group at $\mathcal I^+$, the $BMS$ group (with or with extension depending upon the boundary conditions).  From now on we will add a superscript $BMS^+$ to our notation to remind us where this group is defined. It is not difficult to convince yourself that exactly the same construction can be performed at $\mathcal I^-$ upon switching retarded to advanced coordinates. There is therefore a $BMS^-$ asymptotic symmetry group. Now, acting with symmetries on the initial state should be reflected on the final state. There cannot be independent symmetries acting on both the initial and final states, otherwise the scattering problem would not be defined! The $BMS$ group is therefore defined as a diagonal subgroup of the product $BMS^+ \times BMS^-$. \\

But which diagonal subgroup? Or in other words, how to identify the generators at $\mathcal I^+$ with the ones at $\mathcal I^-$? The crucial clue is that propagating fields around Minkowski spacetime obey universal antipodal matching conditions at spatial infinity. For each bulk field $\Phi^i$, one can define its limit at $\mathcal I^+$ and then take $u \rightarrow -\infty$ which defines the field at $\mathcal I^+_-$. Similarly, one can define the field at  $\mathcal I^-_+$. An antipodal matching condition would mean that 
\begin{equation}
\Phi^i(\theta,\phi)\vert_{\mathcal I^+_-} = \Phi^i(\pi - \theta,\phi + \pi )\vert_{\mathcal I^-_+} 
\end{equation}
Now it turns out that the electromagnetic Li\'enard-Wierchert field describing the retarded electromagnetic field of a uniformly moving source obeys these antipodal matching conditions. The metric of the boosted Kerr black hole is also expected to obey these conditions. Moreover, these conditions are $CPT$ invariant and Lorentz invariant. \\

A complementary perspective comes from perturbative quantum gravity. The $\mathcal{S}$-matrix should be $BMS$ invariant in the sense that acting with BMS charges on the ``in'' state and ``out'' states commute: $Q_{T,R}^+ \mathcal{S} = \mathcal{S} Q_{T,R}^-$. These relationships are the Ward identities of supertranslations and Lorentz transformations or their superrotation extension. Now, as shown in 2014 \cite{He:2014laa} it turns out that the supertranslation Ward identities are identical, after a change of notation, to Weinberg's leading soft graviton theorems derived in 1965 \cite{PhysRev.140.B516}. At subleading order, it was also shown in 2014 \cite{Campiglia:2014yka} that the Diff$(S^2)$ superrotation Ward identities are identical, after rewriting, to the newly found Cachazo-Strominger's subleading soft graviton theorems \cite{Cachazo:2014fwa}.\\

The antipodal matching conditions are compatible with the soft theorems, which validates their range of applicability around Minkowski spacetime. Yet, it has not been derived whether or not the antipodal map is generally valid for any subleading field in the asymptotic expansion close to null infinity, and for spacetimes with other causal structures, such as spacetimes containing a black hole formed from collapse. For the leading order fields, consistent boundary conditions exist which admit antipodal matching boundary conditions both in $4d$ Einstein gravity \cite{Troessaert:2017jcm,Henneaux:2018cst} and $3d$ Einstein gravity \cite{Compere:2017knf}. \\

Assuming that the antipodal matching conditions hold in generality directly leads to conservation laws. Indeed, physical quantities depend upon the fields, so if all relevant fields are antipodally matched, the conserved charges at $\mathcal I^+_-$ and $\mathcal I^-_+$ (either supertranslations or superrotations) are related by the antipodal map symbolically denoted by $\text{AntiPodMap}(\circ)$, 
\begin{equation}
Q\vert_{\mathcal I^+_-} = \text{AntiPodMap}(Q)\vert_{\mathcal I^-_+} 
\end{equation}
Using the conservation laws of these charges, of the form $\partial_u Q = J^u$ and $\partial_v Q = J^v$, we deduce by integration along $u$ and $v$ the conservation laws 
\begin{equation}
\int du \: J^u + Q\vert_{\mathcal I^+_+} = \text{AntiPodMap} \left( \int dv \: J^v + Q\vert_{\mathcal I^-_-} \right). 
\end{equation}
These (supertranslation and superrotation) conservations laws are the \textit{conservation of energy} and \textit{angular momentum at each angle} on the $2$-sphere. \\

This topic is still under investigation, especially in relationship with black holes since it has been conjectured to be relevant if not crucial to resolve the black hole information paradox \cite{Hawking:2016msc,Strominger:2017aeh} (see however \cite{Mirbabayi:2016axw,Bousso:2017dny}) !

\section*{References}
\addcontentsline{toc}{section}{References}

The  {\color{blue}$BMS$ group} of supertranslations and Lorentz transformations of 4-dimensional asymptotically flat spacetimes was described in the founding papers of 1962 \cite{Bondi:1962px,Sachs:1962wk}. The proposed extension by meromorphic superrotations and supertranslations was studied in \cite{deBoer:2003vf,Barnich:2009se,Barnich:2010eb}. The proposed extension to diffeomorphisms on the 2-sphere was derived in \cite{Campiglia:2014yka,Campiglia:2015yka}.\\

The  {\color{blue} displacement memory effect} was independently discovered from the three qualitatively distinct sources of displacement memory: the change of Bondi mass aspect \cite{1974SvA....18...17Z}, gravitational wave flux \cite{Blanchet:1987wq,Blanchet:1992br,Christodoulou:1991cr,Thorne:1992sdb} and null matter radiation \cite{Bieri:2010tq}. The  {\color{blue} spin memory effect} was described in \cite{Pasterski:2015tva}. Experimental prospects include \cite{Lasky:2016knh,Nichols:2017rqr}. \\

The renewal of the topic of $BMS$ symmetries is largely due to the work on  {\color{blue} $BMS$ invariance of scattering}  \cite{Strominger:2013jfa} and the triangle relationship between $BMS$ supertranslation symmetries, {\color{blue} Weinberg' soft graviton theorem} \cite{He:2014laa} and the displacement memory effect \cite{Strominger:2014pwa}. \\

The latest proposal to solve the black hole information paradox using {\color{blue} $BMS$ soft hair} \cite{Strominger:2017aeh}  slightly differs from the first proposal \cite{Hawking:2016msc} (briefly mentioned in \cite{Strominger:2014pwa}). \\

These lectures are also based on several reviews and papers \cite{Barnich:2011mi,Flanagan:2015pxa,Compere:2016jwb,Compere:2016hzt,Strominger:2017zoo}.

	% Lecture 4 : Kerr
\chapter{Rotating black holes}
\label{sec:Kerr}

For this last lecture, we will focus on particular properties of astrophysically realistic $4d$ black holes, which are rotating and uncharged. We will concentrate on stationary asymptotically flat black holes, that admit by definition an asymptotically timelike Killing vector $\partial_t$. The rigidity theorem due to Hawking states that ``stationarity implies axisymmetry'' so these solutions also possess an additional axial Killing vector $\partial_\phi$. Such black holes have necessarily spherical topology. They contain a singularity hidden by an event horizon which is also a Killing horizon. They are exactly described by the {\color{blue} \textbf{Kerr black hole solution}} found in 1963. Quite remarkably, the Kerr black hole solution is also a dynamical attractor: it is the final state of collapse of matter. It is therefore one of the most important analytical solutions of Einstein's equations due to its universality! First of all, we will review the main features of this rich spacetime, before entering in more advanced considerations. \\

In a second step, we will study the maximally spinning limit of the Kerr solution, the {\color{blue} \textbf{extremal Kerr black hole}}. The extremal Kerr black hole lies at the frontier between the regular Kerr black holes and naked singularities which are thought to be unphysical (it would be proven unphysical if one could prove Penrose's cosmic censorship). The third law of black hole mechanics states that it is impossible to spin up a black hole beyond the maximal limit because the extremal black hole has zero Hawking temperature and no physical process can reach absolute zero temperature. If one attempts to send finely-tuned particles or waves into a near-extremal black hole in order to further approach extremality, one realizes that there is a smaller and smaller window of parameters that allows one to do so. On the other hand, if one starts with an extremal black hole, one can simply throw in a massive particle to make the black hole non-extremal. In summary, extremal black holes are finely tuned and (classically) unreachable black holes. \\

The extremal Kerr black hole is a very interesting solution because it admits special near horizon limits. Such limits admit enhanced conformal symmetry that shares features with anti-de Sitter spacetimes where holography and therefore quantum gravity is most understood. The attempts (with successes and failures) to describe the extremal Kerr black hole with holographic techniques is called the {\color{blue} \textbf{\textit{Kerr/CFT} correspondence}} and will be briefly reviewed here. \\

The final part of these lectures will be devoted to the analysis of  {\color{blue} \textbf{gravitational perturbations}} around Kerr geometries. The Kerr black hole is currently under experimental tests by the LIGO/Virgo gravitational wave detectors. Indeed, the final stages of black hole mergers consist in a {\color{blue}\textbf{quasi-normal mode ringing}} of the resulting black hole which is well-described by perturbation theory around the Kerr black hole. Since the Kerr black hole only depends upon 2 parameters, namely the mass and angular momentum, the resonance frequencies (the quasi-normal modes) of the black hole are characteric signatures of Einstein gravity. The emerging experimental science of black hole spectroscopy will soon test the limits of Einstein gravity and look for possible deviations!

\newpage

\section{The Kerr solution --- Review of the main features}

\subsection{Metric in Boyer-Lindquist coordinates}
The Kerr metric describes the most general regular asymptotically flat, stationary and axisymmetric spacetime in $4d$ Einstein gravity. In Boyer-Lindquist coordinates $(t,r,\theta,\phi)$ the metric reads as 
\begin{equation}
ds^2 = - \frac{\Delta}{\Sigma} \left( dt -a \sin^2\theta d\phi \right)^2 + \Sigma \left( \frac{dr^2}{\Delta} + d\theta^2 \right) +\frac{\sin^2\theta}{\Sigma}\left( (r^2+a^2)d\phi - a dt \right)^2
\label{eq:Kerr}
\end{equation}
where 
\begin{equation}
\begin{array}{rcl}
\Delta(r) &\triangleq& r^2 - 2GMr + a^2, \\ 
\Sigma(r,\theta) &\triangleq& r^2 + a^2 \cos^2\theta. 
\end{array}  
\end{equation}
In the following, we will set $G=1$. The metric possesses 2 Killing vectors, $\partial_t$ and $\partial_\phi$, because its components do not depend on these coordinates. Contrary to the Schwarzschild black holes, the Kerr solution is not static, since the time reversal transformation $t \rightarrow -t$ does not preserve the cross-term $dtd\phi$. The mixed term between $t$ and $\phi$ has the effect of ``dragging'' the spacetime along with the rotating body, just as water being dragged along by the surface of an immersed spinning ball. This effect of ``frame dragging'' is very important in the discussion of the physical properties of Kerr black holes, and it can be shown that it is responsible of the gyroscopic precession of test-rotating bodies, called the ``Lense-Thirring effect''.\\

Kerr black holes form a $2$-parameter family of solutions labeled by $M$ and $a$. $M$ is the surface charge associated to $\partial_t$ when evaluated on the sphere at infinity, so it's clearly the total energy (or total mass since we are in the rest frame) of the black hole. The second Killing vector $-\partial_\phi$ gives rise to another conserved surface charge which is $J = aM$, the angular momentum, so the parameter $a=J/M$ is the specific angular momentum.

\subsection{Killing horizon and black hole thermodynamics}

The metric admits a physical singularity at $\Sigma = 0$ because the curvature invariant $R^{\mu\nu\alpha\beta} R_{\mu\nu\alpha\beta}$ blows up at that locus. For $a \neq 0$, it can be shown to be a ring, known as the ring singularity of Kerr. Assuming that Penrose's cosmic censorship principle is true, this singularity must be shielded by a event horizon.\\  

The metric (\ref{eq:Kerr}) admits several coordinate singularities. First at $\theta = 0,\pi$ one has the familiar polar coordinate singularities which can be removed by switching to local cartesian coordinates. Second, at $\Delta (r) = 0$ the first term in \eqref{eq:Kerr} vanishes while the $dr^2$ term blows up. This occurs for 2 values of $r$, denoted by $r_\pm = M \pm \sqrt{M^2-a^2}$, since $\Delta (r)$ is a quadratic function of $r$. The outer value $r=r_+$ is the event horizon $\mathcal H_+$ of the black hole, which we will prove later on. The existence of the event horizon bounds the angular momentum as
\begin{equation}
-M \leq a \leq M. \label{Jb}
\end{equation}
If $a > M$, there is no event horizon and the curvature singularity is naked, which is unphysical, so we assume that the bound \eqref{Jb} holds in Nature.

The horizon $r=r_+$ is in fact a Killing horizon: its null rays are generated by a Killing vector, $\xi = \partial_t + \Omega_H \partial_\phi$. One can check that $\Omega_H = \frac{a}{2Mr_+}$ is such that $\xi^2 = 0$ at $\mathcal H_+$ and  $\nabla_\xi \xi^\mu = \kappa \xi^\mu$ at $\mathcal H_+$ where $\kappa$ is the surface gravity of the black hole (with respect to the unit normalized generator such that $\xi^2 = -1$ at spatial infinity). Thanks to the groundbreaking Hawking  result of 1974 \cite{Hawking:1974sw}, the surface gravity determines the Hawking temperature of the black hole to be
\begin{equation}
T_H = \frac{ \kappa}{2 \pi} = \frac{ (r_+ - M)}{4 \pi M r_+}. 
\end{equation}
Due to the rotation, the Kerr black hole possesses a second Killing horizon, which is called the \textit{inner} horizon defined at the radius $r_- \leq r_+$. This horizon is a Cauchy horizon as we will discuss below. The limiting case $a = M$ is called the \textit{extremal} case. The extremal black hole only possesses one horizon and the Hawking temperature is identically zero. The third law of thermodynamics prevents a thermal system to reach zero temperature. Anagolously, no physical process exists that allows to reach an exactly extremal black hole. \\

The area of the outer horizon divided by $4$ ($4G\hbar$ in $MKSA$ units) has the interpretation of microscopic entropy of the black hole thanks to the famous 1973 Bardeen, Carter and Hawking result \cite{1973CMaPh..31..161B} on black hole thermodynamics combined with Hawking's 1974 result \cite{Hawking:1974sw} on the identification of black hole temperature, with earlier insights from Bekenstein \cite{1972NCimL...4..737B}. The Bekenstein-Hawking entropy of the Kerr black hole is
\begin{equation}
S_{BH} = \frac{A_H}{4} = 2\pi M r_+.
\end{equation} 
For an extremal black hole, $r_+ = M = a$ and $S_{BH}^{ext} = 2\pi M^2 = 2\pi J$. \\

One of the main challenges of a theory of quantum gravity theory is to account for this entropy! It obeys the first law of black hole thermodynamics
\begin{equation}
T_H \delta S_{BH} = \delta M - \Omega_H \delta J
\end{equation} 
and the second law of thermodynamics which states that the entropy of the outer universe plus the black hole entropy always increases. 

\subsection{Ergoregion}

\begin{figure}[h!]
\centering
\begin{tikzpicture}
\draw[white] (-6,-3.5) -- (-6,3.5) -- (6,3.5) -- (6,-3.5) -- cycle;
\begin{polaraxis}[hide axis,anchor=origin,at={(0pt,0pt)}]
\addplot+[mark=none,fill=blue!10,domain=0:360,samples=600,thick] {(1 + sqrt(1 - 0.79*(cos(90-x))^2))}; 
\end{polaraxis}
\draw[black,thick,fill=white] (0,0) circle (2.27);
\draw[black,dashed,thick,->] (0,-2.7) -- (0,3) node[above]{\footnotesize $z$};
\draw[black,thick] (-4,0) -- (4,0);
\coordinate (legend) at (0,-3);
\node[below] at (legend) {\footnotesize $\phi$};
\draw[->] plot [smooth,tension=1] coordinates {($(legend)+(-0.5,0.1)$) ($(legend)+(0,0.05)$) ($(legend)+(0.5,0.1)$)};
\draw[] (0,0) -- (1.75,2.05);
\draw[->] (0,1) to [bend right=-25] (0.6,0.7);
\node[above] at (0.3,1) {\footnotesize $\theta$};
\node[below] at (1,1) {\footnotesize $r$};
%\path[decorate,decoration={text along path,text={\footnotesize Event horizon},text align=center}] (0,0) arc [start angle=90,end angle=180,radius=3];
\path[decorate,decoration={text along path,text={|\footnotesize|Event horizon},text align=center}] (-1.9,0) arc [start angle=180,end angle=90,radius=1.9];
\node[text width=2] at (3.5,0) {\footnotesize Equatorial plane};
\path[decorate,decoration={text along path,text={|\footnotesize\color{blue}|Ergosphere},text align=center}] (-3.18,0.25) arc [start angle=180,end angle=90,radius=2];
\end{tikzpicture}
\caption{Ergoregion of the Kerr black hole ($t$ fixed).}
\label{fig:ergo}
\end{figure}
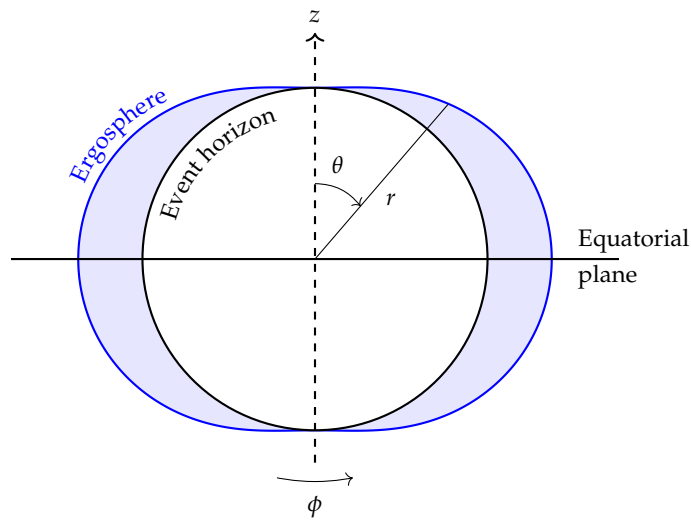

The concept of energy as measured by an asymptotic observer is the conserved quantity associated with $\partial_t$. An interesting feature of the Kerr solution is that $\partial_t$ becomes spacelike beyond a specific surface known as the  \textit{ergosphere}. The metric component $g_{tt} = -(\Delta - a^2 \sin^2 \theta)/\Sigma$ vanishes at the ergosphere radius $r_{erg} = M + \sqrt{M^2 - a^2 \cos^2 \theta}$. The ergosphere lies outside the event horizon and therefore delimits a region called the \textit{ergoregion} which is depicted in Figure \ref{fig:ergo}. Since the energy becomes qualitatively similar to a (spacelike) momentum in that region from the point of view of an asymptotic observer, the energy can take a negative sign locally while remaining globally positive. This leads to many important phenomena of energy extraction from a Kerr black hole, either by particles (the \textit{Penrose process}), waves (the \textit{superradiant effect} \cite{1971JETPL..14..180Z}) or magnetic accretion disks (the \textit{Blandford-Znajek process} \cite{1977MNRAS.179..433B})\dots 

\subsection{Event horizon and Cauchy horizon}
\label{horizons}

\begin{figure}[h!]
  \centering
  \vspace{-10pt}
  \subfloat[Penrose diagram for $r > r_-$]{
  	\begin{tikzpicture}[scale=0.9]
  		\draw[white] (-4.5,-5) -- (-4.5, 5) -- ( 4.5, 5) -- ( 4.5,-5) -- cycle;
		\draw[thick,GrandVert] (-2,-2) -- (-4,0) -- (-2,2);
		\draw[thick,GrandVert] ( 2,-2) -- (4,0) -- ( 2,2);
		\fill[orange!30] (-2,-2) -- (0,-4) -- (2,-2) -- (0,0) -- cycle;
		\fill[orange!10] (-2,2) -- (0,4) -- (2,2) -- (0,0) -- cycle;
		\draw[blue,densely dashed] (-2, 2) -- (0,4) -- ( 2, 2);
		\node[above,blue,rotate=-45] at ( 1,3) {\footnotesize $r=r_-$};
		\node[above,blue,rotate= 45] at (-1,3) {\footnotesize $r=r_-$};
		\draw[blue,densely dashed] (-2,-2) -- (0,-4) -- ( 2,-2);
		\node[below,blue,rotate= 45] at ( 1,-3) {\footnotesize $r=r_-$};
		\node[below,blue,rotate=-45] at (-1,-3) {\footnotesize $r=r_-$};
		\draw[blue] (-2,-2) -- (2, 2);
		\node[above,blue,rotate= 45] at ( 1, 1) {\footnotesize $r=r_+$};
		\draw[blue] (-2, 2) -- (2,-2);
		\node[above,blue,rotate=-45] at (-1, 1) {\footnotesize $r=r_+$};
		\node[above,GrandVert,rotate=-45] at (3, 1) {\footnotesize $\mathcal{I}^+$};
		\node[below,GrandVert,rotate=+45] at (3, -1) {\footnotesize $\mathcal{I}^-$};
		\draw[red] plot[smooth,tension=0.5] coordinates{(2,2) (1,0) (2,-2)};
		\node[red,right] at ( 1, 0) {\footnotesize $r=r_{erg}$};
		\draw[red] plot[smooth,tension=0.5] coordinates{(-2,2) (-1,0) (-2,-2)};
		\node[red,left] at (-1, 0) {\footnotesize $r=r_{erg}$};
		\node[above,GrandVert,rotate=45] at (-3, 1) {\footnotesize $\mathcal{I}^+$};
		\node[below,GrandVert,rotate=-45] at (-3, -1) {\footnotesize $\mathcal{I}^-$};
		\node[below,blue,rotate=-45] at ( 1,-1) {\footnotesize $r=r_+$};
		\node[below,blue,rotate=+45] at (-1,-1) {\footnotesize $r=r_+$};
		\draw[fill=black] (2,2) circle [radius = 2pt];
		\node[above] at (2,2) {\footnotesize $\quad i^+$};
		\draw[fill=black] (2,-2) circle [radius = 2pt];
		\node[below] at (2,-2) {\footnotesize $\quad i^-$};
		\draw[fill=black] (4,0) circle [radius = 2pt];
		\node[right] at (4,0) {\footnotesize $i^0$};
		\draw[fill=black] (-2,2) circle [radius = 2pt];
		\node[above] at (-2,2) {\footnotesize $ i^+ \quad$};
		\draw[fill=black] (-2,-2) circle [radius = 2pt];
		\node[below] at (-2,-2) {\footnotesize $i^-\quad$};
		\draw[fill=black] (-4,0) circle [radius = 2pt];
		\node[left] at (-4,0) {\footnotesize $i^0$};
	\end{tikzpicture}
	\label{fig:PenroseKerrLeft}
   } 
   \subfloat[Penrose diagram for $r < r_-$]{
  	\begin{tikzpicture}[scale=0.9]
  		\draw[white] (-4.5,-5) -- (-4.5, 5) -- ( 4.5, 5) -- ( 4.5,-5) -- cycle;
		\draw[thick,GrandVert] (-2,-2) -- (-4,0) -- (-2,2);
		\draw[thick,GrandVert] ( 2,-2) -- (4,0) -- ( 2,2);
		\fill[orange!10] (-2,-2) -- (0,-4) -- (2,-2) -- (0,0) -- cycle;
		\fill[orange!30] (-2,2) -- (0,4) -- (2,2) -- (0,0) -- cycle;
		\draw[blue,densely dashed] (-2, 2) -- (0,4) -- ( 2, 2);
		\node[above,blue,rotate=-45] at ( 1,3) {\footnotesize $r=r_+$};
		\node[above,blue,rotate= 45] at (-1,3) {\footnotesize $r=r_+$};
		\draw[blue,densely dashed] (-2,-2) -- (0,-4) -- ( 2,-2);
		\node[below,blue,rotate= 45] at ( 1,-3) {\footnotesize $r=r_+$};
		\node[below,blue,rotate=-45] at (-1,-3) {\footnotesize $r=r_+$};
		\draw[blue] (-2,-2) -- (2, 2);
		\node[above,blue,rotate= 45] at ( 1, 1) {\footnotesize $r=r_-$};
		\draw[blue] (-2, 2) -- (2,-2);
		\node[above,blue,rotate=-45] at (-1, 1) {\footnotesize $r=r_-$};
		\node[above,GrandVert,rotate=-45] at (3, 1) {\footnotesize $\mathcal{I}^+$};
		\node[below,GrandVert,rotate=+45] at (3, -1) {\footnotesize $\mathcal{I}^-$};
%		\draw[red] plot[smooth,tension=0.5] coordinates{(2,2) (1,0) (2,-2)};
%		\node[red,right] at ( 1, 0) {\footnotesize $r_{erg}$};
%		\draw[red] plot[smooth,tension=0.5] coordinates{(-2,2) (-1,0) (-2,-2)};
%		\node[red,left] at (-1, 0) {\footnotesize $r_{erg}$};
		\node[above,GrandVert,rotate=45] at (-3, 1) {\footnotesize $\mathcal{I}^+$};
		\node[below,GrandVert,rotate=-45] at (-3, -1) {\footnotesize $\mathcal{I}^-$};
		\node[below,blue,rotate=-45] at ( 1,-1) {\footnotesize $r=r_-$};
		\node[below,blue,rotate=+45] at (-1,-1) {\footnotesize $r=r_-$};
		\draw[very thick,violet] (2,-2) -- (2,2);
		\draw[very thick,violet] (-2,-2) -- (-2,2);
		\node[above,violet,rotate=-90] at (2,0) {\footnotesize Ring singularity};
		\node[above,violet,rotate=+90] at (-2,0) {\footnotesize Ring singularity};
		\draw[fill=black] (2,2) circle [radius = 2pt];
		\node[above] at (2,2) {\footnotesize $\quad i^-$};
		\draw[fill=black] (2,-2) circle [radius = 2pt];
		\node[below] at (2,-2) {\footnotesize $\quad i^+$};
		\draw[fill=black] (4,0) circle [radius = 2pt];
		\node[right] at (4,0) {\footnotesize $i^0$};
		\draw[fill=black] (-2,2) circle [radius = 2pt];
		\node[above] at (-2,2) {\footnotesize $ i^- \quad$};
		\draw[fill=black] (-2,-2) circle [radius = 2pt];
		\node[below] at (-2,-2) {\footnotesize $i^+\quad$};
		\draw[fill=black] (-4,0) circle [radius = 2pt];
		\node[left] at (-4,0) {\footnotesize $i^0$};
	\end{tikzpicture}
	\label{fig:PenroseKerrRight}
   } 
\caption{Penrose diagrams for Kerr spacetime.}
\end{figure}
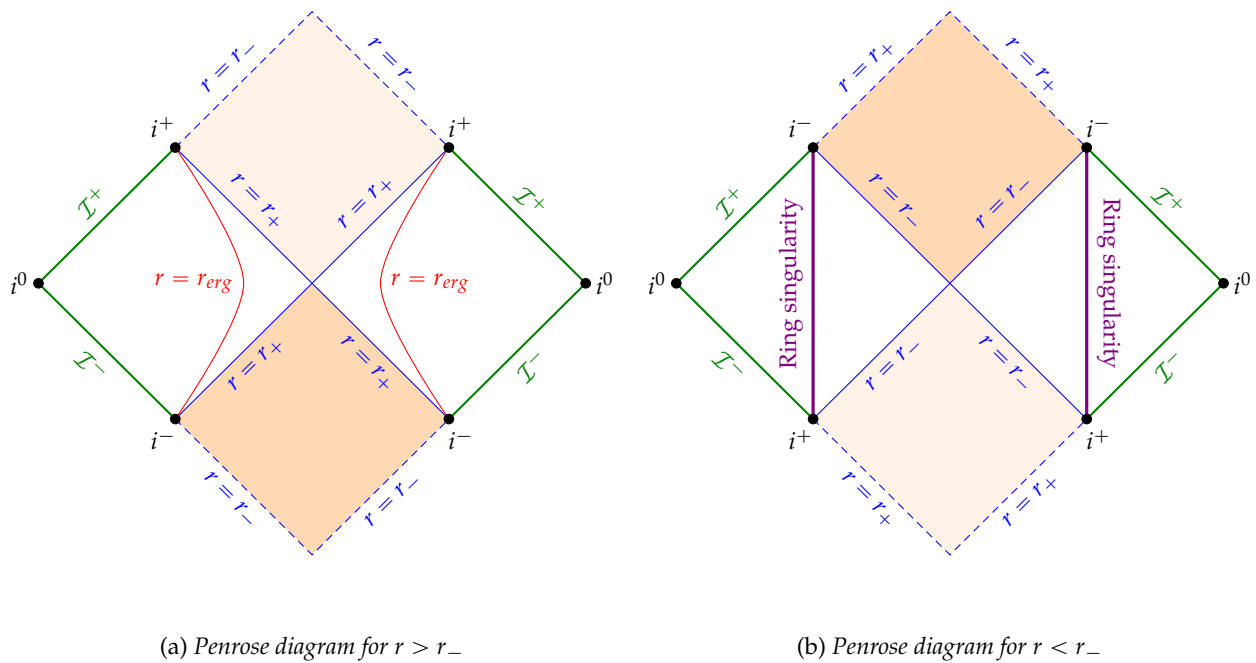

In order to clarify the notions of event horizon and Cauchy horizon we need to clarify the causal structure of the Kerr metric. For that purpose, let us draw the corresponding Penrose diagrams. We only consider the non-extremal case $r_- < r_+$. Extremal diagrams are formally obtained by removing the inner parts between the horizons $r=r_+$ and $r=r_-$ and gluing them. We refer to the excellent Carter lectures \cite{1973blho.conf...57C} for details. The maximal extension of the non-extremal Kerr black hole is obtained by gluing up an infinite sequence of two Penrose diamonds that are depicted in Figures \ref{fig:PenroseKerrLeft} and \ref{fig:PenroseKerrRight}. The orange parts with same opacity are overlapping and must be glued.

Note that these diagrams share some features with the non-extremal \textit{BTZ} ones. The major difference concerns the lateral shapes. For \textit{BTZ}, we get straight vertical lines, which reflect the $AdS_3$ asymptotic geometry of \textit{BTZ}, while for Kerr we get diamonds, which reflect the asymptotic flatness of Kerr! \\

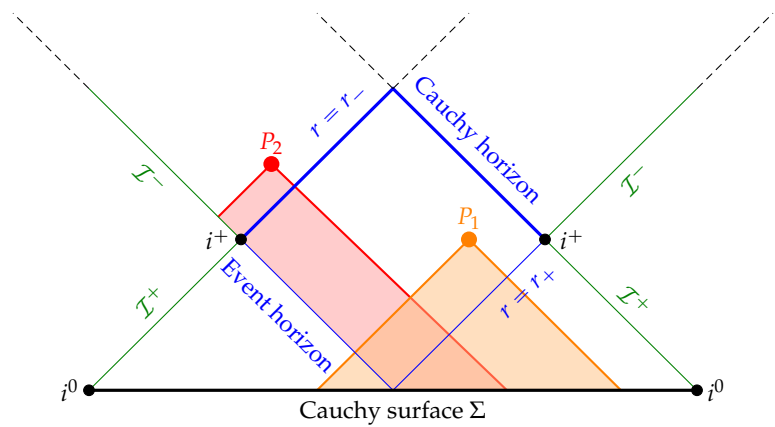
\begin{figure}[h!]
\centering
\vspace{20pt}
\begin{tikzpicture}[scale=1]
	\fill[red!20] (-2.3,2.3) -- (-1.6,3.0) -- (1.5,0) -- (0,0) -- cycle;
	\draw[red,thick] (-2.3,2.3) -- (-1.6,3.0) -- (1.5,0);
	\fill[orange!50,opacity=0.5] (-1,0) -- (1,2) -- (3,0);
	\draw[orange,thick] (-1,0) -- (1,2) -- (3,0);
	\draw[GrandVert] (4,0) -- (2,2) -- (4,4);
	\draw[GrandVert] (-4,0) -- (-2,2) -- (-4,4);
	\draw[very thick,blue] (2,2) -- (0,4) -- (-2,2);
	\draw[densely dashed] ( 4,4) -- ( 5,5);
	\draw[densely dashed] (-4,4) -- (-5,5);
	\draw[densely dashed] ( 0,4) -- ( 1,5);
	\draw[densely dashed] ( 0,4) -- (-1,5);
	\draw[blue] (2,2) -- (0,0) -- (-2,2);
	\draw[very thick,black] (-4,0) -- (4,0);
	\node[above,GrandVert,rotate=-45] at (3, 1) {\footnotesize $\mathcal{I}^+$};
	\node[below,GrandVert,rotate=+45] at (3, 3) {\footnotesize $\mathcal{I}^-$};
	\node[above,GrandVert,rotate=+45] at (-3, 1) {\footnotesize $\mathcal{I}^+$};
	\node[below,GrandVert,rotate=-45] at (-3, 3) {\footnotesize $\mathcal{I}^-$};
	\draw[fill=black] (4,0) circle[radius=2pt];
	\node[right] at (4,0) {\footnotesize $i^0$};
	\draw[fill=black] (-4,0) circle[radius=2pt];
	\node[left] at (-4,0) {\footnotesize $i^0$};
	\draw[fill=black] (2,2) circle[radius=2pt];
	\node[right] at (2,2) {\footnotesize $\: i^+$};
	\draw[fill=black] (-2,2) circle[radius=2pt];
	\node[left] at (-2,2) {\footnotesize $i^+$};
	\node[rotate=+45,blue] at (-0.7,3.7) {\footnotesize $r=r_-$};
	\node[rotate=-45,blue] at ( 1.1,3.3) {\footnotesize Cauchy horizon};
	\fill[red] (-1.6,3.0) circle[radius=3pt];
	\node[above,red] at (-1.6,3.0) {\footnotesize $P_2$};
	\fill[orange] (1,2) circle[radius=3pt];
	\node[above,orange] at (1,2) {\footnotesize $P_1$};
	\node[below] at (0,0) {\footnotesize Cauchy surface $\Sigma$};
	\node[rotate=+45,blue] at (1.8,1.3) {\footnotesize $r=r_+$};
	\node[rotate=-45,blue] at (-1.5,1.0) {\footnotesize Event horizon};
\end{tikzpicture}
\caption{The inner horizon is a Cauchy horizon.}
\label{fig:CauchyHorizon}
\vspace{20pt}
\end{figure}

The physical meaning of the inner horizon $r=r_-$ becomes clear on the Penrose conformal diagram \ref{fig:PenroseKerrLeft}. We will actually show that the surface is a \textit{Cauchy horizon}. Physics usually requires the existence of a Cauchy surface $\Sigma$, which is an hypersurface that all light rays and massive particle trajectories intersect exactly once. Then hyperbolic field equations will determine unambiguously the past and future behaviour of their solutions when initial data is taken on $\Sigma$. Now, the point is that if the region beyond $r=r_-$ is taken into account, no Cauchy hypersurface exists. Indeed, if we pick a point $P_1$ at radius larger than the inner horizon $r=r_-$, the past lightcone of $P_1$ entirely crosses $\Sigma$ (see Figure \ref{fig:CauchyHorizon}), so the physics at $P_1$ is entirely determined by the initial data given on $\Sigma$. Let us now take a point $P_2$ beyond $r=r_-$. After drawing the past lightcone in this point, we see that it also requires information from the other asymptotic region $\mathcal{I}^-$ and so the initial data on $\Sigma$ is not sufficient to determine the event at $P_2$. This prevents the existence of a single Cauchy surface. The surface beyond which the events are no more causally determined by $\Sigma$ is the Cauchy horizon! 

In fact, numerical analysis shows that any perturbation falling into a Kerr black hole leads to a physical curvature singularity around the Cauchy horizon \cite{2000PhRvD..61b4001O}. It is therefore safe to assume that the region beyond the inner horizon is unphysical. Science-fiction movies like to exploit the wormhole which connects the asymptotic region $\mathcal I^-$ of Figure \ref{fig:PenroseKerrRight} to the asymptotic region $\mathcal I^+$ of Figure \ref{fig:PenroseKerrRight}. More precisely, one can imagine a spaceship falling into the black hole (crossing $r=r_+$), passing the Cauchy horizon, avoiding the ring singularity and emerging from the asymptotic flat region on the right-hand side of Figure \ref{fig:PenroseKerrRight}. This scenario is however unphysical because the classical journey will end at the Cauchy horizon where curvature blows up. Only quantum gravity can tell us what happens then... 

\subsection{Linear stability of black holes}

The question of stability of a given spacetime is crucial if one looks at the relevant solutions that one could observe in Nature. Indeed, if we consider a spacetime $(M,g)$ whose metric tensor is perturbed $g_{\mu\nu}\rightarrow g_{\mu\nu} + h_{\mu\nu}$, and if this spacetime is totally disrupted in the long run even by small perturbations, it cannot be a long-lived solution of gravity which will be actually observed! \\

The most simple case to treat is obviously the flat spacetime, whose non-linear stability has been proven by Christodoulou and Klainerman in the nineties \cite{Christodoulou:1993uv} (see also the Lecture \ref{sec:BMS} on asymptotically flat spacetimes). By \textit{non-linear stability}, we mean that, under some weak assumptions, any non-linear perturbation of the Minkowski metric will lead to a spacetime that asymptotes to Minkowski at late times. All perturbations will decay. For some large perturbations, the gravitational field may become so important that an asymptotically flat black hole appears. There is an intermediate, critical behavior, where the perturbation is at the onset of black hole formation known as the Choptuik critical collapse \cite{Choptuik:1992jv}. When a black hole has formed, the stability question can also be addressed but this time in terms of perturbations of a stationary black hole. \\

The linear stability of the static Schwarzschild solution has been proven \cite{PhysRev.108.1063,Vishveshwara:1970cc}. It means that if one imposes a small perturbation $g_{\mu\nu} \rightarrow g_{\mu\nu}+h_{\mu\nu}$ where $h_{\mu\nu}$ obeys the linearised equations of motion, all frequencies of the Fourier modes of $h_{\mu\nu}$ will have a negative imaginary part: all perturbations will decay. A similar statement has been obtained for non-extremal Kerr black holes \cite{Whiting:1988vc,Dafermos:2014cua}: it implies that the Kerr black hole is relevant to describe the late stages of gravitational collapse. \\

Now, something special occurs for extremal black holes. The key point is that the event horizon coincides with the Cauchy horizon. We just saw that a non-extremal black hole has a distinct event horizon located at $r=r_+$ and Cauchy horizon located at $r=r_-$. For an extremal black hole, the Penrose diagram collapses such that $r=r_+=r_-$ is at the same time the event horizon and the Cauchy horizon. We have discussed that the Cauchy horizon is linearly unstable to gravitational perturbations. We can therefore expect that the event horizon will admit unstable modes along the horizon. \\

Indeed, it has been recently shown that rotating and charged black holes in their extremal regime are linearly unstable under respectively gravitational or electromagnetic perturbations \cite{Aretakis:2010gd} ! A generic perturbation of an extremal black hole will produce a non-extremal black hole, and all perturbations will decay consistently with the linear stability results. Some perturbations keep the black hole extremal, and those perturbations are the ones of interest. If one fine-tunes the perturbation to be extremal and leading to an unstable mode, the perturbation leads to infinite gradients at the location of the event horizon, but no curvature singularity appears. The non-linear final state is not a stationary extremal black hole, but a non-stationary extremal black hole \cite{Murata:2013daa}.

\newpage
\section{Extremal rotating black holes}

There are two very different motivations to study the extremal Kerr black hole: 
\begin{itemize}[label=$\rhd$]
\item \textit{From the point of view of quantum gravity}. The extremal Kerr black hole is an intermediate case between the physical but hard to study non-extremal Kerr black hole and unphysical but easier to comprehend supersymmetric (and therefore extremal) black holes in string theory;
\item \textit{From the point of view of observational science}. This black hole is a limit for near-extremal Kerr black holes with very specific observational signatures.
\end{itemize}

The first motivation led to the \textit{Kerr/CFT correspondence} \cite{Guica:2008mu}, which is an attempt to relate the extremal Kerr black hole with a dual $CFT$, with partial successes and failures, see the review \cite{Compere:2012jk}. The second motivation led to several analytical analyses of physical processes around black holes and identifications of ``smoking gun'' observational signatures for either electromagnetic or gravitational wave astronomy. This topic is under active development at the time of writing these lectures (see \cite{Gralla:2017ufe,Compere:2017hsi,Lupsasca:2017exc}). \\

As a physicist, one should ask: are there nearly extremal black holes in Nature? By the third law of black hole thermodynamics, we can never reach extremality. In 1974, Thorne gave a precise bound on how high the spin can be using a specific thin disk accretion disk model \cite{Thorne:1974ve}. The disk can spin up the black hole up to  
\begin{equation}
J \leq 0.998 \: M^2
\end{equation}
where the absorption cross-section of retrograde photons emitted from the disk exceeds the cross-section of prograde photons. If the black hole spins faster, it will accrete too many photons with retrograde orbital motion which will spin down the black hole. However, this bound is only valid for one specific model of disk accretion and it can be beaten!\\

Astronomical observations give encouraging results: the stellar mass black holes known as \textit{GRS} 1905+105 and \textit{Cygnus} X-1 have been claimed to admit $J/M^2 > 98\%$. Also some supermassive black holes at the center of galaxies \textit{MGC}-6-30-15, or 1\textit{H} 0707-495 have also been claimed to have a spin ratio higher than $98 \%$...\\

In theory, it is practical to use the parameter $\lambda$ defined as 
\begin{equation}
\lambda = \sqrt{1-\frac{a^2}{M^2}}
\end{equation}
in order to measure how close one is from extremality. Schwarzschild has $\lambda = 1$ and extremal Kerr has $\lambda = 0$. What we are after is near extremal black holes with $\lambda = 10^{-3}$ or even $10^{-6}$ where the near-extremal features that we are going to describe really express themselves in physical phenomena. A famous science-fiction example of nearly extremal Kerr black hole is ``Gargantua'' in \textit{Interstellar} which, according to Thorne \cite{thorne2014science}, needs to have $\lambda < 10^{-7}$ in order to be consistent with key features of the movie script. Let's now go to the physics.

\subsection{Near horizon geometries}

Let us use from now on the two parameters $M,\lambda$ to denote a generic Kerr black hole. The event horizon and Cauchy horizon lie at radii $r_\pm = M(1\pm \lambda )$. What happens in the limit $\lambda \rightarrow 0$? From the point of view of the asymptotic observer, the geometry becomes the one of extremal Kerr. But from the point of view of an observer close to the black hole horizon, something very different happens. \\ 

We start from the Boyer-Lindquist patch $(t,r,\theta,\phi)$. For an observer close to the horizon, we switch to a coordinate system corotating with the black hole by taking
\begin{equation}
\Phi = \phi - \Omega^{ext}_H t = \phi - \frac{t}{2M}+O(\lambda)\label{NHEKt0}
\end{equation}
up to terms small in the near-extremal limit $\lambda \ll 1$. Moreover, we need to resolve the radius and time which are not good coordinates close to the horizon so we define
\begin{equation}
\left\lbrace
\begin{array}{ccc}
T &=& \frac{t}{2M \kappa} \lambda^{p} \, ; \\ 
R &=& \kappa \frac{r-r_+}{M} \lambda^{-p} \, ; 
\end{array} 
\right.\label{NHEKt}
\end{equation}
where $0< p < 1$ and $\kappa$ is any real normalization. In the limit $\lambda \rightarrow 0$, we can write the Kerr metric as 
\begin{equation}
ds^2 = 2 M^2 \Gamma(\theta ) \left[ -R^2 dT^2 + \frac{dR^2}{R^2} + d\theta^2 + \Lambda^2 (\theta) (d\Phi + R dT)^2 \right] + \mathcal{O}(\lambda^p)
\label{eq:NHEK}
\end{equation}
where $\Gamma(\theta) = (1+\cos^2\theta)/2$ and $\Lambda(\theta) = 2\sin\theta /  (1+\cos^2\theta)$ are two geometrical factors. This metric is known as the \textit{near horizon extremal Kerr} geometry or NHEK geometry. \\ 

Choosing instead $p=1$, we zoom even closer to the horizon and the near-horizon limit changes to 
\begin{equation}
ds^2 = 2 M^2 \Gamma(\theta ) \left[ -R(R+2\kappa) dT^2 + \frac{dR^2}{R(R+2\kappa)} + d\theta^2 + \Lambda^2 (\theta) (d\Phi + (R+\kappa) dT)^2 \right] + \mathcal{O}(\lambda). 
\label{eq:NearNHEK}
\end{equation}
This is the so-called \textit{near-NHEK metric} \cite{Amsel:2009ev,Bredberg:2009pv}. For $p > 1$ or $p<0$, the limit is not well-defined, so that's all we can do. \\

\begin{figure}[h!]
\centering
\begin{tikzpicture}
\draw[white] (-1,-5) -- (-1,5) -- (12,5) -- (12,-5) -- cycle;
\fill[color=gray!50] (0,1) -- (2,1) -- (2,-1) -- (0,-1) -- cycle;
\draw[ultra thick] (0,-1) -- (0,1);
\node[rotate=90,above] at (0,0) {\footnotesize Horizon};
\draw[thick] (0, 1) -- (2, 1);
\draw[thick] (0,-1) -- (2,-1);
\node[] at (1,0) {\footnotesize Near NHEK};
\draw[thick,dashed] (2, 1) -- (3,1);
\draw[thick,dashed] (2,-1) -- (3,-1);
\fill[color=gray!30] (3,1) -- (5,1) -- (5,-1) -- (3,-1) -- cycle;
\draw[thick] (3, 1) -- (5, 1);
\draw[thick] (3,-1) -- (5,-1);
\node[] at (4,0) {\footnotesize NHEK};
\draw[thick,dashed] (5, 1) -- (6,1);
\draw[thick,dashed] (5,-1) -- (6,-1);
\fill[color=gray!10] (6,1) [bend right=10] to (9,2) -- (9,-2) [bend right=10] to (6,-1) -- cycle;
\draw[thick] (6,1) [bend right=10] to (9,2);
\draw[thick] (9,-2) [bend right=10] to (6,-1);
\fill[color=gray!10] (9,2) [bend right=10] to (11,4) -- (11,-4) [bend right=10] to (9,-2) -- cycle;
\draw[thick] (9,2) [bend right=10] to (11,4);
\draw[thick] (11,-4) [bend right=10] to (9,-2);
%\node[] at (7.5,0)% {\footnotesize Kerr geometry};
\node[rotate=-90] at (10,0) {\footnotesize Extremal Kerr};
\end{tikzpicture}
\caption{The three asymptotically decouped regions of Kerr in the near-extremal regime.}
\label{fig:DecoupledKerr}
\end{figure}
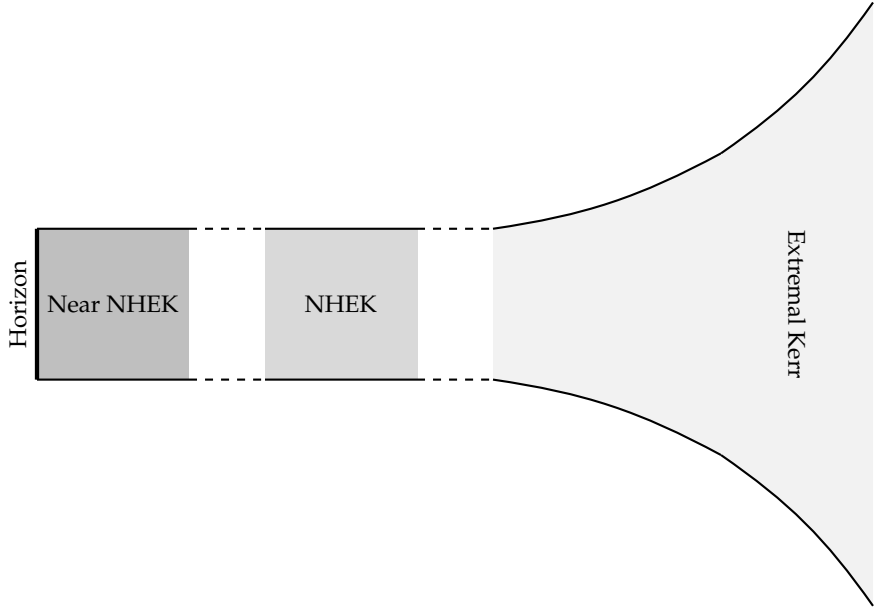

There are therefore 3 different extremal limits from Kerr depending on the type of observer that one is considering! Moreover, these regions formally decouple in the extremal limit, but they never exactly decouple for a nearly extremal black hole. They are summarized in Figure \ref{fig:DecoupledKerr}. Already an interesting consequence of the time definition \eqref{NHEKt} is that the relative gravitational redshift between the (near-)NHEK and extremal Kerr geometries is formally infinite in the limit $\lambda \rightarrow 0$. This leads to very interesting phenomena of high energy collisions in the near-horizon region between near-horizon waves and exterior waves entering the near-horizon region which are highly blueshifted in the near-horizon frame \cite{1975ApJ...196L.107P}. \\

\newpage
Let's now describe the geometry of (near-)NHEK. In fact, NHEK and near-NHEK are diffeomorphic to each other so we only need to discuss the NHEK geometry. The explicit diffeomorphism is 
\begin{align}
R' &= \frac{1}{\kappa}e^{\kappa T} \sqrt{R(R+2\kappa)},\\
T' &= -e^{-\kappa T} \frac{R+\kappa}{\sqrt{R(R+2\kappa)}},\\
\Phi' &= \Phi - \frac{1}{2} \log\frac{R}{R+2\kappa}
\end{align}
as you can check! The NHEK geometry contains the line element $-R^2 dT^2 + dR^2/R^2$ which is exactly $AdS_2$ spacetime in Poincar\'e coordinates. $AdS_2$ spacetime contains $3$ Killing vectors which form a $SL(2,\mathbb R)$ algebra. This symmetry algebra is in fact lifted to a symmetry algebra of the entire NHEK geometry, together with the $U(1)$ with $\partial_\Phi$.  The 4 Killing vectors are $\partial_T$ and $\partial_\Phi$ together with the scale transformation 
\begin{equation}
\xi_3 = T \partial_T - R \partial_R
\end{equation}
and another non-trivial exact symmetry
\begin{equation}
\xi_4 = \left(\frac{1}{2R^2} + \frac{T^2}{2}\right) \partial_T - TR \partial_R - \frac{1}{R} \partial_\Phi .
\end{equation}
The careful reader would have already derived the existence of the symmetry $\xi_3$: rescaling $T$ by $\kappa$ and $R$ by $\kappa^{-1}$ does not change the metric as we saw in \eqref{NHEKt}, which is the finite version of the Killing symmetry! Since $\partial_\Phi$ commutes with the 3 other Killing vectors, the symmetry algebra of the NHEK geometry is 
\begin{align}
SL(2,\mathbb R) \times U(1). 
\end{align}
The existence of conformal $SL(2,\mathbb R)$ symmetry in the near-horizon region of a nearly extremal Kerr black hole implies that physics in the near-horizon region can be described in the language of  critical phenomena. This motivated the recent exploration of magnetospheres, electromagnetic emission, accretion and gravitational wave emission from the NHEK region which all carry critical behavior caused by approximate conformal invariance.  

\subsection{Extremal \textit{BTZ} black holes and their dual \textit{CFT} description}

In $3d$ gravity, we have shown during the second lecture that it exists boundary conditions (Brown-Henneaux \cite{Brown:1986nw}) such that asymptotically $AdS_3$ spacetimes form a phase space whose asymptotic symmetry group is the direct product of two copies of the Virasoro group (which is the infinite-dimensional $2d$ conformal group). This shows that quantum gravity with Brown-Henneaux boundary conditions, if it exists, is a conformal field theory, or in other words can be described in dual terms in the language of a $2d$ $CFT$. (The existence of quantum gravity may require an embedding in string theory.) \\

In particular, extremal $BTZ$ black holes can be understood as particular states of the dual $CFT$, and their entropy can be understood from a microscopic counting in the dual $CFT$. Since this situation is very well understood, we will start by reviewing these results as a starter for describing the attempt at describing the extremal Kerr black hole with a $CFT$. 

\subsubsection{Extremal \textit{BTZ} geometry and near-horizon limit}

The asymptotic boundary cylinder of $AdS_3$ is naturally described by boundary lightcone coordinates $x^\pm = t/\ell \pm \phi$. The extremal $BTZ$ black hole in Fefferman-Graham coordinates $r$, $x^\pm$ is given by 
\begin{equation}
ds^2 = \ell^2 \frac{dr^2}{r^2} - \left(r dx^+ - \ell \frac{4L_-}{r} dx^- \right) \left(r dx^- - \ell \frac{4L_+}{r} dx^+ \right)
\end{equation}
where $L_\pm$ are constant and related to the physical charges by $M\ell = L_+ + L_-$, $J = L_+ - L_-$. Remember that the event horizon lies at $r=0$ in these coordinates. The extremal limit is given by $|J|=M\ell$. Let us choose the branch $J=+M \ell$, or equivalently, $L_- = 0$. The other extremal branch is similar with $L_+$ exchanged with $L_-$. 
In the extremal limit, the line element simply reads as
\begin{equation}
ds^2_{ext} = \ell^2 \frac{dr^2}{r^2} - r^2 dx^+ dx^- + 4J\ell (dx^+)^2. 
\end{equation}
What is the angular velocity in the extremal limit? Let us recall that the vector $\xi = \partial_t + \Omega_H \partial_\phi$ must generate the event horizon. In null coordinates
\begin{equation}
\xi = \frac{1}{\ell} (\partial_+ + \partial_- ) + \Omega_H (\partial_+ - \partial_-) = \left(\frac{1}{\ell} + \Omega_H \right)\partial_+ + \left(\frac{1}{\ell} - \Omega_H \right)\partial_-. 
\end{equation}
But by definition this vector must be null on the horizon ($r=0$), so
\begin{equation}
\left. \xi^2 \right|_{r=0} = g_{\mu\nu}^{ext} \xi^\mu \xi^\nu = g_{++}^{ext} (\xi^+)^2 = 4J\ell \left(\frac{1}{\ell} + \Omega_H^{ext} \right)^2 = 0 \Longrightarrow \Omega_H^{ext} = -\frac{1}{\ell}.
\end{equation}
The generator of the horizon is therefore $\xi =\frac{2}{\ell}\partial_-$. The near-horizon limit can be obtained from the strict extremal solution by introducing a near-horizon coordinate system depending on a small parameter  $\lambda$ running to zero. The adapted change of coordinates is found to be
\begin{equation}
(t,r,\phi) \rightarrow (T,R,\Phi) : \left\lbrace 
\begin{array}{ccl}
t &=& \frac{T}{\lambda} \sqrt{ J \ell} ;\\
r &=& \ell\sqrt{\lambda R}; \\
\phi &=& \Phi + \Omega_H^{ext} \frac{T}{\lambda}  \sqrt{ J \ell}.
\end{array}
\right.
\end{equation}
In the limit $\lambda \rightarrow 0$, the metric becomes
\begin{align}
ds^2 &= \ell^2 \left( \frac{dR^2}{4R^2} - 2 R \sqrt{\frac{J}{\ell}} dT d\Phi + \frac{4J}{\ell } d\Phi^2 \right) \\
&= \frac{\ell^2}{4} \left[ \frac{dR^2}{R^2} - R^2 dT^2 + \frac{16J}{\ell} \left(d\Phi- \sqrt{\frac{\ell}{16J}} R dT \right)^2 \right].
\label{eq:NHEKforBTZ}
\end{align}
The structure of the line element is very similar to the NHEK geometry. Again, we recognize the metric as a combination of  $AdS_2$ with the metric on the $U(1)$ circle (with a non-trivial fibration on  $AdS_2$).  The exact symmetry group is again exactly $SL(2,\mathbb R) \times U(1)$. The Killing vectors $\partial_T$ and $\partial_\Phi$ are again enhanced with the vectors $\xi_3$ and $\xi_4$!

\subsubsection{Chiral zero temperature states and chiral sector of a \textit{CFT}}

Let us now turn to the dual $CFT$ interpretation of extremal $BTZ$ black holes. We remember that the non-extremal $BTZ$ black hole can be described as a thermal high energy ensemble in the dual $CFT$. What about the extremal $BTZ$ black hole? It admits $L_- = 0$ as we saw (or $L_+=0$ if we choose the other chiral branch). It also has zero Hawking temperature which is dual to the temperature of the thermal ensemble in the $CFT$. So the state that is dual to the extremal $BTZ$ black hole is an ensemble of chiral states ($L_- =0$) with zero temperature. \\

Let us see how it fits with the asymptotic symmetry group analysis. We saw that assuming Brown-Henneaux boundary conditions, the asymptotic symmetry group consists of two copies of the Virasoro algebra. What happens when one takes the near-horizon limit of the extremal $BTZ$ black hole? The change of coordinates of the boundary lightcone coordinates is 
\begin{equation}
x^+ = \frac{t}{\ell} + \phi = \Phi \quad ; \quad x^- = \frac{t}{\ell} - \phi =  \frac{2T}{\lambda}\sqrt{J \ell} - \Phi . 
\end{equation}
In the limit $\lambda \rightarrow \infty$, functions of $x^+$ are well-defined but functions of $x^-$ are not. It implies that right-movers are allowed in the near-horizon limit, but left-movers need to be set to the ground state. Therefore, the phase space described in (\ref{eq:EinsteinSolFGBH}) does not admit a near-horizon limit. We need to set $L_-(x^-)= 0$. 
%In fact, the asymptotic symmetries are really ``symplectic symmetries'' acting everywhere in the bulk of spacetime including in the near-horizon region. It is a consequence of the fact that the canonical charges do not depend on the radius and that the canonical Poisson bracket between the charges can be defined at any radius. 
After performing the near-horizon limit, the resulting chiral phase space is described by the metric 
\begin{equation}
ds^2 = \frac{\ell^2}{4} \left[ \frac{dR^2}{R^2} - R^2 dT^2 + \frac{16 L_+ (\Phi)}{\ell} \left(d\Phi- \sqrt{\frac{\ell}{16 L_+ (\Phi)}} R dT \right)^2 \right]. 
\end{equation}
The exact symmetry group $SL(2,\mathbb{R})\times U(1)$ of the near-horizon $BTZ$ geometry is therefore enhanced to the asymptotic symmetry group
\begin{equation}
SL(2,\mathbb{R})\times U(1)_R \rightarrow SL(2,\mathbb{R})_L \times \text{Vir}_R
\end{equation}
since the $U(1)_R$ symmetry refers to $\partial_\Phi$ which is enhanced. (In fact, the $SL(2,\mathbb{R})_L$ symmetry is strictly not present in the asymptotic symmetry group since its charges are all zero and therefore trivial, but since it is a Killing symmetry the factor is usually kept.) The dual field theory corresponding to the phase space of extremal geometries with excitations consistent with the near-horizon limit is therefore the original $CFT$ but amputated to a chiral sector.\\ 

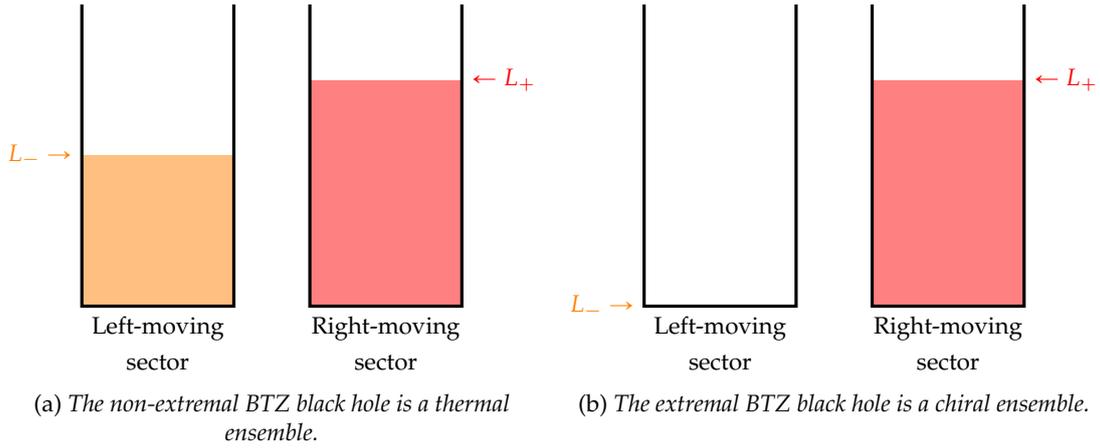
\begin{figure}[h!]
\centering
\subfloat[The non-extremal BTZ black hole is a thermal ensemble.]{
\begin{tikzpicture}
\fill[orange!50] (-2.5,2) node[left,orange] {\footnotesize $L_- \rightarrow$} -- (-2.5,0) -- (-0.5,0) -- (-0.5,2) -- cycle;
\fill[red!50] (2.5,3) node[right,red] {\footnotesize $\leftarrow L_+$} -- (2.5,0) -- (0.5,0) -- (0.5,3) -- cycle;
\draw[very thick] (-2.5,4) -- (-2.5,0) -- (-0.5,0) -- (-0.5,4);
\node[below,text width=2cm,align=center] at (-1.5,0) {\footnotesize Left-moving sector};
\draw[very thick] (2.5,4) -- (2.5,0) -- (0.5,0) -- (0.5,4);
\node[below,text width=2cm,align=center] at (1.5,0) {\footnotesize Right-moving sector};
\end{tikzpicture}
}
\subfloat[The extremal BTZ black hole is a chiral ensemble.]{
\begin{tikzpicture}
\fill[red!50] (2.5,3) node[right,red] {\footnotesize $\leftarrow L_+$} -- (2.5,0) -- (0.5,0) -- (0.5,3) -- cycle;
\node[left,orange] at (-2.5,0) {\footnotesize $L_- \rightarrow$};
\draw[very thick] (-2.5,4) -- (-2.5,0) -- (-0.5,0) -- (-0.5,4);
\node[below,text width=2cm,align=center] at (-1.5,0) {\footnotesize Left-moving sector};
\draw[very thick] (2.5,4) -- (2.5,0) -- (0.5,0) -- (0.5,4);
\node[below,text width=2cm,align=center] at (1.5,0) {\footnotesize Right-moving sector};
\end{tikzpicture}
}
\vspace{10pt}
\caption{$CFT$ thermal and zero temperature ensembles associated to the \textit{BTZ} black hole.}
\label{fig:ChiralCFT}
\end{figure}

The chiral nature of the state dual to extremal $BTZ$ black hole, and the corresponding chiral excitations that still exist in the near-horizon limit are illustrated in Figure \ref{fig:ChiralCFT}. 

\subsubsection{Chiral Cardy formula}

The black hole microscopic entropy counting still works in the extremal limit. Cardy's formula restricted to a chiral sector gives on the $CFT$ side: 
\begin{equation}
S_{CFT}  = 2\pi \sqrt{\frac{c_R L_+}{6}} = \pi \sqrt{\ell L_+}
\end{equation}
which agrees with the Bekenstein-Hawking entropy of the extremal $BTZ$ black hole,
\begin{equation}
S_{ext} = \frac{\pi r_+}{2G} = \ell\pi \frac{M}{G} = \pi \sqrt{\ell L_+}
\end{equation}
where $r_+ = r_- = \ell \sqrt{4GM}$. Here it is crucial that a $2d$ $CFT$ with two chiral sectors exists in order to derive the formula. Another subtetly is that Cardy's formula is strictly valid for $L_\pm \gg c$, so we are outside of its range of validity but the matching still works! \\

This ``unreasonable validity of Cardy's formula'' motivated to look for possible extensions of his range of validity. For example in \cite{Hartman:2014oaa} one uses the fact that there is a mass gap of $1/(8G)= c/12$ between the $AdS_3$ vacuuum and the first $BTZ$ black hole where $c=c_L=c_R=3\ell/(2G)$ is the Brown-Henneaux central charge as we described in the second lecture.  In the dual $CFT$ language it can be translated to the existence of a ``sparse light spectrum of states'' in the energy range $-c/12 < L_+ +L_- <0$. For such $CFTs$, an extended range of applicability of Cardy's formula exists. One first use the standard thermodynamics relations to define a conjugate chemical potential to $L_+$ and $L_-$: 
\begin{equation}
\frac{1}{T_-} = \left( \frac{\partial S}{\partial L_-} \right)_{L_+} \quad \text{ and } \quad \frac{1}{T_+} = \left( \frac{\partial S}{\partial L_+} \right)_{L_-} . 
\end{equation}
The dimensionless chemical potentials $T_+$, $T_-$ are sometimes called by abuse of language ``left and right temperatures''. Using the full Cardy formula we get $L_+ = \frac{\pi^2}{6} c_R T_+^2$, $L_- = \frac{\pi^2}{6} c_L T_-^2$ and the entropy can be written in the canonical ensemble as 
\begin{equation}
S_{CFT} = \frac{\pi^2}{3} (c_L T_L + c_R T_R ) \label{Tcan}
\end{equation}
where the extended range of validity is now $T_L > 1/(2\pi)$, $T_R > 1/(2\pi)$ \cite{Hartman:2014oaa}. For the extremal $BTZ$ black hole we have $T_R \gg 1$ but $T_L = 0$ so we are still outside of the range of validity. More work is yet needed to understand this matching of entropy!

\subsection{The \textit{Kerr/CFT} correspondence}

After defining the near-horizon limit of nearly extremal Kerr and reviewing some relevant background material on the $3d$  $BTZ$ black hole as a warm up, we are now ready to present the \textit{Kerr/CFT} correspondence. In 2009, Guica, Hartman, Song, and Strominger proposed \cite{Guica:2008mu} a new type of \textit{holographic duality} between $4d$ extremal Kerr black holes and $2d$ conformal field theories analogous to a duality between $BTZ$ black holes and $2d$ $CFTs$ that we briefly reviewed. The conjectured duality is based on properties of the near-horizon limit of the extremal Kerr black hole. Even though the original conjecture turned out incorrect, an updated conjecture still holds; the work also generated very interesting developments in holography and still contains a mysterious entropy matching that deserves further research. Let's now review some of these developments following a viewpoint enriched with the subsequent research work \cite{Compere:2012jk}. 

\subsubsection{Virasoro symmetry}

The basis of a $2$-dimensional $CFT$ is its symmetry group consisting of two copies of the Virasoro algebra. But asking for two copies of the Virasoro algebra is asking too much. As we saw for the case of the near-horizon limit of the extremal $BTZ$ black hole we can only hope in the near-horizon limit for only one Virasoro algebra extending the $SL(2,\mathbb R) \times U(1)$ symmetry as $SL(2,\mathbb R) \times \text{Vir}$. The fundamental reason is that any non-extremal excitation prevents the existence of a decoupling geometry such as the near-horizon $BTZ$ or the NHEK geometry.  In the Kerr case, non extremal excitations lead to a non-vanishing coupling between the NHEK region and the asymptotically flat region. This implies that there is no $2d$ $CFT$ describing the Kerr black hole, which invalidates the original \textit{Kerr/CFT} conjecture. There are however some disturbing connections which suggest a relationship with at least part of the structure of a $2d$ $CFT$ \cite{Castro:2010fd,Cvetic:2011dn,Pasterski:2016qvg,Kapec:2016jld}. \\

Since  non-extremal physics does not exist in the near-horizon of an extremal black hole, there is classically no dynamics except non-trivial diffeomorphisms and topology. The NHEK geometry therefore bears much resemblence with the near-horizon $BTZ$ black hole in $3d$ gravity! The original approach of \cite{Guica:2008mu} was inspired from the Brown-Henneaux analysis: one first imposes boundary conditions and study which symmetries preserve the boundary conditions, one checks that the associated charges are finite and integrable and one write the Poisson bracket to derive the charge algebra of asymptotic symmetries. Now, since the boundary conditions contain nothing else than diffeomorphisms and topology, one can just study these two features. \\

The ansatz for the generator of the sought-for Virasoro algebra is the following
\begin{equation}
\chi = \varepsilon (\Phi) \partial_\Phi -  \varepsilon'(\Phi) \left( R \partial_R + \frac{b}{R}\partial_T \right) + \text{subleading terms}
\label{eq:VectorForVirasoro}
\end{equation}
which is built from an arbitrary function on the circle $\varepsilon(\Phi)$. In the original \textit{Kerr/CFT} paper $b$ was set to zero but then there is no smooth classical phase space. Instead $b =1$ leads to a smooth classical phase space \cite{Compere:2015mza,Compere:2015bca}. Since $\Phi$ is $2\pi$-periodic, the function $\varepsilon(\Phi)$ can be mode expanded. We define $L_n \triangleq \chi [\varepsilon(\Phi) = -e^{-in\Phi} ]$. It is straightforward to check that $i[L_m , L_n ] = (m-n) L_{m+n}$ which is the Witt algebra on the circle. The signs are chosen for later convenience (in order to have a positive central charge!). \\

Starting from this ansatz alone, it is possible to exponentiate this generator to generate the finite diffeomorphism of the NHEK geometry depending on one arbitrary function $\psi(\Phi)$ of $\Phi$ and thereby construct a (small) phase space of asymptotically NHEK solutions \cite{Compere:2015bca}. One can then evaluate the surface charge associated with any arbitrary generator \eqref{eq:VectorForVirasoro} for infinitesimal variations on the phase space, which solely amount to vary $\psi$. The resulting charges $\mathcal L_m$ are finite and integrable and therefore the generator \eqref{eq:VectorForVirasoro} is promoted to an asymptotic symmetry. The Poisson algebra of charges is isomorphic to the algebra of asymptotic symmetries up to central terms as follows from the general results derived in the first lecture. One finds 
\begin{equation}
i \{ \mathcal L_m, \mathcal L_n \} = (m-n) \mathcal L_{m+n}+\frac{c}{12}m^3 \delta_{m+n,0}
\end{equation}
with central charge
\begin{equation}
c = 12 J,
\end{equation}
where $J$ is the angular momentum. Moreover, one can check that the asymptotic symmetries act everywhere in the bulk spacetime, which promotes them to \textit{symplectic symmetries}, similarly to $3d$ Einstein gravity. Because there is no finite energy excitation in NHEK, the $SL(2,\mathbb R)$ Killing symmetries are associated with zero charges are therefore trivial. The asymptotic symmetry group therefore consists of one copy of the Virasoro algebra with central charge $c=12J$. \\

\subsubsection{Conjugated chemical potential and Cardy matching}

What it remains to be known to use the (chiral version of the) canonical Cardy formula \eqref{Tcan} is the chemical potential associated with angular momentum. Usually, it is the angular velocity of the black hole, but for an extremal black hole the mass is also function of the angular momentum, so we need to use the definition 
\begin{equation}
\frac{1}{T_\Phi} = \frac{\partial S_{ext}}{\partial J}.
\end{equation}
For extremal Kerr, it gives $T_\Phi = \frac{1}{2\pi}$, which quite annoyingly just lies beyond the extended range of applicability of Cardy's formula. Yet, the entropy of extremal Kerr black hole matches with the chiral canonical Cardy formula
\begin{equation}
S_{ext\, Kerr} = 2\pi J = \frac{\pi^2}{3} c \, T_\Phi
\end{equation}
after using the definitions of $c$ and $T_\Phi$. This is the remarkable entropy matching performed by the \textit{Kerr/CFT} correspondence! In all known examples of extremal black holes including black holes with higher curvature corrections \cite{Azeyanagi:2009wf}, this match was always shown to hold. However, there is no clear $2d$ $CFT$ here and therefore the main hypothesis to derive Cardy's formula does not hold. Yet it matches. This is the ``unreasonable universality of Cardy matching'' which is still a mystery today. \\

\subsubsection{Frolov-Thorne vacuum}

Let us close this section by interpreting the conjugated chemical potential $T_\Phi$. The interpretation can be made in the context of quantum field theories in curved spacetimes. But first let us derive an equivalent formula for $T_\Phi$. So far we defined $T_\Phi \delta S_{ext} = \delta J$.  The first law is $T_H \delta S = \delta M - \Omega_H \delta J$. Let us specialize the first law to extremal variations where $\delta J = \delta (M^2)$ or $\delta M = \frac{1}{2M}\delta J = \Omega_H^{ext} \delta J$. We find $\delta S_{ext} = \frac{\Omega_H^{ext}-\Omega_H}{T_H}\delta J$. Comparing with the definition of $T_\Phi$ we find 
\begin{equation}
\frac{1}{T_\Phi} = \frac{\Omega_H^{ext}-\Omega_H}{T_H} .
\end{equation}

The main result of Hawking in 1974 is that the quantum field state at late times after a black hole has formed by collapse is  described as a thermal ensemble at Hawking temperature $T_H = \frac{\kappa}{2\pi}$ where $\kappa$ is the surface gravity of the black hole. The root of the effect can be attributed to pair creations and annihilations in the quantum vacuum at the vicinity of the horizon. If one of the particles is trapped by the horizon, its conjugated pair cannot annihilate and is emitted by the black hole. The derivation of the thermal nature of the spectrum requires to use the properties of event horizons and the definition of vacuum.\\

In curved spacetimes the definition of vacuum is not unique. It depends on the definition of positive frequency modes, which requires a timelike Killing vector. In Schwarzschild spacetime, there is a globally timelike Killing vector $\partial_t$ which naturally defines the so-called Hartle-Hawking vacuum. This state is regular at the event horizon and thermal. It is described by a density matrix $\rho = \exp ( -\frac{\omega}{T_H})$.  For the Kerr black hole, there is no global timelike Killing vector, due to the ergosphere. For an observer close to the horizon it is natural to define positive frequency modes with respect to the generator of the horizon $\xi = \partial_t + \Omega_H \partial_\phi$. This vector field is timelike in a region bounded by the event horizon and an outer region known as the velocity of light surface where $\xi$ becomes null. The vacuum defined this way is known as the \textit{Frolov-Thorne vacuum}. It is described by a density matrix $\rho = \exp ( -\frac{\omega- \Omega_H m}{T_H}) $ where $\omega$ is the frequency and $m$ the azimuthal number of the wave.\\

In the near-horizon limit, one can rewrite the density matrix in terms of variables adapted to the NHEK coordinates. We first rewrite a wave as 
\begin{equation}
F(r,\theta) \exp(-i\omega t+ i m \phi) = F(R,\theta) \exp(-i\Omega  T+ i M \Phi).
\end{equation}
Using the change of coordinates \eqref{NHEKt0}-\eqref{NHEKt} we find $m=M$ and $\omega = m \Omega_H^{ext} + \lambda \Omega$. It implies that
\begin{equation}
\rho = \exp \left( -\frac{\omega- \Omega_H m}{T_H} \right) = \exp \left( -m \frac{\Omega_H^{ext} - \Omega_H}{T_H} \right)  = \exp \left( -\frac{m}{T_\Phi} \right). 
\end{equation}
This shows that the Frolov-Thorne vacuum is thermally populated in the extremal limit with a ``temperature'' equal to $T_\Phi$. \\

There are however some caveats here that need to be pointed out. While the generator of the horizon $\xi$ is timelike outside the event horizon of a non-extremal Kerr black hole, a singular behavior occurs in the near-horizon extremal limit. In NHEK the vector $\xi = \partial_T$ is timelike only in a polar wedge around the north and south poles, but it is spacelike around the equator (indeed, check the sign of $g_{TT}$ by evaluating $\Lambda(\theta)$ in \eqref{eq:NHEK}). In fact, there is no globally timelike Killing vector in NHEK. So there is no quantum vacuum in NHEK. One should understand the ``NHEK vacuum'' only as a (singular) limit of the Frolov-Thorne vacuum of the near-extremal case. 

%\newpage
\section{Black hole spectroscopy}

On September 14, 2015, gravitational waves were detected for the first time thanks to the twin detectors of the Laser Interferometer Gravitational-wave Observatory (LIGO) \cite{Abbott:2016blz}. This event launched the new era of gravitational wave astronomy. The upcoming direct observations of gravitational physics enable today to test Einstein's theory of gravity in the strong field regime at a precision never reached by other experiments. Third generation detectors, and in particular LISA planned for 2034, will add new precision measurements to be compared with theoretical predictions. This is a unique and very exciting time for gravitational physics. \\

In this lecture, we would like to present the late stages of black hole mergers, after the two bodies have collapsed to form a black hole. The final state of the merger is a Kerr black hole with small perturbations which can be approximated at late stages by linearized perturbations. The mathematical problem of solving linearized Einstein's equations around the Kerr black hole benefited from a crucial contribution from Teukolsky in 1972 during his PhD studies with Kip Thorne. He found a determining set of variables which separates \cite{Teukolsky:1972my}. This allows to reduce the complexity of the problem to solving two coupled ordinary differential equations. In turn, Leaver showed in 1985 how to solve these differential equations to arbitrary precision \cite{Leaver:1985ax}. This allows to deduce the fundamental characteristic decay frequencies (the quasi-normal mode frequencies) of the Kerr black hole numerically up to arbitrary precision. \\

Analoguously to the spectral lines that the hydrogen atom can electromagnetically emit, the Kerr quasi-normal modes tell us what the Kerr black hole can gravitationally radiate. At the time of writing, the observations do not yet allow to check the exact quasi-normal modes frequencies of the Kerr black hole, but the experimental science of black hole spectroscopy will soon start and will allow to check the validity of the Kerr metric!\\

In order to present the founding work of Teukolsky, it is necessary to first introduce the Newman-Penrose formalism and Petrov's classification of the Weyl tensor. 

\subsection{Fundamentals of the Newman-Penrose formalism}

The metric field $g_{\mu\nu}$ is not the appropriate field to decribe fundamental physics which also involves fermions. Coupling Einstein gravity to fermions requires to introduce the more fundamental basis of \textit{tetrads} $e_\mu^a$, $a=0,1,2,3$, such that $e_\mu^a$ is an invertible matrix and $g_{\mu\nu} = \eta_{ab} e^a_\mu e^b_\nu$ where $\eta_{ab}$ is the Minkowski metric of the tangent space at each point. The formulation in terms of tetrads allows to use the \textit{Cartan formalism} where the fundamental ingredient are the one-forms $e^a = e_\mu^a dx^\mu$. We saw in the second Lecture how this Cartan formalism led us to reformulate $3d$ Einstein gravity as a sum of two Chern-Simons theories! \\

The Newman-Penrose formalism is a tetrad formalism with complex tetrads where the tangent space Minkowski metric $\eta_{ab}$ is chosen at each point to be
\begin{equation}
\eta_{ab} = \left( \begin{array}{cccc} 
0& -1 &0&0 \\Ê
-1& 0 &0&0 \\Ê
0& 0 &0&1 \\Ê
0& 0 &1&0 
\end{array}
\right). \label{metric}
\end{equation}
The tetrad frame is chosen to be a set of \textit{$4$ null vectors} $l_\mu,n_\mu,m_\mu,\bar m_\mu$ with
\begin{equation}
g_{\mu\nu} = -l_\mu n_\nu - n_\mu l_\nu + m_\mu \bar{m}_\nu + \bar{m}_\mu m_\nu\, .
\end{equation}
Because they is no complete real basis of null directions in a Lorentzian manifold, two of these vectors have to be \textit{complex}, and since the final metric is real these two complex tetrad are complex conjugated. The two real tetrads label particular ingoing and outgoing null directions, so this formalism is well adapted to describe geometrically the propagation of gravitational waves. \\

The parallel transport must be re-expressed in terms of the general Newman-Penrose tetrad basis, so instead of using the standard $4$ connections $\nabla_\mu$, we consider instead $4$ locally-defined directional covariant derivatives on the flow of the tetrad, which are historically denoted by $D,\Delta,\delta,\bar{\delta}$ and straightforwardly defined as
\begin{equation}
D = l^\mu \nabla_\mu \quad ; \quad \Delta = n^\mu \nabla_\mu \quad ; \quad \delta = m^\mu \nabla_\mu \quad ; \quad \bar{\delta} = \bar{m}^\mu \nabla_\mu\, .
\end{equation}
The Christoffel symbols have $4 \times 6 = 24$ real components, equivalent to 12 complex numbers. In the Newman-Penrose formalism, one defines 12  \textit{complex spin coefficients} which encode the same information. These coefficients have an individual name. Here is the complete list so that you have a precise idea: 
\begin{eqnarray}
	\kappa &=& -m^{\mu}l^{\nu}\nabla_{\nu}l_{\mu} \; ;\qquad 
	\sigma = -m^{\mu}m^{\nu}\nabla_{\nu}l_{\mu} \; ;  \\
	\lambda &=& -n^{\mu}\bar{m}^{\nu}\nabla_{\nu}\bar{m}_{\mu} \; ; \qquad
	\nu = -n^{\mu}n^{\nu}\nabla_{\nu}\bar{m}_{\mu}  \; ;\\
	\rho	 &=& -m^{\mu}\bar{m}^{\nu}\nabla_{\nu}l_{\mu} \; ; \qquad
	\mu	 = -n^{\mu}m^{\nu}\nabla_{\nu}\bar{m}_{\mu} \; ;  \\
	\tau	 &=& -m^{\mu}n^{\nu}\nabla_{\nu}l_{\mu}\; ;  \qquad
	\varpi	 = -n^{\mu}l^{\nu}\nabla_{\nu}\bar{m}_{\mu} \; ; \\
	\epsilon &=& -\frac{1}{2}(n^{\mu}l^{\nu}\nabla_{\nu}l_{\mu} + m^{\mu}l^{\nu}\nabla_{\nu}\bar{m}_{\mu}  )\; ; \\
	\gamma &=& -\frac{1}{2}(n^{\mu}n^{\nu}\nabla_{\nu}l_{\mu} + m^{\mu}n^{\nu}\nabla_{\nu}\bar{m}_{\mu})\; ; \\
	\alpha &=& -\frac{1}{2}(n^{\mu}\bar{m}^{\nu}\nabla_{\nu}l_{\mu}+ m^{\mu}\bar{m}^{\nu}\nabla_{\nu}\bar{m}_{\mu})\; ; \\
	\beta &=& -\frac{1}{2}(n^{\mu}m^{\nu}\nabla_{\nu}l_{\mu} + m^{\mu}m^{\nu}\nabla_{\nu}\bar{m}_{\mu}).
\end{eqnarray}

Now let us examine the formulation of the curvature.  In 4 dimensions, the Riemann tensor has $20$ independent real components, while the Ricci tensor, being symmetric, admits $10$ real components. One can build the traceless part of the Riemann tensor, it is called the Weyl tensor, and it therefore contains $10$ real or $5$ complex independent components. The Weyl tensor is more precisely defined as
\begin{align}
W_{\mu\nu\rho\sigma} = R_{\mu\nu\rho\sigma} - g_{\mu [\rho} R_{\sigma ] \nu} +  g_{\nu [\rho} R_{\sigma ] \mu}+\frac{1}{3}R g_{\mu [\rho} g_{\sigma ] \nu}.
\end{align} 
The basis of $5$ complex scalars $\Psi_i$, $i = 0,\dots, 4$ that allows to label an arbitrary Weyl tensor are called the  \textit{Weyl-Newman-Penrose scalars} and there are defined as 
\begin{align}
\Psi_0 &= W_{\alpha\beta\gamma\delta} l^\alpha m^\beta l^\gamma m^\delta ,\\
\Psi_1 &= W_{\alpha\beta\gamma\delta} l^\alpha n^\beta l^\gamma m^\delta, \\
\Psi_2 &= W_{\alpha\beta\gamma\delta} l^\alpha m^\beta \bar{m}^\gamma n^\delta, \\
\Psi_3 &= W_{\alpha\beta\gamma\delta} l^\alpha n^\beta \bar{m}^\gamma n^\delta, \\
\Psi_4 &= W_{\alpha\beta\gamma\delta} n^\alpha \bar{m}^\beta n^\gamma \bar{m}^\delta.
\end{align} 
These $5$ scalars are invariant under diffeomorphisms but depend on the choice of the tetrad basis. Let us define $3$ antisymmetric bivectors
\begin{equation}
X_{\mu\nu} = -2 n_{[\mu}\bar m_{\nu]} \quad ; \quad Y_{\mu\nu} = 2 l_{[\mu}m_{\nu]} \quad ; \quad Z_{\mu\nu} = 2 m_{[\mu}\bar{m}_{\nu]} - 2 l_{[\mu} n_{\nu]}.
\end{equation}
One can show as an exercise that the Weyl tensor can indeed be written as a combination of the 5 Weyl-Newman-Penrose scalars in the tetrad basis. The exact formula is 
\begin{align}
W_{\alpha\beta\gamma\delta} = &\Psi_0 \: X_{\alpha\beta} X_{\gamma\delta} + \Psi_1 \left( X_{\alpha\beta} Z_{\gamma\delta} + Z_{\alpha\beta} X_{\gamma\delta} \right) + \Psi_2 \left( Y_{\alpha\beta} X_{\gamma\delta} + X_{\alpha\beta} Y_{\gamma\delta} + Z_{\alpha\beta}Z_{\gamma\delta} \right)\nonumber \\
&+ \Psi_3 \left( Y_{\alpha\beta} Z_{\gamma\delta} + Z_{\alpha\beta} Y_{\gamma\delta} \right) + \Psi_4 \: Y_{\alpha\beta} Y_{\gamma\delta}+ c.c.\label{Weyle}
\end{align}

Since the Weyl-Newman-Penrose scalars are not invariant under a change of tetrad, it is important to discuss the exact ambiguity. One can perform a local Lorentz transformation at each spacetime point, which rotates the tetrad. Such a transformation is labelled by 6 real numbers at each point. It is useful to categorize these 6 local Lorentz rotations in 3 types:
\begin{itemize}
\item Rotations of type I which leave $l^\mu$ unchanged ($a \in \mathbb C$);
\begin{align}
l^\mu \mapsto l^\mu,\qquad n^\mu \mapsto n^\mu +a^* m^\mu +a \bar m^\mu + aa^* l^\mu ,\qquad m^\mu \mapsto m^\mu + a l^\mu\qquad \bar m^\mu \mapsto \bar m^\mu + a^* l^\mu. 
\end{align}
\item Rotations of type II which leave $n^\mu$ unchanged  ($b \in \mathbb C$);
\begin{align}
\hspace{-0.5cm}n^\mu \mapsto n^\mu,\qquad l^\mu \mapsto l^\mu +b^* m^\mu +b \bar m^\mu + bb^* n^\mu ,\qquad m^\mu \mapsto m^\mu + b n^\mu, \qquad m^\mu \mapsto m^\mu + b^* n^\mu. 
\end{align}
\item Rotations of type III which leave the directions of $l^\mu$ and $n^\mu$ unchanged and rotate $m^\mu$ by an angle in the $m^\mu,\bar m^\mu$ plane ($A,\theta \in \mathbb R$);
\begin{align}
l^\mu \mapsto A^{-1}l^\mu,\qquad n^\mu \mapsto A n^\mu, \qquad m^\mu \mapsto e^{i \theta} m^\mu\qquad \bar m^\mu \mapsto e^{-i \theta} \bar m^\mu . 
\end{align}
\end{itemize}
You can check that under these transformations, the tetrads preserve the same orthonomality conditions. In other words,  the metric on the tangent space \eqref{metric} is preserved. There are also discrete flips, such as exchanging $l^\mu \leftrightarrows n^\mu$. After some work involving the evaluation of the Weyl tensor contracted with each of the possible combinations of tetrads, one can find the transformation laws of the Weyl-Newman-Penrose scalars under each rotation of type I, II and III: 
\begin{eqnarray}
&\Psi_0 \mapsto \Psi_0, \nonumber\\
&\Psi_1 \mapsto \Psi_1+a^* \Psi_0, \nonumber\\
&\Psi_2 \mapsto \Psi_2+2 a^* \Psi_1+(a^*)^2 \Psi_0,\label{R1} \\
&\Psi_3 \mapsto \Psi_3+3 a^* \Psi_2 + 3 (a^*)^2 \Psi_1+(a^*)^3 \Psi_0, \nonumber\\
&\Psi_4 \mapsto \Psi_4+4 a^* \Psi_3 + 6 (a^*)^2 \Psi_2 + 4 (a^*)^3 \Psi_1+(a^*)^4 \Psi_0; \nonumber
\end{eqnarray}
\begin{eqnarray}
&\Psi_0 \mapsto \Psi_0+4 b \Psi_1 + 6 b^2 \Psi_2 + 4 b^3 \Psi_3 + b^4 \Psi_4, \nonumber\\
&\Psi_1 \mapsto \Psi_1+3 b \Psi_2 + 3 b^2 \Psi_3 + b^3 \Psi_4, \nonumber\\
&\Psi_2 \mapsto \Psi_2+2 b \Psi_3 + b^2 \Psi_4 , \label{R2}\\
&\Psi_3 \mapsto \Psi_3+b \Psi_4, \nonumber\\
&\Psi_4 \mapsto \Psi_4; \nonumber
\end{eqnarray}
\begin{eqnarray}
&\Psi_0 \mapsto A^2 e^{-2 i \theta }\Psi_0, \nonumber\\
&\Psi_1 \mapsto A^{-1} e^{i \theta}\Psi_1, \nonumber\\
&\Psi_2 \mapsto \Psi_2,\label{R3} \\
&\Psi_3 \mapsto A e^{-i \theta}\Psi_3, \nonumber\\
&\Psi_4 \mapsto A^2 e^{-2 i \theta}\Psi_4. \nonumber
\end{eqnarray}

\subsection{Fundamentals of Petrov's classification}
\label{sec:Petrov}

Let us now review the classification that Petrov obtained in 1954 \cite{1954UZKGU.114...55P}\footnote{As an anecdote on ULB connections, it is amusing to notice that this theory was also independently developed at ULB by G\'eh\'eniau in 1957 (who supervised the PhD of M. Henneaux 23 years later).}, using the Newman-Penrose formalism. This classification is taylor made for 4 spacetime dimensions. A higher dimensional classification also exists but was established only 50 years later  \cite{Coley:2004jv}. \\

We would like to get an algebraic, coordinate-independent, classification of solutions to Einstein field equations. The non-homogeneous solutions ($T_{\mu\nu} \neq 0$) are dependent on the matter content through the Ricci tensor $R_{\mu\nu}$ which is the trace part of the Riemann tensor. The traceless part of the Riemann tensor, the Weyl tensor, is left unconstrained by the matter fields and therefore represents the purely gravitational field. For vacuum solutions ($T_{\mu\nu} = 0$), the Ricci tensor is always zero even though the gravitational field (the Weyl tensor) could be non-trivial. The classification of the Weyl tensor is therefore useful to categorize both homogeneous and non-homogeneous solutions to Einstein's equations. Note that the Weyl tensor is conformally invariant and therefore may classify at most conjugacy classes of metrics differing by a global conformal Weyl factor (for example $AdS_4$ and Minkowski will belong to the same class). \\

Petrov classified the Weyl tensor by the number of degenerate local eigenvalues and (antisymmetric) eigenbivectors of the Weyl tensor. The eigenvalue equation reads as 
\begin{equation}
W^{\mu\nu}_{\phantom{\mu\nu}\alpha\beta} X^{\alpha\beta} = \lambda \: X^{\mu\nu}. 
\end{equation}
A non-trivial result due to Penrose in 1960 shows that solving this eigenvalue problem is equivalent to classify spacetimes according to the degeneracy of \textit{principal null directions} of the Weyl tensor. Such directions are spanned by null vectors $k^\mu$ obeying 
\begin{equation}
k_{[\alpha} W_{\beta]\gamma\delta [\rho} k_{\sigma ]} k^\gamma k^\delta = 0.
\end{equation}
Yet another equivalent formulation of the classification is the following.  We have just seen that with respect to a chosen tetrad, the Weyl tensor is completely determined by the five Weyl-Newman-Penrose scalars. The third formulation of the classification consists in determining how many of these scalars can be made to vanish for a given spacetime by choosing a suitable orientation of the tetrad frame. \\

Let us concentrate on this third formulation. Given a metric together with a Newman-Penrose frame $l^\mu,n^\mu,m^\mu,\bar{m}^\mu$, we can deduce the 5 Weyl components $\Psi_i$. Let us assume that $\Psi_4 \neq 0$. If this is not the case, we can perform a rotation of the type I \eqref{R1} to make it non zero, as long as the spacetime is not conformally flat in which case all the Weyl tensor vanishes. Now consider a rotation of type II \eqref{R2} with complex parameter $b$. It is clear that $\Psi_0$ can be made to vanish if $b$ is a root of the equation 
\begin{equation}
\Psi_0+4 b \Psi_1 + 6 b^2 \Psi_2 + 4 b^3 \Psi_3 + b^4 \Psi_4 = 0. \label{eqb}
\end{equation}
This equation has always exactly 4 roots, and the corresponding directions of $l^\mu$, namely $ l^\mu +b^* m^\mu +b \bar m^\mu + bb^* n^\mu$ are called the \textit{principal null directions} of the Weyl tensor. Indeed, contracting the tetrad decomposition of the Weyl tensor \eqref{Weyle} with multiple instances of $l^\mu$, one obtains after using the orthonormality condition \eqref{metric} of the tetrad: 
\begin{equation}
l_{[\alpha} W_{\beta]\gamma\delta [\rho} l_{\sigma ]} l^\gamma l^\delta = \Psi_0 l_{[\alpha} \bar m_{\beta]} l_{[\rho} \bar m_{\sigma]}+ \Psi^*_0 l_{[\alpha} m_{\beta]} l_{[\rho} m_{\sigma]} .
\end{equation}
After the type II frame rotation that we just defined by solving \eqref{eqb}, we have a vanishing new $\Psi_0 = 0$, and therefore the new null direction $l^\mu$ is indeed a principal null direction in the sense of Penrose. If one or two more roots coincide, the Weyl tensor is said to be algebraically special; otherwise it is said to be algebraically general. The various ways in which the roots coincide or are distinct lead to Petrov's classification.\\

\begin{figure}[p]
\centering
\subfloat[Characterisation of Petrov types. $k^\mu$ is always the most degenerate principal null direction (\textit{p.n.d.}). Adapted from \cite{Stephani:2003tm}.]{
\renewcommand{\arraystretch}{2}
\begin{tabular}{cccc}
\toprule
\textit{Petrov type} & \textit{Multiplicity of p.n.d.} & \textit{Vanishing Weyl components} & \textit{Criterion on $W_{\alpha\rho\sigma\beta}$} \\ 
\hline 
$I$ & $(1,1,1,1)$ & $\Psi_0 = 0$ & $k_{[\gamma} W_{\alpha ] \rho\sigma [\beta} k_{\delta ]} k^\rho k^\sigma = 0$ \\
$II$ & $(2,1,1)$ & $\Psi_0 = \Psi_1 = 0$ & $ k_{[\gamma} W_{\alpha ] \rho\sigma \beta} k^\rho k^\sigma = 0$ \\
$D$ & $(2,2)$ & $\Psi_0 = \Psi_1 = 0$ & $ k_{[\gamma} W_{\alpha ] \rho\sigma \beta} k^\rho k^\sigma = 0$ \\
$III$ & $(3,1)$ & $\Psi_0 = \Psi_1 = \Psi_2 = 0$ & $ k_{[\gamma} W_{\alpha ] \rho\sigma \beta} k^\rho = 0$ \\
$N$ & $(4)$ & $\Psi_0 = \Psi_1 = \Psi_2 = \Psi_3 = 0$ & $W_{\alpha\rho\sigma\beta} k^\alpha = 0$ \\
$O$ & $\emptyset$ & $\Psi_i = 0,\: \forall i$ & $W_{\alpha\rho\sigma\beta} = 0$ \\ 
\bottomrule 
\end{tabular}
} \\
\vspace{40pt}
\subfloat[The Penrose graph summarizing the degeneracy growth in Petrov's classification. Each arrow indicates one additional degeneracy. Adapted from \cite{Stephani:2003tm}.]{
\begin{tikzpicture}[scale=1.25]
\draw[black] (-1,-1) -- (-1,5) -- (5,5) -- (5,-1) -- cycle;
\node[circle=10pt,draw=black,thick] (III) at (0,0) {$III$};
\node[circle=10pt,draw=black,thick] (N)   at (2,0) {$N$};
\node[circle=10pt,draw=black,thick] (O)   at (4,0) {$O$};
\node[circle=10pt,draw=black,thick] (II)  at (2,2) {$II$};
\node[circle=10pt,draw=black,thick] (D)   at (4,2) {$D$};
\node[circle=10pt,draw=black,thick] (I)   at (4,4) {$I$}; 
\draw (I) edge [-latex,thick] (II);
\draw (I) edge [double, -latex,double distance=1pt,thick] (D);
\draw (II) edge [-latex,thick] (D);
\draw (II) edge [-latex,thick] (III);
%\draw (D) edge [-latex,thick] (N);
\draw (III) edge [-latex,thick] (N);
\draw (II) edge [double, -latex,double distance=1pt,thick] (N);
\draw (N) edge [ -latex,thick] (O);
\draw (D) edge [ -latex,thick] (O);
\end{tikzpicture}
}
\vspace{40pt}
\caption{Summary of Petrov's algebraic classification.}
\label{fig:Petrov}
\end{figure}
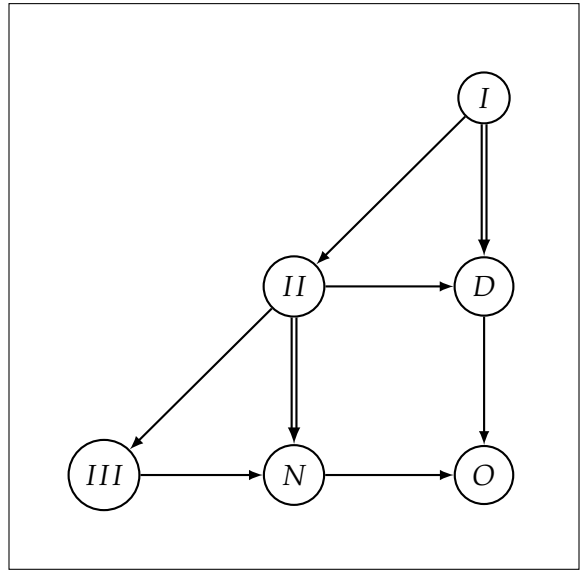

There are in total 6 types of Weyl tensors, which is summarized in Table \ref{fig:Petrov}(a).  Type $I$ is the most general one with 4 distinct principal null directions, and type $O$ is the totally degenerate case where the Weyl tensor is vanishing. The Kerr spacetime is type $D$ with two distinct but doubly degenerate principal null directions. The so-called \textit{Goldberg-Sachs theorem} implies that for any vacuum type $D$ spacetime such as  the Kerr black hole, the principal null directions are shear-free geodesic congruences. The fact that only $\Psi_2$ is non-vanishing for Kerr and the fact that its null principal directions are shear-free geodesic congruences make the Newman-Penrose formalism particularly adapted to describe the gravitational physics of Kerr! Let us now turn on to this physics...

\newpage

\subsection{Quasi-normal mode ringing of Kerr}

In this section, we will analyze the linear perturbations of the Kerr black hole. Generally, one can consider a theory of matter fields $\Phi_M^i$ minimally coupled to Einstein gravity without cosmological constant, whose action can be written as
\begin{equation}
S= \frac{1}{16\pi G} \int d^4 x \: \rmg \: R + \int d^4 x \: \rmg \: L[\Phi_M^i,g_{\mu\nu}]
\end{equation}
where $L[\Phi_M^i,g_{\mu\nu}]$ is the lagrangian density of matter fields collectively denoted by $\Phi_M^i$. The variation of $S$ with respect to the metric field $g_{\mu\nu}$ gives rise to Einstein's equations $G_{\mu\nu} = 8\pi G T_{\mu\nu}$, and the variation of the matter fields supplements it by the equations of motion for $\Phi_M^i$. We define the \textit{background field} as the Kerr black hole solution of this system of differential equations, denoted by $\bar{g}_{\mu\nu}$ and $\bar{\Phi}^i_M = 0$. We consider perturbations around that background: $g_{\mu\nu}=\bar{g}_{\mu\nu} + h_{\mu\nu}, \Phi_M^i = \bar{\Phi}_M^i + \delta\Phi_M^i$. If the deviations are small, we can approximate them by the linearized perturbations around the background field. The linearized field equations are a set of partial second-order differential equations which depend upon the background geometry. \\

The Kerr black hole $\bar{g}_{\mu\nu}$ admits two Killing isometries, $\partial_t$ and $\partial_\phi$ in the traditional Boyer-Lindquist coordinates. As a result, no explicit $t$ or $\phi$ dependence is present in the equations, and one can Fourier transform them to impose the equations on the Fourier modes of both metric and matter variables. In other words, one Fourier expands all the metric and matter fields collectively denoted as $\Phi^i(t,r,\theta,\phi) $ as
\begin{equation}
\Phi^i(t,r,\theta,\phi) = \frac{1}{2\pi} \int d\omega e^{-i\omega t} \sum_{m\in \mathbb{Z}} e^{im\phi} F^i(r,\theta)
\end{equation}
and all equations reduce to equations in terms of $F^i(r,\theta)$ which depend upon $r,\theta$, but also $M,J$ (the Kerr black hole parameters) and $m,\omega$ (the perturbation parameters). \\

To complete the problem, we still need to impose the physical boundary conditions. By definition of the event horizon, all modes need to be ingoing at $r=r_+$ so we demand that 
\begin{equation}
e^{-i\omega t + im\phi} F(r,\theta) \xrightarrow{r\rightarrow r_+} e^{-i\omega v_\star +im\phi_\star} F(\theta)
\end{equation}
where $r_\star$ is the \textit{tortoise coordinate}, $v_\star = t+r_\star$ the advanced time and $\phi_\star$ the angular coordinate which define the regular ingoing Eddington-Finkelstein coordinates $v_\star,r_\star,\theta,\phi_\star$ (or in other words, which resolves the geometry near the horizon). For the Kerr black hole, one has 
\begin{equation}
dr_\star \triangleq \frac{r^2+a^2}{\Delta} dr, \qquad d\phi_* \triangleq d\phi + \frac{a}{\Delta}dr. 
\end{equation}
We also impose that there is no ingoing field from $\cI^-$, which defines a ``purely outgoing mode at null infinity $\cI^+$''. This boundary condition prevents the stimulated emission of the black hole, and selects the intrinsic emission. It reads as
\begin{equation}
e^{-i\omega t + im\phi} F(r,\theta) \xrightarrow{r\rightarrow \infty, \, u\text{ fixed}} e^{-i\omega u+im\phi} \tilde F(\theta)
\end{equation}
where $u=t-r$ is the asymptotically flat retarded time.\\

Providing these boundary conditions, the solutions of the linearized equations of motion will turn out to belong to discrete sets of solutions, since one can show that the frequency $\omega$ will be quantized $\omega_{lmN}$ in terms of 3 integer numbers. The numbers $l,m$ are nothing else than (deformed) spherical harmonic numbers and $N=0,1,2,\dots$ is known as the \textit{overtone}.  Here the system is dissipative since waves escape either at null infinity or inside the event horizon. The frequencies are therefore complex: a positive imaginary part means a growing mode (an unstable mode) and a negative imaginary part means a damping/decaying mode. The proof of linear stability of Kerr amounts to show that all imaginary parts are negative! The presence of a family of discrete frequencies characterizing the Kerr black hole is very similar to the spectral lines of the hydrogen atom and therefore the science of quasi-normal modes is often called \textit{black hole spectroscopy}, as announced in the introduction. Let us now derive all this! 

\subsubsection{Separation of variables and Teukolsky master equation}
The question is: how to deal with partial differential equations involving $(r,\theta)$ in a Kerr background? The answer is: find how to separate them! Let us first give a hint that the equations of spins 0, 1 and 2 are separable. \\

We consider a causal geodesic that travels outside the Kerr black hole $x^\mu (\lambda)$. The tangent vector $\dot{x}^\mu \triangleq \frac{dx^\mu}{d\lambda}$ obeys $g_{\mu\nu} \dot{x}^\mu\dot{x}^\nu = 0$ or $-1$. Both stationarity and axisymmetry guarantee that the geodesic motion has two dynamical invariants, which are simply the energy of the particle moving on the geodesic $e =-\mu k^\mu g_{\mu\nu} \dot{x}^\nu$ ($k = \partial_t$, $\mu$ is the rest mass) and its orbital angular momentum $j = \mu k^\mu g_{\mu\nu} \dot{x}^\nu$ ($k = \partial_\phi$). In addition to this, Carter found in 1968 \cite{Carter:1968rr} that there is in fact another conserved quantity on the flow of causal geodesics, quadratic in the momentum, called in his name the \textit{Carter constant} $Q$. It takes the following form in Boyer-Lindquist coordinates
\begin{equation}\label{defQ}
Q = \mu^2 {\dot x}_\theta^2 + \cos^2 \theta \left[ a^2 (\mu^2 -e^2) + \left( \frac{j}{\sin \theta} \right)^2 \right].
\end{equation} 
One often uses the related constant $K = Q + (j-a e)^2$ which is always non-negative. The existence of 4 first order equations for $x^\mu(\lambda)$ allows to analytically solve the geodesic problem in Kerr, though in terms of elliptic integrals \cite{Fujita:2009bp}. The conservation of $Q$ is related to a hidden symmetry, which is of higher order than Killing isometries. For the Kerr background, one can construct a symmetric tensor $K_{\mu\nu}$ which satisfies a close cousin of the Killing equation
\begin{equation}
\nabla_{(\lambda} K_{\mu\nu )} = 0. 
\end{equation}
For this reason, $K_{\mu\nu}$ is said to be a \textit{Killing tensor}, and is non-trivial in the sense that it is not simply the tensor product of the two Killing vectors always discovered. \\
%One can actually show that 
%\begin{equation}
%K_{\mu\nu} = 2 \Sigma l_{(\mu} n_{\nu )} + r^2 g_{\mu\nu}
%\end{equation}
%where $l_\mu$ and $n_\nu$ are the two (double) principal null directions of the Weyl tensor (recall that the Kerr vacuum is Petrov type $D$!)
%\begin{equation}
%l^\mu = \frac{1}{\Delta}\left(r^2+a^2,\Delta,0,a \right) \quad \text{ and } \quad n^\mu = \frac{1}{2\Sigma}\left( r^2+a^2, -\Delta,0,a \right)
%\label{eq:KerrTetralLandM}
%\end{equation}
%An explicit computation gives $Q = \mu^2 K_{\mu\nu} \dot{x}^\mu \dot{x}^\nu$. Remark that in static limit, we have $j_\theta = 0$ (the orbit is planar), and we can perform a rotation to put this orbit in the equatorial plane $\theta = \pi/2$ so $q$ reduces simply to the square of the orbital kinetic momentum $q = j_z^2$. 

The contraction between $K_{\mu\nu}$ or $g_{\mu\nu}$ with the $4$-velocity of any particle gives a dynamical invariant, so we can define an analogue operator to the Kerr d'Alembertian $\Box = g^{\mu\nu} \nabla_\mu \nabla_\nu$ by performing the contraction $\Box_K = K^{\mu\nu} \nabla_\mu \nabla_\nu$. We have the fundamental property that these two second-order differential operators on the manifold actually \textit{commute} $[\Box,\Box_K] \equiv 0$, as you can check as an exercise. As shown by Carter \cite{Carter:1977pq}, this property is at the origin of the separability of the scalar wave equation in the Kerr background. While it is not known whether there is a similar symmetry-based structure that allows the separability of the spin 2 perturbation (the linearized Einstein perturbation), it turns out to hold as we will now show! \\

The proof of separability came from Teukolsky in 1972 \cite{Teukolsky:1972my} who found the right combination of the metric perturbation which separates. He employed the Newman-Penrose formalism and Petrov's classification. Providing that the Kerr spacetime is Petrov type $D$, one can find 2 preferential null orthogonal directions to build the Newman-Penrose basis, which are the degenerate principal null directions of the Weyl tensor.\\

The two (double) principal null directions of the Weyl tensor are
\begin{equation}
l^\mu = \frac{1}{\Delta}\left(r^2+a^2,\Delta,0,a \right) \quad \text{ and } \quad n^\mu = \frac{1}{2\Sigma}\left( r^2+a^2, -\Delta,0,a \right).
\label{eq:KerrTetralLandM}
\end{equation}
One completes the basis with the complex vector
\begin{equation}
m^\mu = \frac{1}{\sqrt{2}(r+i a \cos \theta)} \left( i a \sin \theta, 0, 1, \frac{i}{\sin\theta}\right)
\end{equation}
to get the so-called \textit{Kinnersley tetrad} \cite{Kinnersley:1969zza}. With respect to this choice of null basis, the only non-vanishing Weyl component is
\begin{equation}
\Psi_2 = -\frac{M}{(r-i a \cos\theta)^3}.
\end{equation}
We therefore proved that the Kerr black hole is Type D, as stated in Section \ref{sec:Petrov}: all effects of gravitation are caused by the single Weyl scalar $\Psi_2$ in the Kinnersley tetrad. 
A modification $g_{\mu\nu} \rightarrow g_{\mu\nu} + h_{\mu\nu}$ of the Kerr metric amounts to perturb $\Psi_i$ by $\delta \Psi_i$. In fact, if one can solve only for $\Psi_0$ or $\Psi_4$. The other Weyl perturbations can be then deduced, up to the additional information on how $M$ and $J$ are perturbed. The entire perturbation $h_{\mu\nu}$ could then be reconstructed from all variations $\delta \Psi_i$.\\

The point is that the equation for $(r-i a \cos \theta)^4 \delta \Psi_4$ is separable (equal to a function of $\theta$ times a function of $r$). Teukolsky realised later that the equation for $\delta \Psi_0$ is also separable. In fact, there is an identity relating $\delta\Psi_0$ and $\delta\Psi_4$ so the equations are equivalent. Physically, the Weyl scalars $\delta \Psi_0$ and $\delta \Psi_4$ both describe the two polarization modes of the gravitational waves. At the linear level, there are only three possible perturbations: gravitational waves, changes of the gravitational potentials (change of $M,J$)\footnote{To be precise, there are 2 additional gravitational potentials that are usually discarded because considered unphysical: the \textit{NUT charge}, a sort of magnetic analogue to the mass but which generates closed timelike curves, and the \textit{acceleration parameter}, which introduces conical wire singularities.} or changes of coordinates (which can belong to the asymptotic symmetry group and therefore be non-trivial). It is thus easy to understand that we can reconstruct all $\delta\Psi_i$ from either $\delta\Psi_0$ or $\delta\Psi_4$ together with the knowledge of how $M$ and $J$ change since these variables are diffeomorphic invariant. Since the perturbations of spin 0 or 1 also separate, it is convenient to write the master equation which separates all spins in a single notation. This is called the \textit{Teukolsky master equation}!
\begin{align}
&\left[ \frac{(r^2+a^2)^2}{\Delta} - a^2\sin^2\theta \right] \frac{\partial^2 \psi}{\partial t^2} + \frac{4Mar}{\Delta} \frac{\partial^2\psi}{\partial t \partial\phi} + \left[ \frac{a^2}{\Delta} - \frac{1}{\sin^2\theta}\right] \frac{\partial^2\psi}{\partial \phi^2} \nonumber\\
&- \Delta^{-s} \frac{\partial}{\partial r} \left(\Delta^{s+1} \frac{\partial\psi}{\partial r} \right) - \frac{1}{\sin\theta} \frac{\partial}{\partial\theta} \left(\sin \theta \frac{\partial\psi}{\partial\theta}\right) - 2s \left[ \frac{a(r-M)}{\Delta} + \frac{i \cos \theta}{\sin^2\theta} \right] \frac{\partial\psi}{\partial\phi} \\
&- 2s \left[ \frac{M(r^2-a^2)}{\Delta} -r-ia \cos\theta\right] \frac{\partial\psi}{\partial t} + (s^2 \cot^2 \theta - s ) \psi = T \nonumber
\end{align}
where $T$ is a source term if we couple gravity to matter. For the spin 0 case, $\psi$ is just the original scalar field. For the spin 2 case, $\psi$ is either $\delta\Psi_0$ or $(r-ia\cos\theta)^4\delta\Psi_4$.
The general solution of the Teukolsky master equation takes the separable form 
\begin{equation}
\psi(t,r,\theta,\phi) = \frac{1}{2\pi} \int d\omega e^{-i\omega t} \sum_{l=|s|}^{\infty} \sum_{m=-l}^{+l} e^{im\phi} R^s_{lm\omega}(r) S^s_{lm\omega}(\cos\theta).
\end{equation}
The equation for $S^s_{lm\omega}(\cos\theta)$ is called the \textit{spin weighted spheroidal harmonic} equation
\begin{equation}
\left[ \frac{d}{d x}(1-x^2)\frac{d}{d x} \right] S^s_{lm\omega}(x) + \left[ a^2\omega^2 x^2 - 2a\omega sx  + \mathcal E^s_{lm\omega} - \frac{m^2 + 2 m s x + s^2}{1-x^2} \right] S^s_{lm\omega}(x) = 0
\label{eq:EqSpheroidal}
\end{equation}
where $x = \cos\theta$ and $\mathcal E^s_{lm\omega}$ is the separation constant. When $a = 0$, the dependence in $\omega$ drops out and the functions $S^s_{lm}(\cos\theta)$ reduce to spin-weighted spherical harmonics $Y^s_{lm}(\theta,\phi) = S^s_{lm}(\cos\theta) e^{im\phi}$ after inclusion of the Fourier $\phi$ factor. In this case, the angular separation constants $\mathcal E^s_{lm\omega}=\mathcal E^s_{lm} $ are known analytically to be $\mathcal E^s_{lm} = l(l+1)$.\\

The radial equation is the \textit{radial Teukolsky equation}:
\begin{align}
	\Delta^{-s} \frac{\partial}{\partial {r}}(\Delta^{s+1}\frac{\partial  R_{lm {\omega}}}{\partial {r}}) - V({r}) R_{lm {\omega}}({r}) = T_{l m {\omega}}({r})	\label{eqn:RadialTeukolsky}
\end{align}
with source $T_{l m {\omega}}({r})$ and potential
\begin{eqnarray}
	V({r})  &=& -\frac{(K_{m  \omega})^2-2si({r}-M)K_{m \omega}}{\Delta}- 4si {\omega} {r}+ \lambda_{\ell m \omega} ,\nonumber\\
	K_{m \omega}  &\triangleq & ({r}^2 + a^2){\omega} -ma, \\
	\lambda_{\ell m \omega} & \triangleq & \mathcal E_{lm\omega} -2am{\omega}+a^2{\omega}^2-s(s+1).\nonumber
\end{eqnarray}
When $a=0$, the $m$ dependence drops out. This is a consequence of $SO(3)$ symmetry.

\subsubsection{Solving the angular equation}

The solutions of (\ref{eq:EqSpheroidal}) are the so-called \textit{spin-weighted spheroidal harmonics} $S^s_{lm\omega} (x)$ which form, for each value of the spin $s$, an orthogonal system of functions on the interval $[-1,1]$ (recall that $x = \cos\theta$)
\begin{equation}
\int_{-1}^{+1} \: dx \: S^s_{lm\omega} (x) S^s_{l'm'\omega'} (x) = \delta_{ll'}\delta_{mm'}\delta(\omega - \omega'). 
\end{equation}
They are only defined for values of integers parameters $l,m$ in the range $l \geq |s|$ and $|m| \leq l$, otherwise they are simply identically zero. The case $s=0$ reduces to the spheroidal harmonics which are well-known, even by symbolic computation softwares such as \textit{Mathematica}\textsuperscript{TM}. \\

An analytic representation of the solutions for $s>0$ was found for first time by Leaver \cite{Leaver:1985ax}, thanks to the Frobenius method. Let us now review this result.  The equation (\ref{eq:EqSpheroidal}) has three singular points, among which two regular singular points at $x = \pm 1$, and an irregular singularity at $x = \infty$ if we extend analytically the domain of definition of the coordinate $x$. Boundary conditions are imposed such that solutions to (\ref{eq:EqSpheroidal}) are finite at the regular singular points. The local behaviour around these points can be worked out easily. One finds that $S_{lm}(x) \sim x^k$, with $k = \pm \frac{1}{2}|m+s|$ at $x=+1$ and $k = \pm \frac{1}{2}|m-s|$ at $x=-1$. The boundary conditions allow us to discard the negative exponents. When $x$ runs to infinity, the equation can be integrated and the boundary condition is fixed analytically from the other two singular points. It leads to $S_{lm}(x) \sim \exp(ia\omega x)$. So the Frobenius method gives a natural ansatz: one can consider the product of the solutions at singular points times a Taylor series around $x=-1$ 
\begin{equation}
S_{lm} (x) = e^{ia\omega x} (1+x)^{\frac{1}{2}|m+s|} (1-x)^{\frac{1}{2}|m-s|} \sum_{n=0}^{+\infty} c_n (1+x)^n.
\end{equation}
Substituting this ansatz in the angular equation, and equating all coefficients multiplying the same monomials $x^k$, for any $k\in\mathbb{N}$, we find
\begin{equation}
\left\lbrace
\begin{array}{l}
\alpha_0 c_1 + \beta_0 c_0 = 0 \, ;\\
\alpha_n c_{n+1} + \beta_n c_n + \gamma_n c_{n-1} = 0, \: \forall\: n \in \mathbb{N}_0\; ,
\end{array}
\right.
\end{equation}
which spans a 3-term recurrence on the coefficients $c_n$. The coefficients $\lbrace \alpha_n,\beta_n,\gamma_n\rbrace$ are independent of $x$ (or $\theta$) and their explicit forms can be found in \cite{Leaver:1985ax}. If we define $r_n = c_{n+1}/c_n$, the second equality becomes $\alpha_n r_n + \beta_n + \gamma_n/r_{n-1} = 0$ from which we can extract
\begin{equation}
r_{n-1} = \frac{-\gamma_n}{\beta_n + \alpha_n r_n} = \frac{-\gamma_n}{\beta_n + \frac{\alpha_n (-\gamma_{n+1})}{\beta_{n+1}+\alpha_{n+1}r_{n+1}}}  \cdots\label{eq:CtdFrac1}
\end{equation}
It is a continuous fraction which expresses the ratio $r_n$ in terms of the next one $r_{n+1}$. It converges if $r_n \xrightarrow{n\rightarrow \infty} 0$, and it turns out to be the case! 
%For any set of parameters $a,m,\omega,s$, the separation constants $A_{lm}^s$ are roots of the continued fraction equation
%\begin{equation}
%\beta_0 - \frac{\alpha_0 \gamma_1}{\beta_1 - \frac{\alpha_1\gamma_2}{\beta_2 - \cdots}} = 0
%\label{eq:CtdFrac1}
%\end{equation}
%Since the $\gamma_n$ are linear in $a$, they vanish identically in the static limit $a=0$, and the recursion will stop whenever $A_{lm}^s$ is such that $\beta_N = 0$ for some $N$. This will happen when $A_{lm}^s = N(N+1)-s(s+1)$, so when $l$ reaches $N$. 
When $a=0$, $\omega$ completely disappears of all the equations, so the spheroidal equation can be solved numerically to arbitrary precision, which gives at the same time $S_{lm}^s (\cos \theta)$ and the constants $\mathcal E_{lm}^s$. Otherwise the frequency impacts the recursion, so one has to solve \textit{simultaneously} the spheroidal equation and the radial equation, whose quantized solutions will give the proper frequencies of the system. 

\subsubsection{Solving the radial equation}
The radial equation has been solved formally by means of a similar method (Frobenius expansion and continued fraction) by Leaver, also in \cite{Leaver:1985ax}. The singular points of the radial equation for $R_{lm}(r)$ are the two roots of $\Delta$, which are the radial position of the event horizon $r = r_+$ and Cauchy horizon $r = r_-$, and $r=\infty$. The point $r=r_+$ is a regular singularity, and the behaviour of the solution in its neighbourhood is fixed by the ingoing boundary condition. On the other hand, the point at infinity $r=\infty$ is an essential singularity, where we impose purely outgoing boundary conditions as a definition of quasi-normal modes. The problem consists at solving an ordinary differential equation together with boundary conditions are two separated locations: it is a boundary value problem, which you might be familiar with in the context of the quantum description of the hydrogen atom.
% In fact, the solutions are trapped in the potential $V(r)$ which is convex and admits an unique minimum around the last photon orbit\footnote{The last photon orbit defines the smallest radius $r_L$ under which it is impossible to find a photon travelling a stable circular orbit. For $r<r_L$, the null circular geodesics are unstable, and a photon cannot remains on it! For Schwarzschild metric, $r_L = 3GM$. For Kerr, it varies with the rotation parameter $a$. When $a=M$ (extremal regime), the last photon orbit is at $r=M$.}. 
Around $r = r_+$, one can integrate the radial equation to find the two solutions $R_{lm}(r) \sim r^{i\sigma_+}$ and $R_{lm}(r) \sim r^{-s-i\sigma_+}$ where $\sigma_+ \triangleq (\omega r_+ - am)/\sqrt{1-4a^2}$. The ingoing boundary condition rules out the first behaviour. Similarly, we can proceed near infinity to get $R_{lm}(r)$ equal to a combinaison of $r^{-1-i\omega} e^{-i\omega r}$ which is ingoing and then forbidden by the boundary conditions, and $R_{lm}(r) \sim r^{-1-2s+i\omega} e^{i\omega r}$ which is outgoing and accepted. After working out the contribution at the second regular singular point $r=r_-$, the Frobenius ansatz can be written as
\begin{equation}
R_{lm}(r) = e^{i\omega r} (r-r_-)^{-1-s+i\omega+i\sigma_+} (r-r_+)^{-s-i\sigma_+} \sum_{n=0}^{+\infty} d_n \left(\frac{r-r_+}{r-r_-}\right)^n.
\end{equation}
Injecting it into the radial Teukolsky equation gives a second set of recursion relations
\begin{equation}
\left\lbrace
\begin{array}{l}
\alpha'_0 d_1 + \beta'_0 d_0 = 0\; ; \\
\alpha'_n d_{n+1} + \beta'_n d_n + \gamma'_n d_{n-1} = 0, \: \forall\: n \in \mathbb{N}_0\; .
\end{array}
\right.
\end{equation}
The complex coefficients $\lbrace \alpha'_n,\beta'_n,\gamma'_n\rbrace$ again do not depend on $r$, but well in $\omega$, $a$, and the separation constants $\mathcal E_{lm\omega}^s$. For a given set of parameters $a,m,l,s$, the frequency $\omega$ is a root of the continued fraction equation
\begin{equation}
\beta'_0 - \frac{\alpha'_0 \gamma'_1}{\beta'_1 - \frac{\alpha'_1\gamma'_2}{\beta'_2 - \cdots}} = 0.
\label{eq:CtdFrac2}
\end{equation}
Equations (\ref{eq:CtdFrac1}) and (\ref{eq:CtdFrac2}) are two equations for the unknown variables $\mathcal E_{lm\omega}^s$ and $\omega$. They may be solved simultaneously by non-linear root-search algorithms. The solutions of coupled recurrences are found to be a set a quantized modes of frequency $\omega_{lmN}$ where $N$ is the previously announced overtone number ($N\in \mathbb{N}$). The proof of linear stability of the Kerr black hole in Einstein gravity is then obtained after checking (numerically) that indeed $Im(\omega_{lmN})<0$ for all modes. 

\subsubsection{Geometric interpretation in the eikonal limit}

Let us briefly mention that some analytical quasi-normal modes solutions have been derived, see the review \cite{Berti:2009kk} and references therein. Although the spectrum can be numerically calculated to arbitrary precision, these analytic solutions give us more insight. As as example, when the multipolar index $l$ runs to infinity (the so-called eikonal regime or geometric optics approximation), it is possible to reformulate the Teukolsky radial and angular equations to put them into the form $\varepsilon f''(z) + U(z)f(z) = 0$, where $\varepsilon \ll 1$ and $z$ is either $x,r$. The solution of such an equation can be approximatively computed thanks to \textit{WKB} expansions $f(z) \sim \exp (S_0/\varepsilon + S_1 + \varepsilon S_2 +\cdots )$. This allows to relate the corresponding wave solutions to null geodesics as pioneered by Press in 1971 \cite{1971ApJ...170L.105P} and derived for Kerr much more recently \cite{Yang:2012he}!  Each quasi-normal mode solution $(\omega_{lmN},l,m)$ corresponds to a specific null geodesic with certain conserved quantities $(e,j,Q)$ (remember the definition \eqref{defQ}) according to the following dictionary: 
%The region in which the wave propagates is identical to the region in which geodesics with these conserved quantities can propagate. In addition, for each point in this region, there is one (and only one) geodesic passing through it. So conversely, given any set of conserved parameter $(e,j_z,q)$, one will be able to find  a wave solution that exists in the region in which the geodesics travel. Not all such sets of conserved quantities correspond to quasi-normal modes, however, because they may not satisfy the correct boundary conditions we have fixed. More explicitly:
\begin{itemize}[label=$\rhd$]
\item $Re(\omega_{lmN}) = e$. The wave frequency is the same as the energy of the corresponding null ray;
\item $m = j$. The azimuthal number corresponds to the $z$-axis projection of the orbital momentum $\vec{j}$ of the null ray. The latter is quantized to get a standing wave in the azimuthal direction $\phi$;
\item $Re(\mathcal E_{lm\omega})$ is related to a combination of $Q$ and $m^2$. The separation constants $\mathcal E_{lm\omega}$ are complex. Their real part is linked to the Carter constant $Q$ and the square of the axial orbital kinetic momentum $j^2$. The angular momentum eigenvalue in the colatitudinal direction also gets quantized in order to get standing waves in the $\theta$ direction. 

\item The decay rate $Im(\omega_{lmN}) = \gamma_L$, where $\gamma_L$ is a \textit{Lyapunov exponent} of the orbit, which indicates a rate of instability of the null geodesic congruence. 

\end{itemize}

\subsubsection{Schwarzschild spectroscopy}

\begin{figure}[h!]
\subfloat[Quasi-normal modes frequencies for gravitational perturbations ($s=2$).]{
\includegraphics[width=0.50\textwidth]{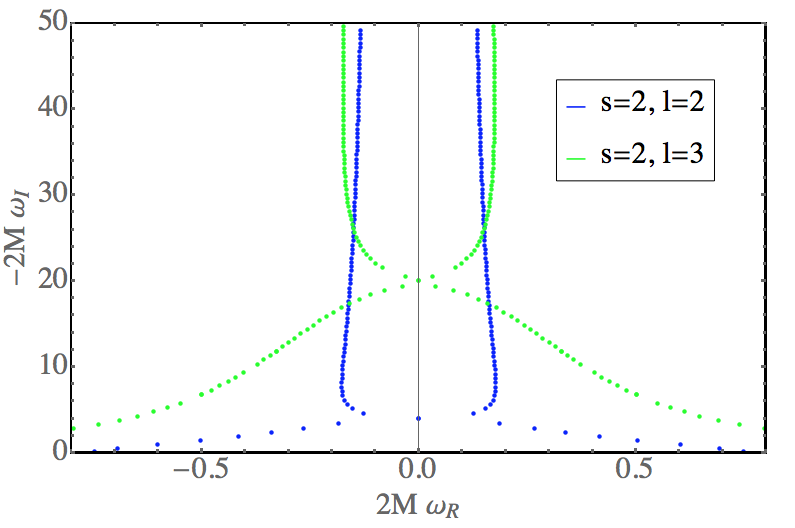}
\label{fig:SchwQNM1}
}
\subfloat[Comparison of quasi-normal modes fundamental spectra $l=|s|$ for scalar, vector, and gravitational perturbations ($s=0,1,2$).]{
\includegraphics[width=0.49\textwidth]{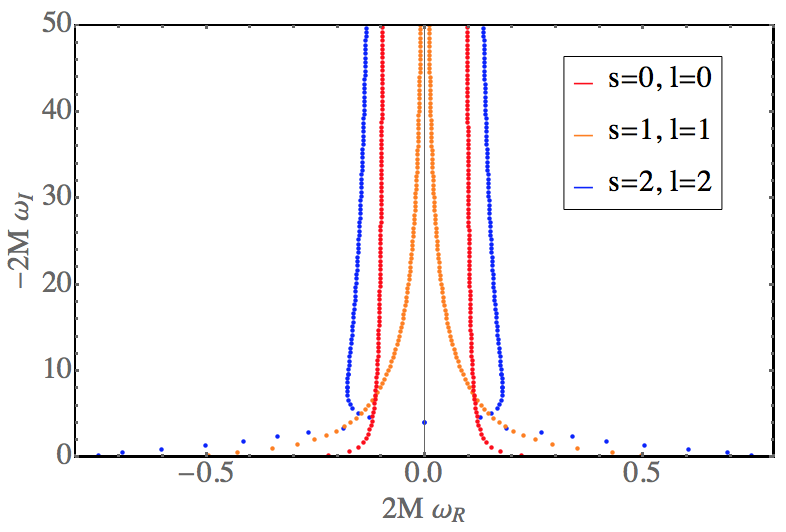}
\label{fig:SchwQNM2}
}
\caption{Quasi-normal modes for Schwarzschild black holes. {\footnotesize Reproduced from online data \cite{Berti:web} and private communication of E. Berti. Original figure published in \cite{Berti:2009kk} $\copyright$ IOP Publishing. Reproduced with permission. All rights reserved.}}
\end{figure}

We present here some concrete numerical results. Let us begin by the Schwarzschild black hole (hence with $a=0$), see Figure \ref{fig:SchwQNM2}.

The graphs are symmetric under real frequency reversal: each positive frequency $Re(\omega )$ mode is associated with a 
negative $Re(\omega )$ mode with the same imaginary part. All imaginary parts are negative (this is the property of linear stability). There are several branches for each $l$, with increasing overtone number $N$. There is no dependency in $m$ since it factors out of the radial equation (this is due to $SO(3)$ symmetry).\\

 The first feature worth to notice is that for gravitational perturbations ($s=2$) of static black holes there are algebraically special modes that have exactly $Re(\omega )=0$. %These are indicated with an arrow in Figure \ref{fig:SchwQNM1}. 
 They are given by $M\omega \approx \pm i (l-1)l(l+1)(l+2)/12$. Such real frequency modes do not exist for other spin $s$ perturbations. For each $l$, we will call the modes below these algebraically special modes as the weakly damped modes, and above, the high damped ones. This qualitative distinction is peculiar to gravitational perturbations: for other fields $s\neq 2$, one approaches monotonically the asymptotic high-damped regime. Here are some additional properties:
\begin{itemize}[label=$\rhd$]
\item The most weakly damped modes are the most relevant ones for experimental detection since they give the leading signal. ln black hole mergers, these modes appear in the ringdown at late stages after the merger. They have not yet been detected at the time of writing. The dominant $s=2$ mode is the $l=2$, $N=0$ mode with frequency $M\omega = 0.3737-0.0890i$.  

\item In the large overtone limit $N\rightarrow\infty$, the modes are highly damped. The asymptotic expansion of the frequency reads as $\omega \approx T_H \ln 3 - i 2\pi T_H (N+\frac{1}{2}) + \mathcal{O}(N^{-1/2})$ for gravitational perturbations. The appearence of Hawking's temperature suggests a microscopic interpretation of these modes.

\item In the large $l$ limit, one has the eikonal regime or geometric optics regime where the quasi-normal modes can be mapped to null geodesics, as we already explained. The leading eikonal approximation is $\omega = \Omega_\gamma l - i(N+1/2) \lambda_\gamma$, where $\Omega_\gamma$ is the angular frequency of a photon orbiting a geodesic with same conserved charges as the quasi-normal mode, and $\gamma_L = -i(N+1/2) \lambda_\gamma$ is the Lyapunov exponent related to the instability frequency $\lambda_\gamma$ of the orbit.
\end{itemize}
 
 \subsubsection{Kerr spectroscopy}

Let us finally discuss the rotating Kerr black hole. The quasi-normal mode spectrum has a rich and complex structure. The $SO(3)$ symmetry is now slightly broken, so the spectrum depends upon $m$ in addition to $l$. This leads to a sort of Zeeman effect (which is the splitting of quantum states of electrons around atoms due to an external magnetic field) which splits the spectrum of frequencies.  \\

The weakly damped modes are again the most relevant for gravitational wave detection. The $N=0$, $l=2$, $m=0$ mode is the least damped with $M\omega_{020} \approx 0.4437-0.0739 (1-a/M)^{0.3350}$, as obtained by numerical interpolation. There seems to be no known analytic formula in the highly damped regime. The spacing of the imaginary part of the frequency is not the constant $2\pi T_H$ but now grows with $a/M$. The eikonal regime has been described earlier; \\

Let us conclude with a brief comment about quasi-normal modes around extremal rotating black holes. It was shown recently  \cite{Yang:2013uba} that the spectrum bifurcates in the near-extremal limit into the so-called ``zero-damped'' and ``damped''  quasi-normal modes. The zero-damped quasi-normal modes as their name indicates have $Im(\omega)\rightarrow 0$ in the extremal limit while the damped ones keep a non-zero imaginary part. This new physical feature arises from the occurence of the near-horizon region of near-extremal Kerr black hole. The zero-damped quasi-normal modes are emitted from the near-horizon region, while the damped ones from the ``far'' extremal Kerr region. \\

Here we conclude this journey into the marvelous world of rotating black holes. We hope that the material will trigger the curiosity of the reader to dig in further into these very rich topic full of discoveries! 

\newpage
\section*{References}
\addcontentsline{toc}{section}{References}

Many excellent references exist on the {\color{blue} Kerr black hole}, probably many of which I missed reading. For a review of the coordinates patches of Kerr and its Penrose diagrams, I would recommend the lecture notes of Carter of 1973 \cite{1973blho.conf...57C}. For the thermodynamics of Kerr, I particularly like the review of Wald \cite{Wald:1999vt}. For clearly and concisely reviewing many classical features of Kerr and also the Hawking radiation, I recommend the lecture notes of Townsend of 1997 \cite{Townsend:1997ku}. A modern review of the Kerr metric which includes a discussion of perturbation theory and hidden symmetries (together with a historical review) is given by Teukolsky in 2014 \cite{Teukolsky:2014vca}. It largely inspired this lecture.   \\

The study of the {\color{blue} near-horizon region of extremal Kerr black holes} started by the work of Bardeen, Teukolsky and Press in 1972 \cite{Bardeen:1972fi} but since it didn't have direct astrophysical interest it was not until 1999 that the actual decoupled near-horizon region was discovered by Bardeen and Horowitz \cite{Bardeen:1999px}. Recent interest for this limit is partly due to the {\color{blue} \textit{Kerr/CFT} correspondence} in 2009 \cite{Guica:2008mu}. The lectures are mostly based on my review (updated in Dec 2016) \cite{Compere:2012jk}. In particular, some of the developments of \cite{Bredberg:2009pv,Balasubramanian:2009bg,Amsel:2009ev,Dias:2009ex,Compere:2015mza,Compere:2015bca,Compere:2015knw} were covered. \\

The {\color{blue} quasi-normal modes} of the Kerr black hole are reviewed in \cite{Berti:2009kk,Konoplya:2011qq}. A classic reference on the Newman-Penrose formalism is \cite{Chandrasekhar:1985kt} but unfortunately it uses the overall minus signature convention for the metric. I mostly followed the appendix of \cite{Merlin:2016boc} that contains many useful expressions in overall plus signature. For a review of Frolov's classification, I would recommend the corresponding introductory chapter of the book \cite{Stephani:2003tm}. For more details on the mysterious relationship between hidden symmetries and separability, see the reviews \cite{Kalnins:1999ycc,Frolov:2017kze}. A partial database of explicit values of quasi-normal modes can be found in \cite{Berti:web,Berti:2009kk}. For a recent high accuracy study, see \cite{Cook:2014cta}.

\vspace{1cm}
\vfill

\section*{Acknowledgments}
G.C. thanks the organizers of the Amsterdam-Brussels-Geneva-Paris Doctoral School and the International Solvay Institutes for the opportunity to give these lectures. This work was partly supported by the ERC Starting Grant 335146 ``HoloBHC".  G.C. is a Research Associate and A.F. is Research Fellow of the Fonds de la Recherche Scientifique F.R.S.-FNRS (Belgium).

	\blanc
	
	% Bibliographie

%\Bibliographie{mybib2}

\providecommand{\href}[2]{#2}\begingroup\raggedright\endgroup

%============================================================

% D - Tables des illustrations et des théorèmes
% ---------------------------------------------

	% Tables des illustrations
	%	\listoffigures % Insertion de la table des illustrations
	%	\addcontentsline{toc}{chapter}{Table des illustrations}
	% Table des théorèmes et définitions
	%	\ListOfThmAndDef

%============================================================

	% Fin du document
	
		\end{document}